\documentclass[a4paper,11pt]{article}
\pdfoutput=1 % if your are submitting a pdflatex (i.e. if you have
\usepackage{jcappub} % for details on the use of the package, please
\usepackage[T1]{fontenc} % if needed
\usepackage{mathrsfs}
\usepackage{amsmath}
\usepackage{upgreek}

\title{\boldmath Observational features of deformed Schwarzschild black holes illuminated by an anisotropic accretion disk}

\author[a]{Dan Li,}
\author[a,1]{Shiyang Hu,\note{Corresponding author.}}
\author[b]{Chen Deng,}
\author[c]{and Xin Wu}

\affiliation[a]{School of Mathematics and Physics, University of South China,\\ Hengyang, 421001 People's Republic of China}
\affiliation[b]{School of Astronomy and Space Science, Nanjing University,\\ Nanjing, 210023 People's Republic of China}
\affiliation[c]{School of Mathematics, Physics and Statistics, Shanghai University of Engineering Science,\\ Shanghai, 201620 People's Republic of China}

\emailAdd{danli@usc.edu.cn}
\emailAdd{husy\_arcturus@163.com}
\emailAdd{dengchen@smail.nju.edu.cn}
\emailAdd{wuxin\_1134@sina.com}

\abstract{The projection effect of an anisotropic accretion disk causes its electromagnetic radiation to depend on the emission angle. Although this dependency has the potential to influence the observational characteristics of black holes, it has not received sufficient attention. In this paper, we employ a relativistic ray-tracing algorithm to numerically simulate $86$ GHz and $230$ GHz images of deformed Schwarzschild black holes illuminated by an equatorial anisotropic accretion disk, aiming to reveal the observational signatures of the target black hole and the impact of the projection effect on the images. The study demonstrates that while the introduction of the projection effect does not alter the profiles of the black hole's inner shadow and critical curve, it significantly suppresses the specific intensity of light rays, particularly in direct emission, thereby reducing image brightness. The extent of this reduction depends on both the observation inclination and frequency. This phenomenon aids in the extraction of geometric information from higher-order subrings in the image. Furthermore, we find that increasing the deformation parameter enhances the brightness of the deformed Schwarzschild black hole image, accompanied by a reduction in the size of the critical curve and inner shadow. This relationship establishes a connection between the intrinsic properties of deformed Schwarzschild black holes and their observational characteristics, providing a reliable tool for testing the no-hair theorem and gravitational theories. Specifically, we propose a novel method for constraining parameters based on the silhouette of the inner shadow, which holds promise for extension to any spherically symmetric black hole in other gravitational theories.}

\begin{document}
\maketitle
\flushbottom

\section{Introduction}
In general relativity, black holes are precisely described by only three parameters--mass, charge, and spin--according to the famous no-hair theorem. However, general relativity exhibits notable limitations in addressing the problem of black hole singularities and the phenomenon of cosmic accelerated expansion. These limitations have prompted the scientific community to question the universality of general relativity, leading to the development of numerous modified theories of gravity \cite{Nojiri and Odintsov (2007),Clifton et al. (2012)}. These alternative theories often introduce additional fields to the Einstein-Hilbert action based on specific requirements, inevitably resulting in black holes characterized by ``hair'' parameters beyond mass, charge, and spin. It is worth emphasizing that current astrophysical observations have not yet excluded the possibility of the existence of such hairy black holes \cite{Yunes and Siemens (2013),Berti et al. (2015),Mizuno et al. (2018),Gralla (2021),Vagnozzi et al. (2023)}. Moreover, these black holes play a crucial role in the deeper understanding and advancement of gravitational theories, and thus warrant significant attention.

Interestingly, an unusual class of black holes exists within modified theories of gravity that does not originate from a specific action but is instead derived through the parameterization of classical black holes in general relativity. Notable examples include the deformed Schwarzschild black hole and its rotating counterpart \cite{Johannsen and Psaltis (2011)}, the Kerr-like black hole \cite{Johannsen (2013)}, the Cardoso-Pani-Rico metric \cite{Cardoso et al. (2014)}, the Rezzolla-Zhidenko spacetime \cite{Rezzolla and Zhidenko (2014)} and its derivation \cite{Rezzolla et al. (2016)}, and the parameterized Schwarzschild black hole \cite{Tian and Zhu (2019)}. Among these, the deformed Schwarzschild black hole stands out as one of the pioneering studies in this direction. It is derived by introducing a deformation function related to $M/r$ into the Schwarzschild metric, where $M$ and $r$ represent the black hole's mass and the radial coordinate, respectively. It is important to emphasize that the deformed Schwarzschild black hole is not a vacuum solution of the Einstein field equations, and its corresponding modified field equations and action remain unknown. Nevertheless, it serves as an effective tool for testing the no-hair theorem and exploring gravitational theories. Specifically, if the deformation function were confirmed through astronomical observations, the no-hair theorem would be automatically disproven. Consequently, the deformed Schwarzschild black hole has garnered significant attention within the scientific community \cite{Rayimbaev (2016),Yi and Wu (2020),Magalhaes et al. (2020),Zhang et al. (2021),Huang et al. (2022),Xu et al. (2024)}. However, the observational characteristics of this black hole and the impact of the deformation function on its image remain unclear, a gap this paper aims to fill.

The investigation of black hole images is closely linked to actual observations. Recently, the Event Horizon Telescope (EHT) collaboration successfully captured images of the supermassive black holes at the centers of the M87 galaxy and the Milky Way using Very Long Baseline Interferometry (VLBI) with an Earth-sized aperture \cite{Akiyama et al. (2019),Akiyama et al. (2022)}. This globally acclaimed achievement not only provided strong evidence for the existence of black holes but also opened a new avenue for studying gravity and high-energy physics. It is important to highlight that this milestone was underpinned by theoretical simulations of black hole images, the development of an image template library, and the subsequent matching of these templates with actual observations. Therefore, it is crucial to comprehensively consider the various accretion environments that may exist in the vicinity of a black hole when simulating its images to ensure reliable observational predictions. In other words, the potential impact of the accretion environment on image features must be a central focus in black hole image simulations.

In the case of a black hole surrounded by a spherically symmetric accretion flow, the authors of \cite{Narayan et al. (2019)} proposed that the properties of the accretion flow--such as the emission region and dynamical characteristics--affect only the brightness of the black hole image and do not influence the geometric information of the observable shadow. In other words, the boundary of the black hole shadow coincides with the projection of the critical photon orbit, which is determined solely by the intrinsic properties of the black hole and serves as a fingerprint of spacetime. Similarly, Zeng et al. numerically simulated images of four-dimensional Gauss--Bonnet black holes surrounded by a spherically symmetric accretion flow and examined the impact of the specific emissivity of the accreting matter on these images \cite{Zeng et al. (2020)}. They found that altering the profile of the specific emissivity changes only the brightness of the image, without affecting the size of the observable black hole shadow. These findings are consistent with those from other studies \cite{Guo et al. (2021),Hu et al. (2022),He et al. (2022a),Heydari-Fard et al. (2023)}, confirming that the details of spherically symmetric accretion flows do not impact the size of the observable black hole shadow.

However, astrophysical black holes, particularly the supermassive black holes at the center of galaxies that are potential targets for EHT imaging, are often surrounded by an accretion disk. This suggests that the conclusions drawn from scenarios involving spherically symmetric accretion flows may not be universally applicable and highlights the need to investigate the impact of accretion disks on black hole images. In this context, Gralla and colleagues pioneered the investigation of how an optically and geometrically thin equatorial accretion disk influences the qualitative observational features of Schwarzschild black holes \cite{Gralla et al. (2019)}. They were the first to define the projection of the critical photon orbit onto the observation plane as the ``critical curve''. Their research revealed that the sharp bright ring in the image is composed of a series of subrings, and that the image asymptotically approaches the critical curve as the order of subring increases. Additionally, numerical simulations indicated that the critical curve is unlikely to correspond to the boundary of the observable black hole shadow, especially when the inner edge of the accretion disk extends to the event horizon, which confines the observable shadow to a smaller region. This phenomenon has been extensively studied in a flood of subsequent research \cite{Hu et al. (2022),Zeng and Zhang (2020),Peng et al. (2021),Li and He (2021),Zeng et al. (2022),He et al. (2022b),Guo et al. (2022),Meng et al. (2023),Wang et al. (2023),Yang et al. (2023),Gao et al. (2023),Li et al. (2024),Sui et al. (2024)}, and this type of shadow has been termed the ``inner shadow'' by the authors of \cite{Chael et al. (2021)}. Moreover, the authors of \cite{Hu et al. (2024)} numerically simulated images of a Horndeski black hole with a tilted thin accretion disk and identified a degeneracy between the disk inclination and the observation angle. They introduced the concept of an effective observation angle to describe the combined effect of the accretion disk tilt and observation inclination on image features. Their study also noted that the precession of a tilted accretion disk could cause the rotation of bright spots and the inner shadow in the image, thus establishing a link between the dynamics of tilted accretion disk and the observed signatures in black hole images.

The details of the accretion disk's emission--including the spatial position of radiating particles, the plasma's temperature, density, pressure, and the ion-electron temperature ratio--have the potential to alter the characteristics of black hole images. In addition, it is important to note that the radiation from accreting material also exhibits a dependence on the emission angle, with the specific intensity of the radiation varying according to the angle between the light ray and the disk's normal \cite{Thorne (1974)}. This phenomenon is often referred to as the ``projection effect'' of the accretion disk on radiation. While the projection effect has been extensively considered in numerical simulations with respect to the iron line profile of accretion disks \cite{Laor (1991),Svoboda et al. (2009),Dauser et al. (2010),Dauser et al. (2013),Garcia et al. (2014)}, it has received relatively scant attention in the context of black hole image simulations \cite{Tian and Zhu (2019),Bambi (2012)}. In particular, there remains significant potential for exploring how the projection effect influences black hole images. In response, this paper presents a comprehensive analysis of the spacetime properties of deformed Schwarzschild black holes, employs a ray-tracing algorithm to simulate images of these black holes surrounded by equatorial accretion disks with and without considering projection effect, elucidates the impact of deformation function on the observable characteristics of deformed Schwarzschild black holes, and meticulously examines the projection effect of anisotropic accretion disk on image features.

The remainder of this paper is organized as follows. In section 2, we provide a brief review of the metric describing deformed Schwarzschild black holes, followed by the derivation of geodesic equations governing the motion of time-like and null-like particles. Additionally, we examine the influence of the deformation parameter on the critical curve. In section 3, we employ a ray-tracing method to simulate images of targeted black holes at $86$ and $230$ GHz bands, both with and without considering the projection effect of the accretion disk. Moreover, we meticulously scrutinize the impacts of parameters on the redshift factor and the appearance of the inner shadow, and propose a novel method to constrain system parameters using the geometric information of the inner shadow. Finally, the conclusions are presented in section 4. Throughout this paper, we employ geometric units, where fundamental physical constants such as the speed of light $c$, the gravitational constant $G$, and the black hole mass $M$ are set to unity. Additionally, we adopt the metric signature $(-,+,+,+)$ to describe the spacetime geometry.
\section{Geodesics and critical curve of deformed Schwarzschild black holes}
In this section, we first introduce the metric of deformed Schwarzschild black holes and reveal the impact of the deformation parameter on spacetime geometry by means of embedding diagrams. Then, we derive the canonical equation of light rays and illustrate the radii of the photon ring and the critical curve across a range of deformation parameter values. Finally, we provide information on the circular orbits and plunging orbits of massive particles, which are essential for the ray-tracing method outlined in the following section.
\subsection{Deformed Schwarzschild black holes}
Considering a possible deviation from the Schwarzschild geometry, the authors of \cite{Johannsen and Psaltis (2011)} proposed a parameterized static spherically symmetric black hole based on the work in \cite{Yunes and Stein (2011)}, which can be expressed in Schwarzschild coordinates as \cite{Rayimbaev (2016),Yi and Wu (2020),Magalhaes et al. (2020),Zhang et al. (2021),Huang et al. (2022),Xu et al. (2024)}
\begin{equation}\label{1}
\textrm{d}s^{2}=g_{tt}\textrm{d}t^{2}+g_{rr}\textrm{d}r^2+g_{\theta \theta}\textrm{d}\theta^{2}+g_{\varphi \varphi}\textrm{d}\varphi^{2},
\end{equation}
in which $g_{\mu\nu}$ represents the covariant metric tensor with nonzero components
\begin{eqnarray}\label{2}
g_{tt} &=& -f(r)\left[1+h(r)\right], \quad g_{rr}=\frac{1+h(r)}{f(r)}, \nonumber \\
g_{\theta \theta} &=& r^{2}, \quad g_{\varphi \varphi}=r^{2}\sin^{2}\theta.
\end{eqnarray}
The metric potential and the deformation function are defined in geometric units as
\begin{equation}\label{3}
f(r)=1-\frac{2}{r}
\end{equation}
and
\begin{equation}\label{4}
h(r)=\sum_{k=0}^{\infty}\varepsilon_{k}\left(\frac{1}{r}\right)^{k},
\end{equation}
respectively, with $\varepsilon_{k}$ denotes the deformation parameter. Although theoretically, the deformation function is of infinite order, the deformation parameters are constrained by the current astrophysical observations. Specifically, parameters $\varepsilon_{0}$ and $\varepsilon_{1}$ must vanish to ensure the asymptotic flatness of the deformed Schwarzschild spacetime and its rotating counterpart \cite{Johannsen and Psaltis (2011)}. According to the Lunar Laser Ranging experiment, parameter $\varepsilon_{2}$ satisfies $|\varepsilon_{2}| \leq 4.6 \times 10^{-4}$ and can safely be approximated as $0$ \cite{Johannsen and Psaltis (2011),Williams et al. (2004)}. Furthermore, it is conceivable that the higher-order deformation terms ($k\geq 4$) have a negligible impact on the black hole spacetime. Therefore, the present research only focuses on the case where $k = 3$. Consequently, we rearrange the deformation function as
\begin{equation}\label{5}
h(r)=\frac{\varepsilon}{r^{3}},
\end{equation}
where $\varepsilon=\varepsilon_{3}$. Clearly, the degree to which the deformed Schwarzschild black hole deviates from the Schwarzschild black hole is determined by the deformation parameter $\varepsilon$. In particular, the black hole is prolate for $h(r) > 0$, but becomes oblate for $h(r) < 0$. When the deformation function $h(r) = 0$, the metric reduces to Schwarzschild case.
\begin{figure}%[tbph]
\center{
\includegraphics[width=6cm]{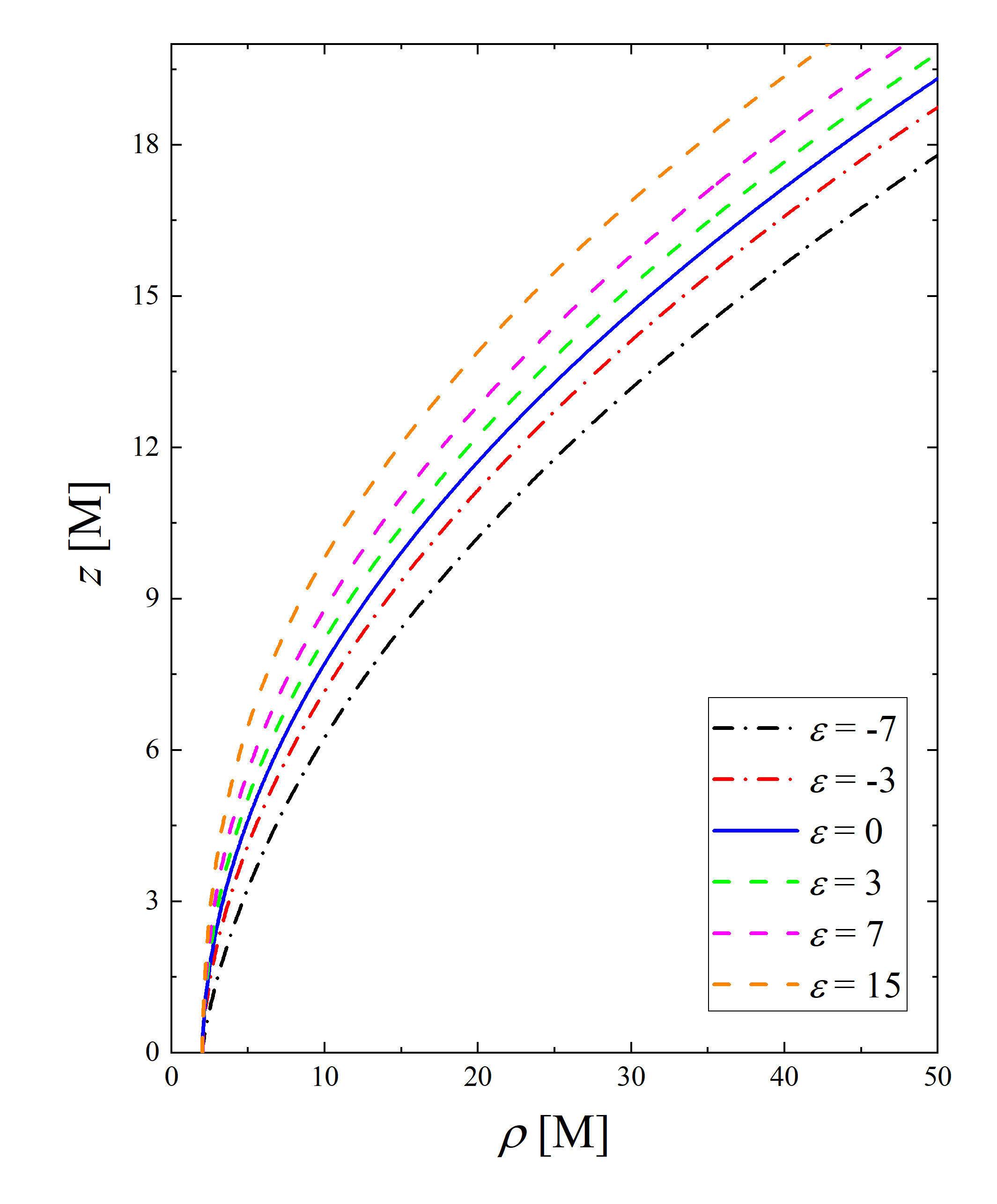}
\caption{Sections of the embedding diagram for the deformed Schwarzschild spacetime with different values of deformation parameter. Dashed-dot lines represent cases with negative $\varepsilon$, while dashed lines represent those with positive $\varepsilon$. Evidently, the slope of the curve increases with the deformation parameter, indicating a transition of the spacetime morphology from oblate to prolate.}}\label{fig1}
\end{figure}

It is worth mentioning that the deformation occurs inside the event horizon and is hardly measurable, but the impact of the deformation on the spacetime structure can be observed through the embedding diagram. For the sake of conciseness and coherence in the paper, we present the method for drawing the embedding diagram in the Appendix. Figure 1 depicts the profile of the deformed Schwarzschild spacetime embedding diagram for various $\varepsilon$, with all curves starting from the event horizon at $r_{\textrm{eh}}=2$\footnote{Several authors have confirmed that the event horizon radius of the deformed Schwarzschild black hole remains consistent with that of the Schwarzschild black hole when the deformation parameter satisfies $\varepsilon > -8$ \cite{Magalhaes et al. (2020),Xu et al. (2024)}.}. It is observed that as the deformation parameter increases, the curves converge towards the polar region, indicating a transition of the black hole morphology from oblate to prolate. Additionally, the slope of the curves correlates positively with the deformation parameter, suggesting an inverse relationship between the strength of the black hole's gravitational field and the deformation parameter.
\subsection{Null-like geodesics and the critical curve}
The motion of light rays in deformed Schwarzschild spacetime satisfies the Lagrangian formula
\begin{eqnarray}\label{6}
\mathscr{L} &=& \frac{1}{2} g_{\mu\nu} \dot{x}^{\mu}\dot{x}^{\nu} \nonumber \\
&=& \frac{1}{2} \left( g_{tt} \dot{t}^{2}+g_{rr} \dot{r}^{2}+g_{\theta \theta}\dot{\theta}^{2}+g_{\varphi \varphi}\dot{\varphi}^{2} \right),
\end{eqnarray}
where the four-velocity $\dot{x}^{\mu}$ is a derivative of the generalized coordinate $x^{\mu}$ with respect to an affine parameter $\lambda$. According to the Euler-Lagrange equation $p_{\mu}=\partial \mathscr{L}/\partial \dot{x}^{\mu}$, we have photon's covariant four-momentum
\begin{equation}\label{7}
p_{t}=g_{tt}\dot{t}=-E_{\textrm{n}},
\end{equation}
\begin{equation}\label{8}
p_{r}=g_{rr}\dot{r},
\end{equation}
\begin{equation}\label{9}
p_{\theta}=g_{\theta \theta}\dot{\theta},
\end{equation}
and
\begin{equation}\label{10}
p_{\varphi}=g_{\varphi \varphi}\dot{\varphi}=L_{\textrm{n}}.
\end{equation}
Here, $E_{\textrm{n}}$ and $L_{\textrm{n}}$ are two motion constants during photon propagation due to time-translational and rotational invariance, corresponding to the specific energy and specific angular momentum of the photon, respectively. Next, we would like to introduce the Hamiltonian $\mathscr{H}$ that governs the motion of particles, which can be derived via the Legendre transformation and is expressed as
\begin{equation}\label{11}
\mathscr{H} = p_{\mu}\dot{x}^{\mu}-\mathscr{L} = \frac{1}{2}g^{\mu\nu}p_{\mu}p_{\nu},
\end{equation}
where $g^{\mu\nu}$ denotes the contravariant metric tensor and satisfies $g^{\mu\nu} = 1/g_{\mu\nu}$ in spherically symmetric spacetime. Note that $\mathscr{H}=0$ for null-like particles, while $\mathscr{H}=-1/2$ applies to massive particles. Then, the propagation of light rays can be calculated using the canonical equations,
\begin{equation}\label{12}
\dot{x^{\mu}}=\frac{\partial \mathscr{H}}{\partial p_{\mu}}, \quad \dot{p_{\mu}} = -\frac{\partial \mathscr{H}}{\partial x^{\mu}}.
\end{equation}

\begin{table*}
\begin{center}
\small \caption{Dependence of the radii of photon's unstable circular orbits $r_{\textrm{p}}$ and the critical curve $b_{\textrm{p}}$ on the deformation parameter $\varepsilon$. The values of $r_{\textrm{p}}$ and $b_{\textrm{p}}$ monotonically decrease with increasing $\varepsilon$, consistent with the findings in \cite{Magalhaes et al. (2020)}.} \label{t1}
\begin{tabular}{ccccccccccc}\hline
$\varepsilon$ & $-7$ & $-5$ & $-3$ & $-1$ & $0$ & $1$ & $4$ & $7$ & $10$ & $15$  \\
\hline
$r_{\textrm{p}}$ & $3.4474$ & $3.3139$ & $3.1820$ & $3.0575$ & $3.0000$ & $2.9466$ & $2.8142$ & $2.7211$ & $2.6572$ & $2.5900$\\
\hline
$b_{\textrm{p}}$ & $5.8429$ & $5.6666$ & $5.4823$ & $5.2923$ & $5.1962$ & $5.1000$ & $4.8175$ & $4.5537$ & $4.3154$ & $3.9754$\\
\hline
\end{tabular}
\end{center}
\end{table*}

Photons can be trapped in unstable circular orbits (critical photon orbits) around a black hole, where these orbits satisfy the condition of $p_{r}=\dot{p_{r}}=0$. Hence, according to eq. \eqref{11}, we define the effective potential for photon motions, read as\footnote{Note that due to the spherical symmetry of the deformed Schwarzschild spacetime, the propagation of light rays within any plane can safely be extended to the entire spacetime, allowing us to set $\theta=\pi/2$ and $p_{\theta}=\dot{p_{\theta}}=0$.}
\begin{equation}\label{13}
V_{\textrm{eff}}^{\textrm{n}} = \frac{E_{\textrm{n}}}{L_{\textrm{n}}} = \frac{\sqrt{f(r)\left[1+h(r)\right]}}{r}.
\end{equation}
The unstable circular orbits of photons correspond to the local maximum of the effective potential, and their radius, labeled as $r_{\textrm{p}}$, can be numerically calculated using the relation $\partial V_{\textrm{eff}}^{\textrm{n}}/\partial r = 0$. It is worth emphasizing that the circular orbits of photons are unstable; once perturbed, photons either fall into the event horizon of the deformed Schwarzschild black hole or escape to infinity, thereby potentially forming a closed curve in the observation plane. This profile is termed as ``critical curve'' in \cite{Gralla et al. (2019)}, which serves as the boundary of the traditional concept known as the ``black hole shadow''. Once the value of $r_{\textrm{p}}$ is determined, the radius of the critical curve, $b_{\textrm{p}}$, can be obtained using eq. \eqref{13}, expressed as
\begin{equation}\label{14}
b_{\textrm{p}}=\frac{r_{\textrm{p}}}{\sqrt{f(r_{\textrm{p}})\left[1+h(r_{\textrm{p}})\right]}}.
\end{equation}

Table $1$ lists the values of $r_{\textrm{p}}$ and $b_{\textrm{p}}$ of deformed Schwarzschild black holes for various $\varepsilon$. It is evident that as the deformation parameter increases, the critical photon orbit moves closer to the central object, accompanied by a shrinking of the critical curve. This trend not only confirms a negative correlation between the deformation parameter and the gravitational field of deformed Schwarzschild black holes but also provides a basis for testing the no-hair theorem of black holes. Furthermore, it is noteworthy that decreasing the mass of the deformed Schwarzschild black hole can also lead to such a trend, indicating a degeneracy between the deformation parameter and the black hole mass.
\subsection{Time-like geodesics}
The dynamics of the accreting material play a crucial role in the simulation of black hole images. In this study, we assume that the light ``illuminating'' the deformed Schwarzschild black hole is provided by an optically thin, geometrically thin accretion disk located on the equatorial plane. This accretion disk consists of two parts: the Keplerian region where $r_{\textrm{s}} \geq r_{\textrm{ISCO}}$, and the plunging region where $r_{\textrm{eh}} < r_{\textrm{s}} < r_{\textrm{ISCO}}$, in which $r_{\textrm{s}}$ and $r_{\textrm{ISCO}}$ represent the radial coordinate of the accreting material and the radius of the innermost stable circular orbit (ISCO) of timelike particles, respectively. We then discuss the motion behavior of accreting particles in each region.

In the Keplerian region, particles move in a stable circular orbit around the central body, and the specific energy $E_{\textrm{k}}$ and specific angular momentum $L_{\textrm{k}}$ of the particles are read as \cite{Liu et al. (2021)}
\begin{equation}\label{15}
E_{\textrm{k}} = \frac{-g_{tt}}{\sqrt{-g_{tt}-\frac{1}{2}r_{\textrm{s}}(-g_{tt})'}},
\end{equation}
\begin{equation}\label{16}
L_{\textrm{k}} = \frac{r_{\textrm{s}}\sqrt{r_{\textrm{s}}(-g_{tt})'}}{\sqrt{-2g_{tt}-r_{\textrm{s}}(-g_{tt})'}}.
\end{equation}
Here, ``$()'$'' denotes the first-order partial derivative with respect to $r$, and the value of $g_{tt}$ is evaluated at $r = r_{\textrm{s}}$. Notably, the motion of time-like particles is also governed by the Lagrangian \eqref{6} or Hamiltonian \eqref{11}. Therefore, we can further derive the four-velocity of the particle in the Keplerian region, expressed as
\begin{equation}\label{17}
\dot{t_{\textrm{k}}} = \frac{E_{\textrm{k}}}{-g_{tt}},
\end{equation}
\begin{equation}\label{18}
\dot{\varphi_{\textrm{k}}} = \frac{L_{\textrm{k}}}{r_{\textrm{s}}^{2}},
\end{equation}
and
\begin{equation}\label{19}
\dot{r_{\textrm{k}}} = \dot{\theta_{\textrm{k}}} = 0
\end{equation}
since the particle is moving in a stable circular orbit on the equatorial plane.

In astrophysics, the viscosity of the accretion disk causes the accreting material to transfer angular momentum outward as it orbits the black hole. Consequently, particles in the Keplerian region gradually migrate closer to the ISCO. Once particles cross the ISCO, they accelerate and spiral toward the black hole until they cross the event horizon. The dynamics of particles during this process have attracted considerable scientific attention and have been confirmed to correlate with astrophysical observations \cite{Chael et al. (2021),Mummery and Balbus (2022),Mummery et al. (2024)}. To simplify the calculations, we adopt the strategy outlined in the literature \cite{Hu et al. (2024),Hou et al. (2022)}, specifically assuming that the specific energy $E_{\textrm{p}}$ and specific angular momentum $L_{\textrm{p}}$ of particles in the plunging region are identical to those at the ISCO. Thus, we have
\begin{equation}\label{20}
E_{\textrm{p}} = E_{\textrm{ISCO}} = \frac{-g_{tt}}{\sqrt{-g_{tt}-\frac{1}{2}r_{\textrm{ISCO}}(-g_{tt})'}},
\end{equation}
\begin{equation}\label{21}
L_{\textrm{p}} = L_{\textrm{ISCO}} = \frac{r_{\textrm{ISCO}}\sqrt{r_{\textrm{ISCO}}(-g_{tt})'}}{\sqrt{-2g_{tt}-r_{\textrm{ISCO}}(-g_{tt})'}},
\end{equation}
where the value of $g_{tt}$ is evaluated at $r = r_{\textrm{ISCO}}$. The value of  $r_{\textrm{ISCO}}$ can be determined by the iteration relation \cite{Gao et al. (2023)}
\begin{equation}\label{22}
r_{\textrm{ISCO}} = \frac{-3g_{tt}(-g_{tt})'}{2(-g_{tt})'^{2}+g_{tt}(-g_{tt})''},
\end{equation}
in which ``$()''$'' denotes the second-order partial derivative with respect to $r$. Similar to the Keplerian region, we have the four-velocity of the particle in the plunging region:
\begin{equation}\label{23}
\dot{t_{\textrm{p}}} = \frac{E_{\textrm{p}}}{-g_{tt}},
\end{equation}
\begin{equation}\label{24}
\dot{\varphi_{\textrm{p}}} = \frac{L_{\textrm{p}}}{r_{\textrm{s}}^{2}},
\end{equation}
\begin{equation}\label{25}
\dot{\theta_{\textrm{p}}} = 0,
\end{equation}
and
\begin{equation}\label{26}
\dot{r_{\textrm{p}}} = -\sqrt{\frac{1}{g_{rr}}\left(-1-\frac{E_{\textrm{p}}^{2}}{g_{tt}}-\frac{L_{\textrm{p}}^{2}}{r_{\textrm{s}}^{2}}\right)}.
\end{equation}
Here, the values of $g_{tt}$ and $g_{rr}$ are evaluated at $r = r_{\textrm{s}}$, eq. \eqref{26} is derived from the condition $g_{\mu\nu}\dot{x}^{\mu}\dot{x}^{\nu}=-1$, where ``$-$'' indicates that the particle is spiralling towards the black hole.
\section{Observational features of deformed Schwarzschild black holes}
In this section, we first introduce the projection effect of an anisotropic accretion disk as well as the ray-tracing algorithm. Then, we numerically simulate the images of deformed Schwarzschild black holes at $86$ GHz and $230$ GHz for different parameter spaces to analyze the impact of the accretion disk projection effect on the images. Finally, we focus on the influence of parameters on the inner shadow and propose a novel method for constraining black hole using the geometric information of the inner shadow.
\begin{figure}%[tbph]
\center{
\includegraphics[width=8cm]{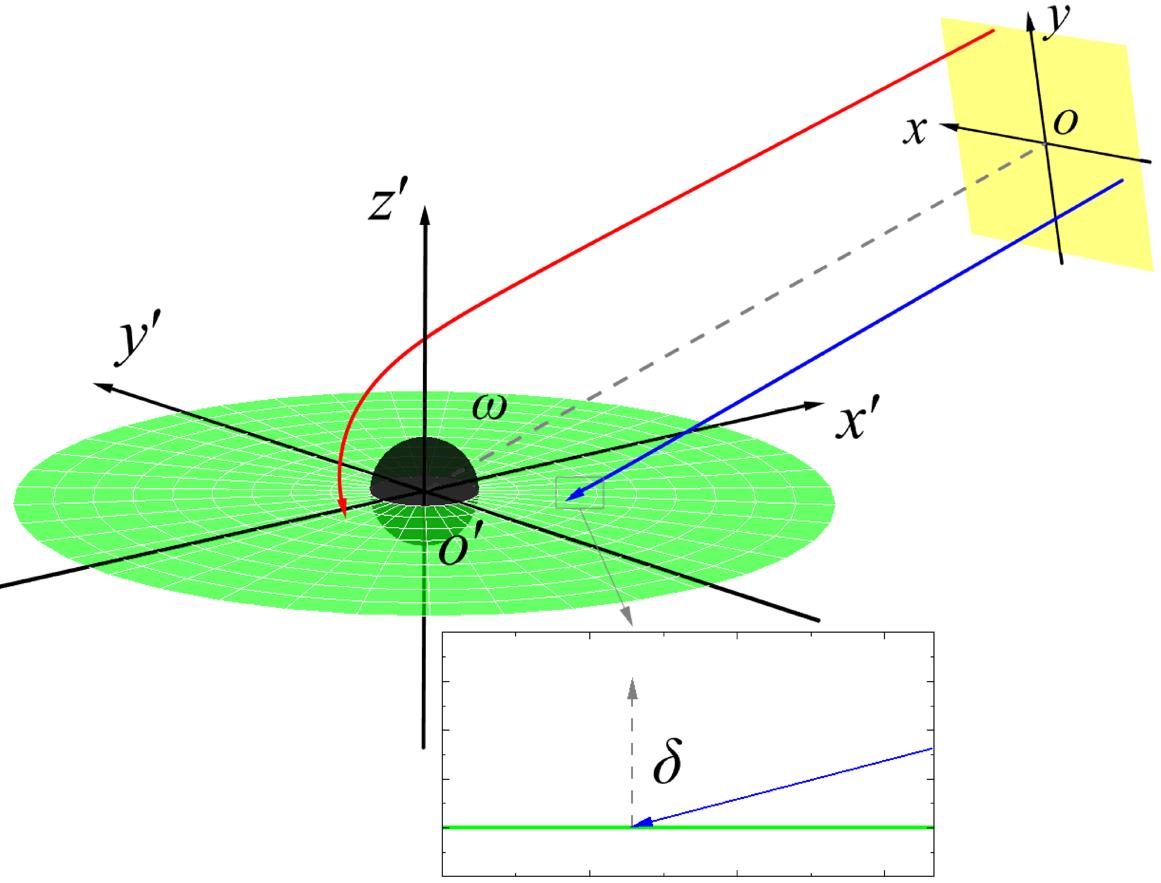}
\caption{Schematic diagram of the backward ray-tracing method. The deformed Schwarzschild black hole and the accretion disk are represented by the black sphere and the green disk, respectively. The yellow solid block represents the observation plane, with $\omega$ being the observation inclination. The rays emitted from the accretion disk and reaching the observation plane (i.e., red line and blue line) can be inversely viewed as being emanated from the observation plane and hitting the accretion disk (reversibility of light). Note that these rays, in general, do not intersect the accretion disk perpendicularly but rather at an angle $\delta$ with the disk normal (see zoomed-in inset). Clearly, the radiation from an anisotropic accretion disk is determined by $\delta$.}}\label{fig2}
\end{figure}
\subsection{Ray-tracing scheme}
The propagation of light bridges the details of the accreting material and the black hole image. Performing forward ray tracing from the accretion flow to ascertain which rays reach the observation plane is a computationally intensive process. Fortunately, by leveraging the principle of light path reversibility, we can trace light rays backward in time from the observation plane and identify those that intersect with the accretion disk and the event horizon. Rays intersecting the accretion disk contribute to the image brightness, while those intersecting the event horizon delineate the observable shadow of the black hole.

Figure 2 presents a schematic diagram of this backward ray-tracing process. In this diagram, the spacetime of the black hole and the observer's local frame are represented by the Cartesian coordinate systems $(x^{\prime},y^{\prime}, z^{\prime})$ and $(x,y,z)$, respectively, where the $z$-axis of the latter aligns with $\overline{oo^{\prime}}$ and points towards the black hole singularity. The green disk and yellow block denote the anisotropic accretion disk and the observation plane, respectively. Notably, the pixel coordinates within the observation plane satisfy $z=0$. The angle $\omega$ between $\overline{oo^{\prime}}$ and $\overline{o^{\prime}z^{\prime}}$ represents the observation inclination, ranging from $[0,\pi/2]$. The red and blue solid lines illustrate two representative rays emitted perpendicular to the observation plane. It is evident that the intersection points of the photons emitted from different pixels with the accretion disk vary. These photons can also traverse the accretion disk multiple times under the influence of the black hole's gravitational field, accumulating their ``brightness'' \cite{Gralla et al. (2019)}. By calculating the total specific intensity of the rays corresponding to each pixel, the black hole image can be reconstructed.

It is noteworthy that, in general, the wavevector of a photon does not hit the accretion disk perpendicularly but rather at an angle $\delta$ relative to the normal of the disk plane (see zoomed-in inset in figure 2). For an anisotropic accretion disk, it is evident that $\delta$ influences the disk's emission and has the potential to modify the characteristics of the black hole's image. This phenomenon is referred to as the ``projection effect'' of the anisotropic accretion disk on radiation. Let us denote the radiation along the normal to the disk as $\mathscr{F}$. Then, the radiation at an angle $\delta$ to the normal can be represented as $\kappa\mathscr{F}$, where $\kappa$ is a function of $\delta$. While Bambi proposed $\kappa = 1/2 + (3/4)\cos\delta$ in \cite{Bambi (2012)}, we adopt a more straightforward approach in this study, setting $\kappa = \cos\delta$. The influence of $\kappa$ on the black hole image features will be examined in detail in our forthcoming work.

We now have a clear understanding of the principles of ray-tracing and the projection effect. The next task is to use a numerical integrator to trace the fate of light rays emitted from each pixel on the observation plane. Typically, this operation needs to be performed in the local frame of the black hole, which means we need to determine the initial conditions of the rays, $(x^{\mu},p_{\mu})=(t,r,\theta,\varphi,p_{t},p_{r},p_{\theta},p_{\varphi})$, in the local coordinate system $(x^{\prime},y^{\prime}, z^{\prime})$. As shown in figure 2, the observer's local coordinate system can be aligned with the black hole's local coordinate system through multiple rotations and translations, thus yielding \cite{Younsi et al. (2016),Pu et al. (2016),Lin et al. (2022)}
\begin{eqnarray}\label{27}
x^{\prime} &=& (r_{\textrm{obs}}-z)\sin\omega-y\cos\omega, \nonumber \\
y^{\prime} &=& x, \nonumber \\
z^{\prime} &=& (r_{\textrm{obs}}-z)\cos\omega+y\sin\omega.
\end{eqnarray}
Here, $r_{\textrm{obs}}$ denotes the distance between the deformed Schwarzschild black hole and the observer. Further, we have the initial position of the photon in Schwarzschild coordinates,
\begin{eqnarray}\label{28}
r &=& \sqrt{x^{\prime 2}+y^{\prime 2}+z^{\prime 2}}, \nonumber \\
\theta &=& \arccos\left(\frac{z^{\prime}}{r}\right), \nonumber \\
\varphi &=& \textrm{atan}2\left(y^{\prime},x^{\prime}\right).
\end{eqnarray}
Assuming that the observation distance $r_{\textrm{obs}}$ is sufficiently large (i.e., $r_{\textrm{obs}}=1000$), the observer's spacetime can be approximated as flat. Consequently, it is reasonable to assume that the light rays are emitted perpendicularly from the observation plane. Therefore, the initial velocity of the photon measured in the $(x,y,z)$ coordinate system can be represented as $(\dot{x},\dot{y},\dot{z})=(0,0,1)$. Utilizing the differential form of eq. \eqref{27}, we have
\begin{eqnarray}\label{29}
\dot{x}^{\prime} &=& -\sin\omega, \nonumber \\
\dot{y}^{\prime} &=& 0, \nonumber \\
\dot{z}^{\prime} &=& -\cos\omega.
\end{eqnarray}
Substituting this result into the differential form of eq. \eqref{28}, we obtain the initial velocity of the photon in Schwarzschild coordinates,
\begin{eqnarray}\label{30}
\dot{r} &=& -\sin\theta\sin\omega\cos\varphi-\cos\theta\cos\omega, \nonumber \\
\dot{\theta} &=& \frac{1}{r}\left(\sin\theta\cos\omega-\cos\theta\sin\omega\cos\varphi\right), \nonumber \\
\dot{\varphi} &=& \frac{\sin\omega\sin\varphi}{r\sin\theta}.
\end{eqnarray}
Recalling eqs. \eqref{8}-\eqref{10}, solving for the initial conjugate momentum of the photon in the local frame of the black hole is straightforward\footnote{The constant of motion $p_{t}$ can be obtained from the Hamiltonian constraint \eqref{11}.}.

We now have the initial conditions of the photon corresponding to each pixel $(x,y,0)$ within the observation plane. Using a fifth- and sixth-order Runge-Kutta-Fehlberg integrator with variable step sizes to integrate eq. \eqref{12}, we can simulate the propagation of light rays and obtain the intersection points with the accretion disk as well as the angle $\delta$.
\subsection{Redshift factors}
The relative motion between the accreting particles and the observer introduces a Doppler effect, which can significantly impact the characteristics of the black hole image. For a stationary observer, with four-velocity $u^{\mu}_{\textrm{o}}=(1,0,0,0)$, the redshift factor of the light ray is given by
\begin{equation}\label{31}
g = \frac{p_{\mu}u^{\mu}_{\textrm{o}}}{p_{\nu}u^{\nu}_{\textrm{e}}} = \frac{p_{t}}{p_{\nu}u^{\nu}_{\textrm{e}}}.
\end{equation}
Here, $p_{\mu}$ is the photon's conjugate momentum, which can be directly read from the ray-tracing; $u^{\nu}_{\textrm{e}}$ is the four-velocity of the emitting source. The four-velocity of particles in the Keplerian region and the plunging region can be calculated using eqs. \eqref{17}-\eqref{19} and eqs. \eqref{23}-\eqref{26}, respectively.
\begin{figure*}%[tbph]
\center{
\includegraphics[width=2.9cm]{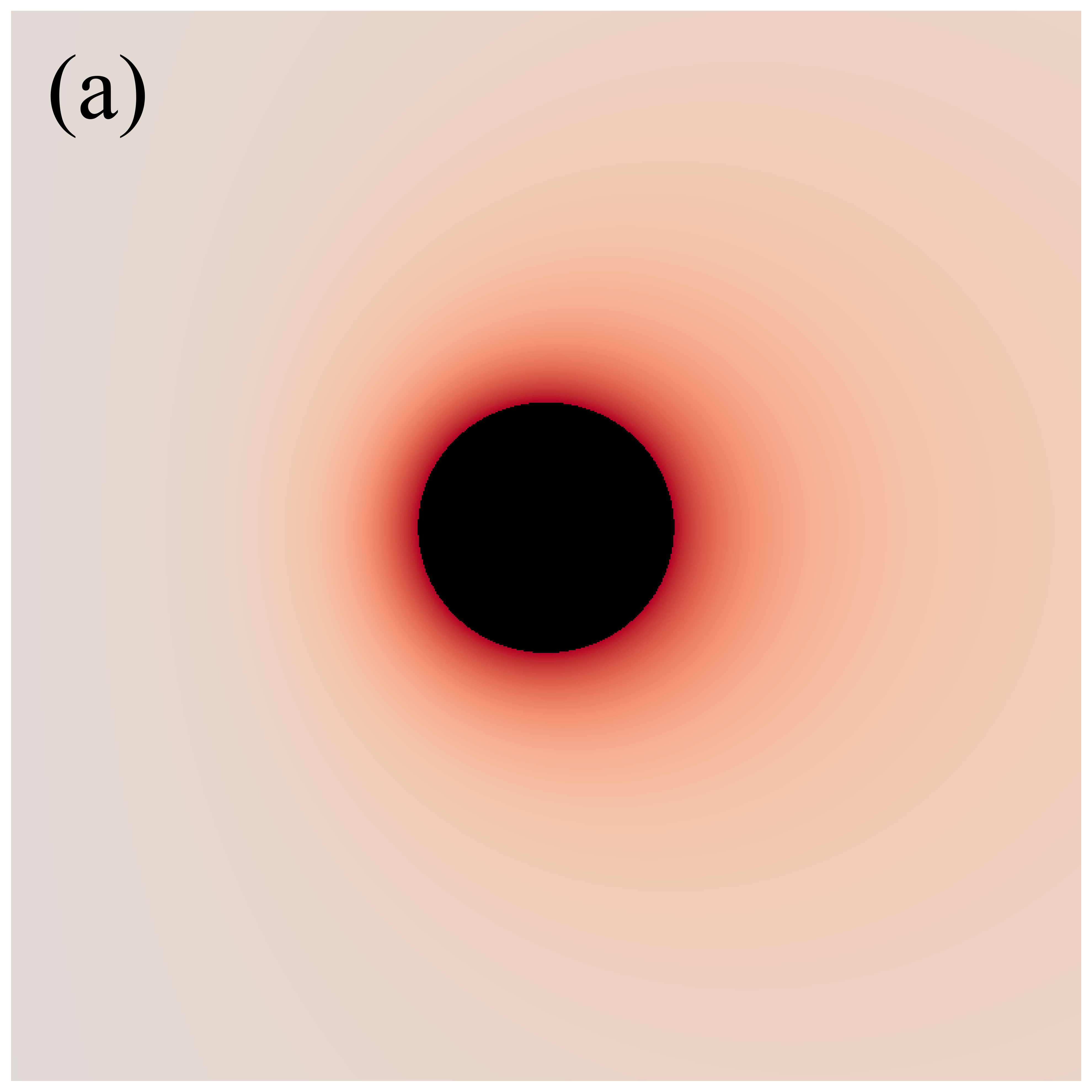}
\includegraphics[width=2.9cm]{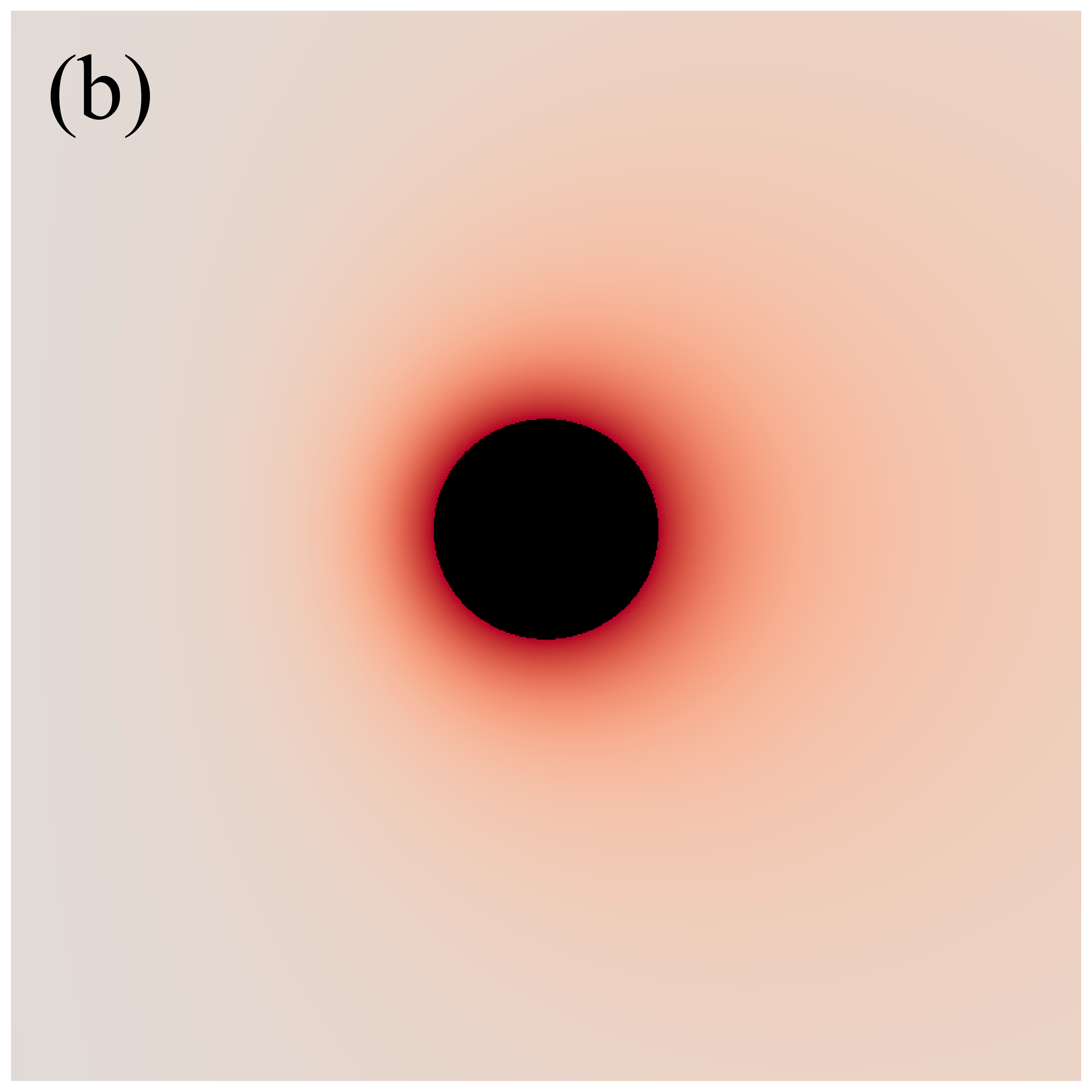}
\includegraphics[width=2.9cm]{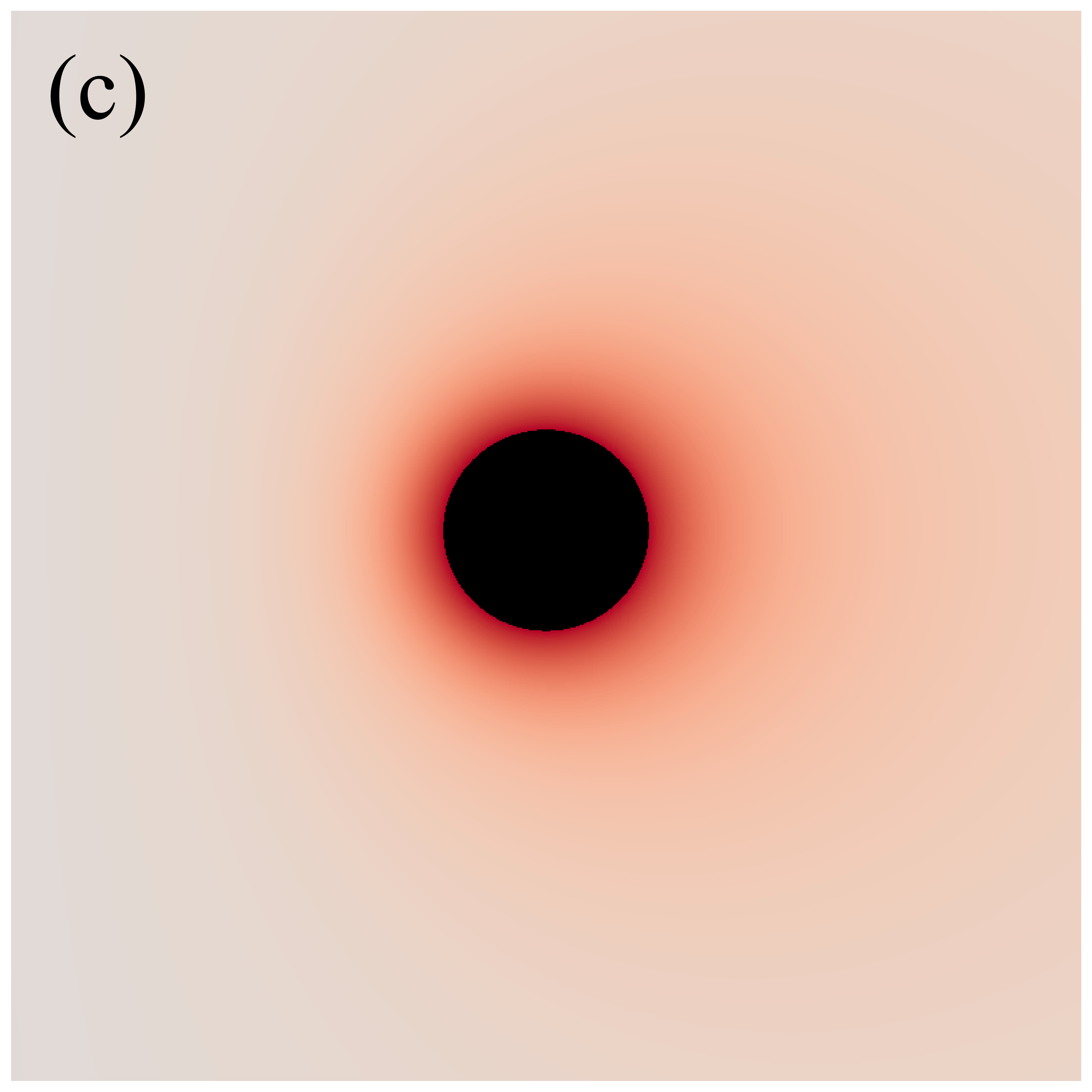}
\includegraphics[width=2.9cm]{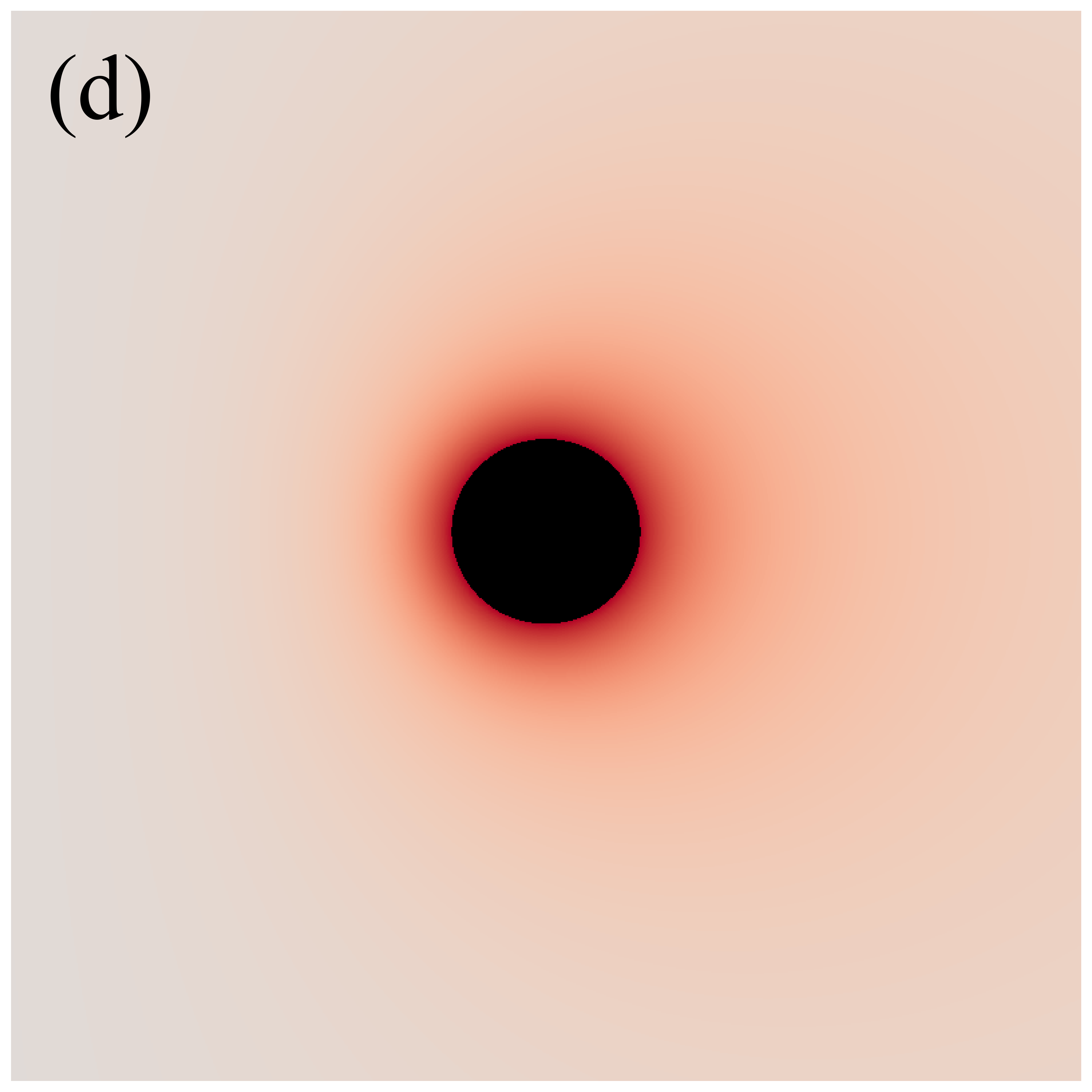}
\includegraphics[width=2.9cm]{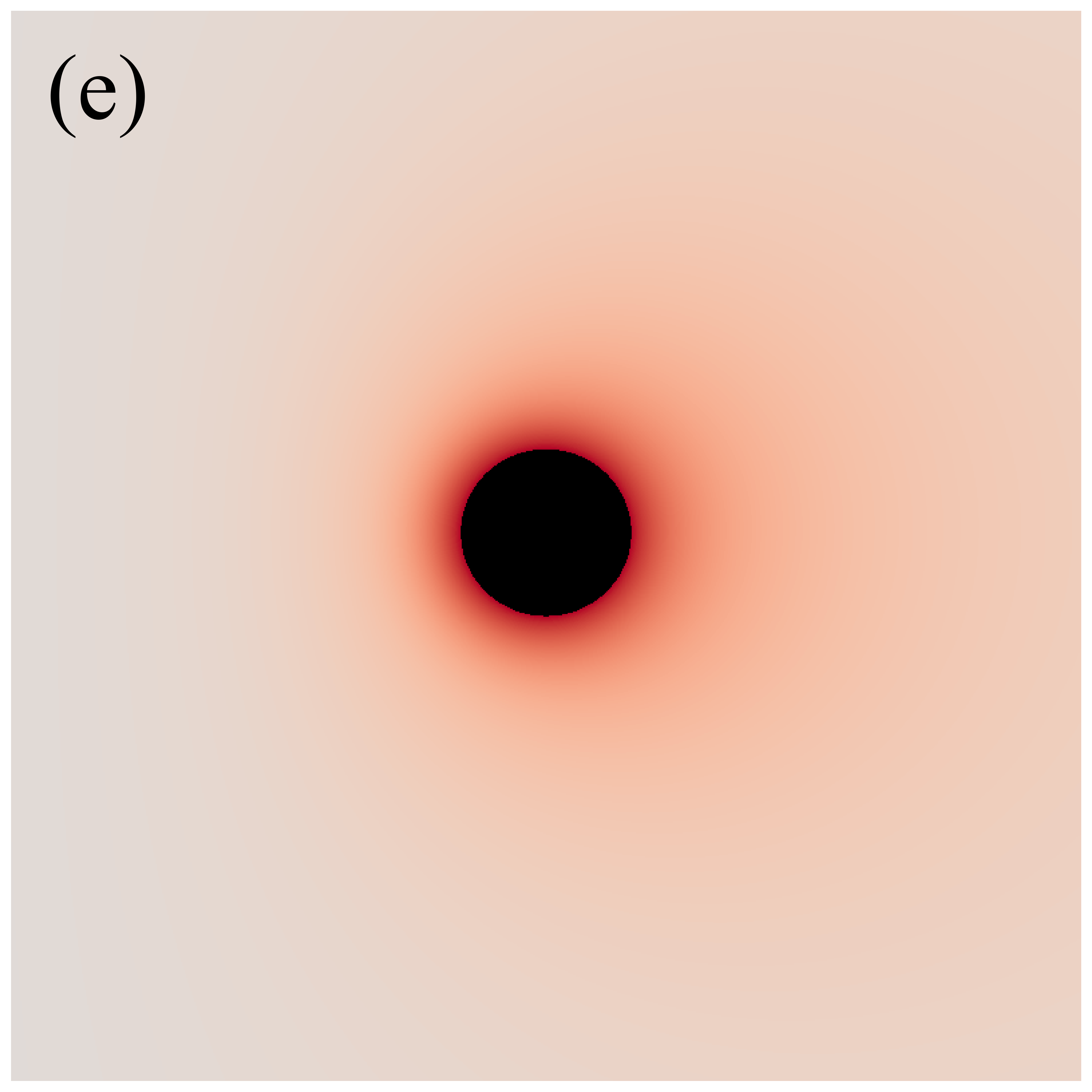}
\includegraphics[width=2.9cm]{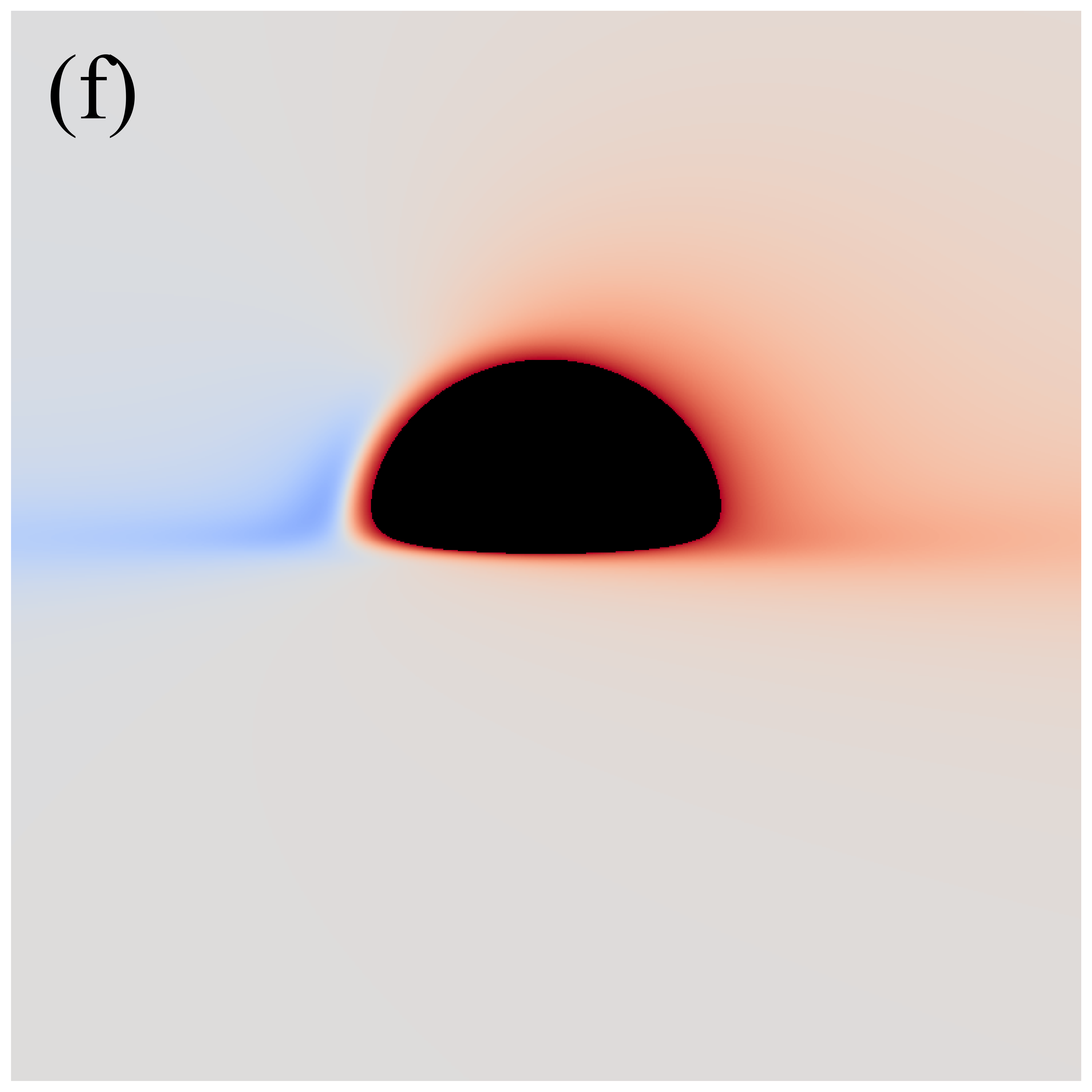}
\includegraphics[width=2.9cm]{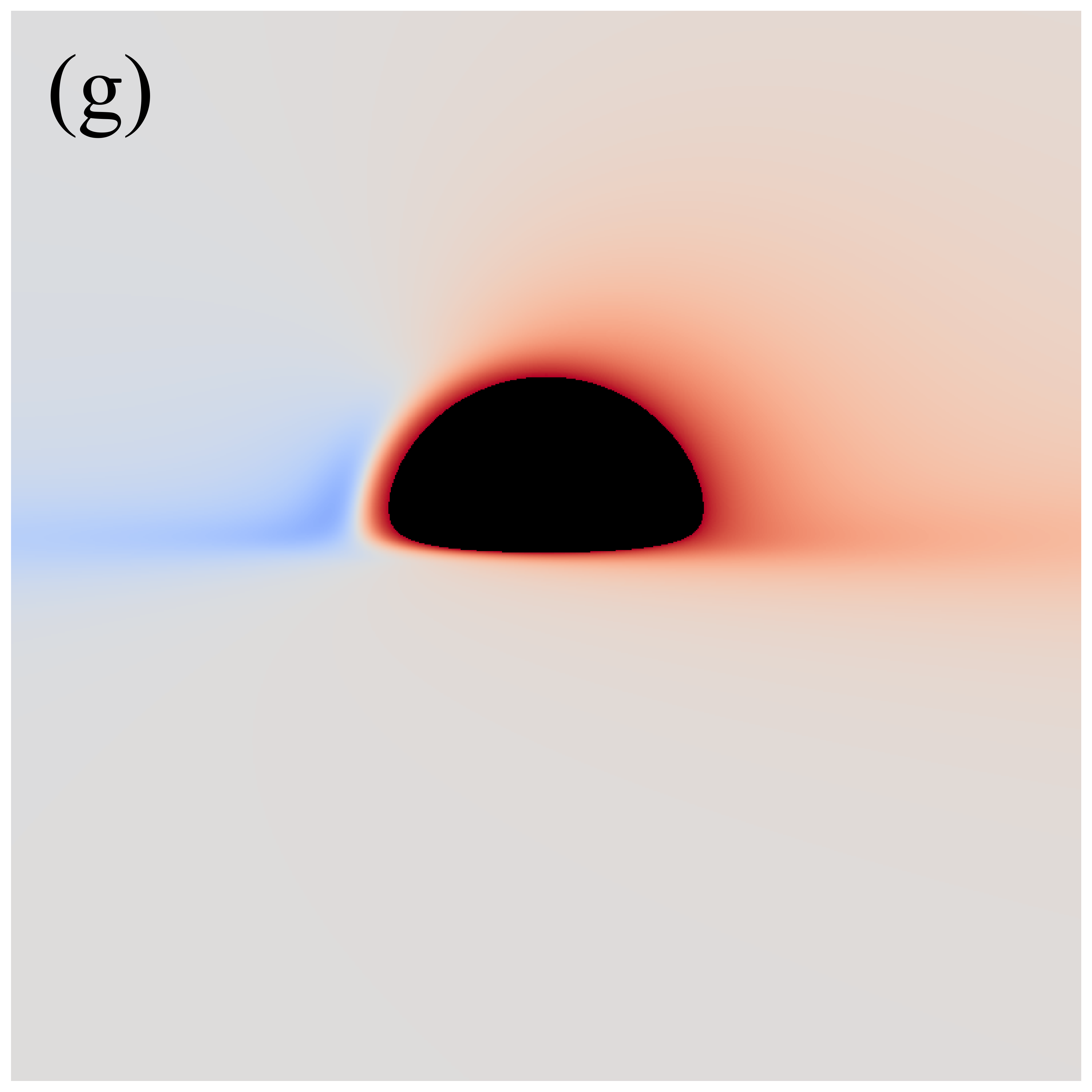}
\includegraphics[width=2.9cm]{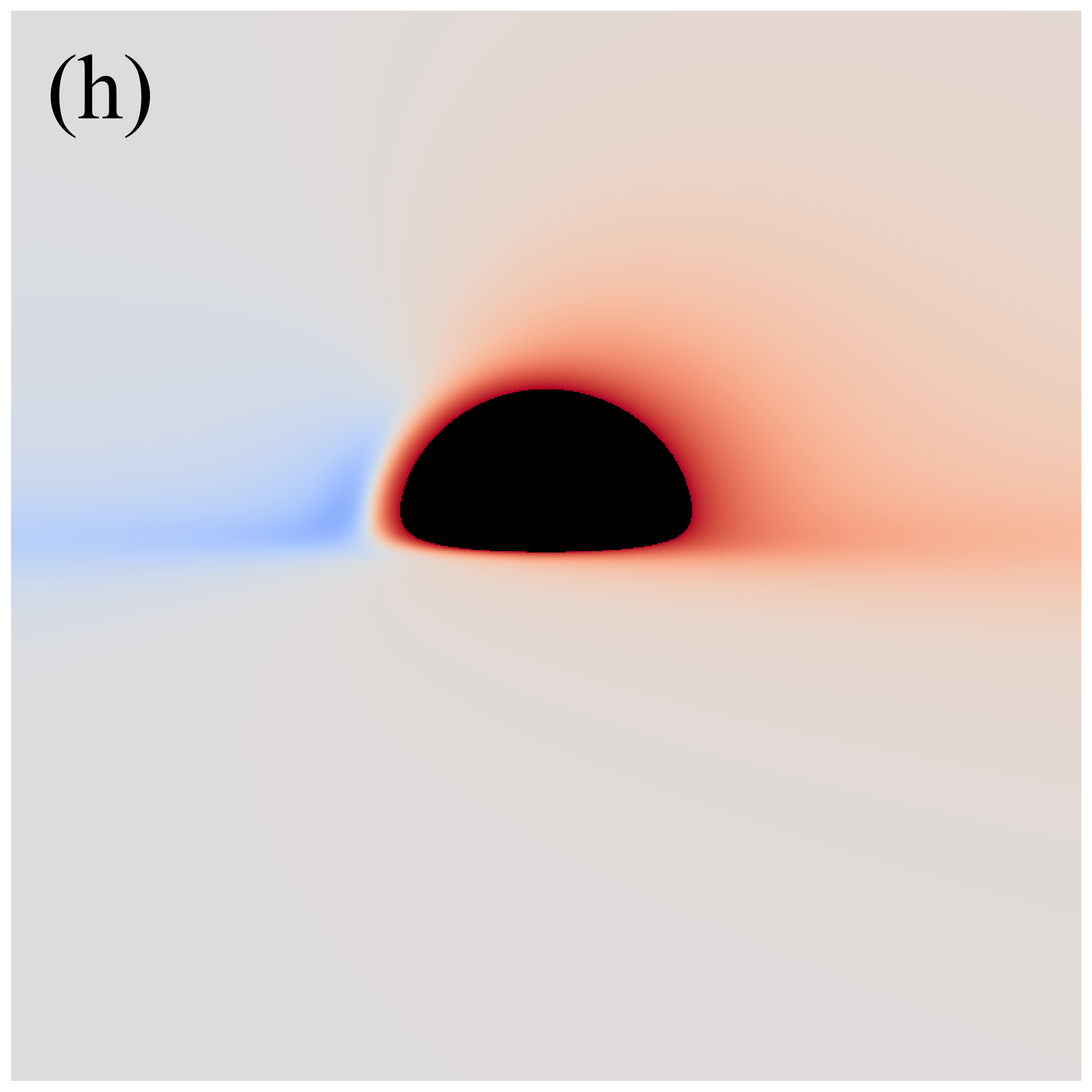}
\includegraphics[width=2.9cm]{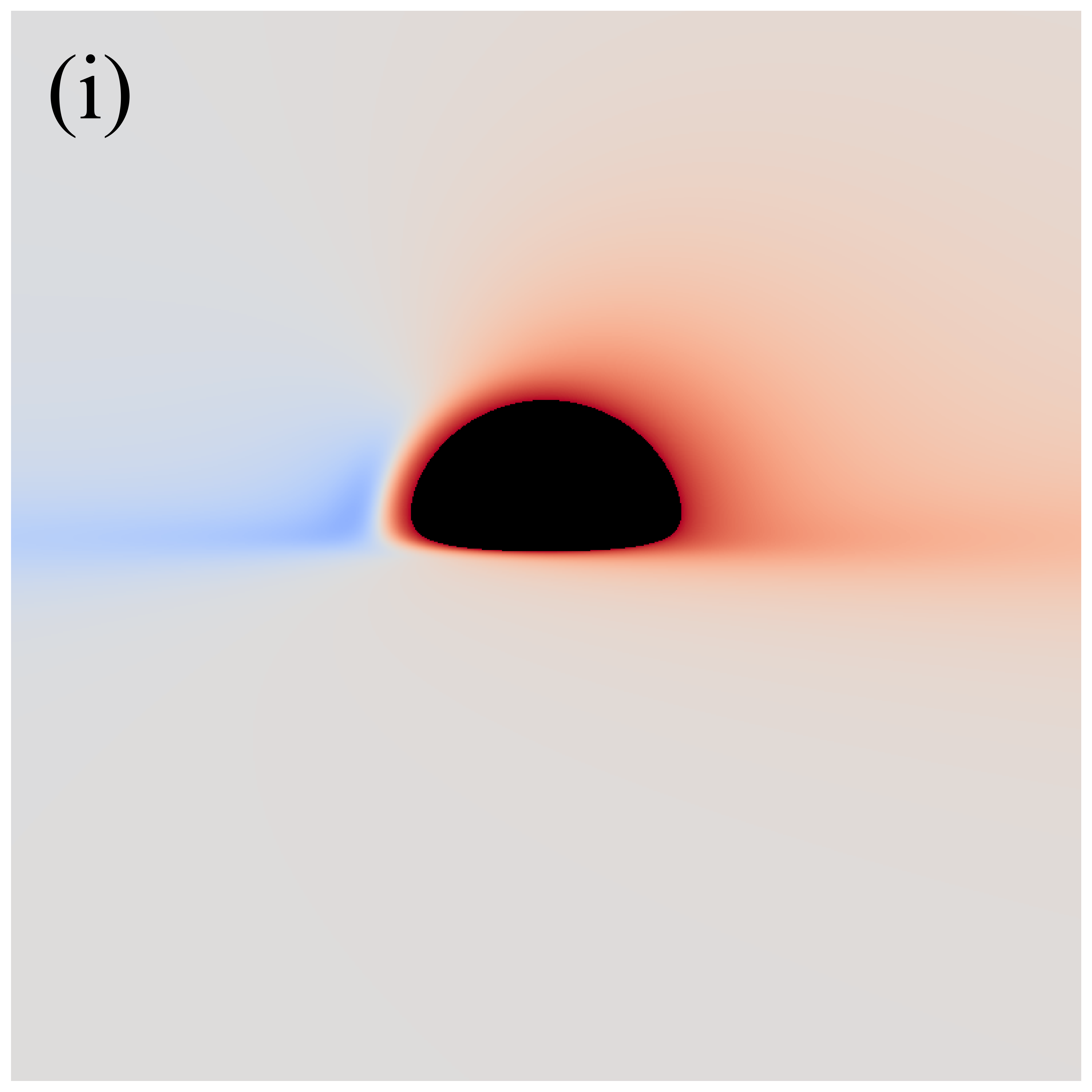}
\includegraphics[width=2.9cm]{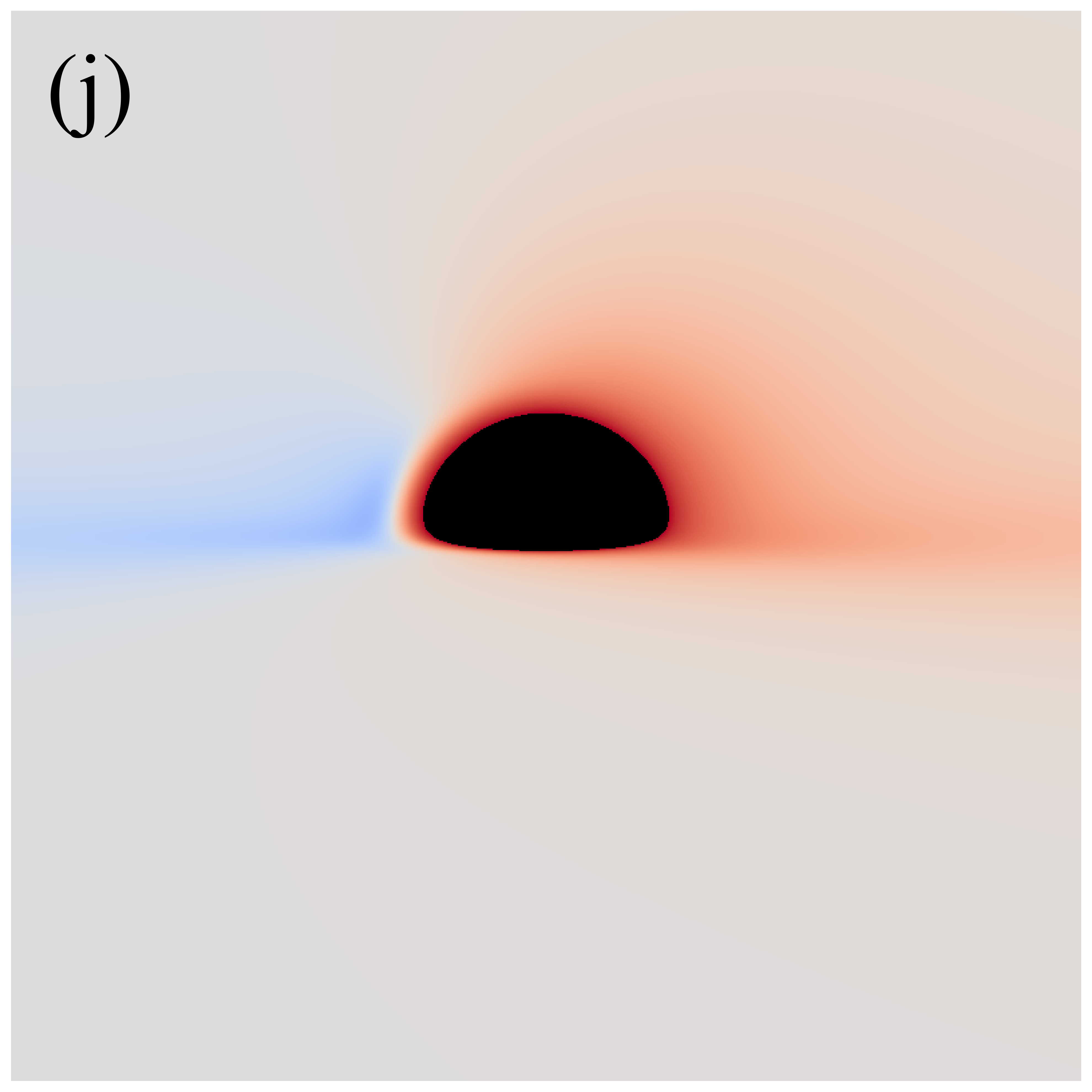}
\caption{Distributions of the redshift factor $g$ in direct images of deformed Schwarzschild black holes surrounded by an equatorial accretion disk for different values of the deformation parameter $\varepsilon$ and observation angle $\omega$. Each panel has a field of view set to $30 \times 30$ M with a resolution of $800 \times 800$ pixels. From left to right in each row, $\varepsilon$ takes values of $-7$, $-3$, $0$, $3$, and $7$. The first row corresponds to $\omega = 17^{\circ}$, and the second row corresponds to $\omega = 85^{\circ}$. Each redshift factor is visualized by means of a continuous, linear color spectrum, with red indicating redshift and blue indicating blueshift. It can be deduced that at smaller observation angles, blueshift signals are difficult to identify; whereas at larger observation angles, both blueshift and redshift factors are suppressed with the enhancement of $\varepsilon$.}}\label{fig3}
\end{figure*}

Figure 3 shows the distributions of redshift factors in the direct images of deformed Schwarzschild black holes for $\varepsilon = -7$, $-3$, $0$, $3$, and $7$ from left to right. The first and second rows correspond to observation angles of $17^{\circ}$ and $85^{\circ}$, respectively. Each redshift factor is visualized using a continuous, linear color map, where red represents redshift $(g < 1)$ and blue is associated with blueshift $(g > 1)$. Each panel's central black region represents the black hole's inner shadow, with its boundary defined by the projection of the event horizon \cite{Chael et al. (2021)}. The size of the inner shadow is modulated by the deformation parameter, while its shape is affected by the observation inclination. Surrounding the inner shadow is a prominent ``red ring'', composed of light emitted by particles within the plunging region, resulting in a significant redshift. At low observation inclinations, blueshift features are almost indiscernible. This is because, at smaller observation angles, the projection of the accretion disk's motion along the line of sight is negligible, resulting in gravitational redshift becoming the dominant effect. At high observation inclinations, the blueshift and redshift regions predominantly appear on the west and east sides of the image, respectively, determined by the direction of the accretion disk's rotation. Additionally, we find that the redshift and blueshift are suppressed with increasing deformation parameter but are enhanced with increasing observation inclination.
\begin{figure*}%[tbph]
\center{
\includegraphics[width=6cm]{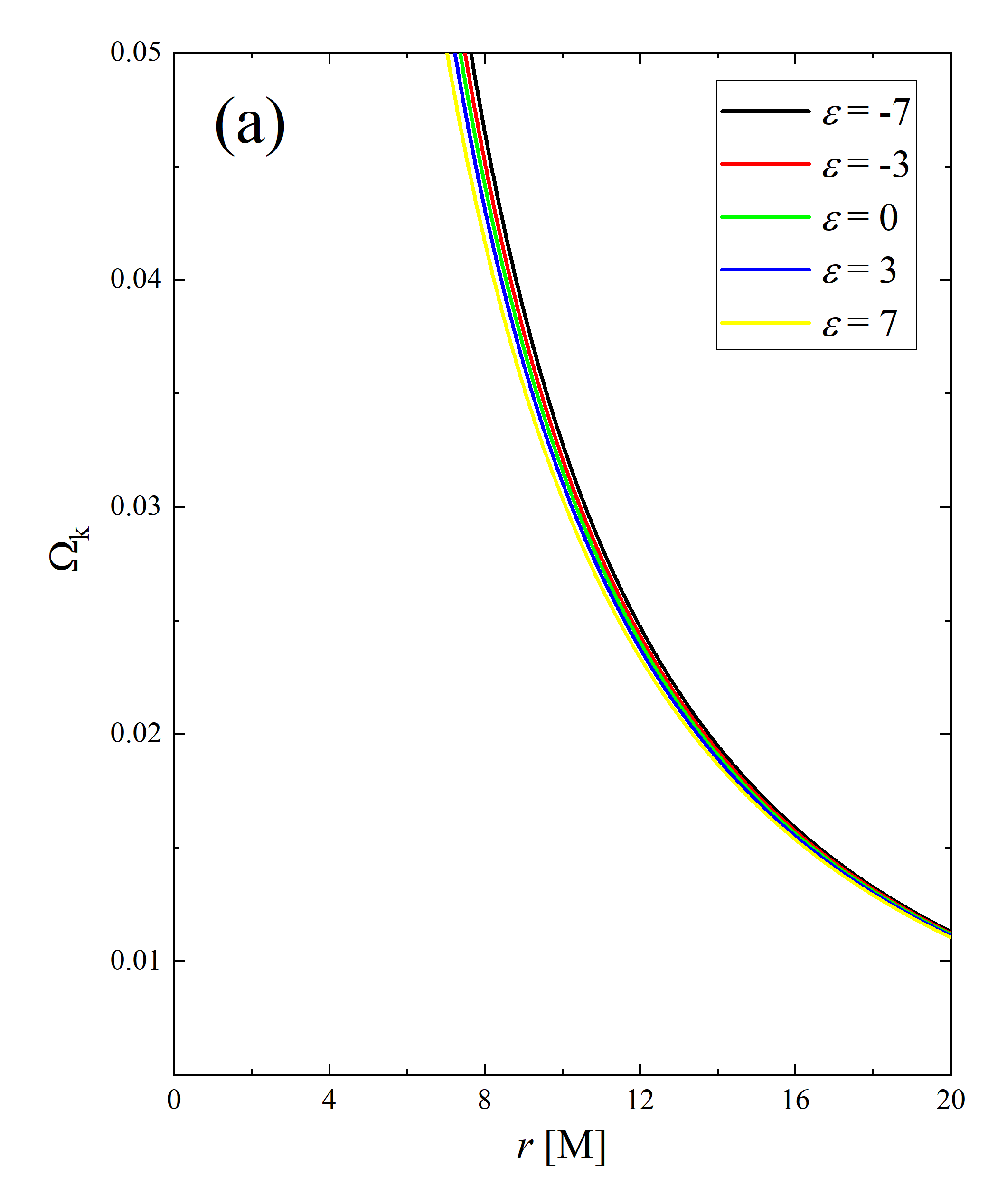}
\includegraphics[width=6cm]{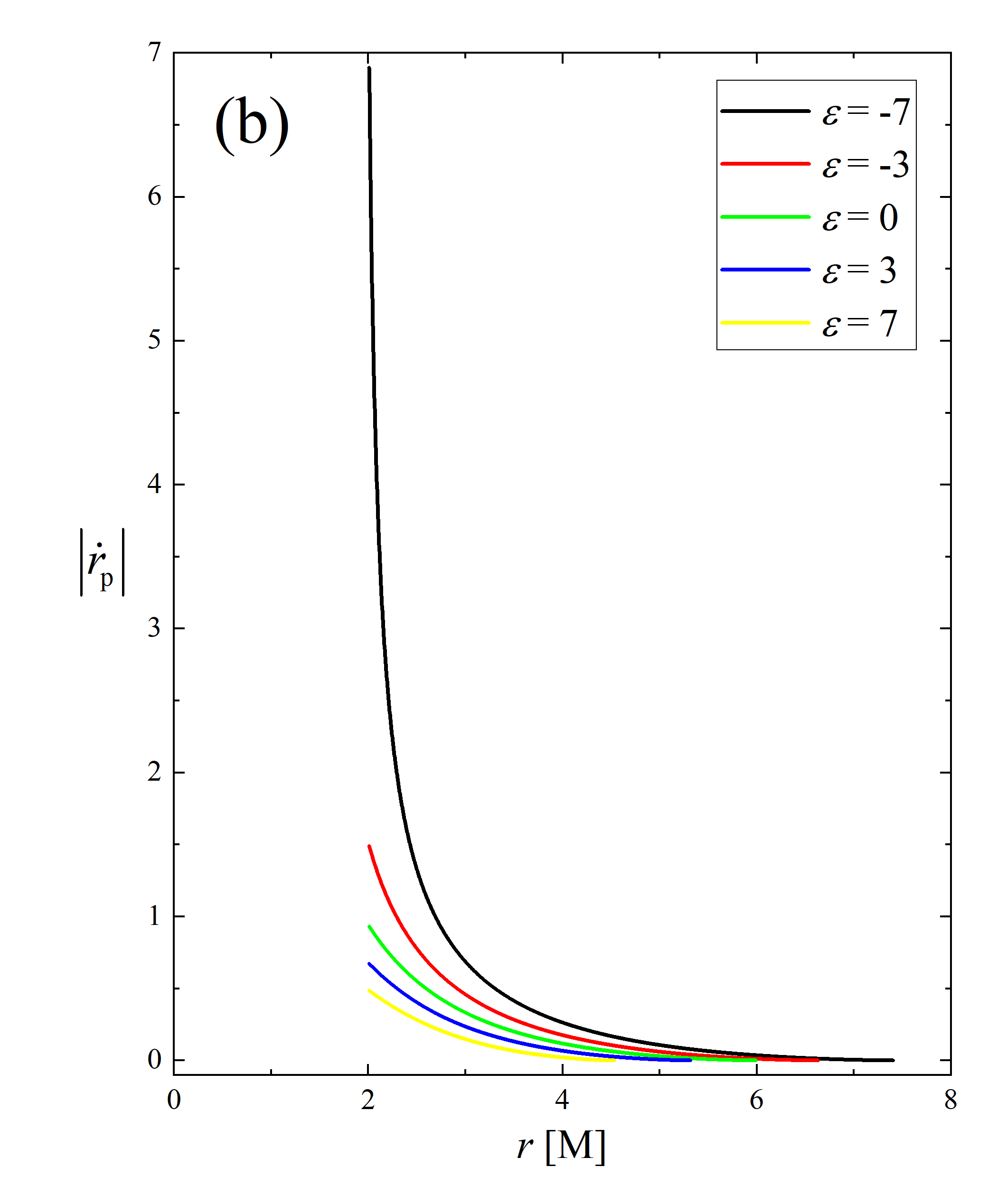}
\caption{Variations in angular velocity $\Omega_{\textrm{k}}$ and the absolute value of radial velocity $|\dot{r}_{\textrm{p}}|$ of the emitting source for different values of the deformation parameter in the Keplerian region (a) and the plunging region (b). Both the values of $\Omega_{\textrm{k}}$ and $|\dot{r}_{\textrm{p}}|$ decrease with the increase of $\varepsilon$, which well explains the dependence of the redshift factor on the deformation parameter demonstrated in figure 3.}}\label{fig4}
\end{figure*}

The impact of observation inclination on the distribution of the redshift factors is quite intuitive: at larger observation inclinations, the line of sight is closer to the disk plane, and the component of the accreting material's velocity along the line of sight increases, thereby enhancing the Doppler effect. However, the mechanism by which the deformation parameter affects the redshift factor is not immediately apparent. We plot figure 4 to elucidate the intrinsic nature of the changes in the redshift factor due to the deformation parameter. The left and right panels respectively show the evolution of the angular velocity $(\Omega_{\textrm{k}} = \dot{\varphi}_{\textrm{k}}/\dot{t}_{\textrm{k}})$ in the Keplerian region and the absolute value of radial velocity $|\dot{r}_{\textrm{p}}|$ in the plunging region of the accreting particles as a function of radius for different $\varepsilon$. It can be confirmed that the values of $\Omega_{\textrm{k}}$ and $|\dot{r}_{\textrm{p}}|$ of the oblate deformed Schwarzschild black holes $(\varepsilon < 0)$ are greater than those of the prolate deformed Schwarzschild black holes $(\varepsilon > 0)$. This implies that with increasing $\varepsilon$, the angular (radial) velocity of particles in the Keplerian (plunging) region decrease, leading to a reduction in the Doppler effect on the light rays.

\begin{figure*}%[tbph]
\center{
\includegraphics[width=2.9cm]{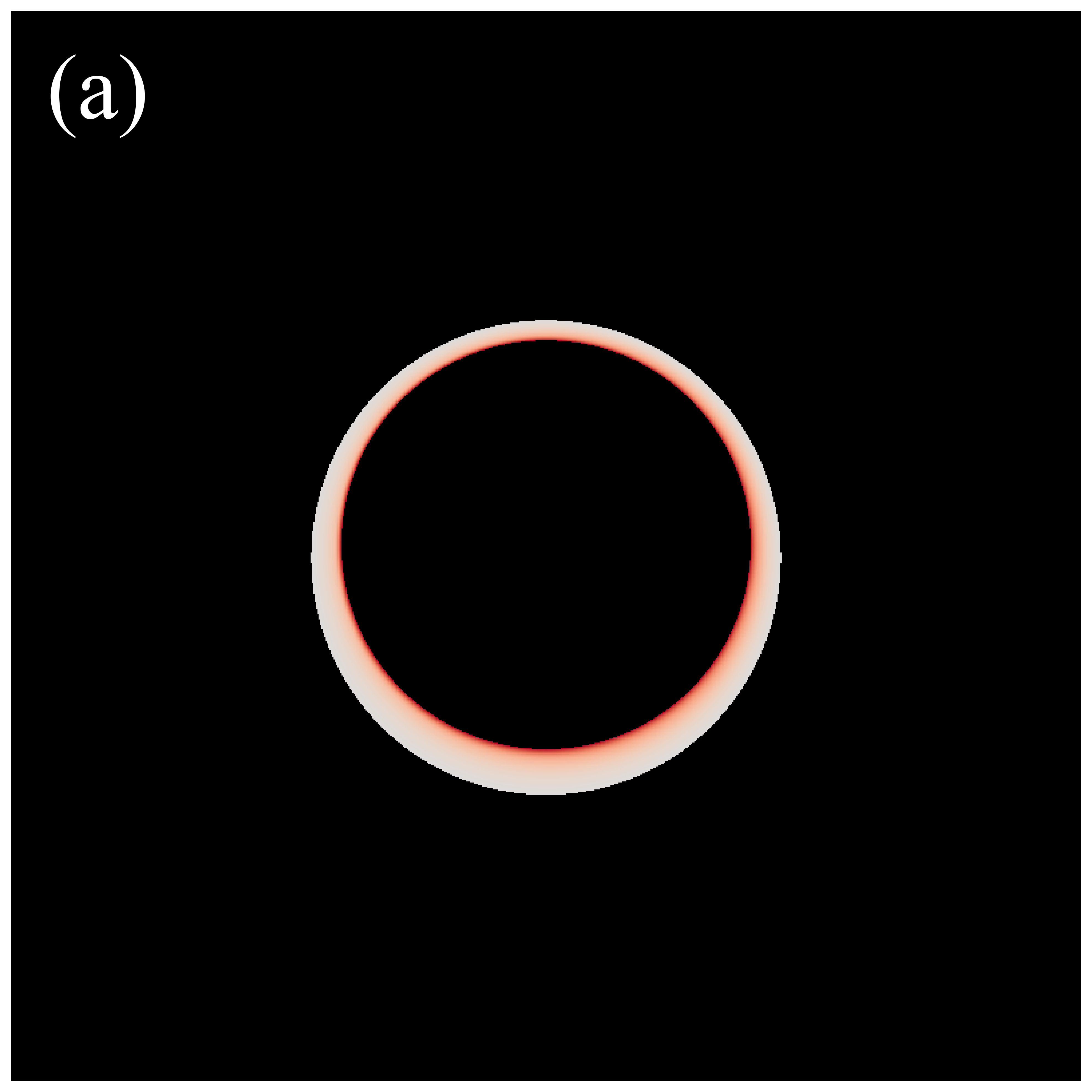}
\includegraphics[width=2.9cm]{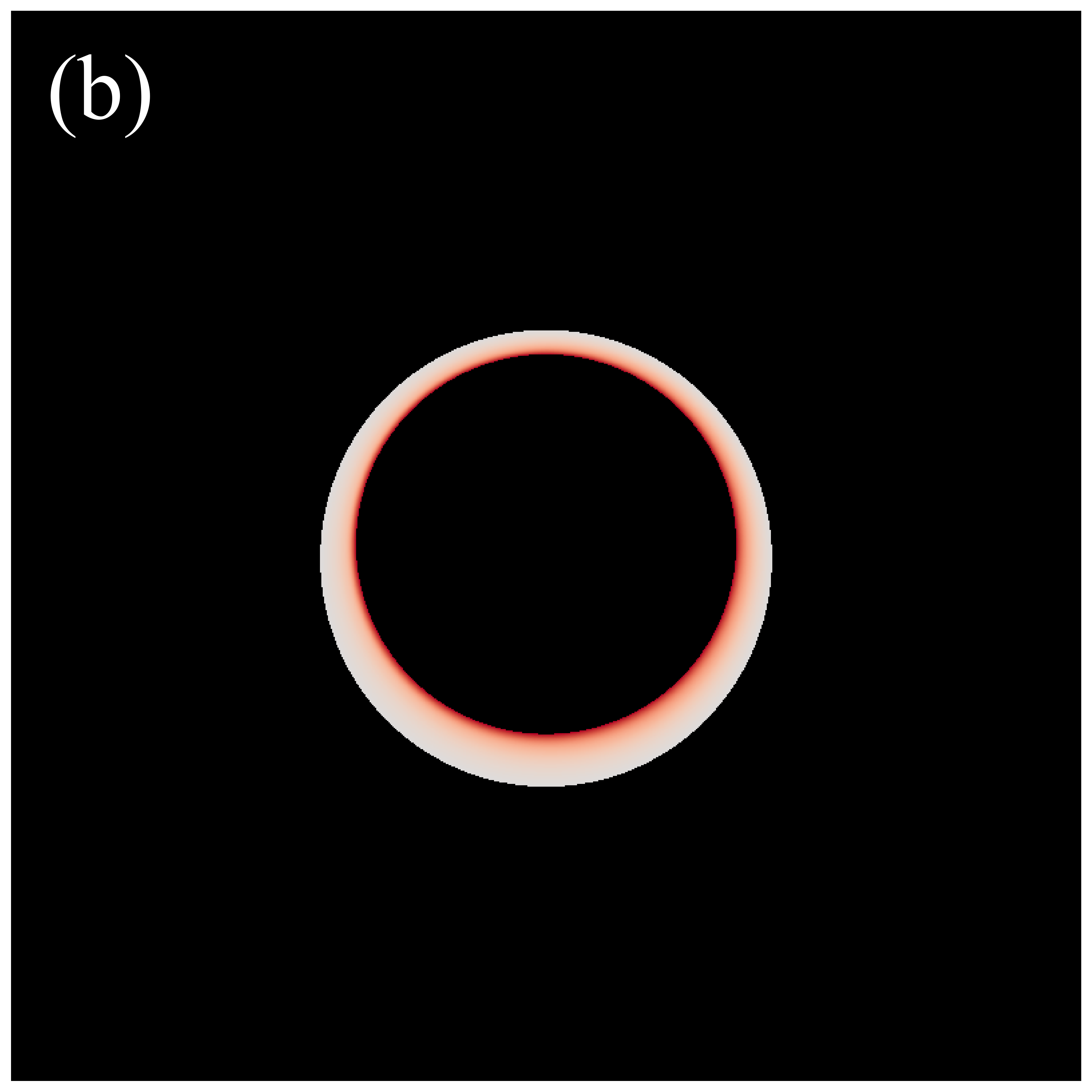}
\includegraphics[width=2.9cm]{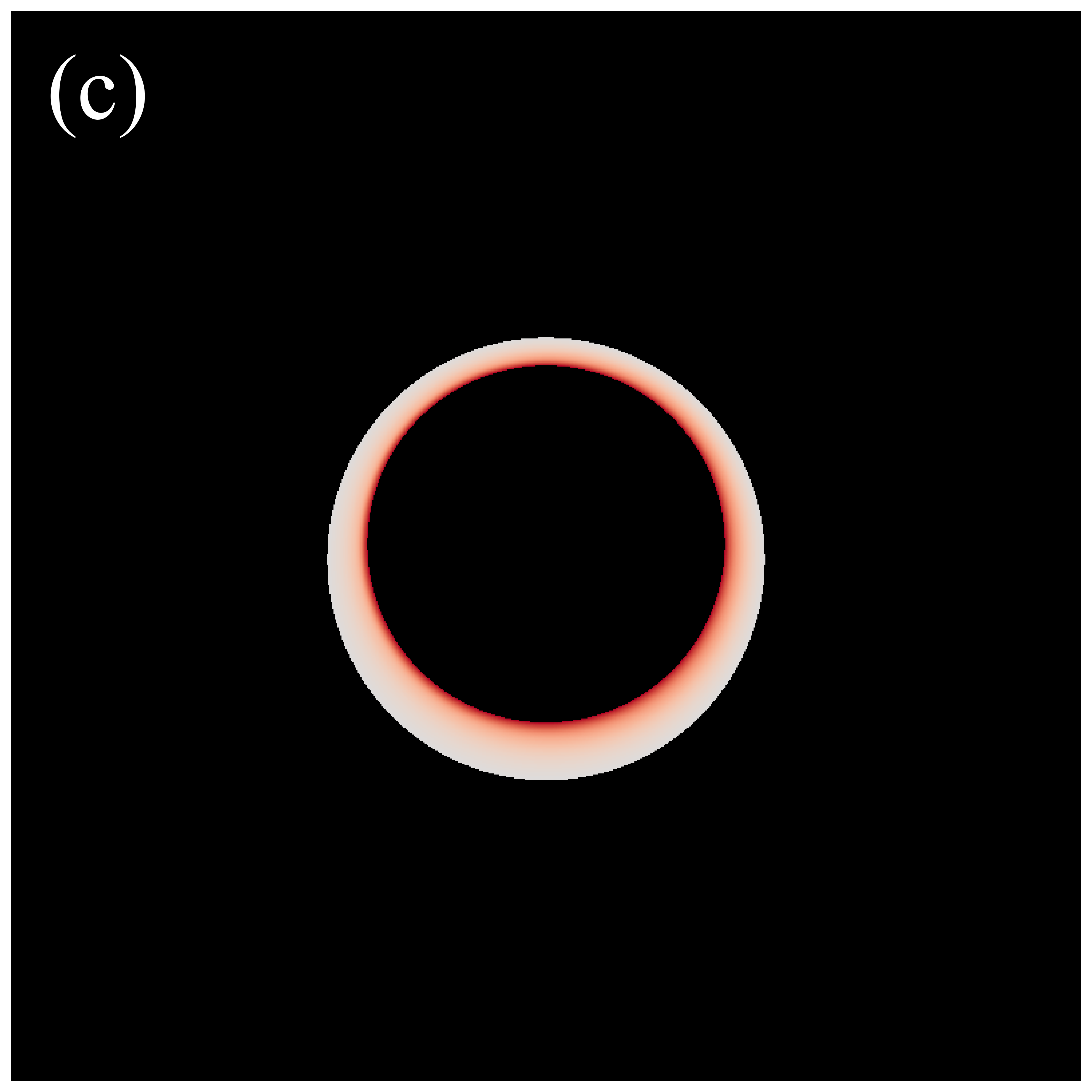}
\includegraphics[width=2.9cm]{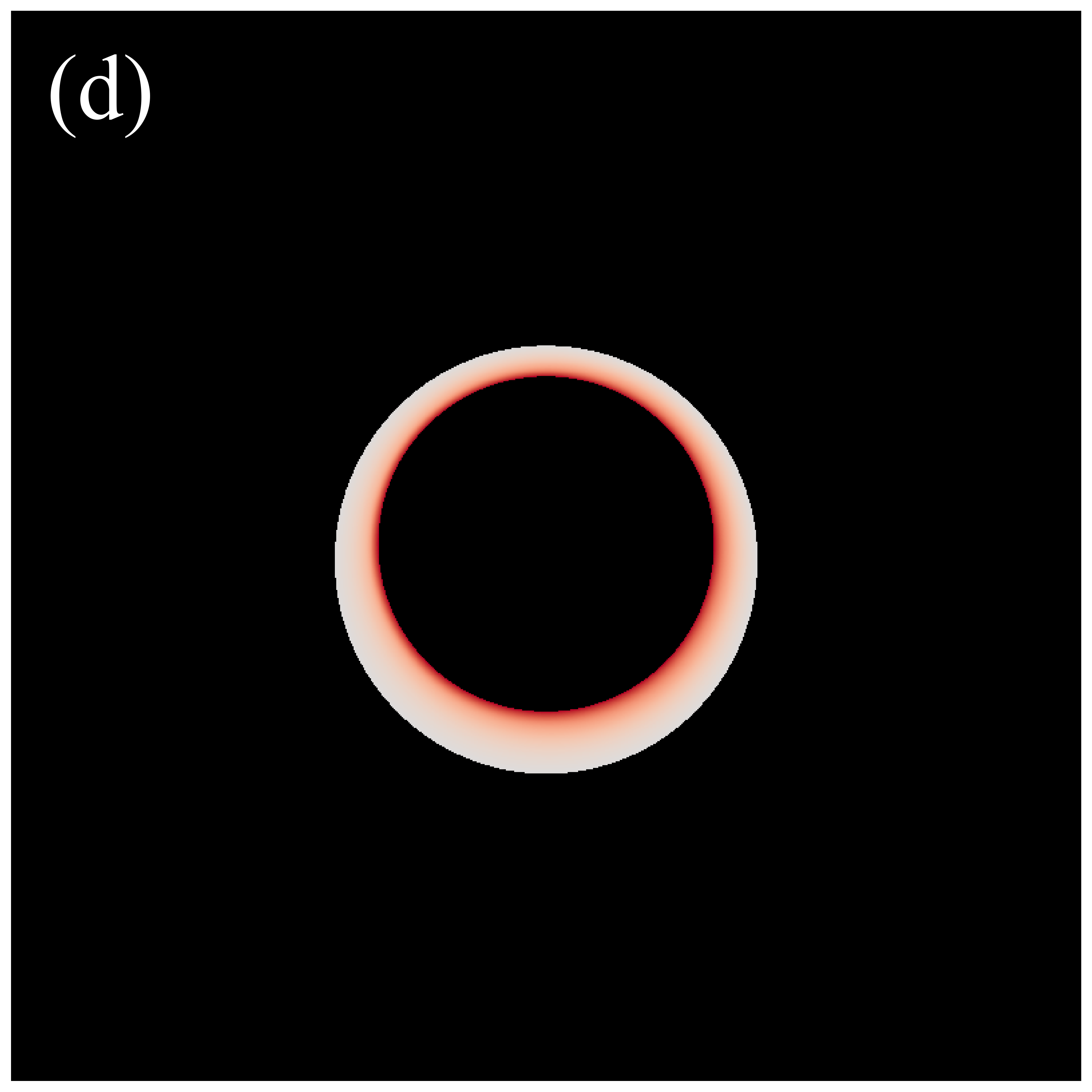}
\includegraphics[width=2.9cm]{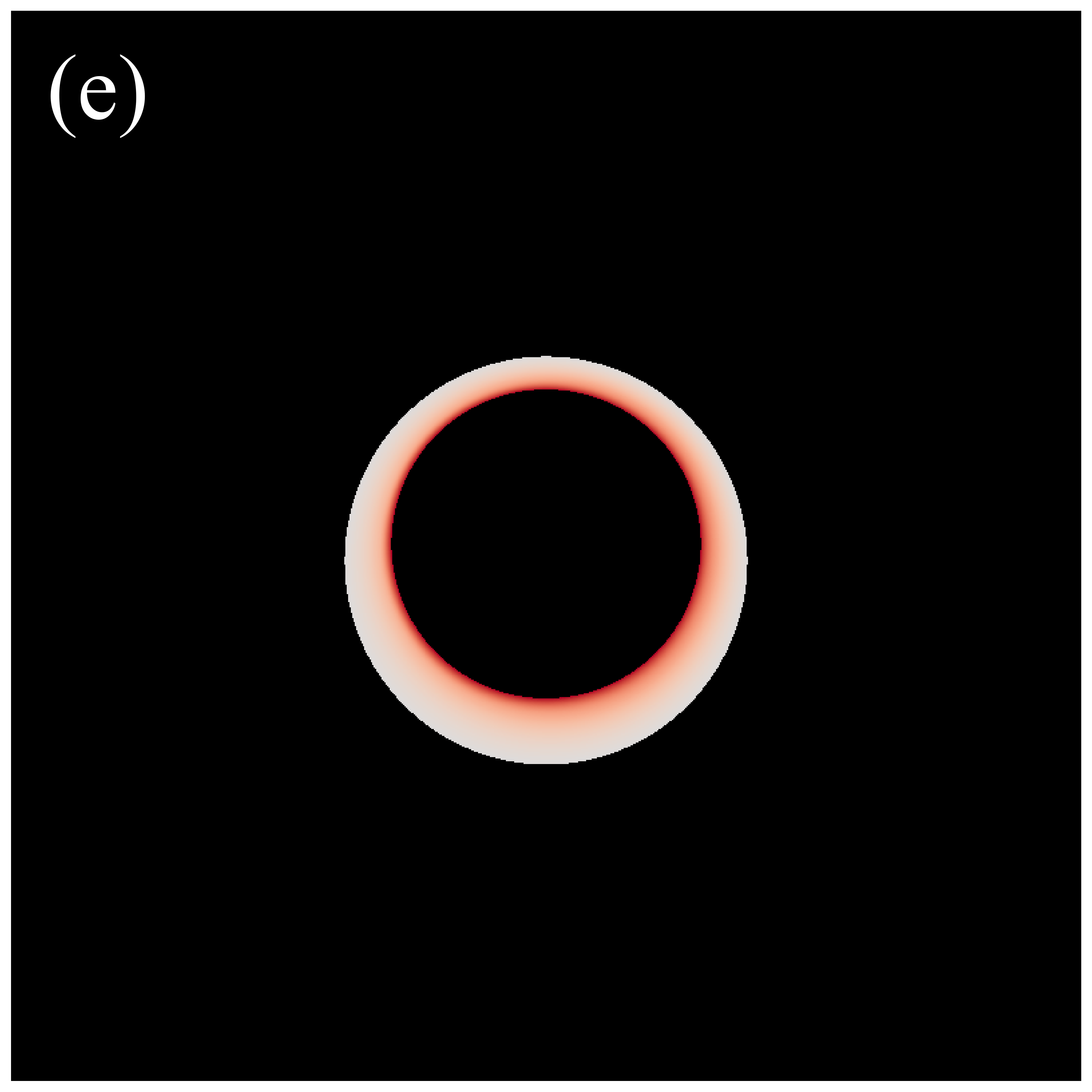}
\includegraphics[width=2.9cm]{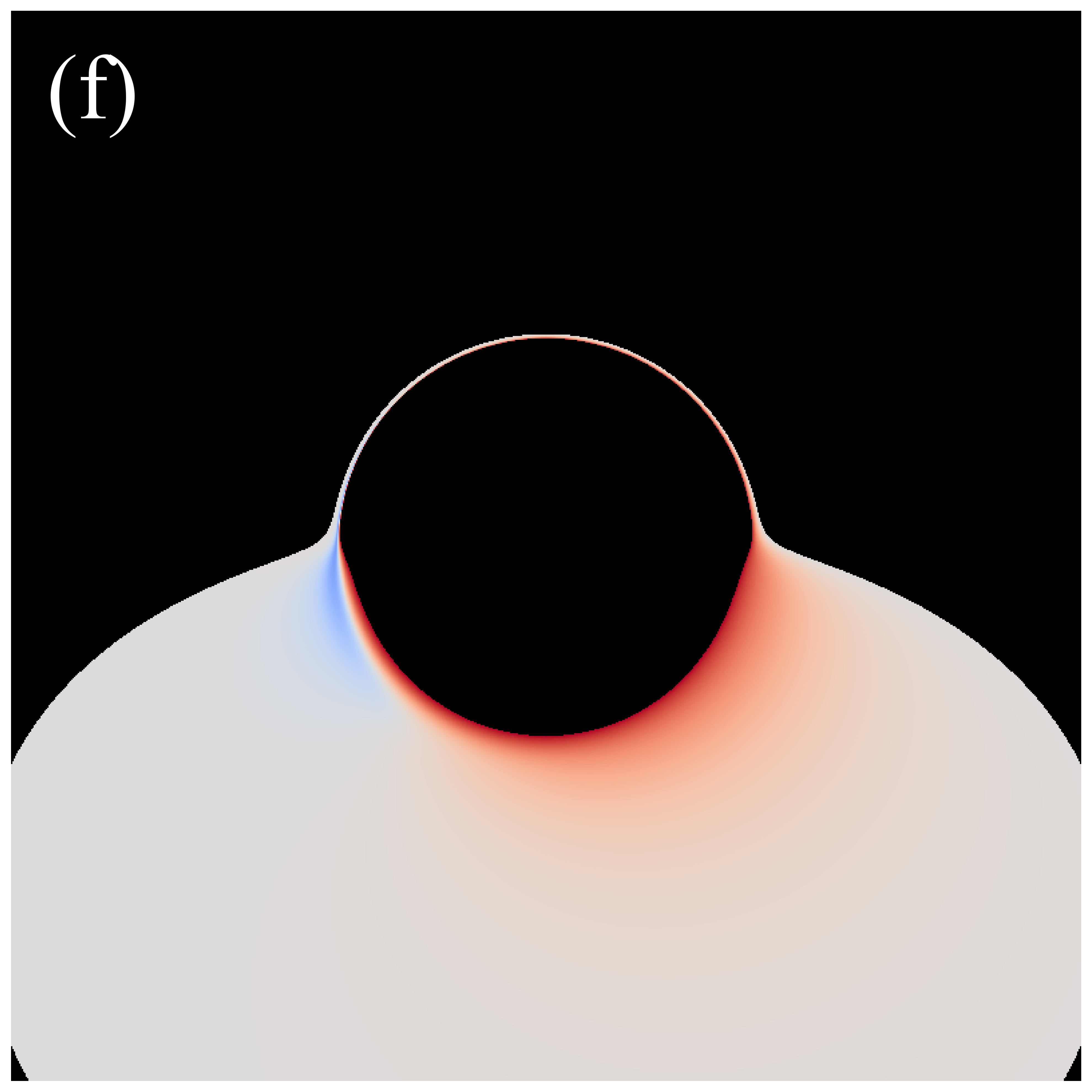}
\includegraphics[width=2.9cm]{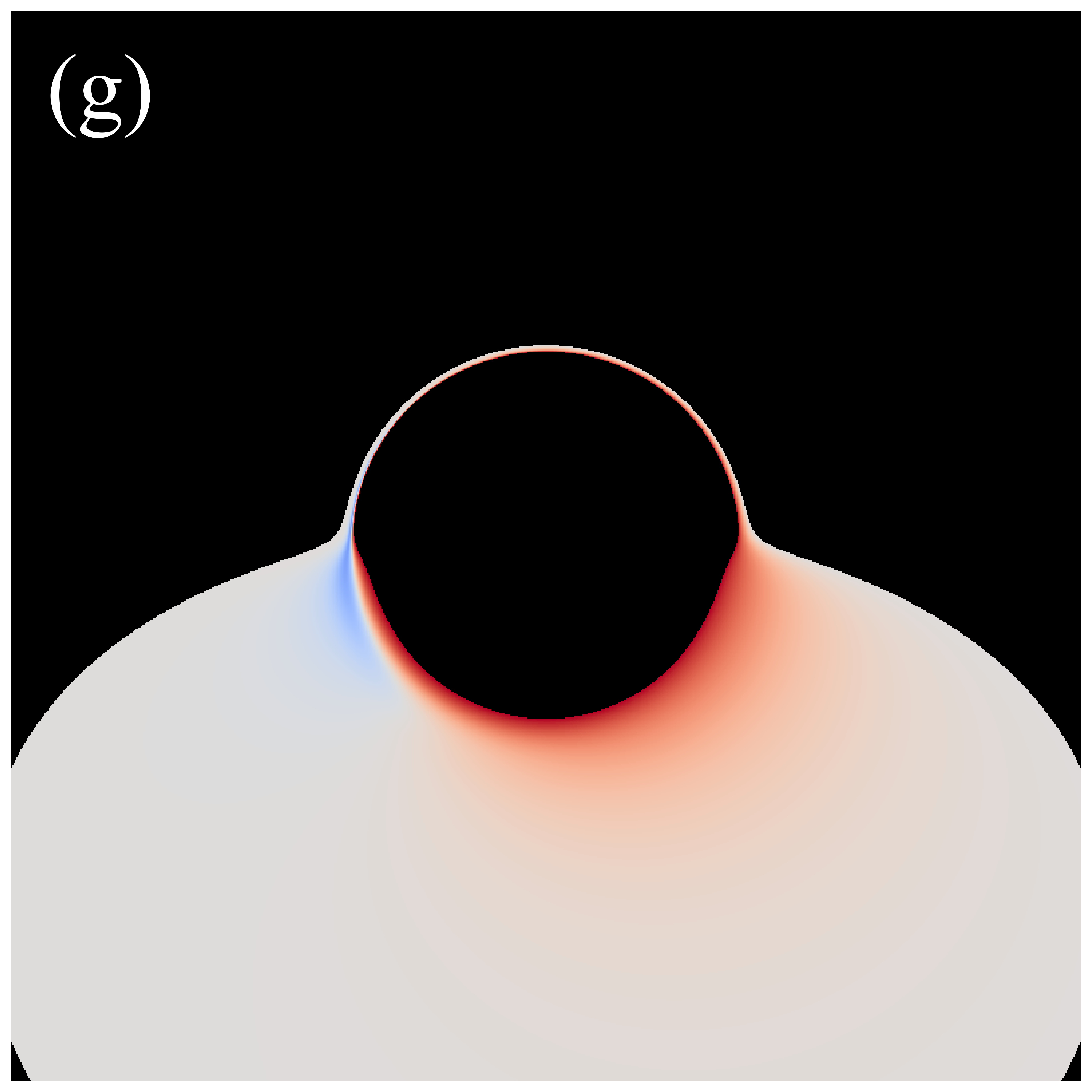}
\includegraphics[width=2.9cm]{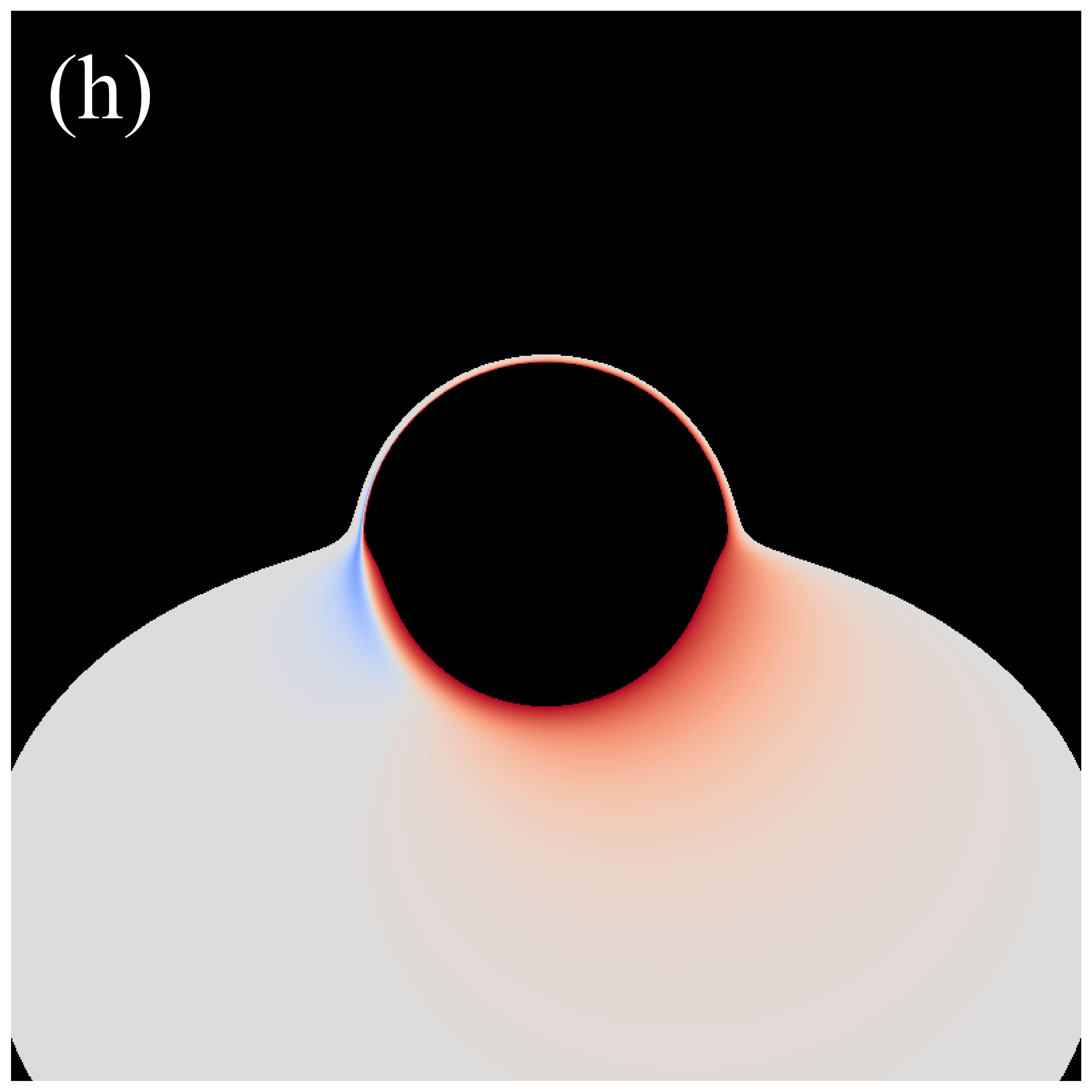}
\includegraphics[width=2.9cm]{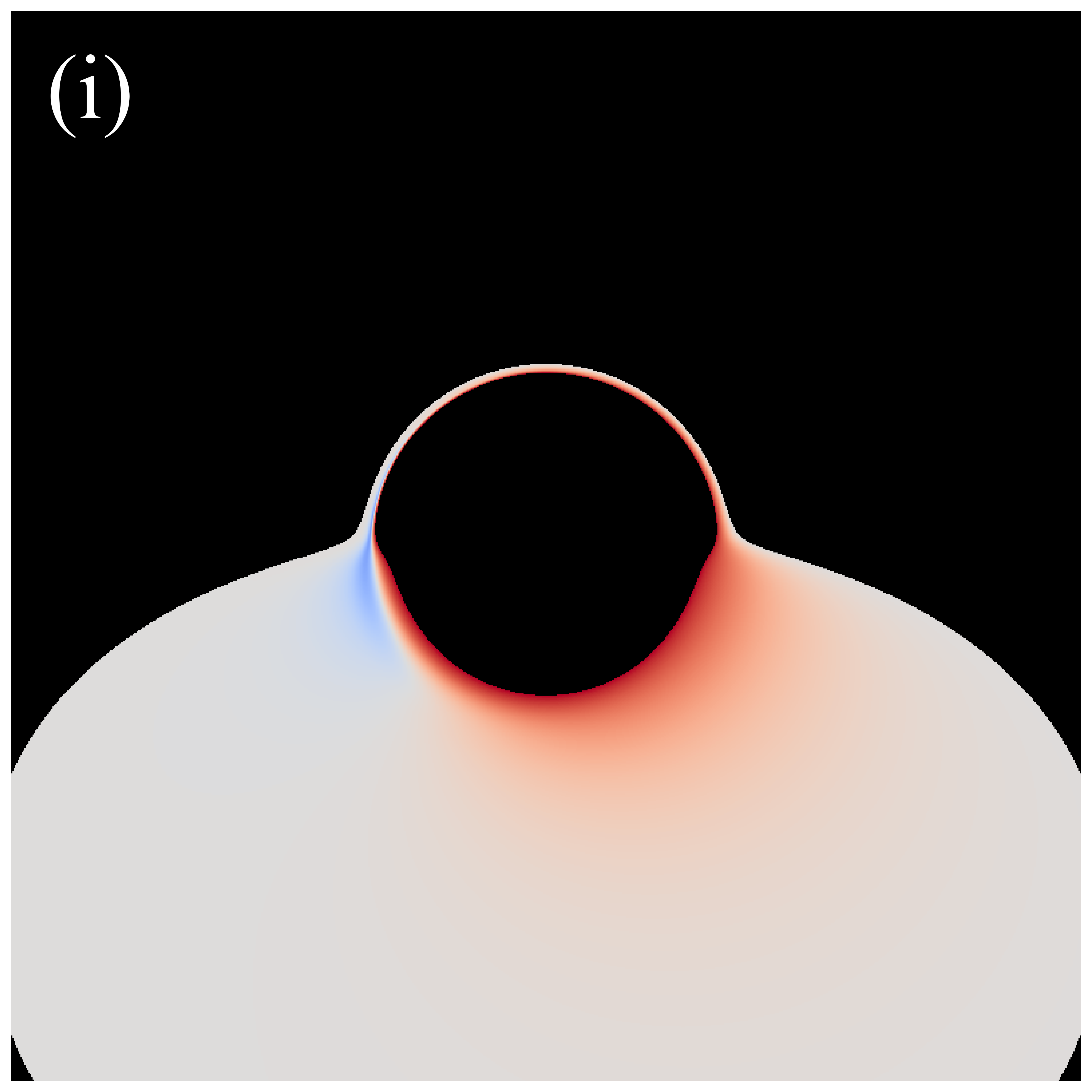}
\includegraphics[width=2.9cm]{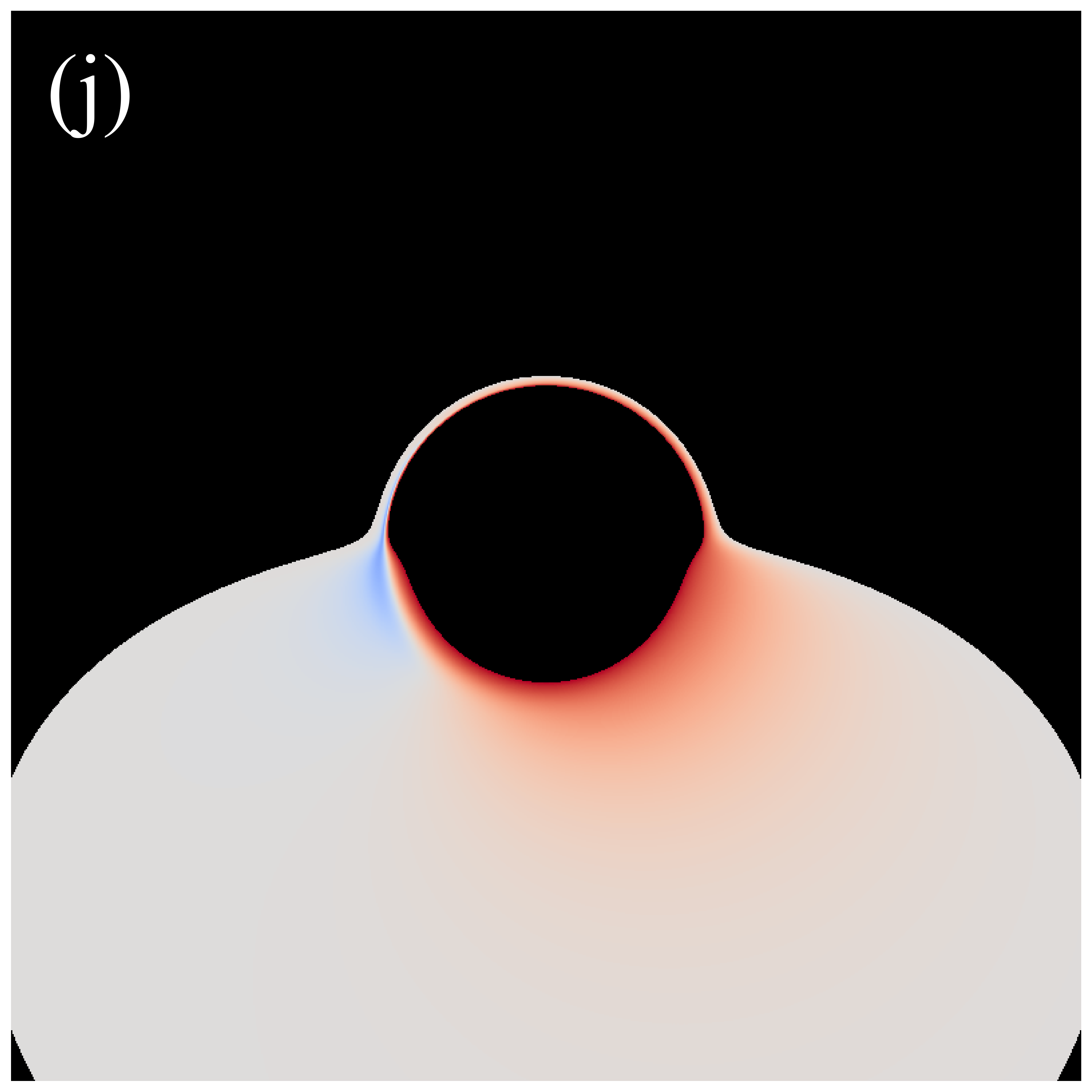}
\caption{Similar to figure 3, but for the secondary images.}}\label{fig5}
\end{figure*}
We also investigate the distribution of the redshift factors in the secondary images, as shown in figure 5. Similar to the results in figure 3, increasing the observation angle introduces blueshift. However, in this scenario, the influence of the deformation parameter on the redshift factor is difficult to discern. Simultaneously, we find that the central dark region in the secondary image is larger than that in the direct image. This is because the deflection angle of the light rays forming the secondary image is larger, making the image closer to the critical curve \cite{Gralla et al. (2019)}.
\subsection{Deformed Schwarzschild black hole images with and without considering projection effect}
To make the simulated images of black holes observationally meaningful, it is crucial to set the radiation of the accretion disk properly. Theoretically, the radiation is related to the pressure, temperature, density, and spatial distribution of the plasma, and the relevant calculations involve general relativistic magnetohydrodynamic (GRMHD) simulations and integrating the radiative transfer equation. Undoubtedly, this is an extremely complex and tedious process. Interestingly, in recent years, some authors have derived analytical expressions for the radiation of accretion disks by fitting time-averaged images from GRMHD simulations \cite{Chael et al. (2021),Gralla et al. (2020)}. These profiles not only serve as a powerful tool for qualitatively revealing black hole observational features but also significantly simplify the computational process.
\begin{figure*}%[tbph]
\center{
\includegraphics[width=3.5cm]{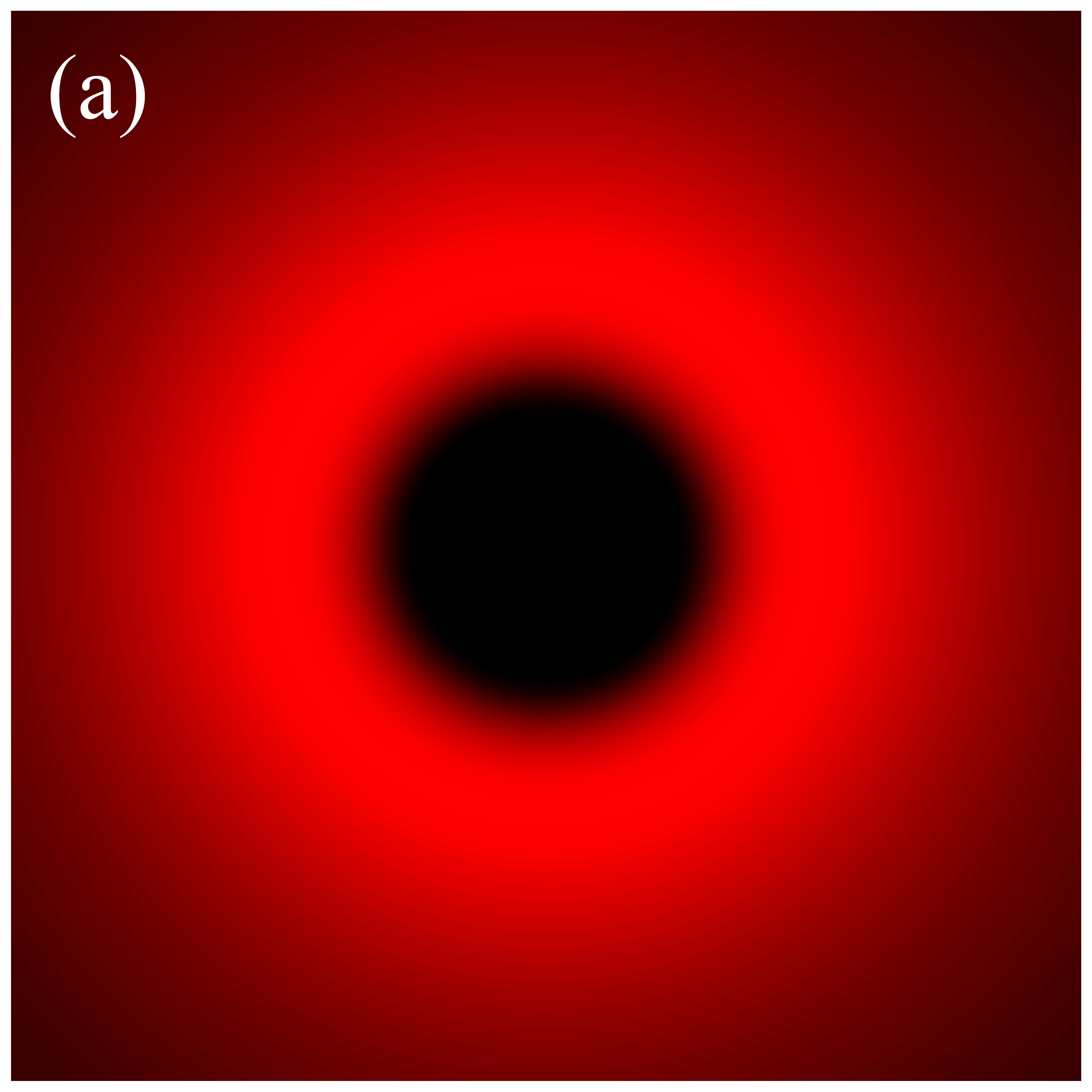}
\includegraphics[width=3.5cm]{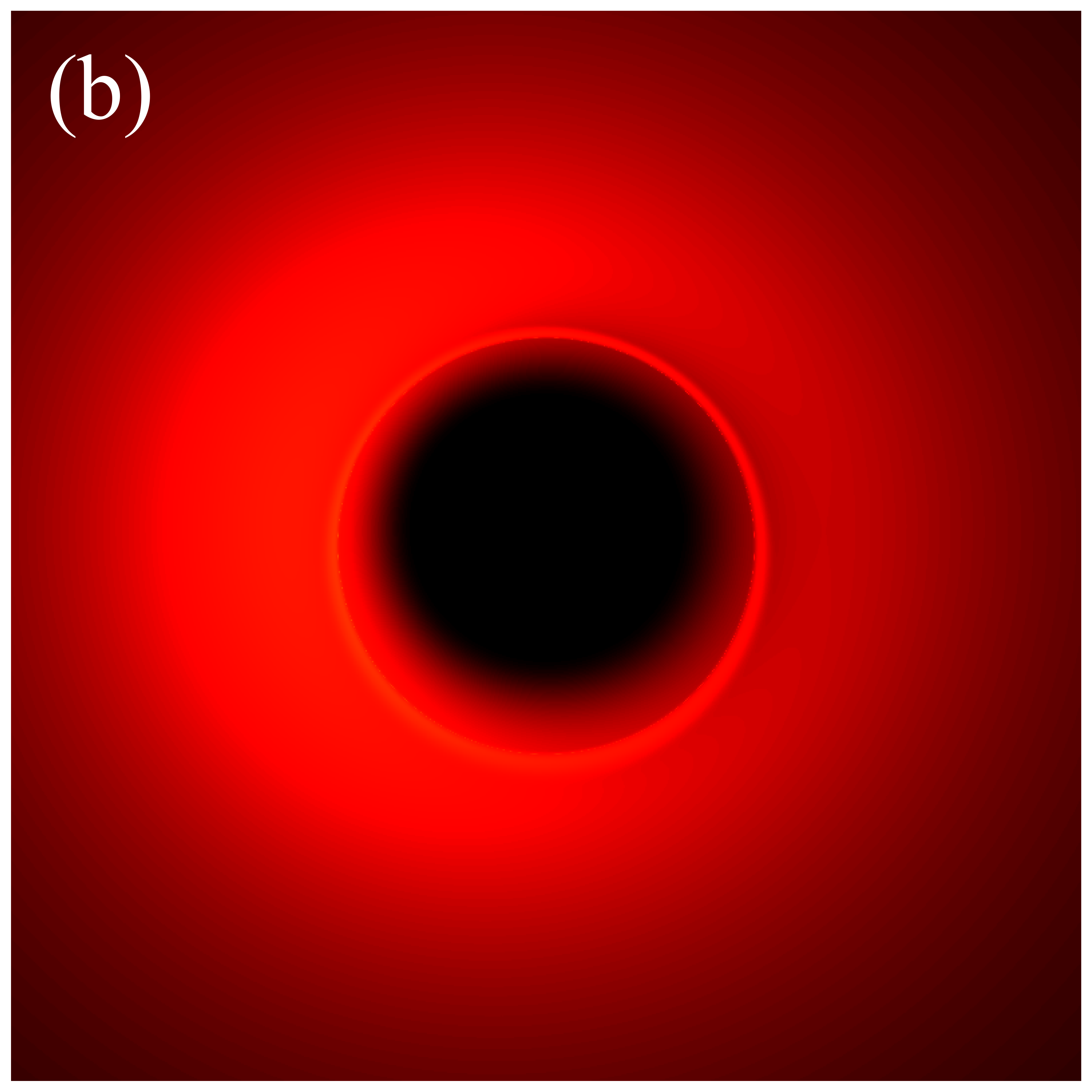}
\includegraphics[width=3.5cm]{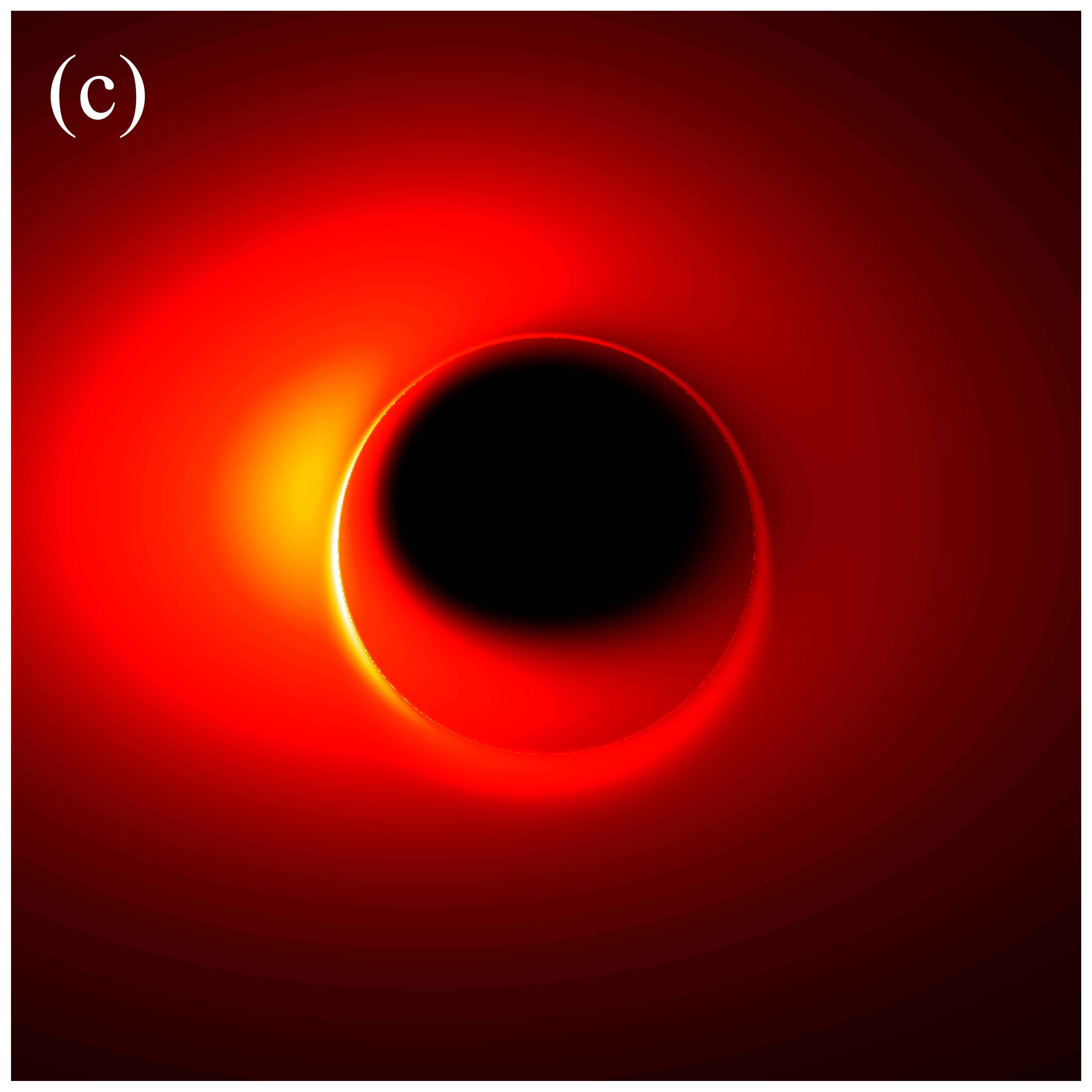}
\includegraphics[width=3.5cm]{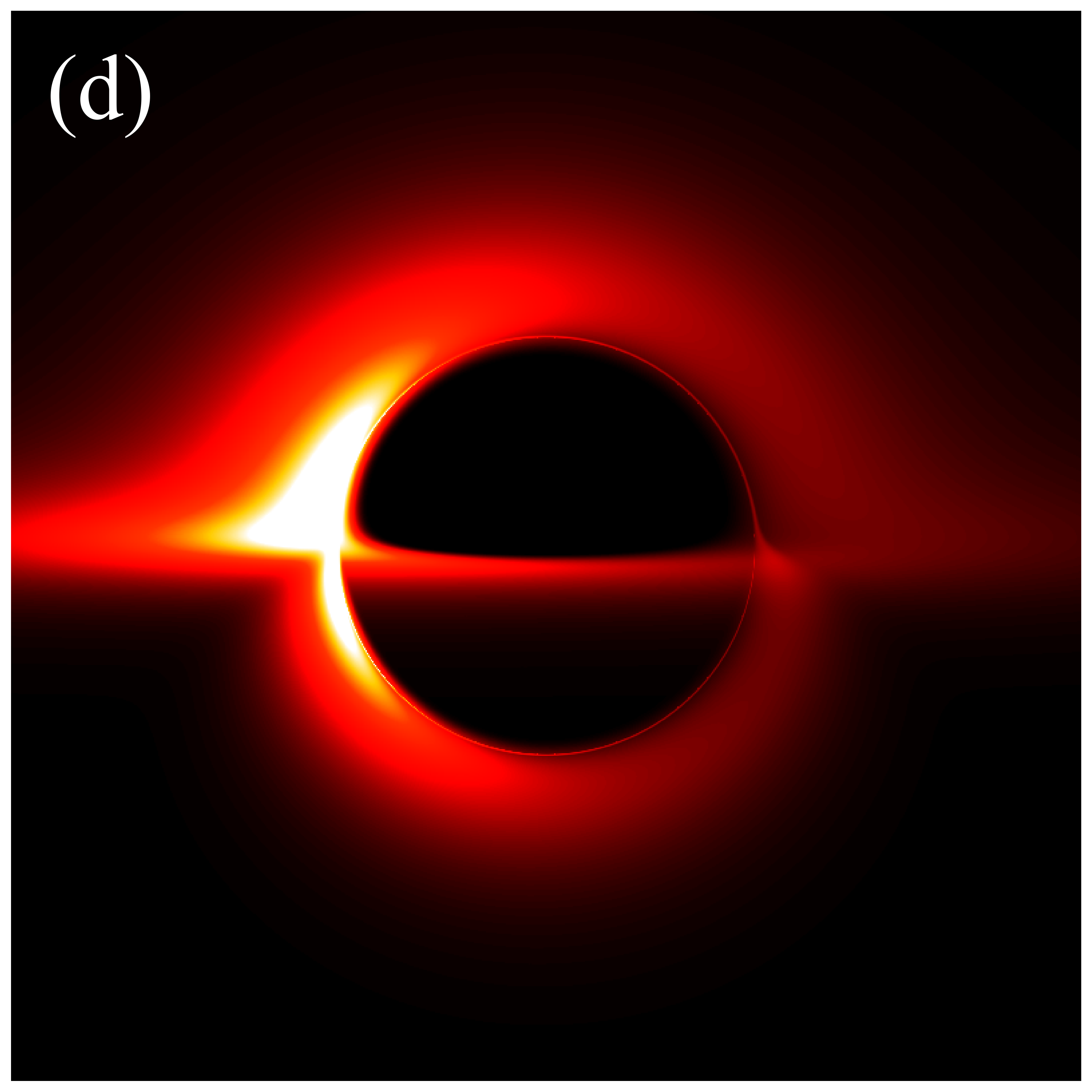}
\includegraphics[width=3.5cm]{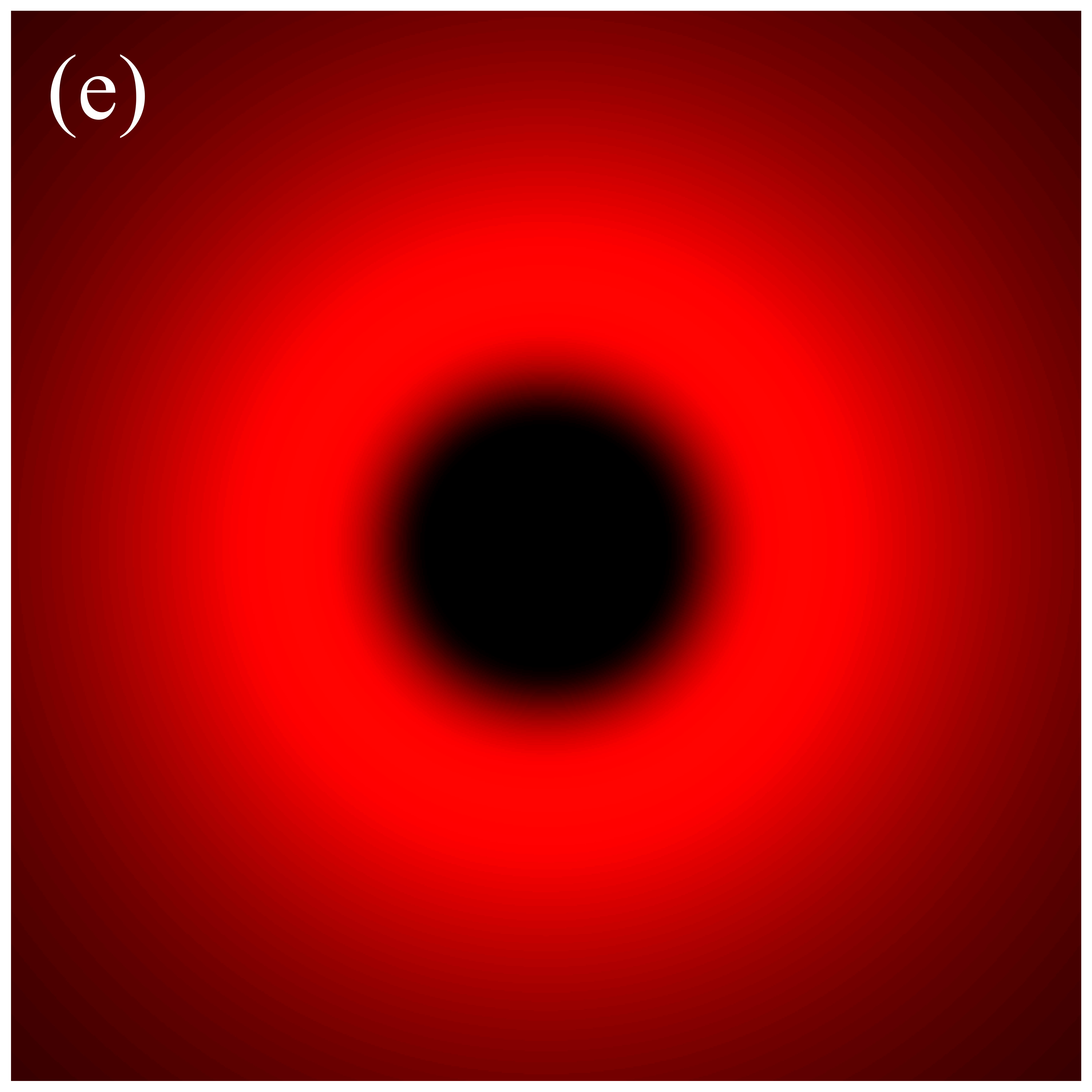}
\includegraphics[width=3.5cm]{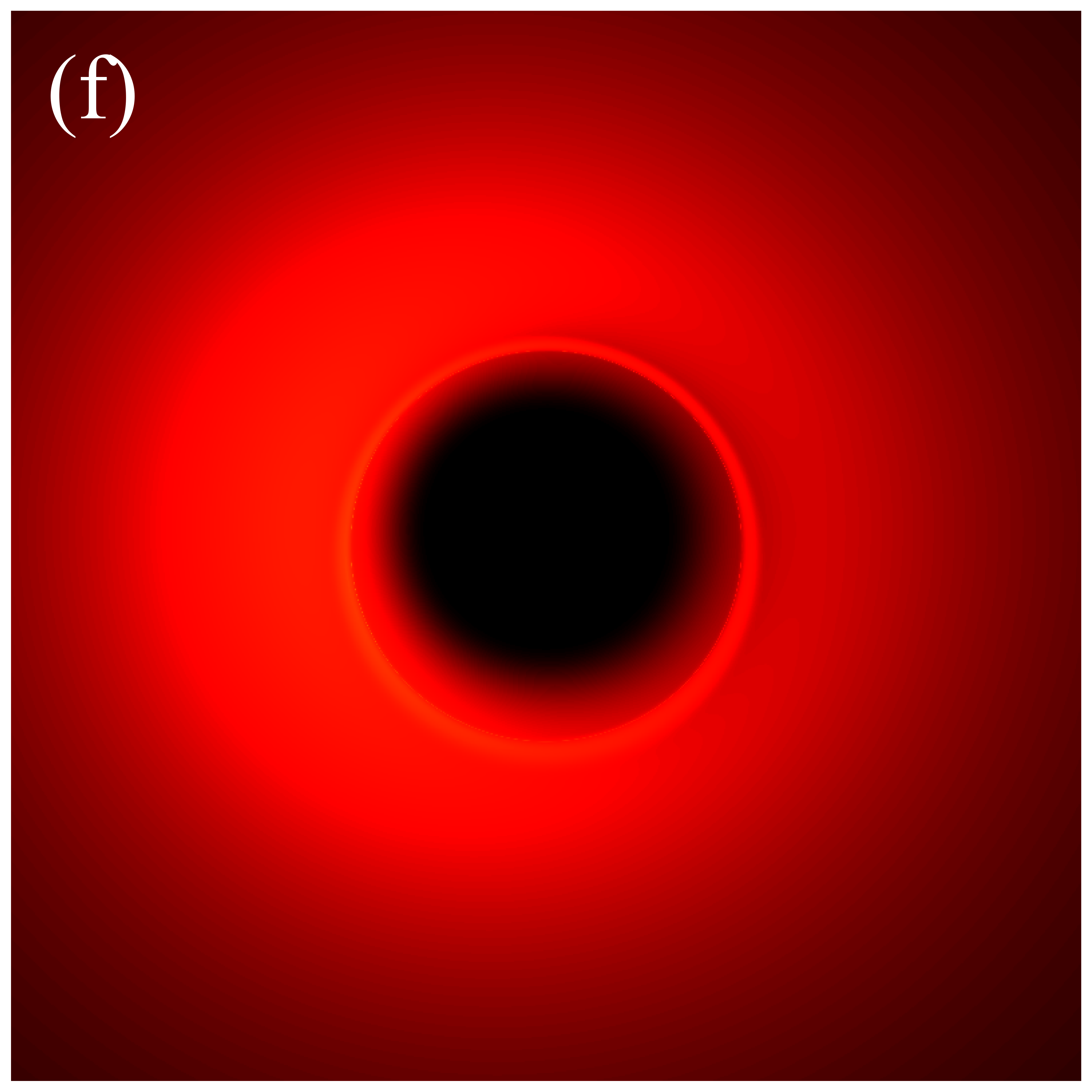}
\includegraphics[width=3.5cm]{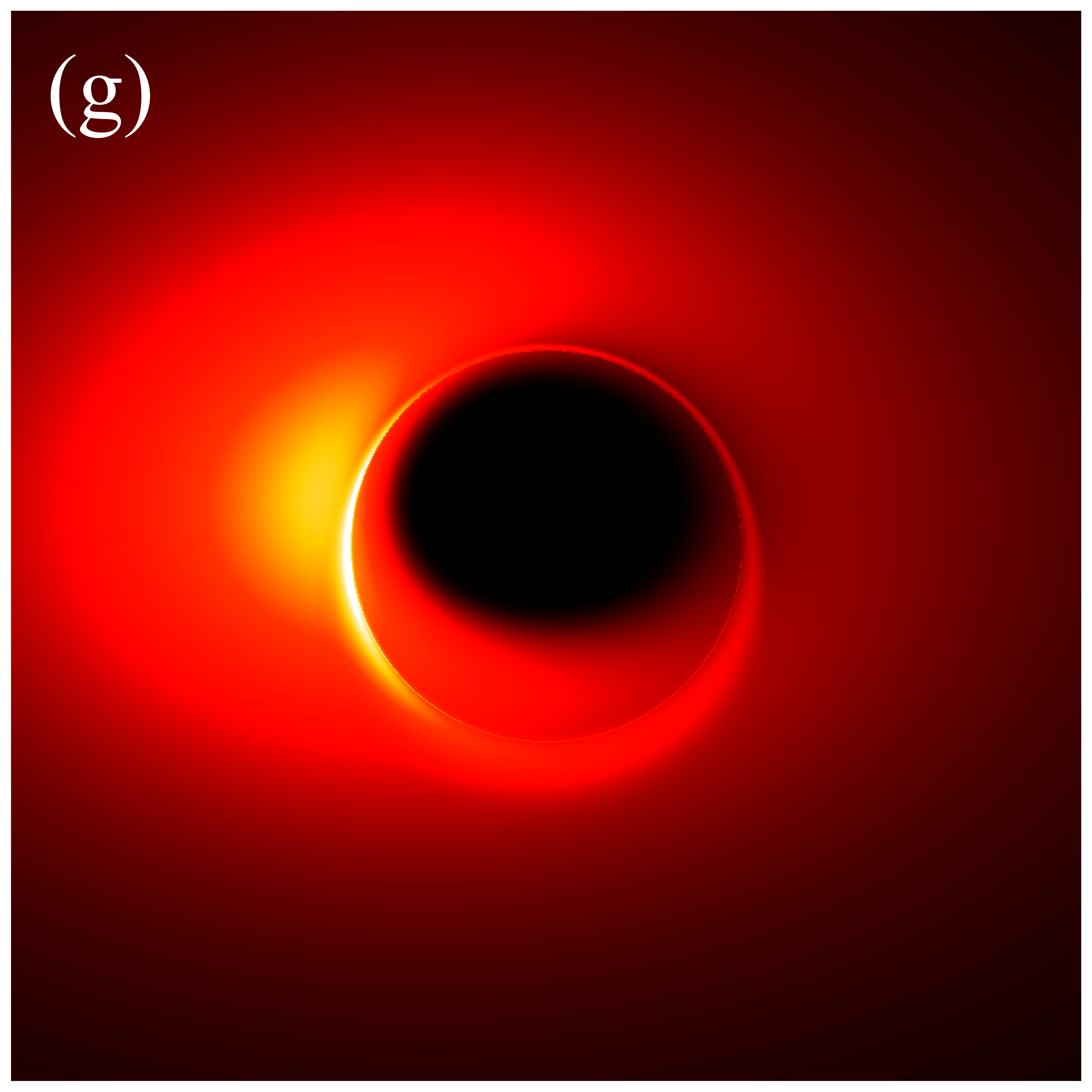}
\includegraphics[width=3.5cm]{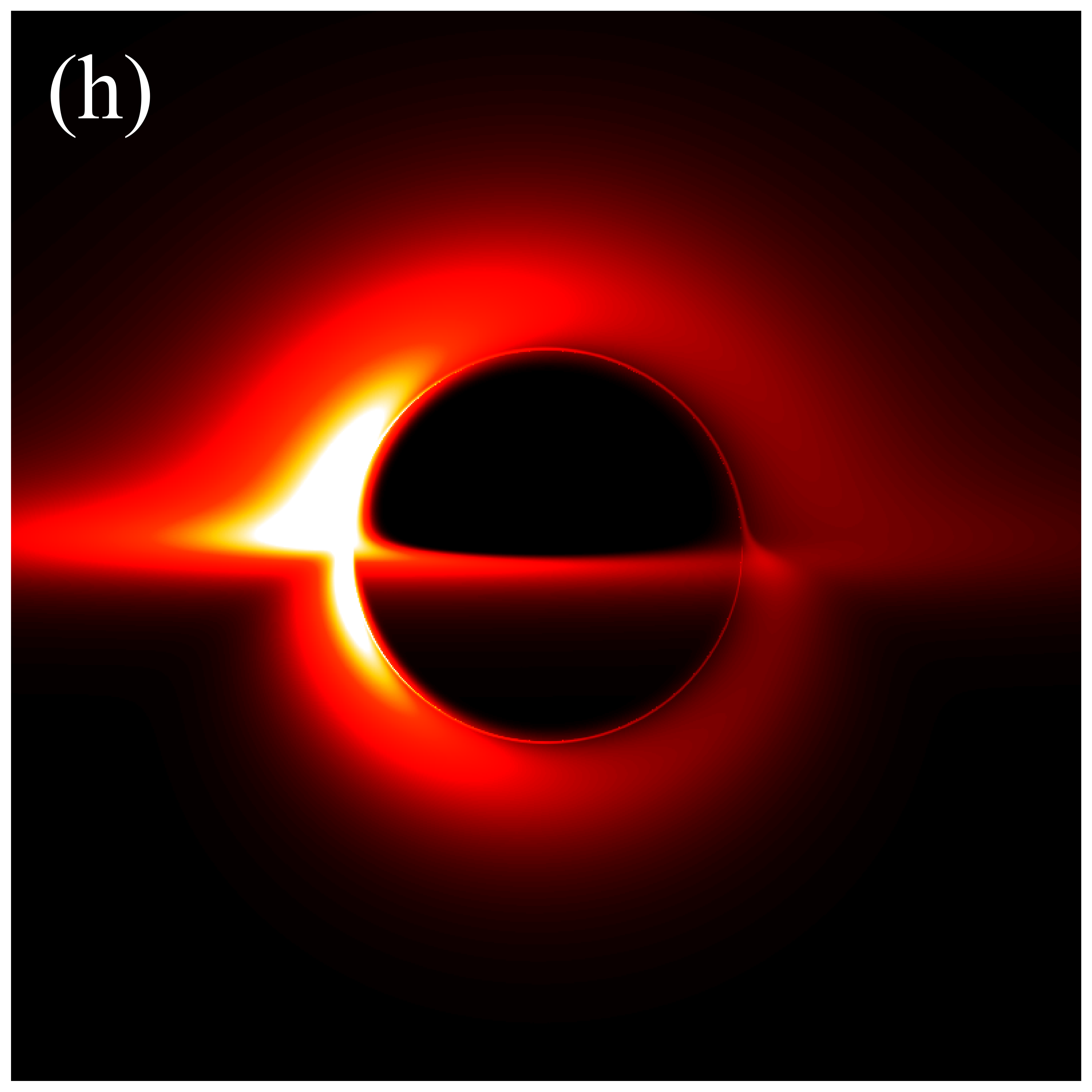}
\includegraphics[width=3.5cm]{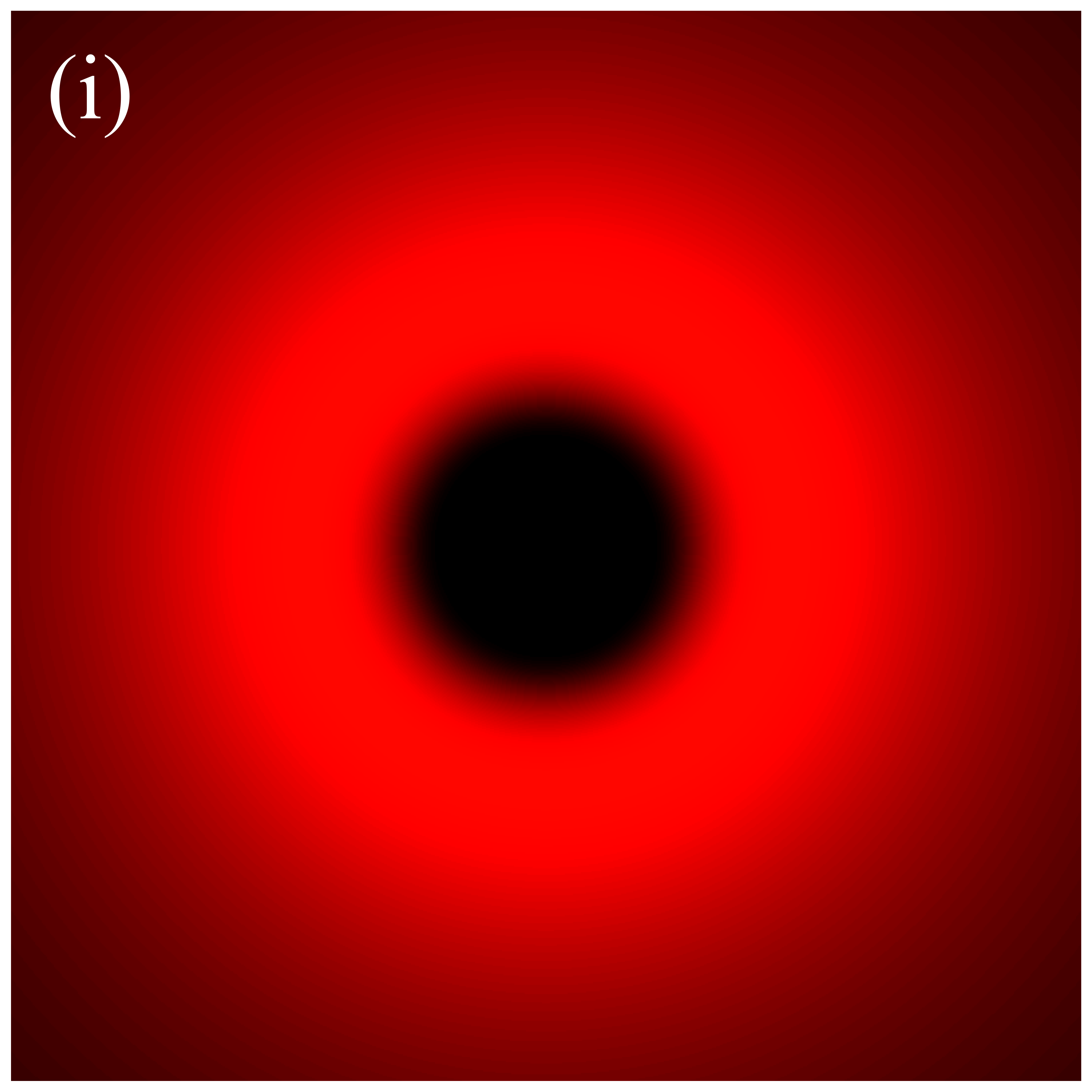}
\includegraphics[width=3.5cm]{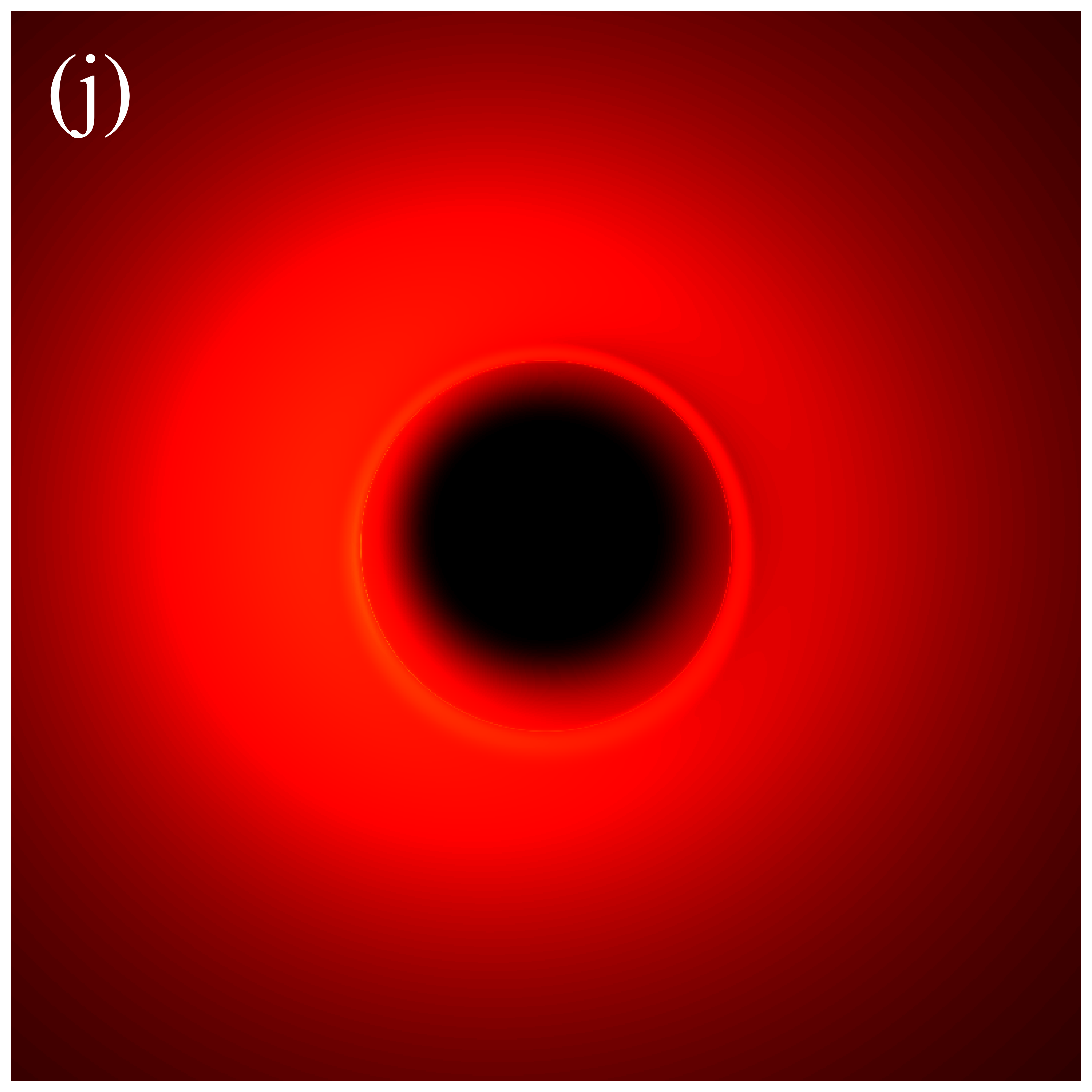}
\includegraphics[width=3.5cm]{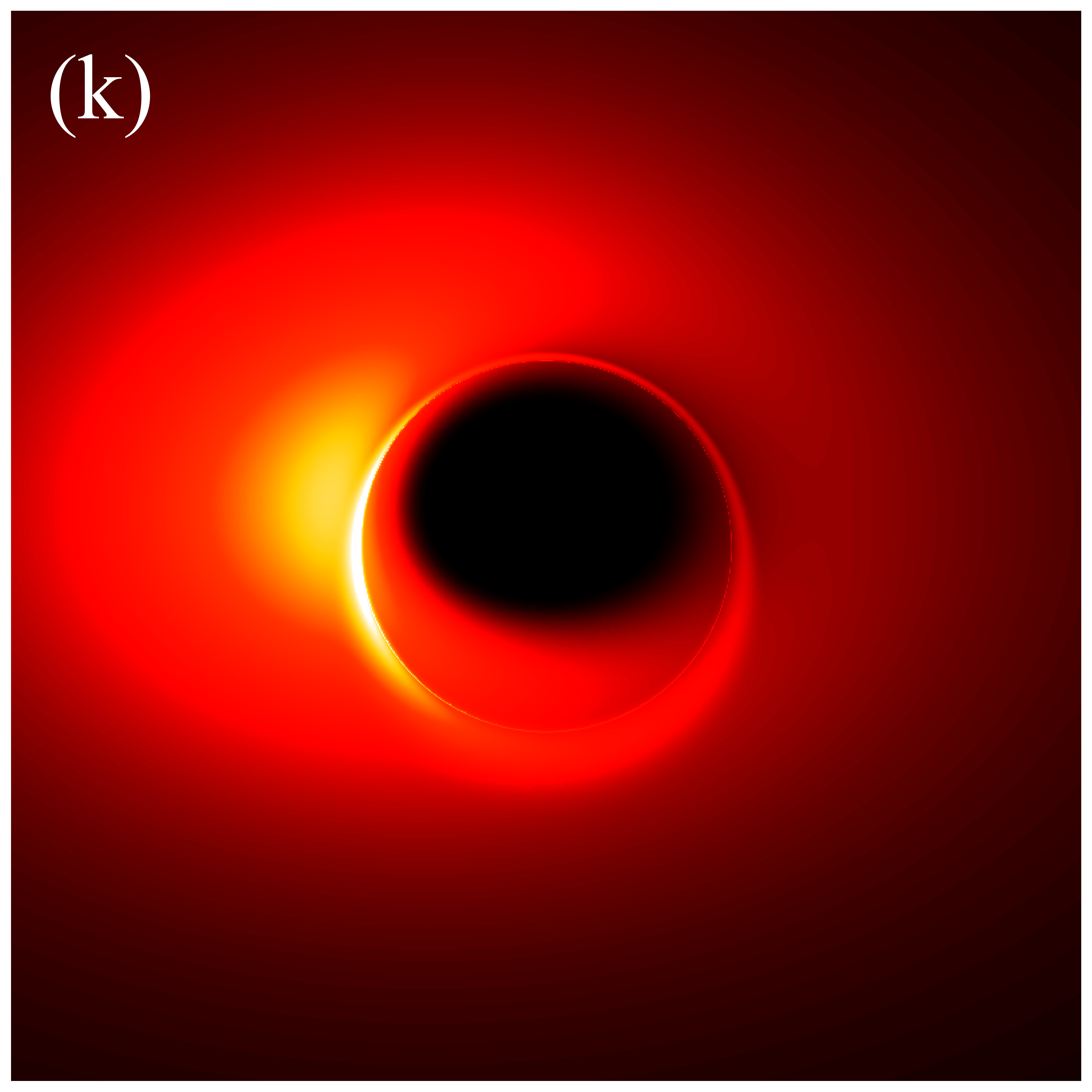}
\includegraphics[width=3.5cm]{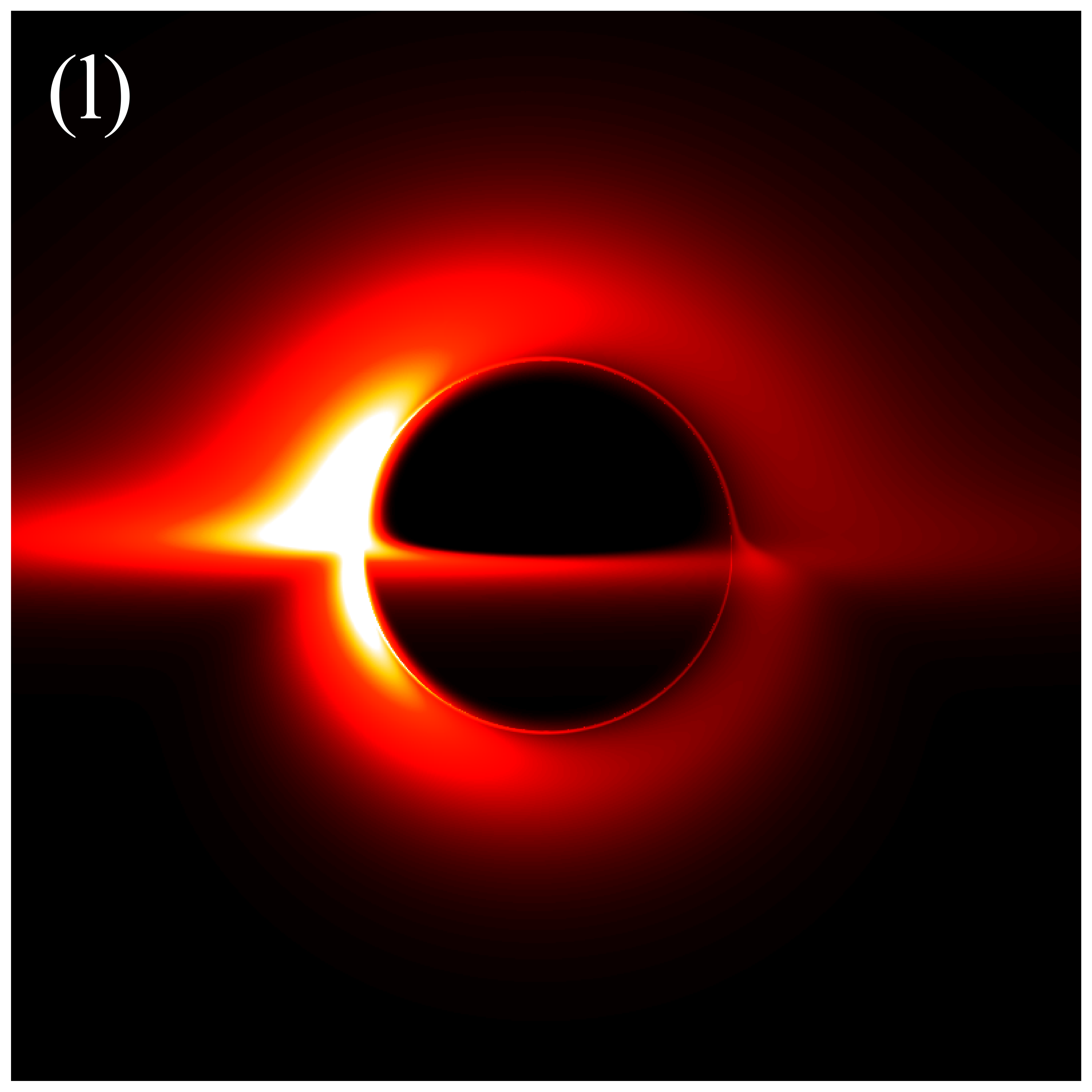}
\includegraphics[width=3.5cm]{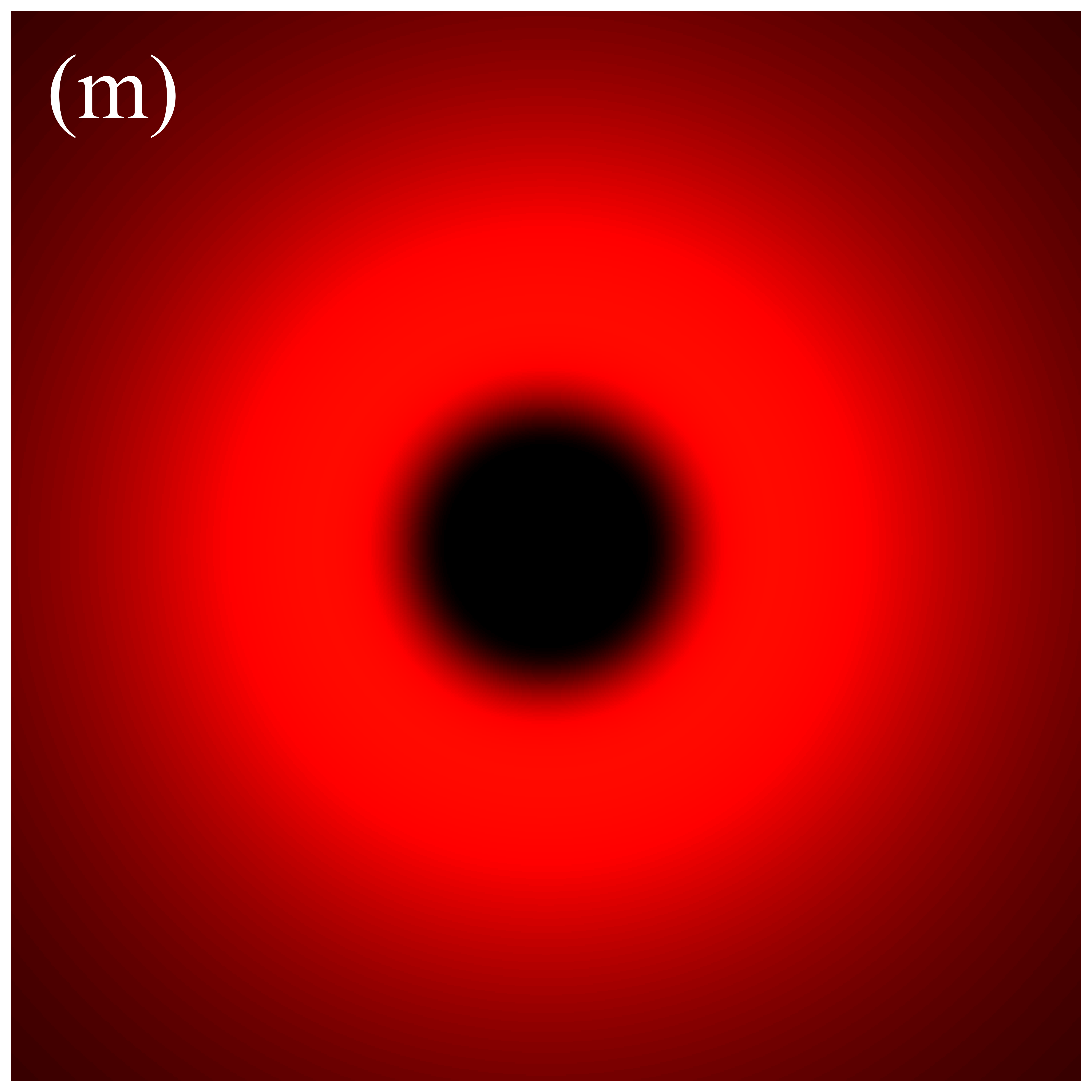}
\includegraphics[width=3.5cm]{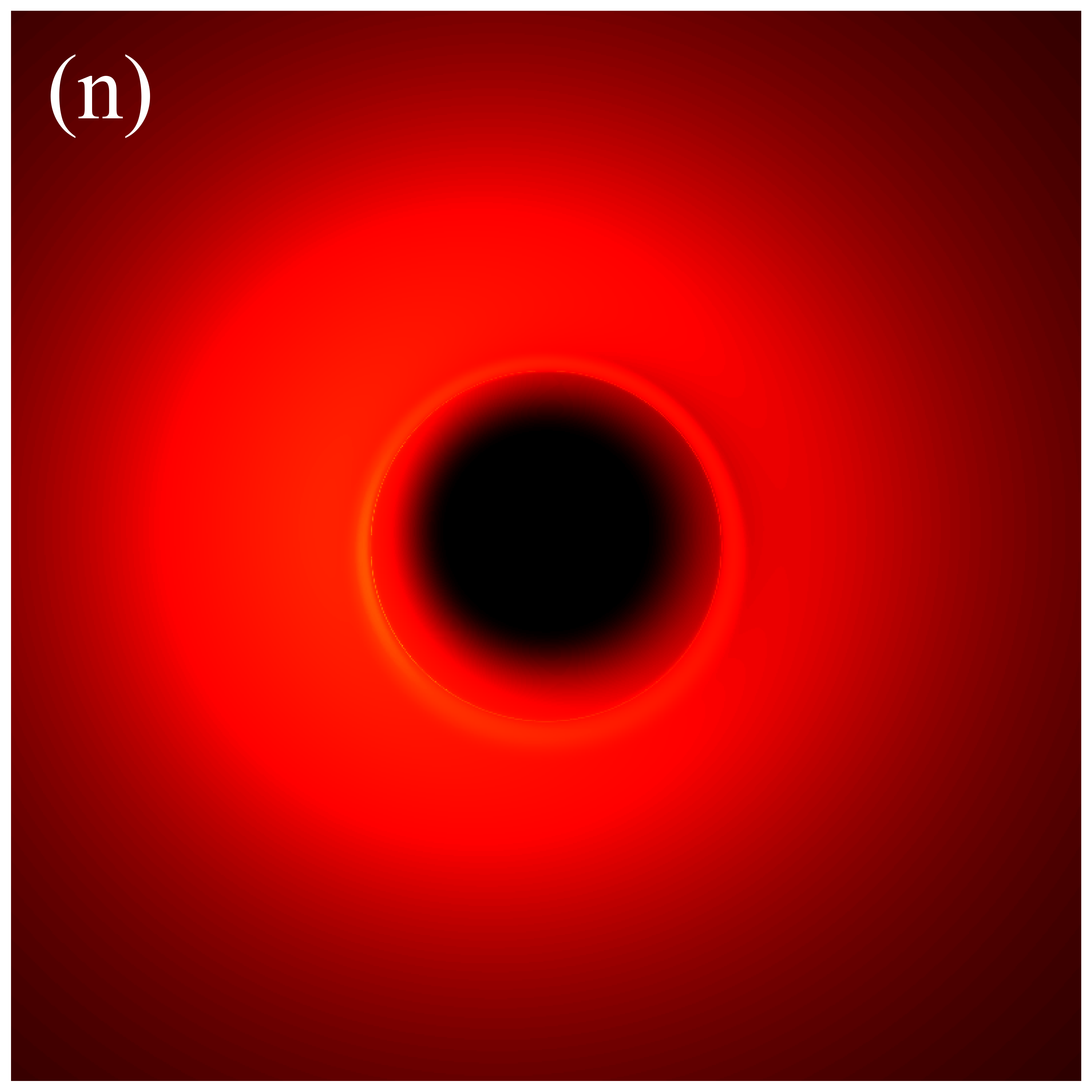}
\includegraphics[width=3.5cm]{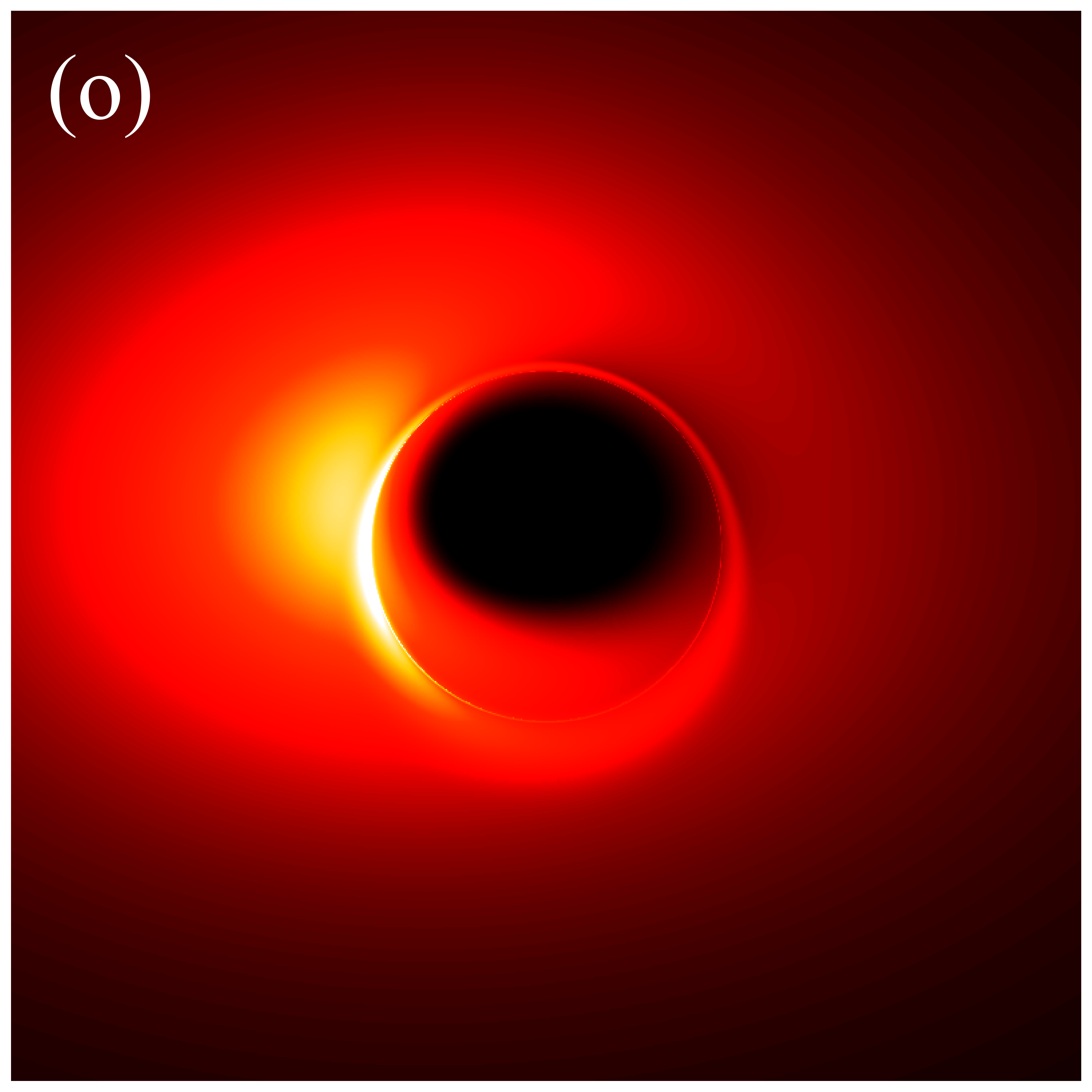}
\includegraphics[width=3.5cm]{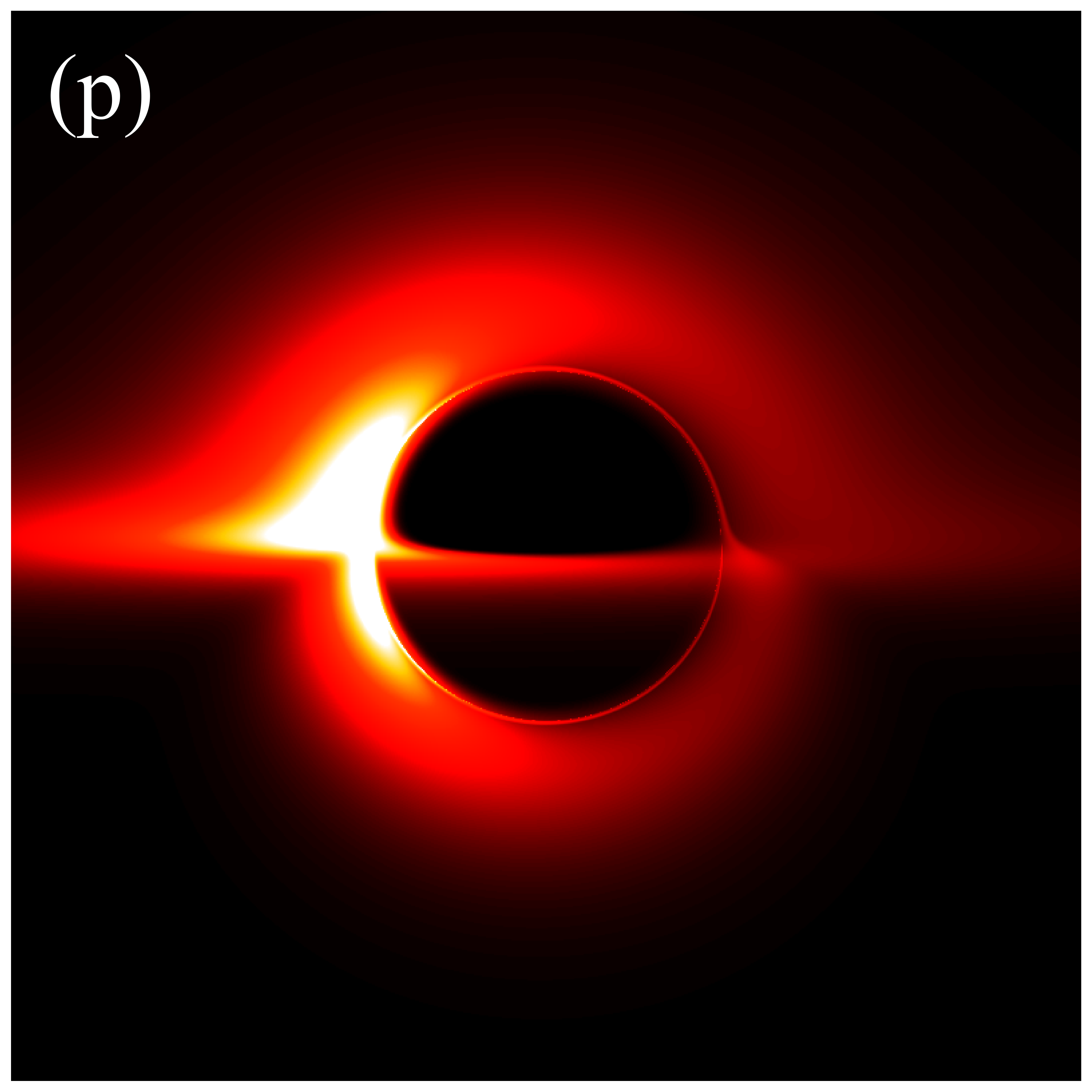}
\includegraphics[width=3.5cm]{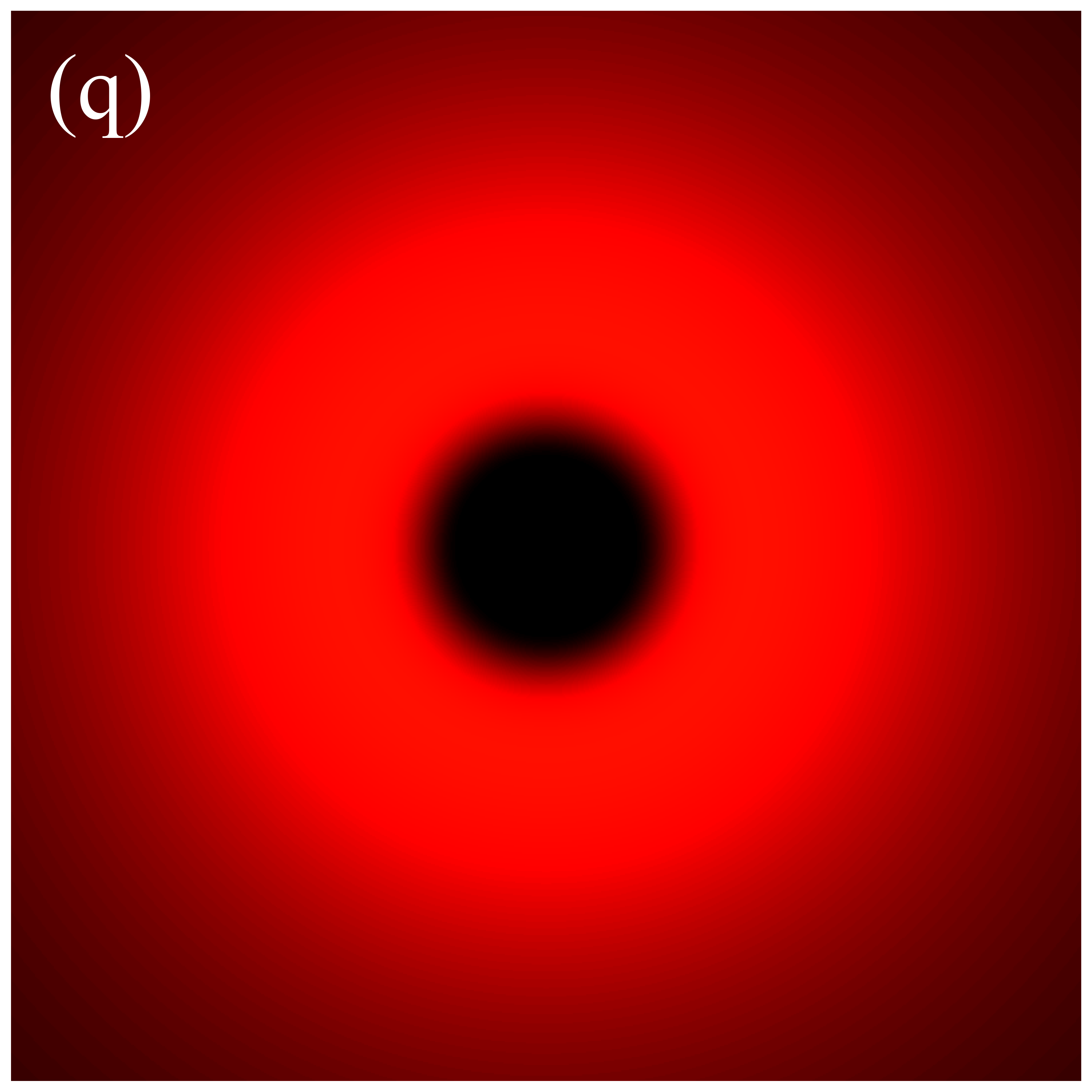}
\includegraphics[width=3.5cm]{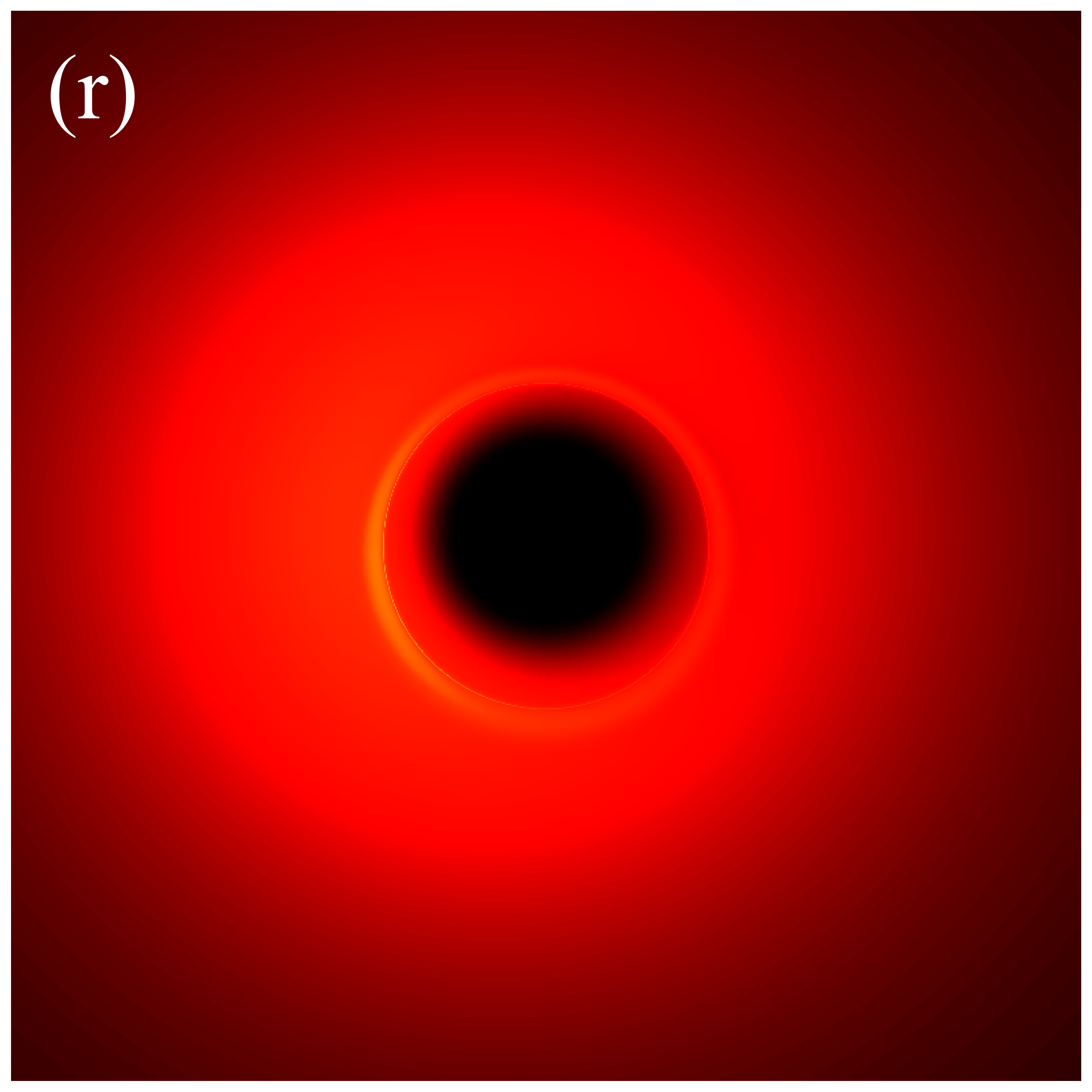}
\includegraphics[width=3.5cm]{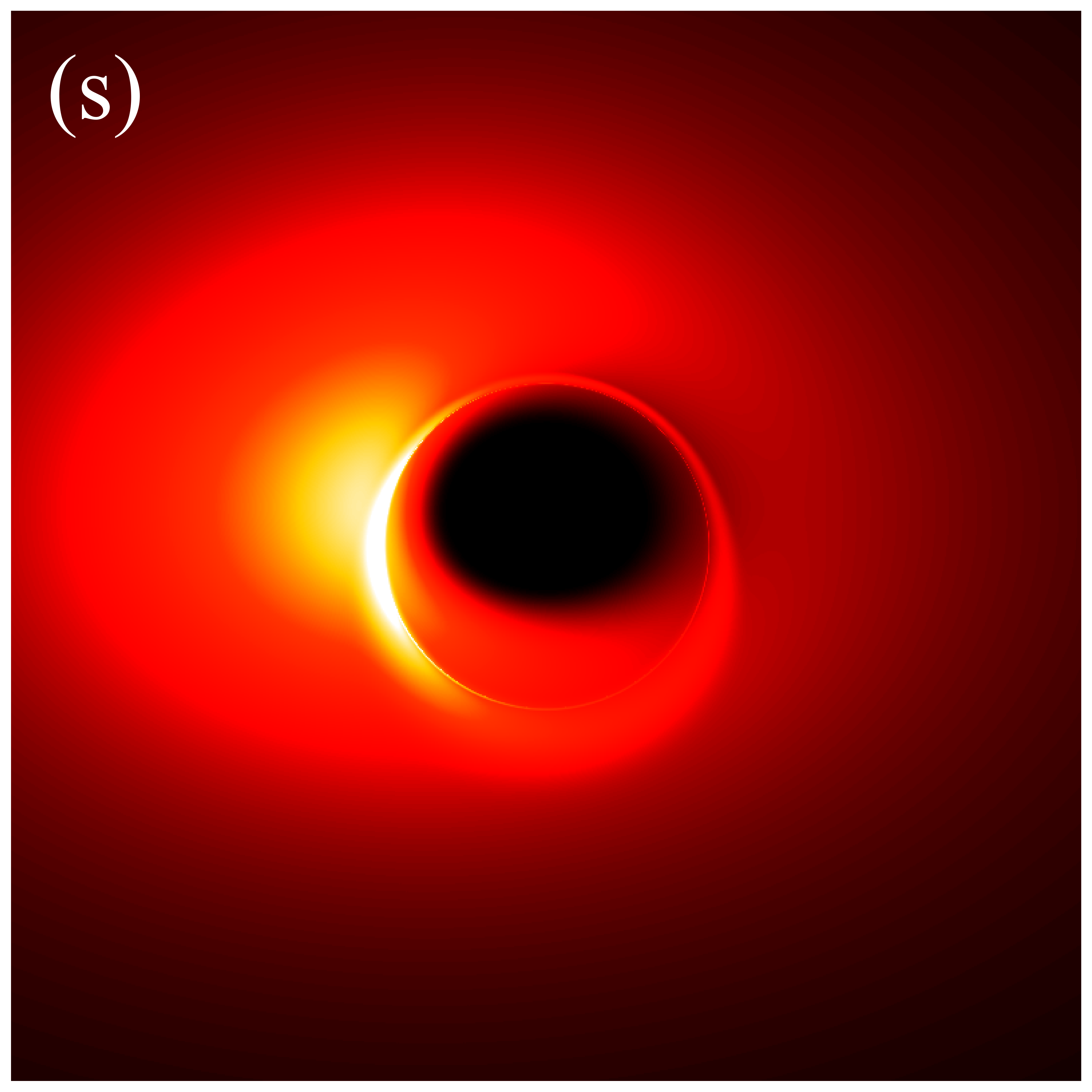}
\includegraphics[width=3.5cm]{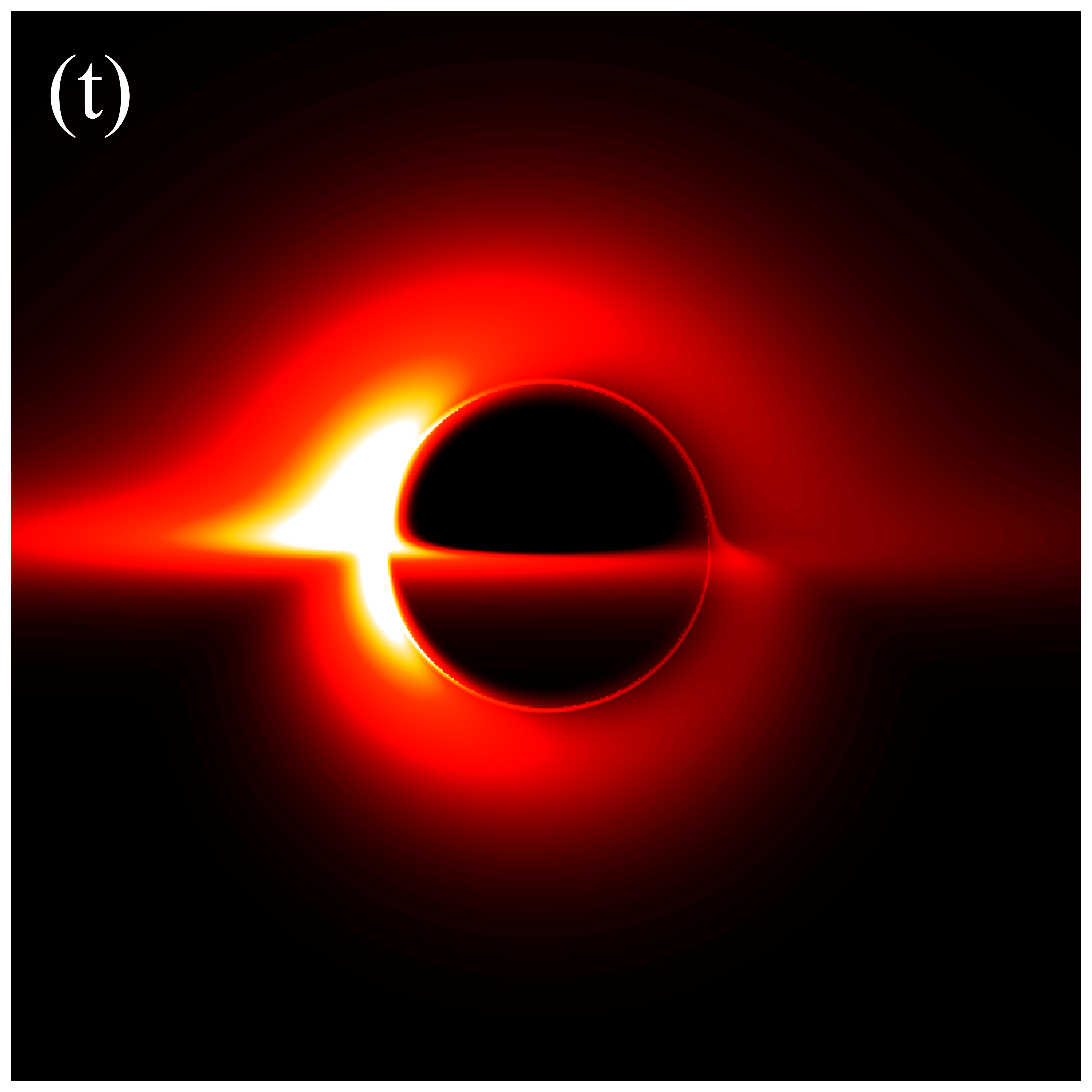}
\caption{Images of deformed Schwarzschild black holes at $86$ GHz across different parameter spaces. From left to right, the observation angles are $0^{\circ}$, $17^{\circ}$, $50^{\circ}$, and $85^{\circ}$; from top to bottom, the values of deformation parameter $\varepsilon$ are $-7$, $-3$, $0$, $3$, and $7$. Each row maintains a consistent deformation parameter, while each column represents a fixed observation angle. The image brightness is depicted by means of a linear gamma color scale with an index of $\gamma = 0.7$, where black and white are associated with the minimum and maximum values of $I_{\textrm{obs}}^{\gamma}$, within the range of $[0,0.9]$, respectively.}}\label{fig6}
\end{figure*}

In this study, we adopt the analytical accretion disk model established in the literature \cite{Chael et al. (2021)}, where the disk emission is a function of the source coordinate $r_{\textrm{s}}$ and is expressed in log-space as
\begin{equation}\label{32}
\log\left[\mathscr{F}\left(r_{\textrm{s}}\right)\right] = p_{1}\log\left(\frac{r_{\textrm{s}}}{r_{\textrm{eh}}}\right) + p_{2}\left[\log\left(\frac{r_{\textrm{s}}}{r_{\textrm{eh}}}\right)\right]^{2}.
\end{equation}
Here, $p_{1}$ and $p_{2}$ are parameters associated with the observation frequency. Specifically, when $p_{1}=0$ and $p_{2}=-3/4$, the model corresponds to $86$ GHz, and when $p_{1}=-2$ and $p_{2}=-1/2$, it matches $230$ GHz. The observed intensity of the light ray corresponding to each pixel within the observation plane is then given by
\begin{equation}\label{33}
I_{\textrm{obs}} = \sum^{N_{\textrm{max}}}_{n=1}C_{n}\mathscr{F}\left(r_{\textrm{s}}\right)\kappa g^{3},
\end{equation}
where $n$ denotes the number of intersections of the light ray with the accretion disk, and $N_{\textrm{max}}$ represents the maximum number of times the light ray can pass through the disk. The coefficient $C_{n}$, referred to as the ``fudge factor'', accounts for the varying contributions of high-order rings to the image brightness. Here, we set $C_{1}=1$ and $C_{n}=2/3$ for $n > 1$ to ensure that model \eqref{33} aligns with GRMHD simulations. It is important to note that, compared to equation (16) in the literature \cite{Chael et al. (2021)}, our model incorporates an additional projection effect $\kappa$ arising from the anisotropic accretion disk. As previously mentioned, we set $\kappa = \cos\delta$, where $\delta$ is obtained from the ray-tracing. When $\kappa=1$, model \eqref{33} degenerates into the case of an isotropic accretion disk.
\begin{figure*}%[tbph]
\center{
\includegraphics[width=3.5cm]{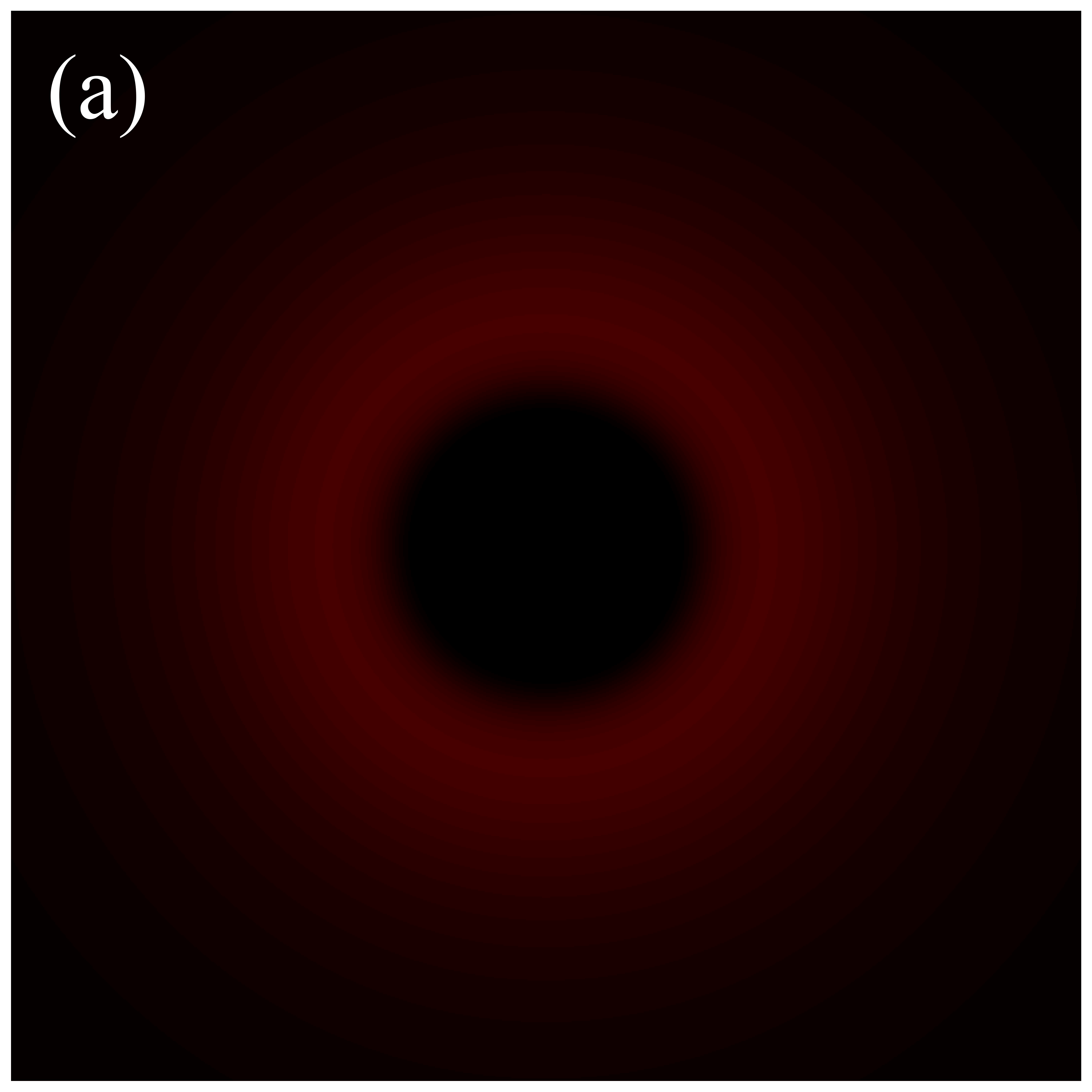}
\includegraphics[width=3.5cm]{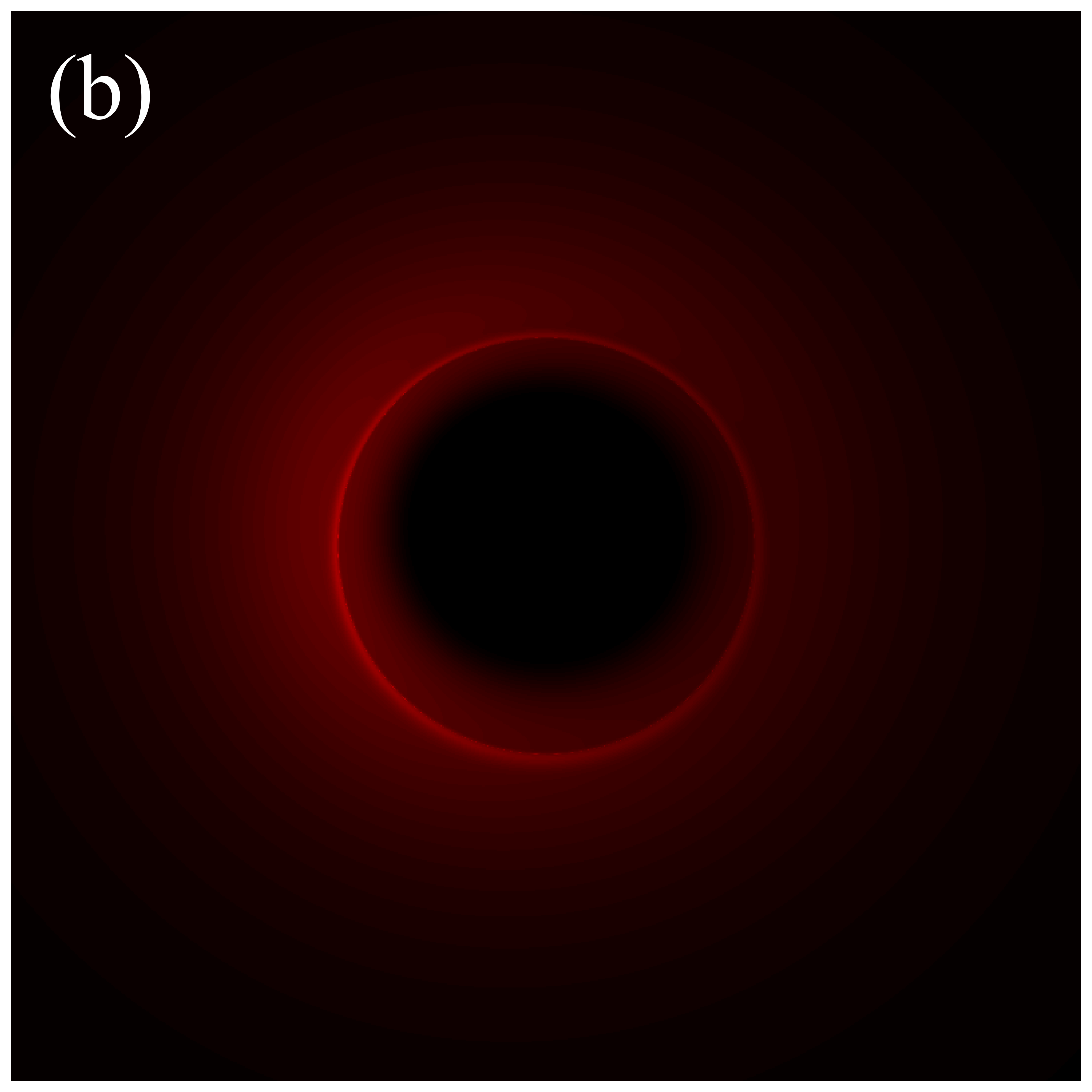}
\includegraphics[width=3.5cm]{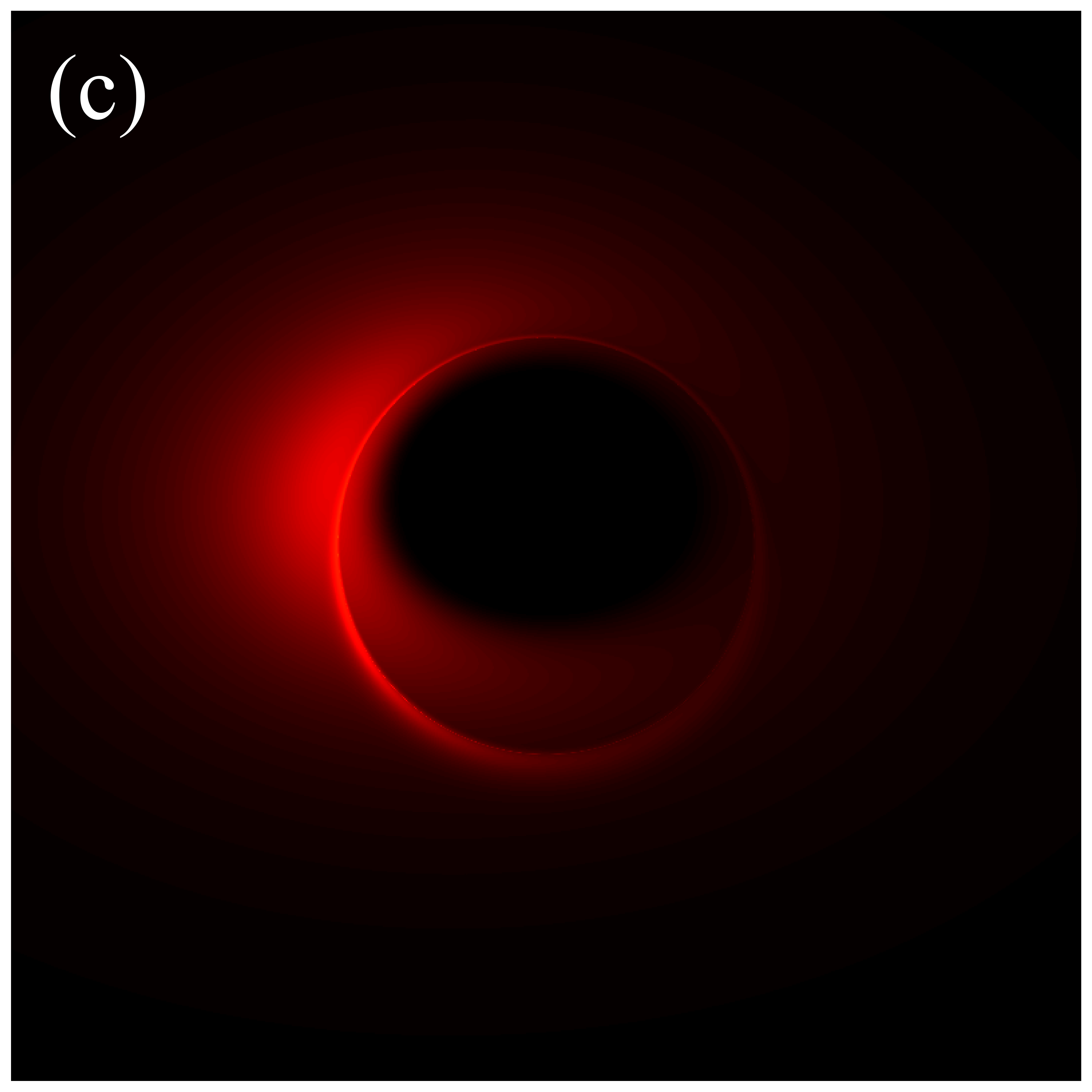}
\includegraphics[width=3.5cm]{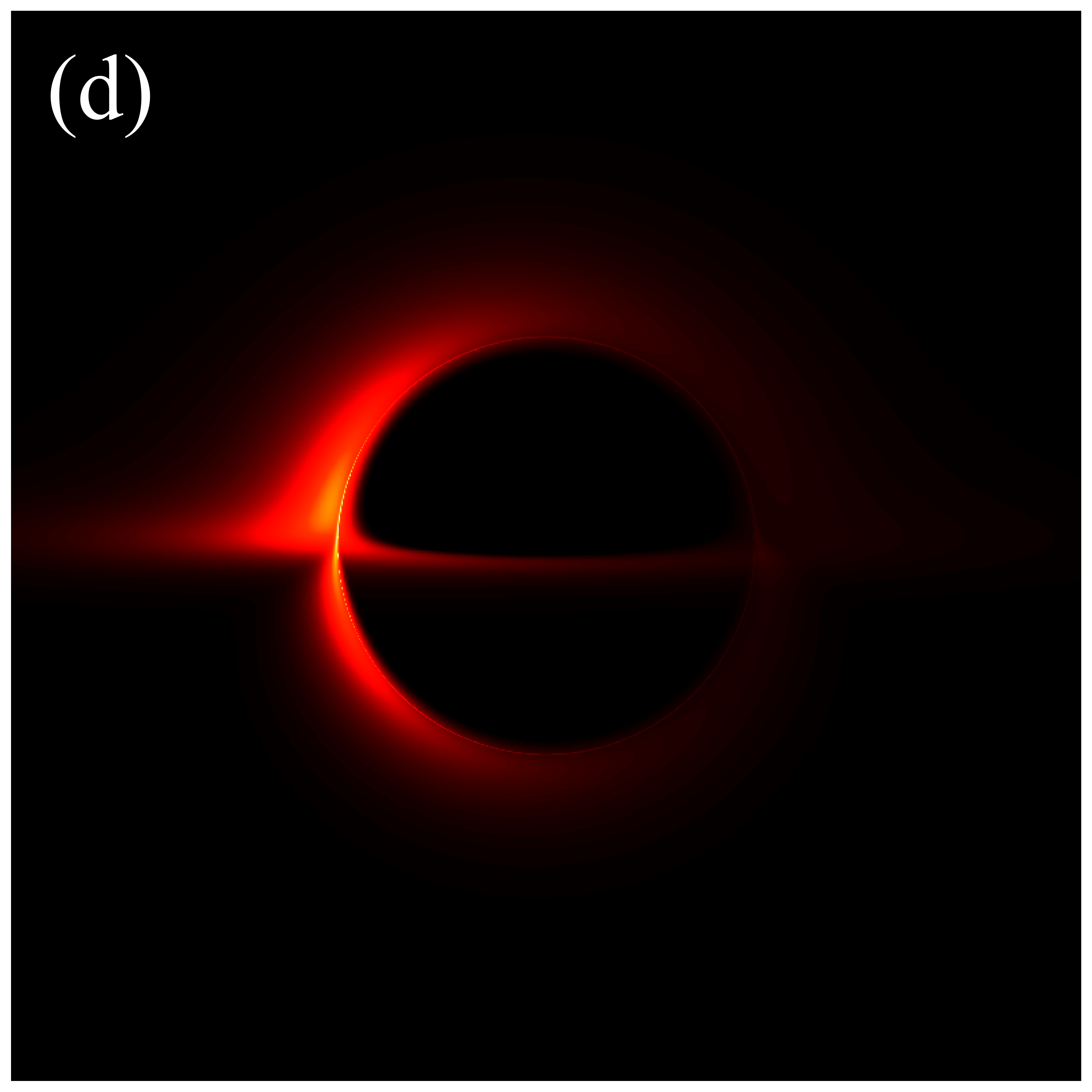}
\includegraphics[width=3.5cm]{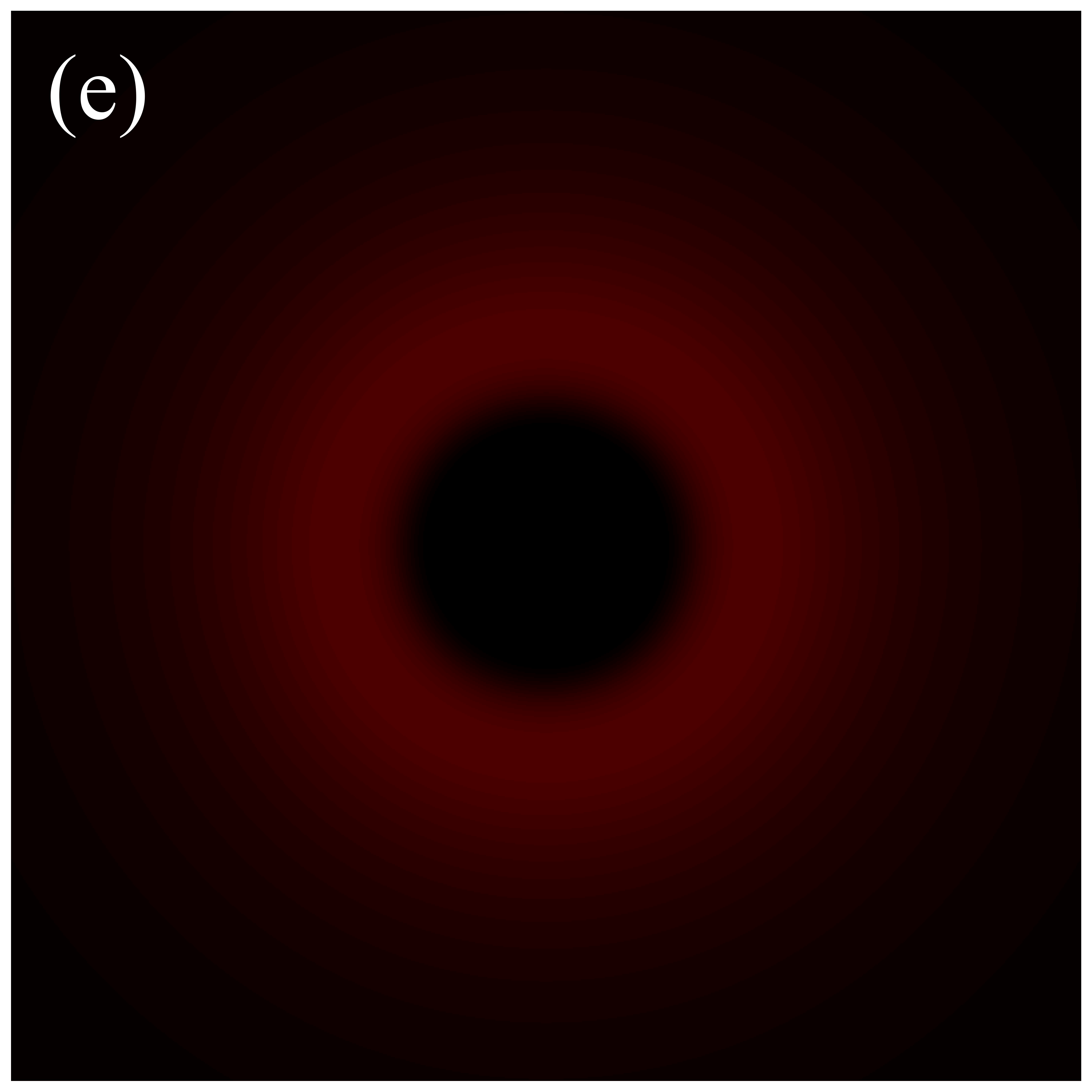}
\includegraphics[width=3.5cm]{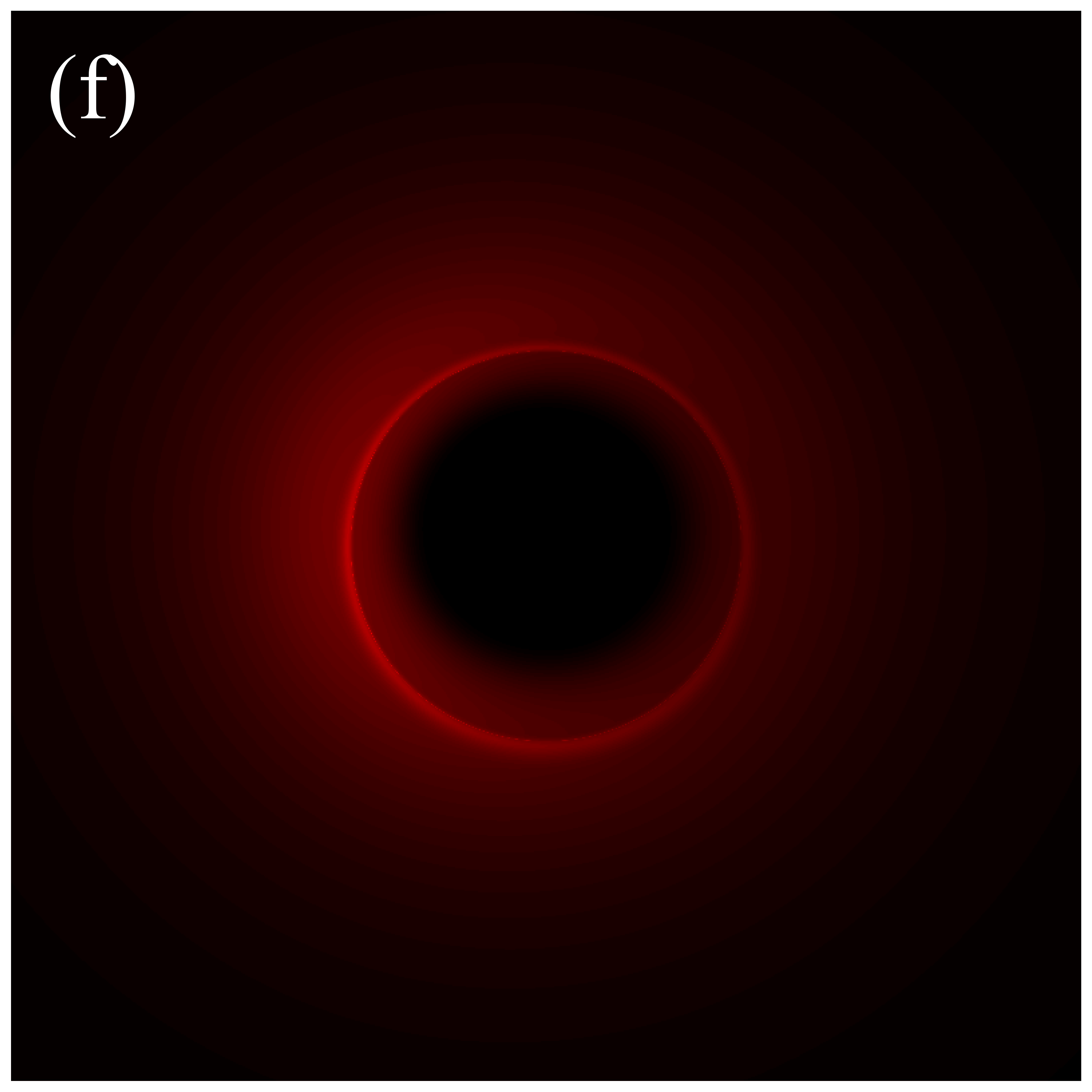}
\includegraphics[width=3.5cm]{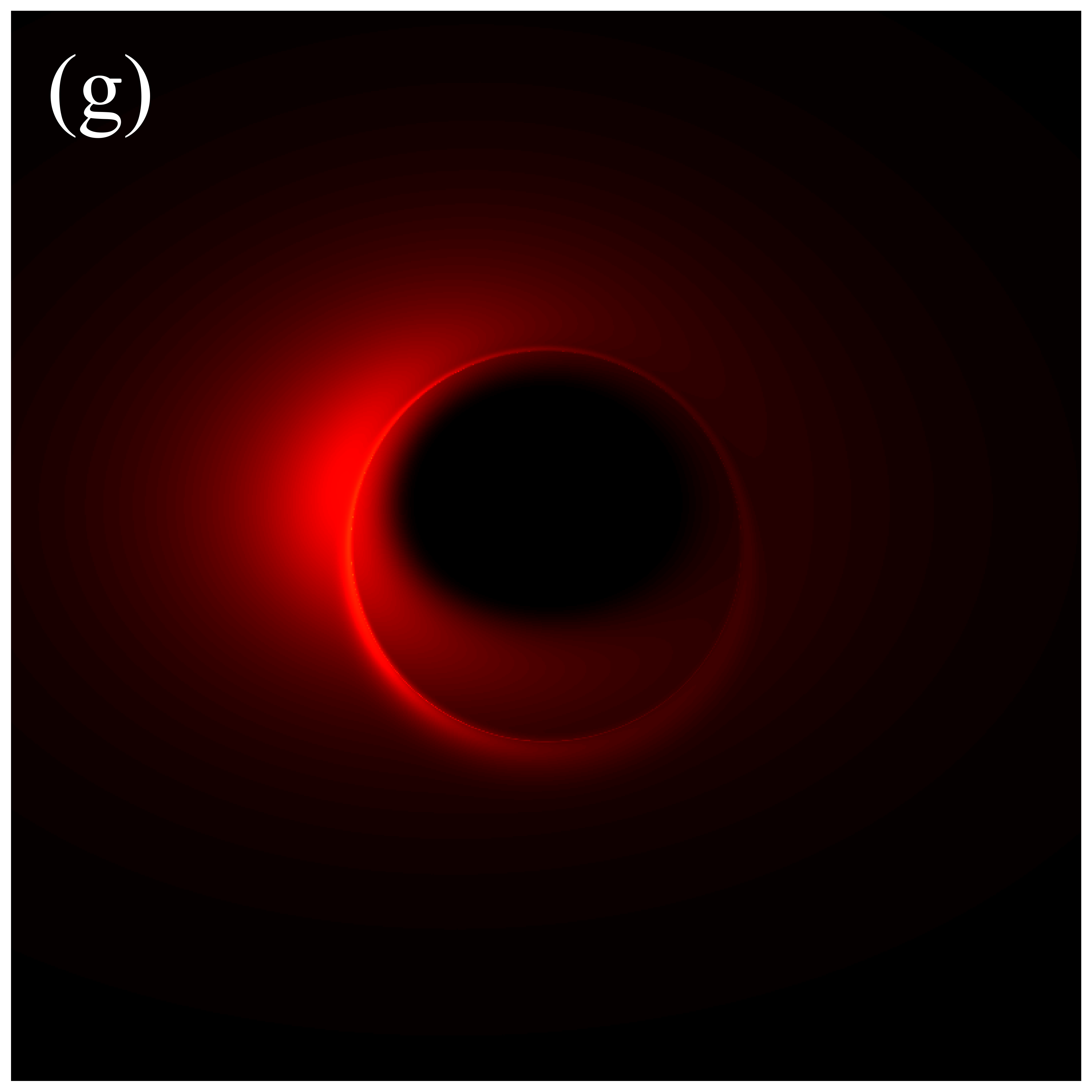}
\includegraphics[width=3.5cm]{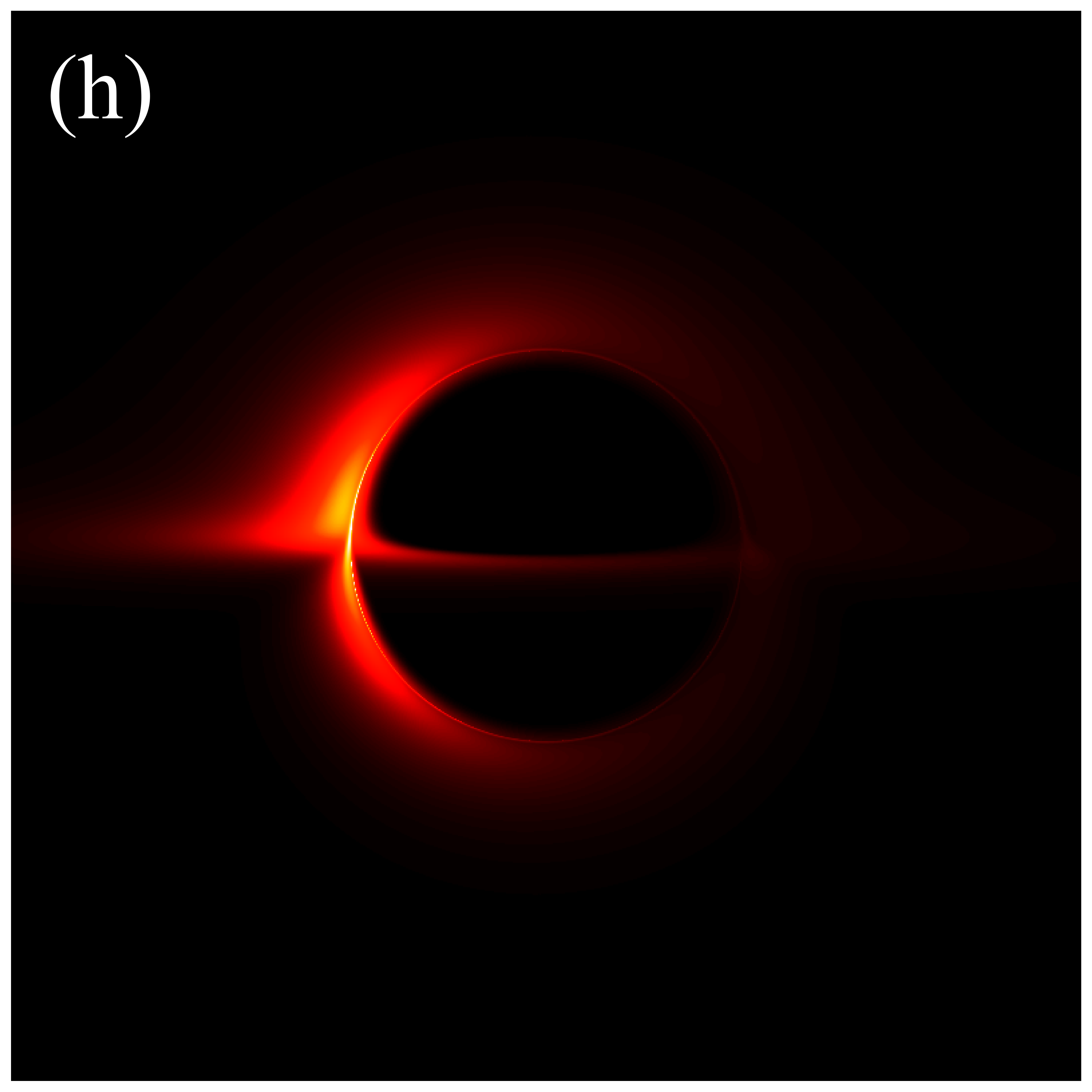}
\includegraphics[width=3.5cm]{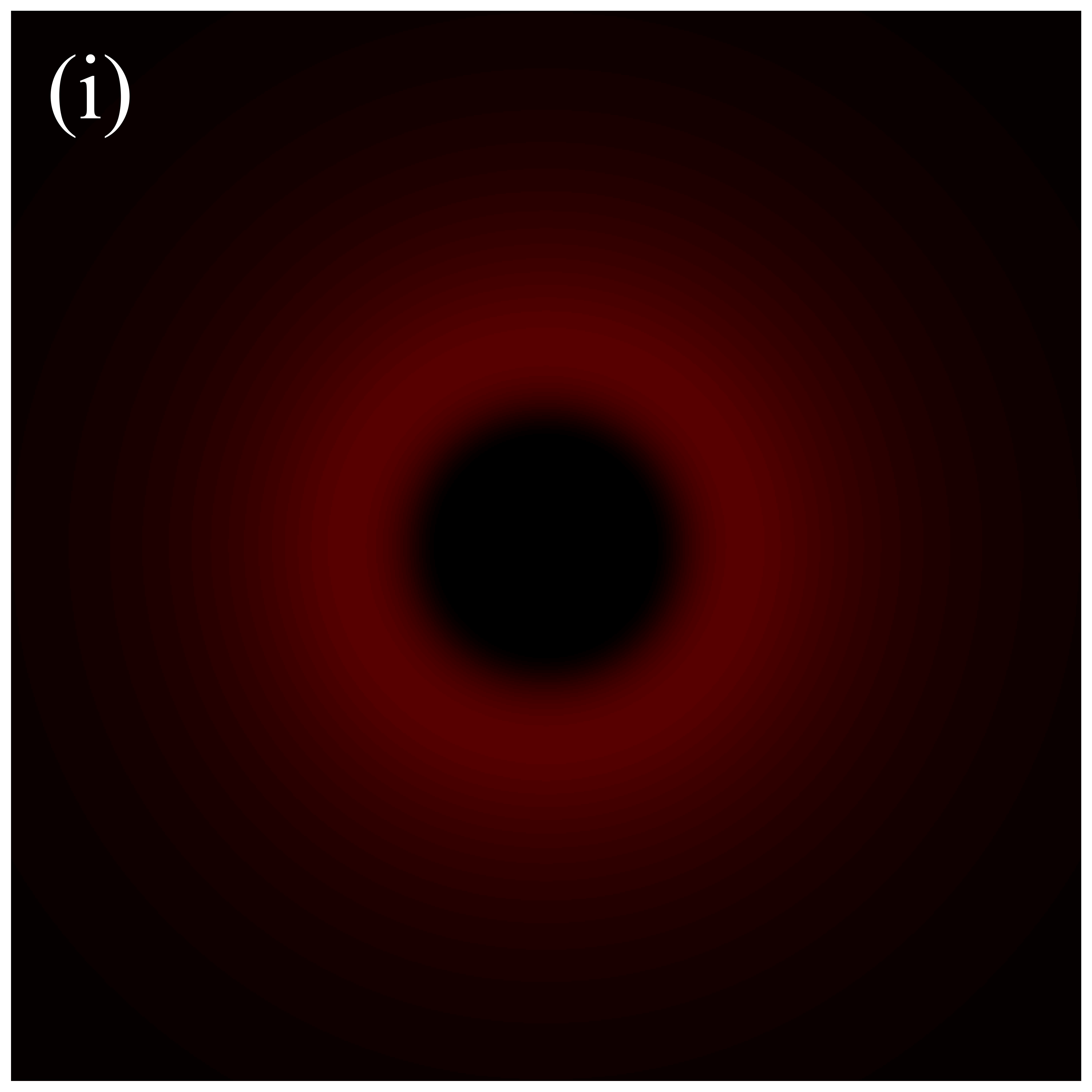}
\includegraphics[width=3.5cm]{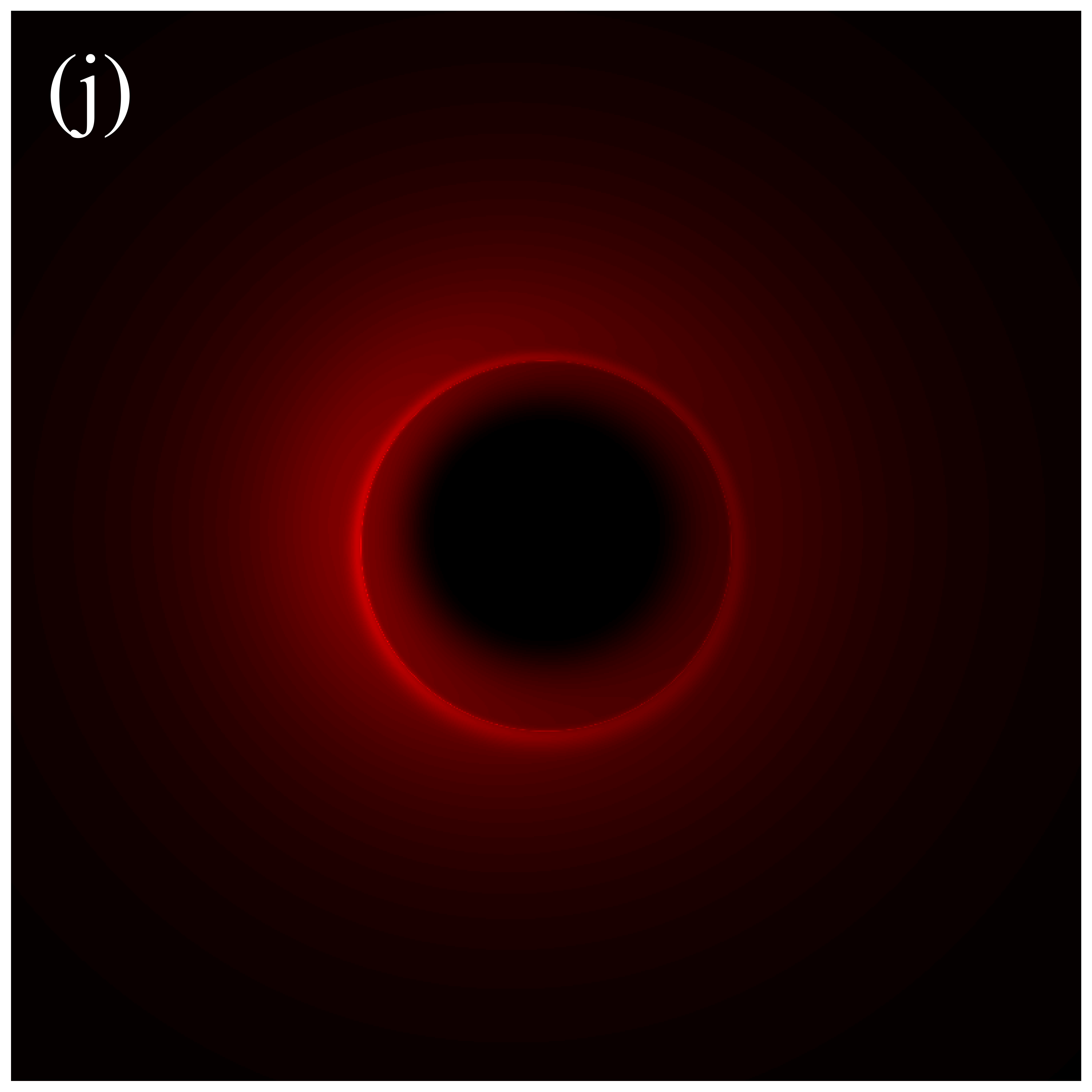}
\includegraphics[width=3.5cm]{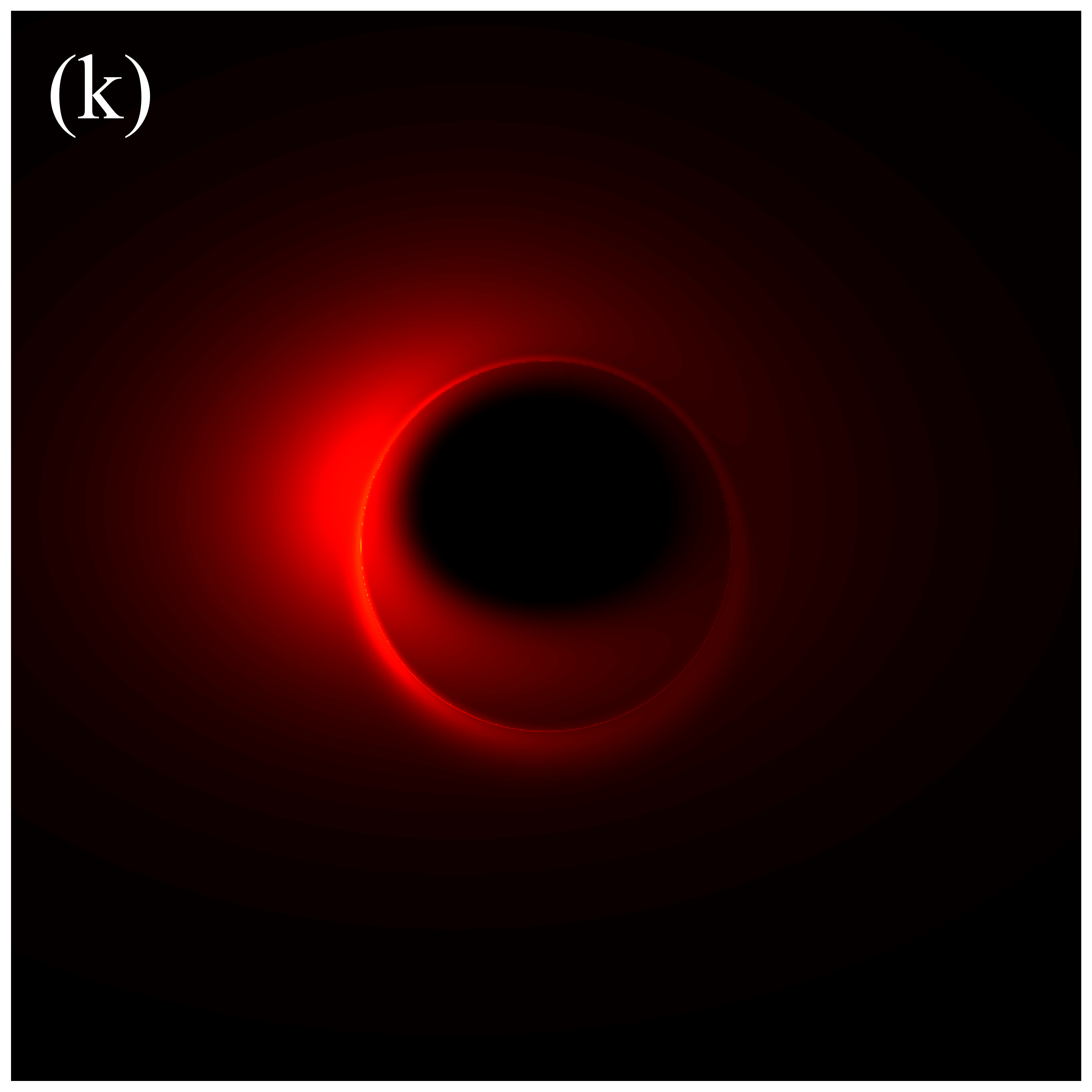}
\includegraphics[width=3.5cm]{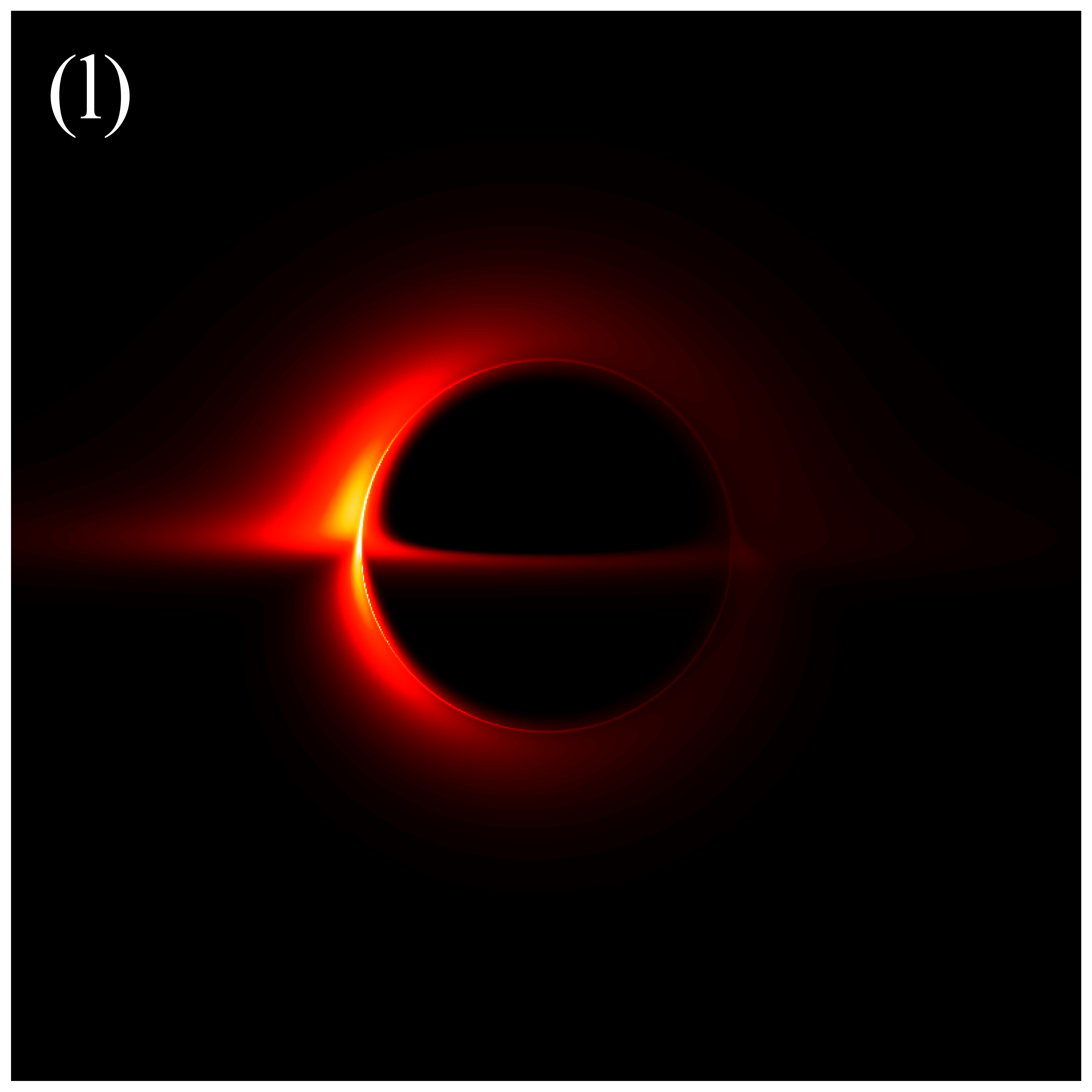}
\includegraphics[width=3.5cm]{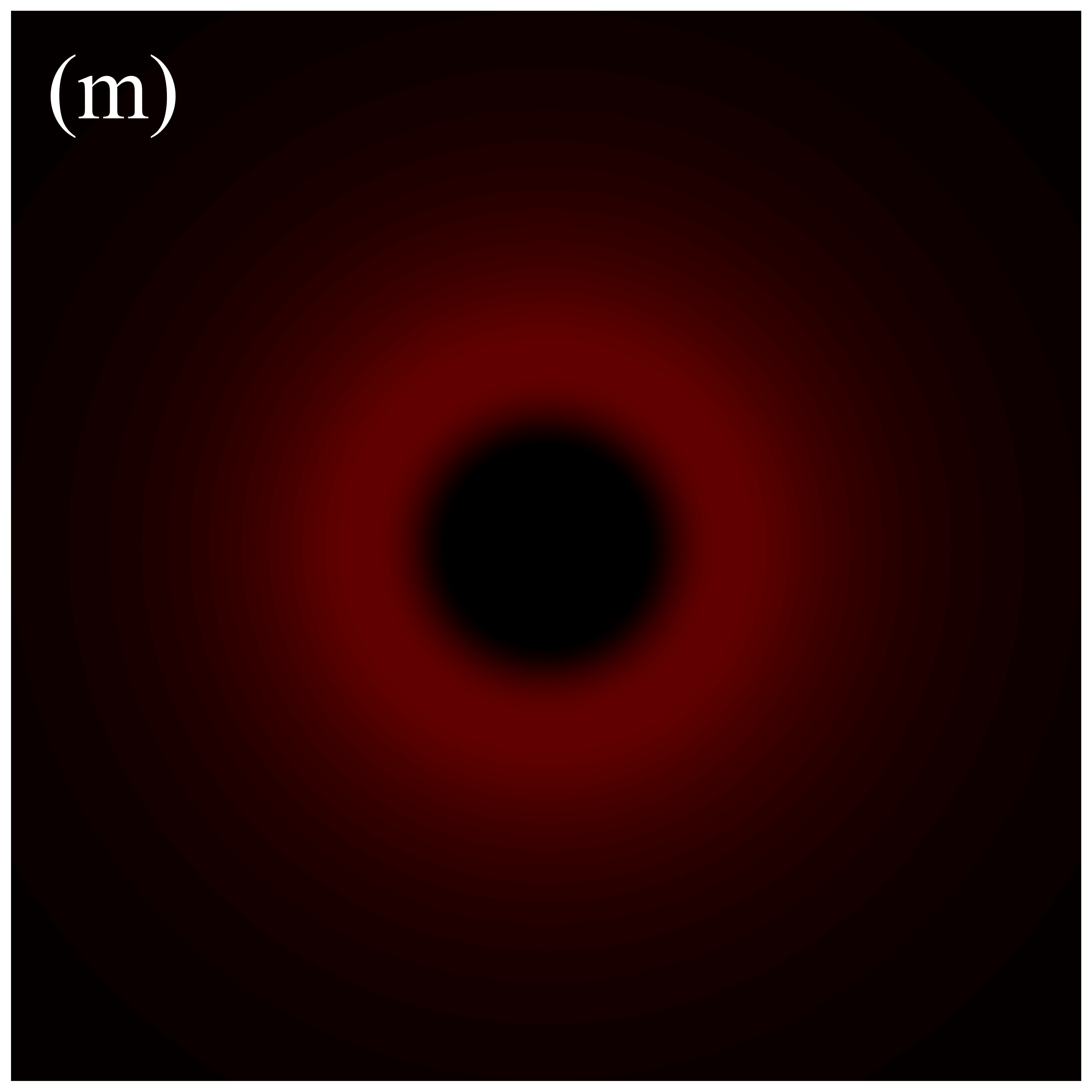}
\includegraphics[width=3.5cm]{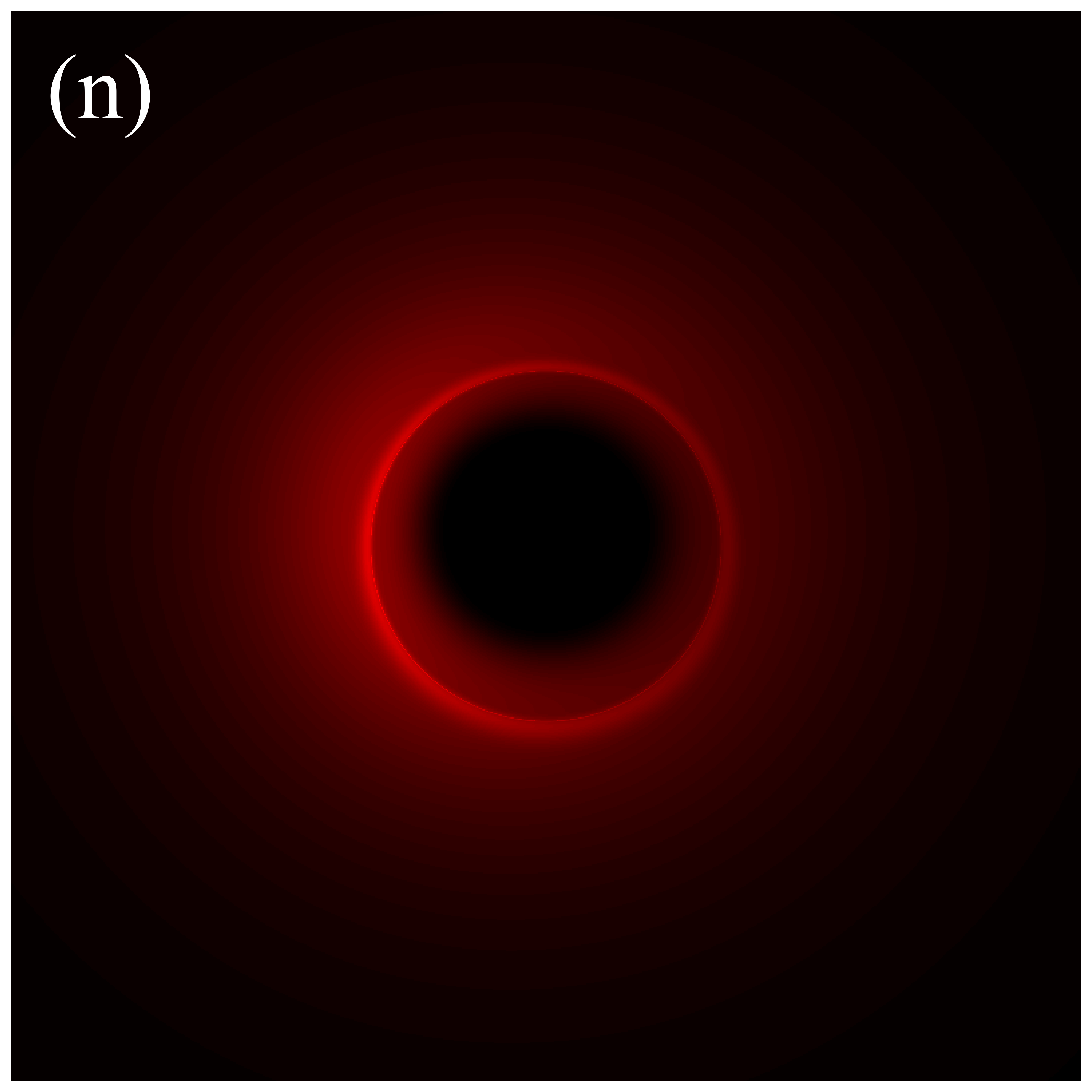}
\includegraphics[width=3.5cm]{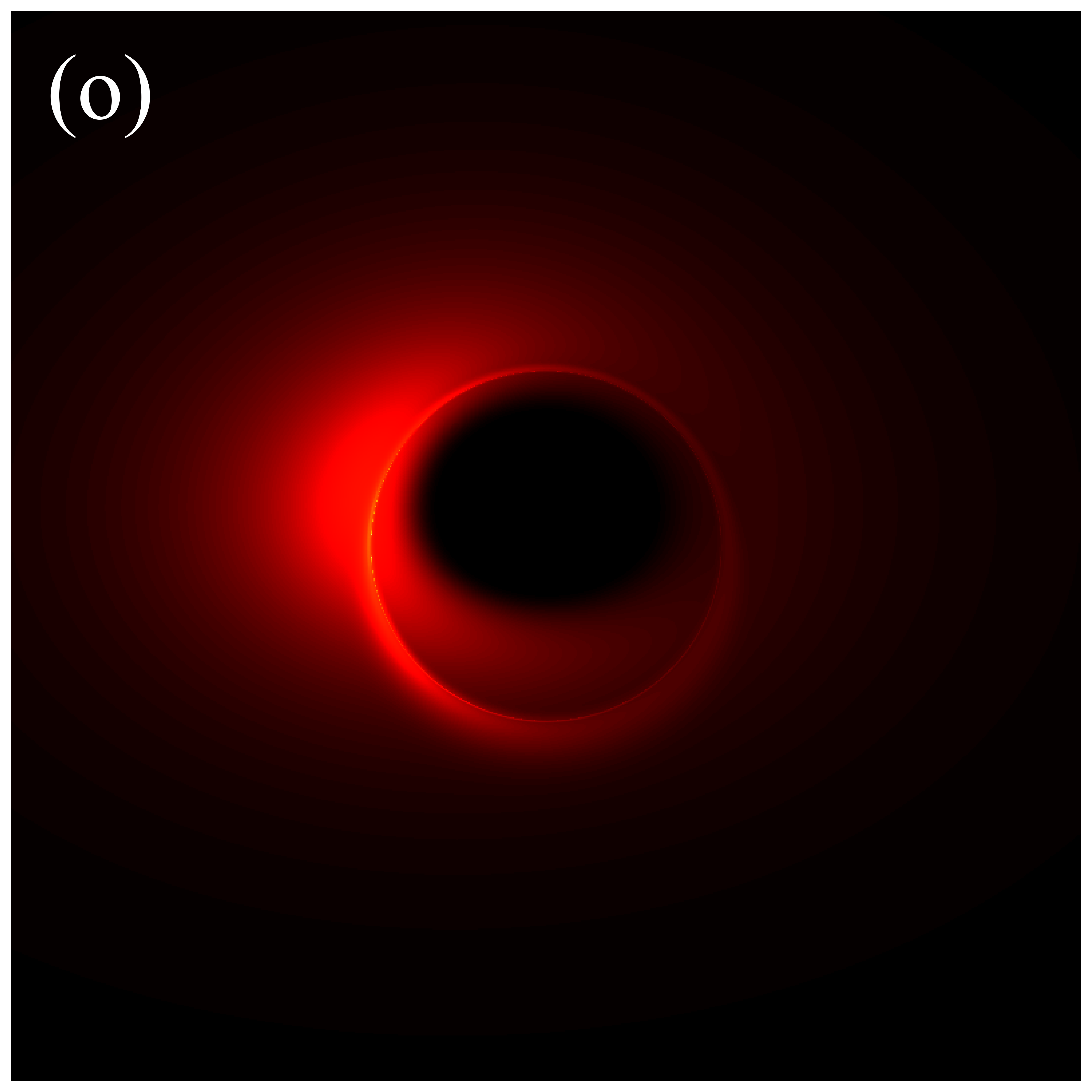}
\includegraphics[width=3.5cm]{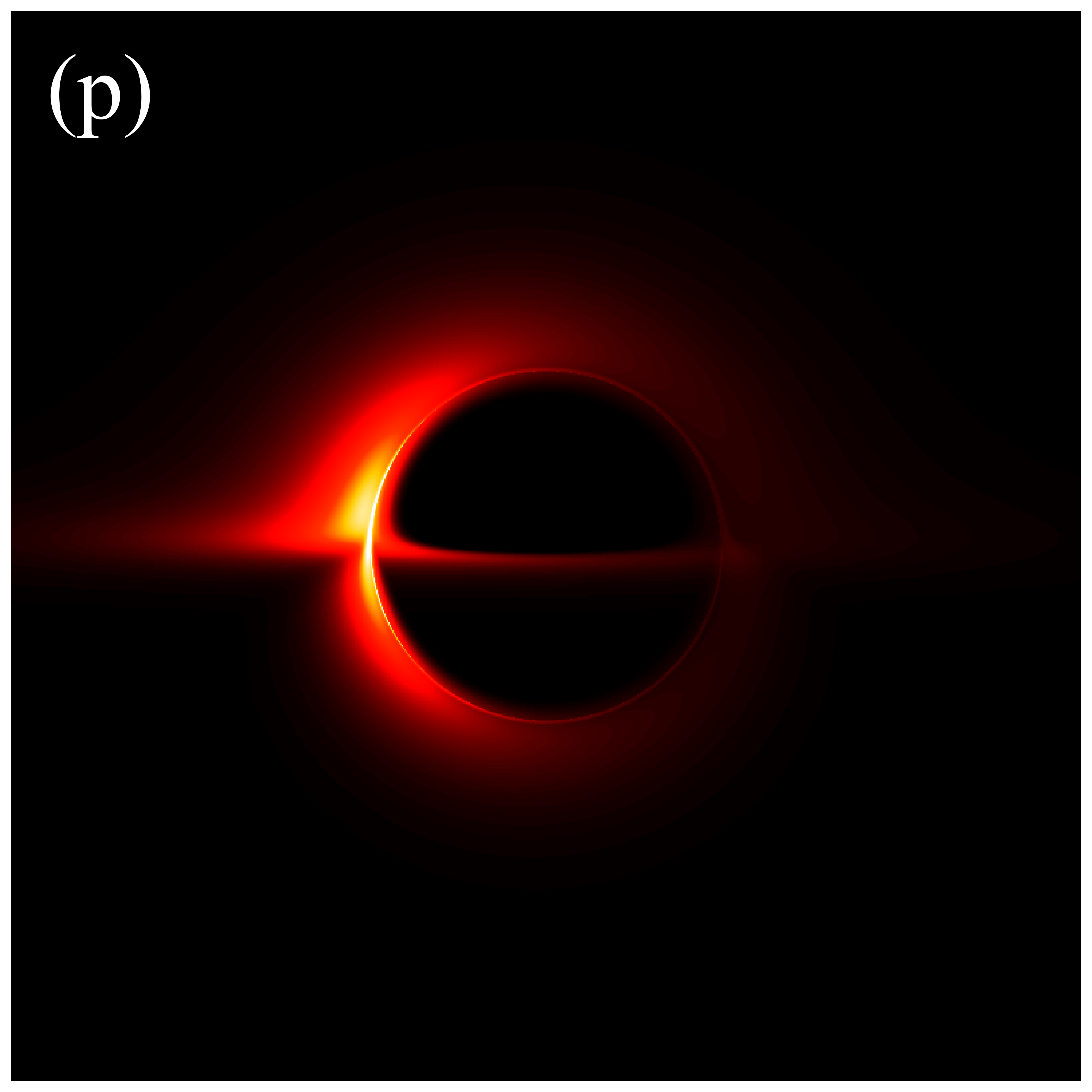}
\includegraphics[width=3.5cm]{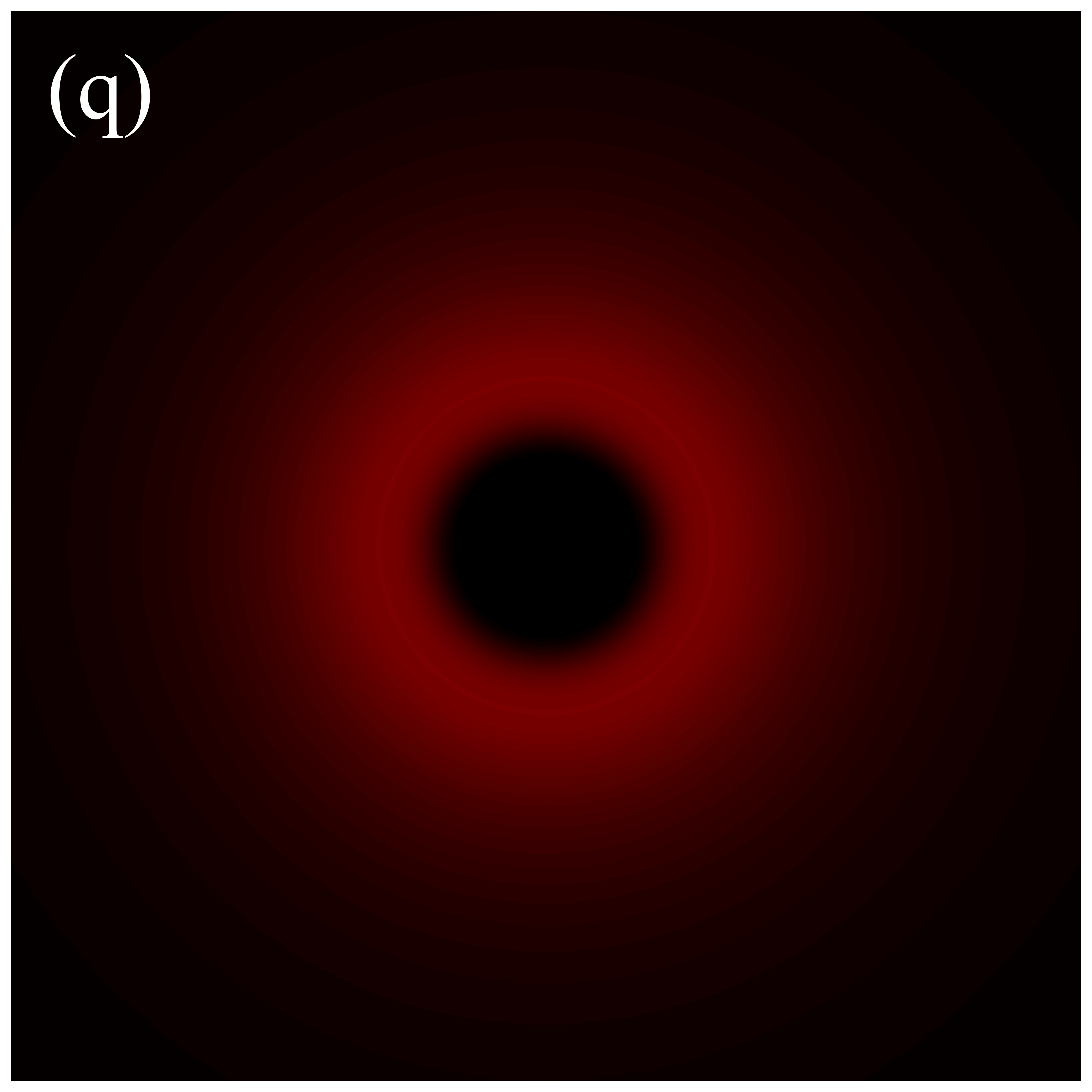}
\includegraphics[width=3.5cm]{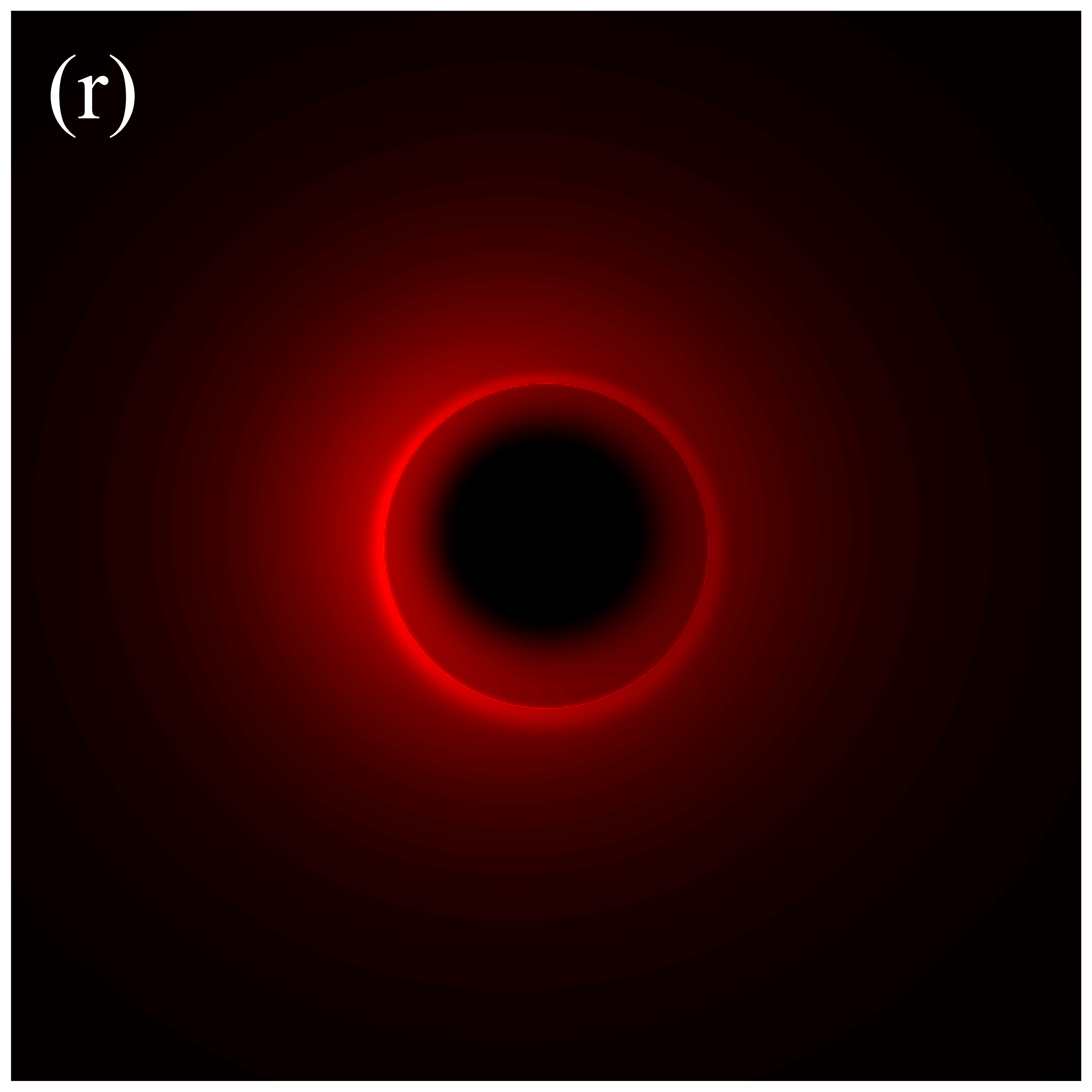}
\includegraphics[width=3.5cm]{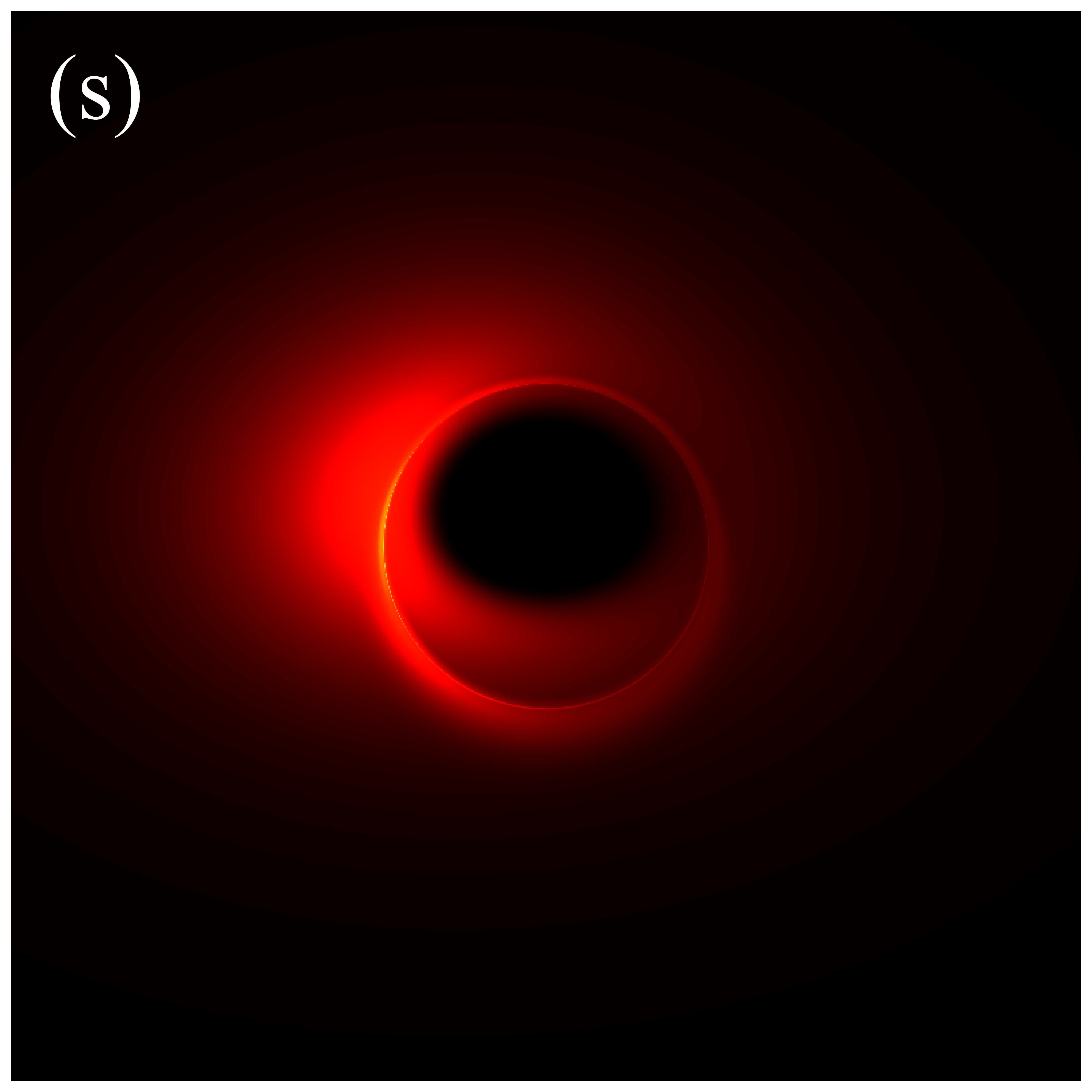}
\includegraphics[width=3.5cm]{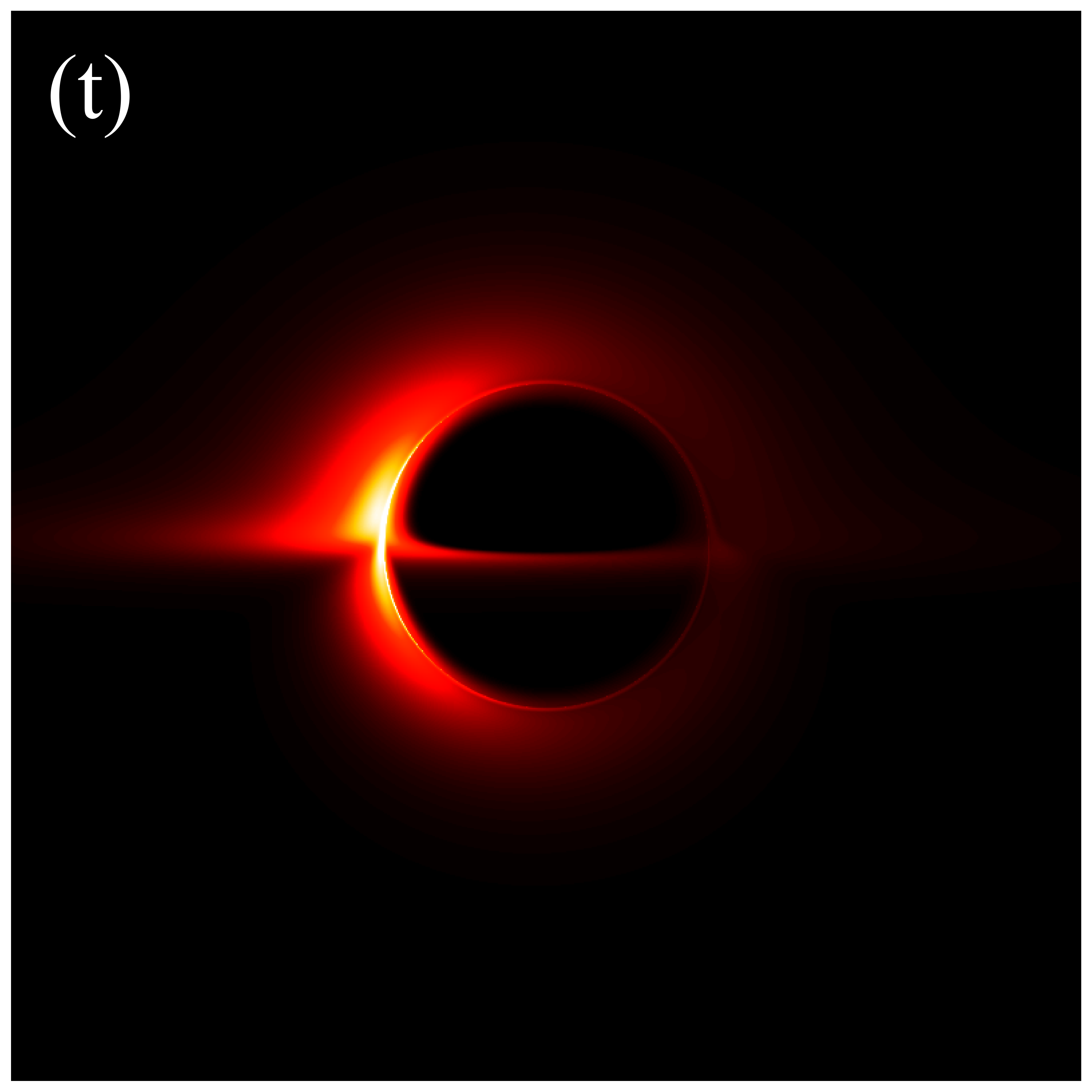}
\caption{Similar to figure 6, but for $230$ GHz band.}}\label{fig7}
\end{figure*}

In the subsequent simulations, we fix the observation distance at $r_{\textrm{obs}}=1000$ M, the field of view at $30 \times 30$ M, and the image resolution at $800 \times 800$ pixels. Figure 6 presents the images of deformed Schwarzschild black holes with an isotropic accretion disk ($\kappa=1$) at an observation frequency of $86$ GHz, across varying deformation parameters and inclination angles. When the observation inclination is $0^{\circ}$ (leftmost column), the black hole image exhibits central symmetry and can be viewed as a series of concentric circles with varying radii. This symmetry arises because, when the line of sight is perpendicular to the disk plane, the motion of the accretion disk has no component along the line of sight, resulting in a redshift factor $g$ that lacks a Doppler component and includes only gravitational redshift. As non-zero inclination angles are introduced, the image develops an east-west asymmetry that becomes increasingly pronounced with larger inclination angles. In particular, at high inclination angles (e.g., $\omega=50^{\circ}$, $85^{\circ}$), crescent-shaped and eyebrow-like bright spots emerge on the left side of the image, with their brightness and size increasing with $\omega$. This is attributed to the Doppler effect becoming more significant as the inclination angle increases. Additionally, It is observed that increasing the deformation parameter slightly enhances the image brightness. Furthermore, the inner shadow of the black hole is influenced by both the deformation parameter and the inclination angle. Specifically, as the inclination angle increases, the shape of the inner shadow transitions from circular to elliptical, eventually appearing arched at $\omega=85^{\circ}$; the size of the inner shadow decreases noticeably with increasing deformation parameter. It is important to emphasize that in figure 6, the critical curve is barely discernible, with its size decreasing as the deformation parameter increases. This implies that in the $86$ GHz simulations, the contributions of higher-order subrings are easily obscured by the image background, presenting certain challenge for using these images to test gravitational theories.

Figure 7 presents the images of deformed Schwarzschild black holes surrounded by an isotropic accretion disk at $230$ GHz, across different deformation parameters and observation inclinations. The effects of the inclination angle and deformation parameter on the image features are consistent with those observed in figure 6: increasing the deformation parameter reduces the sizes of both the inner shadow and the critical curve, while enhancing the image brightness; increasing the inclination angle modifies the shape of the inner shadow and accentuates the east-west asymmetry in the image. However, there are two fundamental distinctions between the images at $86$ GHz and $230$ GHz. Firstly, the brightness of the $230$ GHz images is significantly lower than that of the $86$ GHz images. Secondary, in the $230$ GHz images, a clear, thin, and complete critical curve is easily discernible, aligning with the results listed in Table 1. The reason for this difference is that the accretion disk has a higher optical depth for low-frequency light, which causes the contribution of higher-order rings to the image to be washed out \cite{Chael et al. (2021)}. In other words, simulations at higher frequencies are more likely to preserve the features of higher-order images, thereby enhancing the ability to test gravitational theories.
\begin{figure*}%[tbph]
\center{
\includegraphics[width=2.9cm]{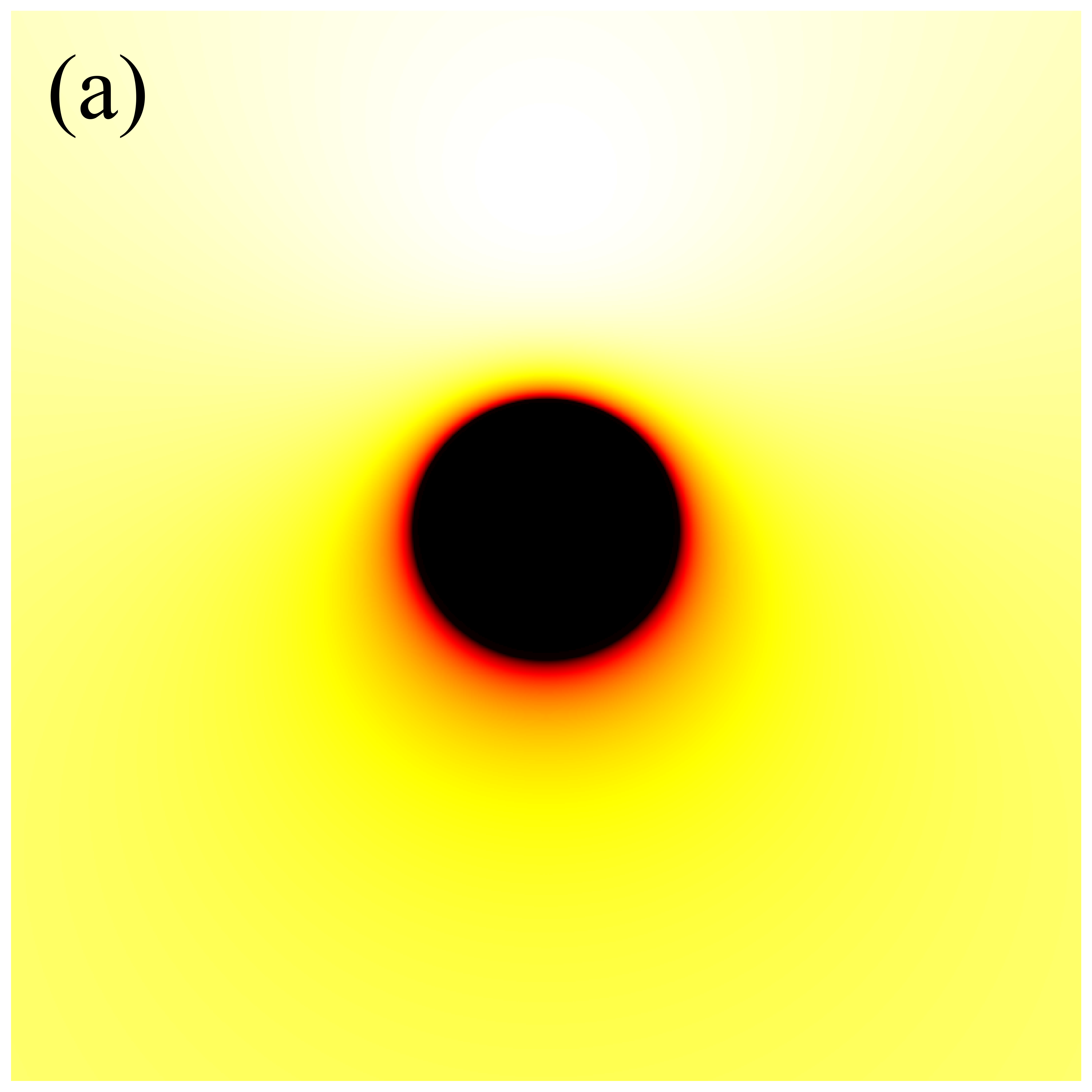}
\includegraphics[width=2.9cm]{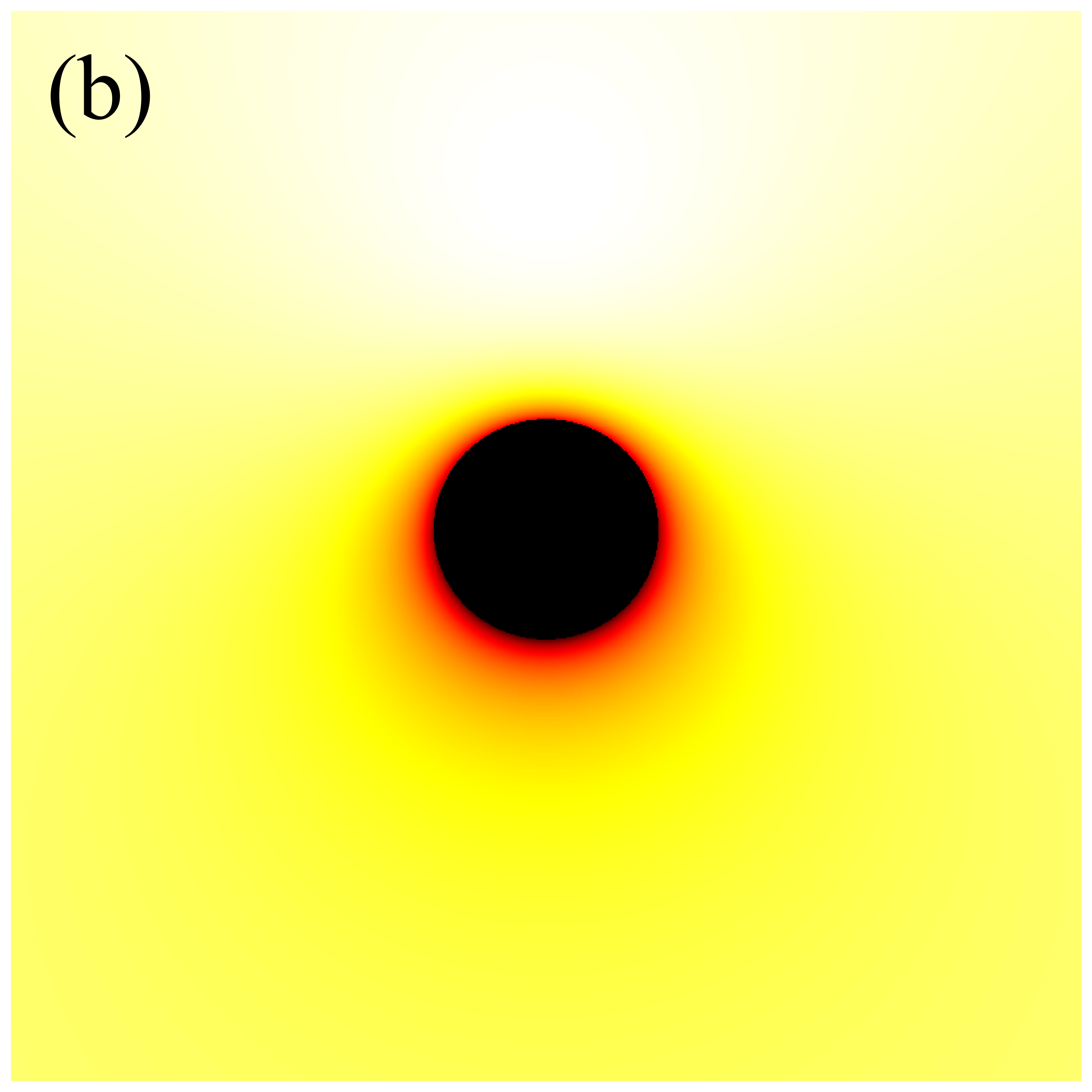}
\includegraphics[width=2.9cm]{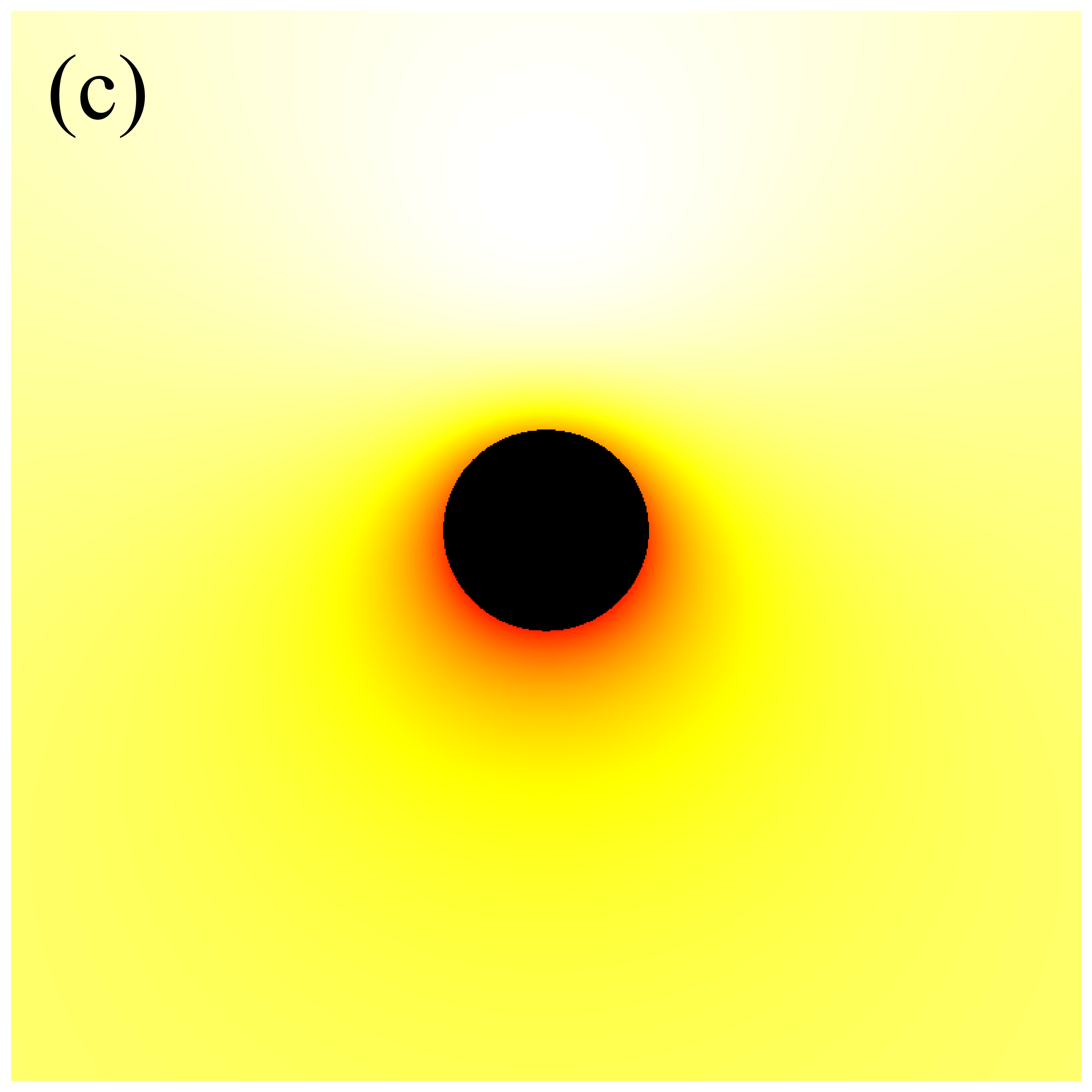}
\includegraphics[width=2.9cm]{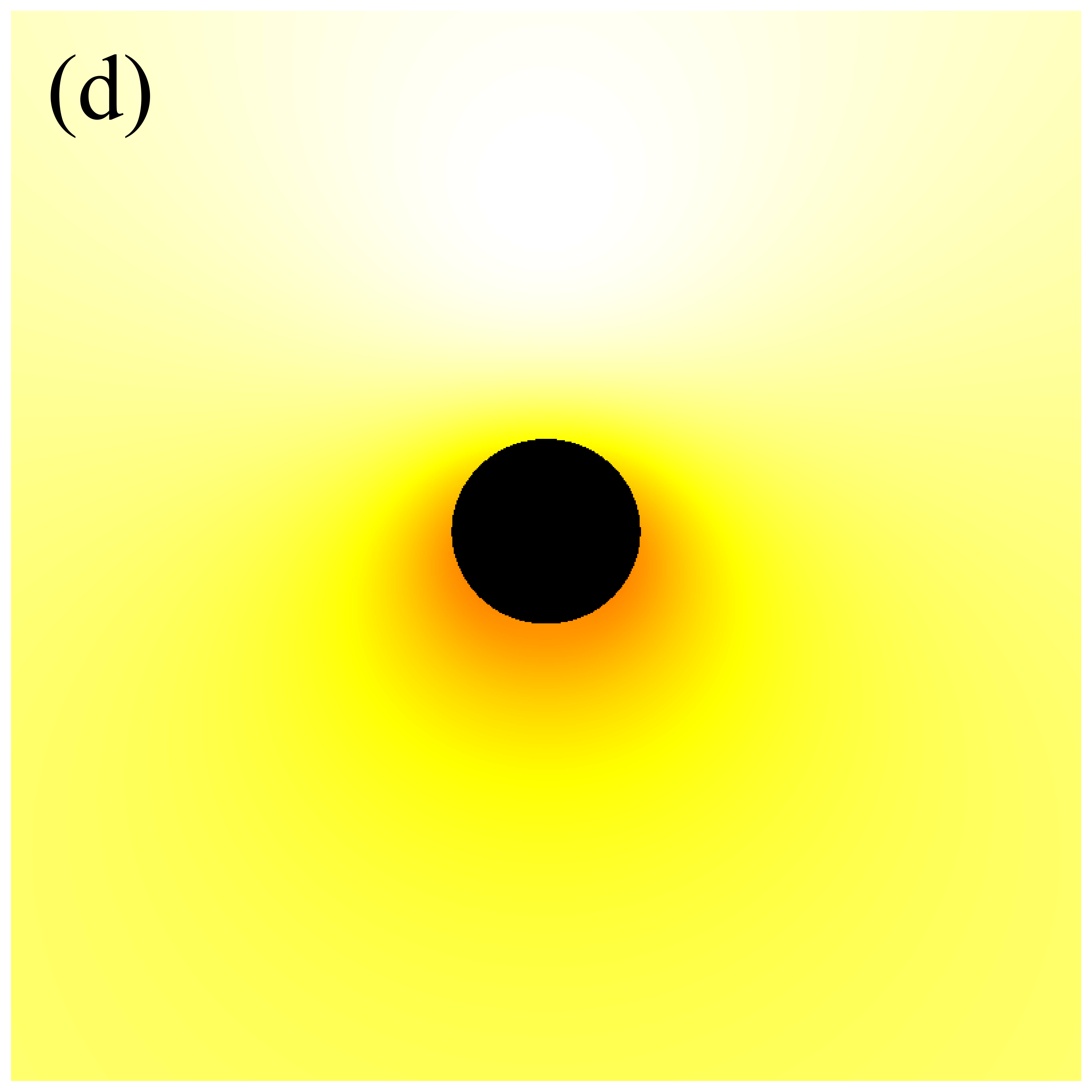}
\includegraphics[width=2.9cm]{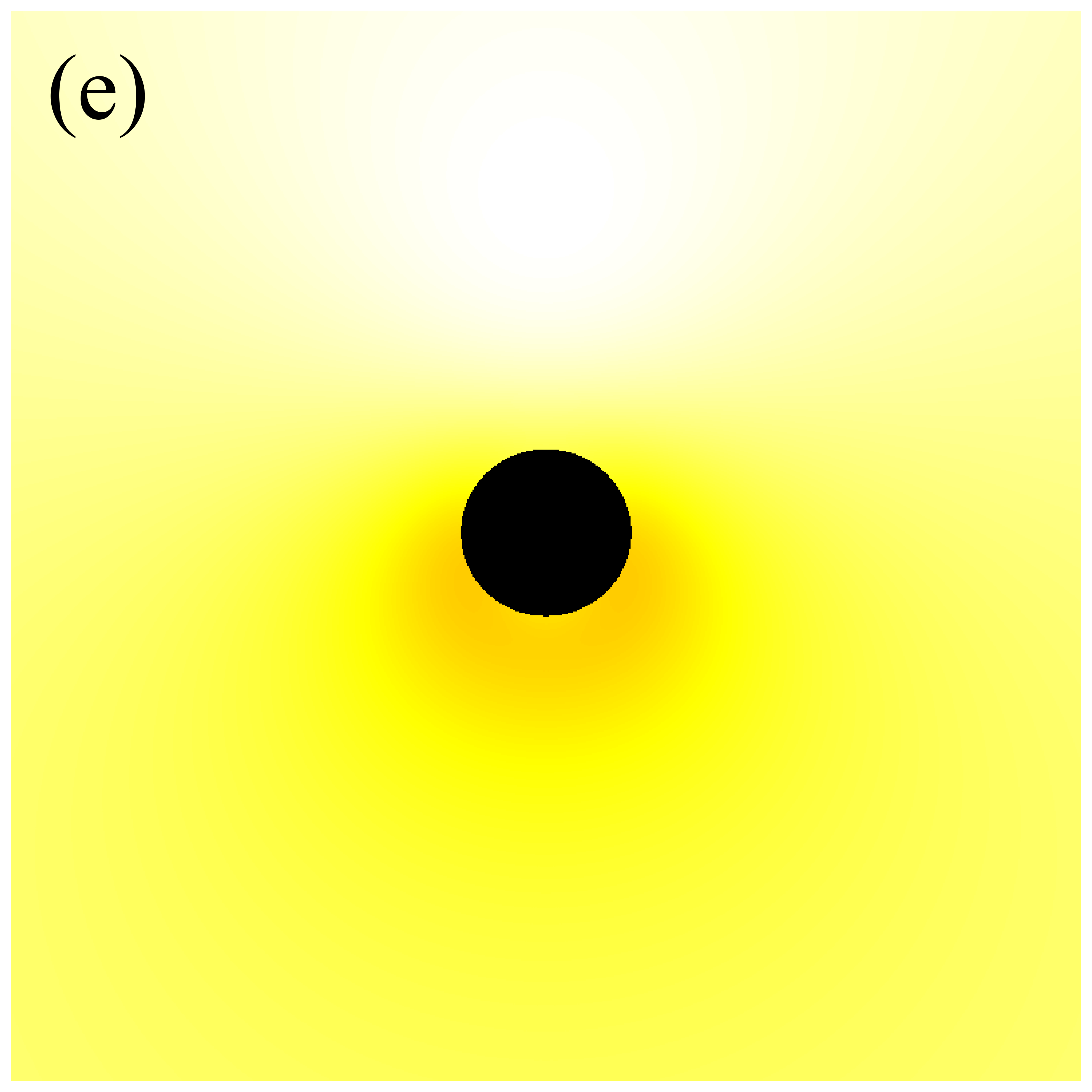}
\includegraphics[width=2.9cm]{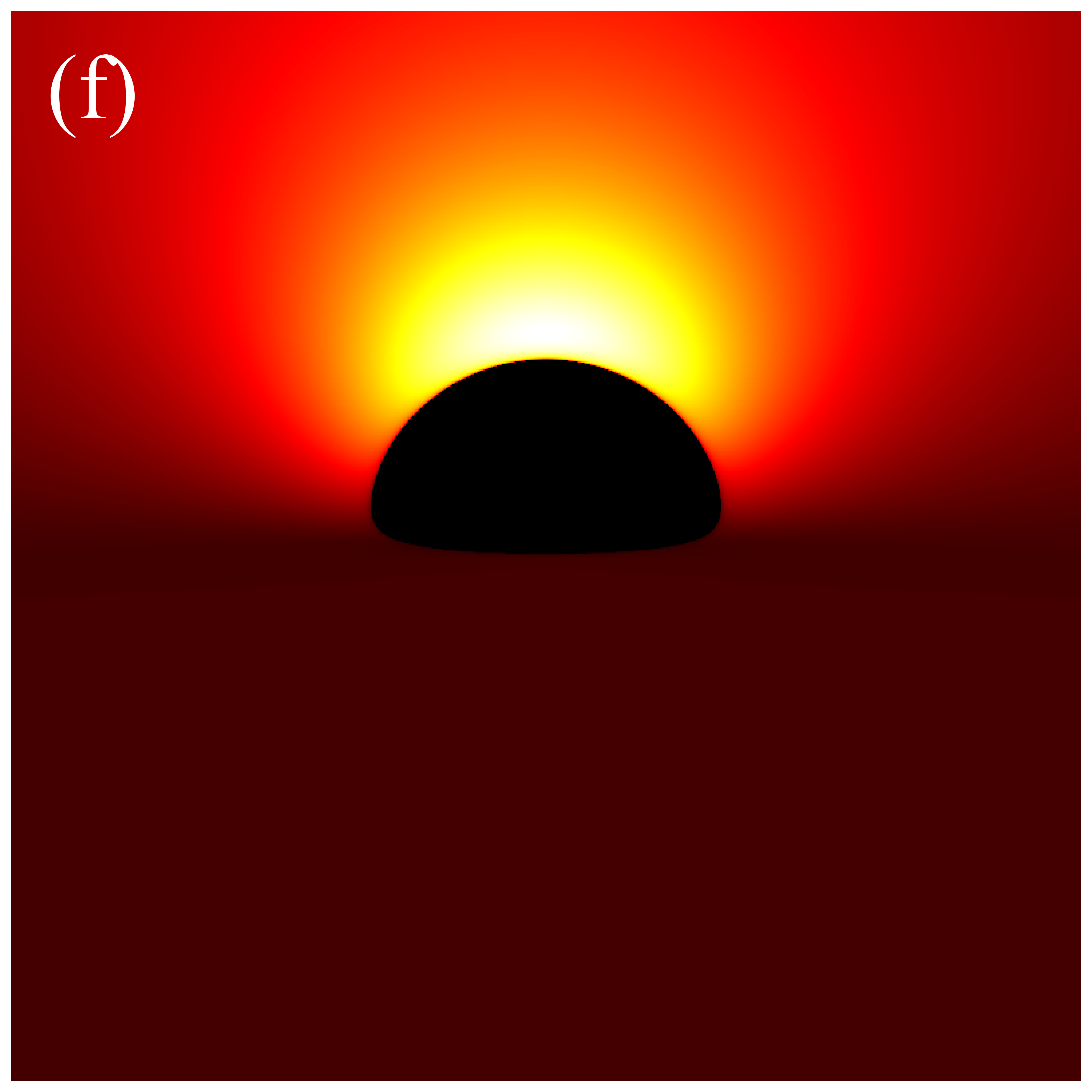}
\includegraphics[width=2.9cm]{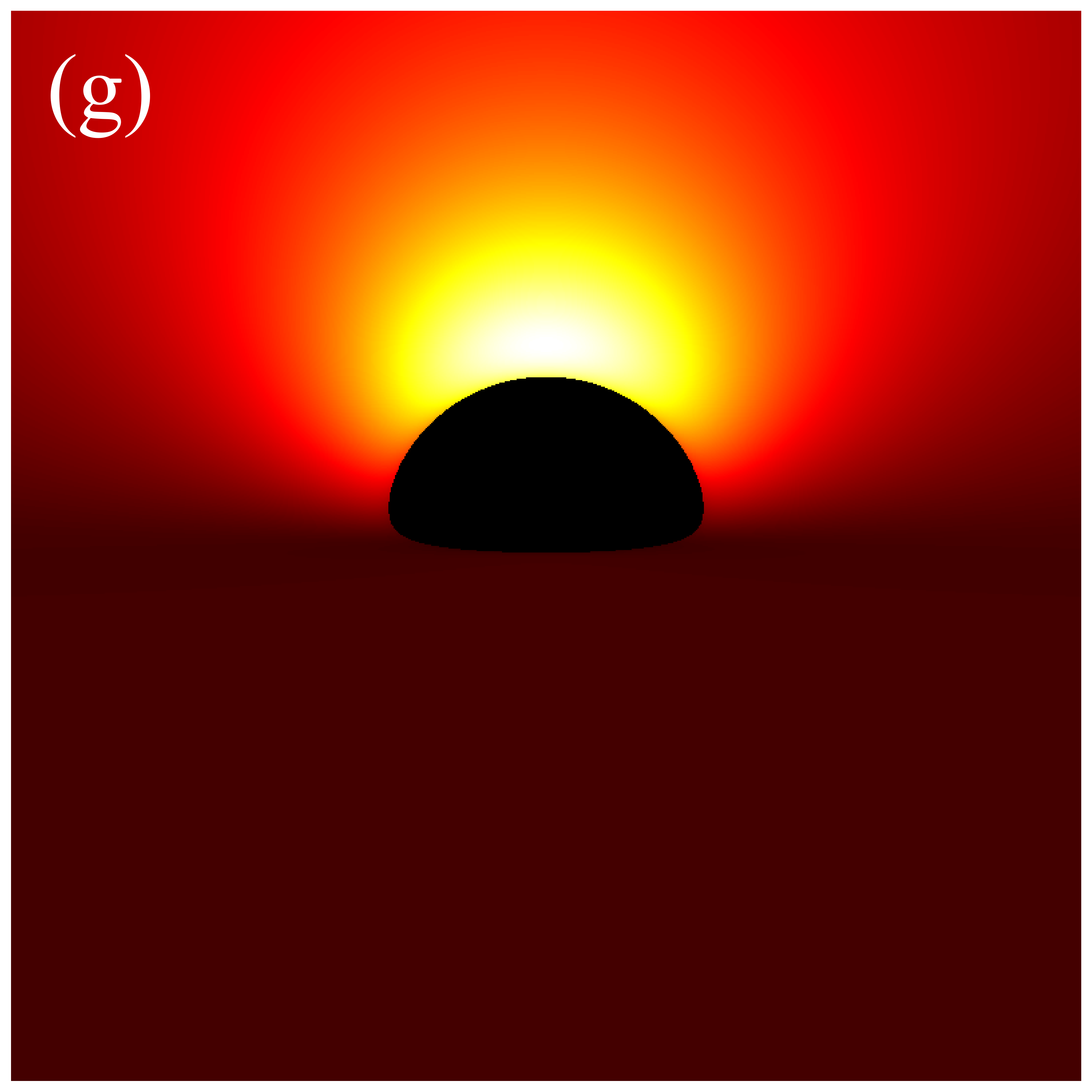}
\includegraphics[width=2.9cm]{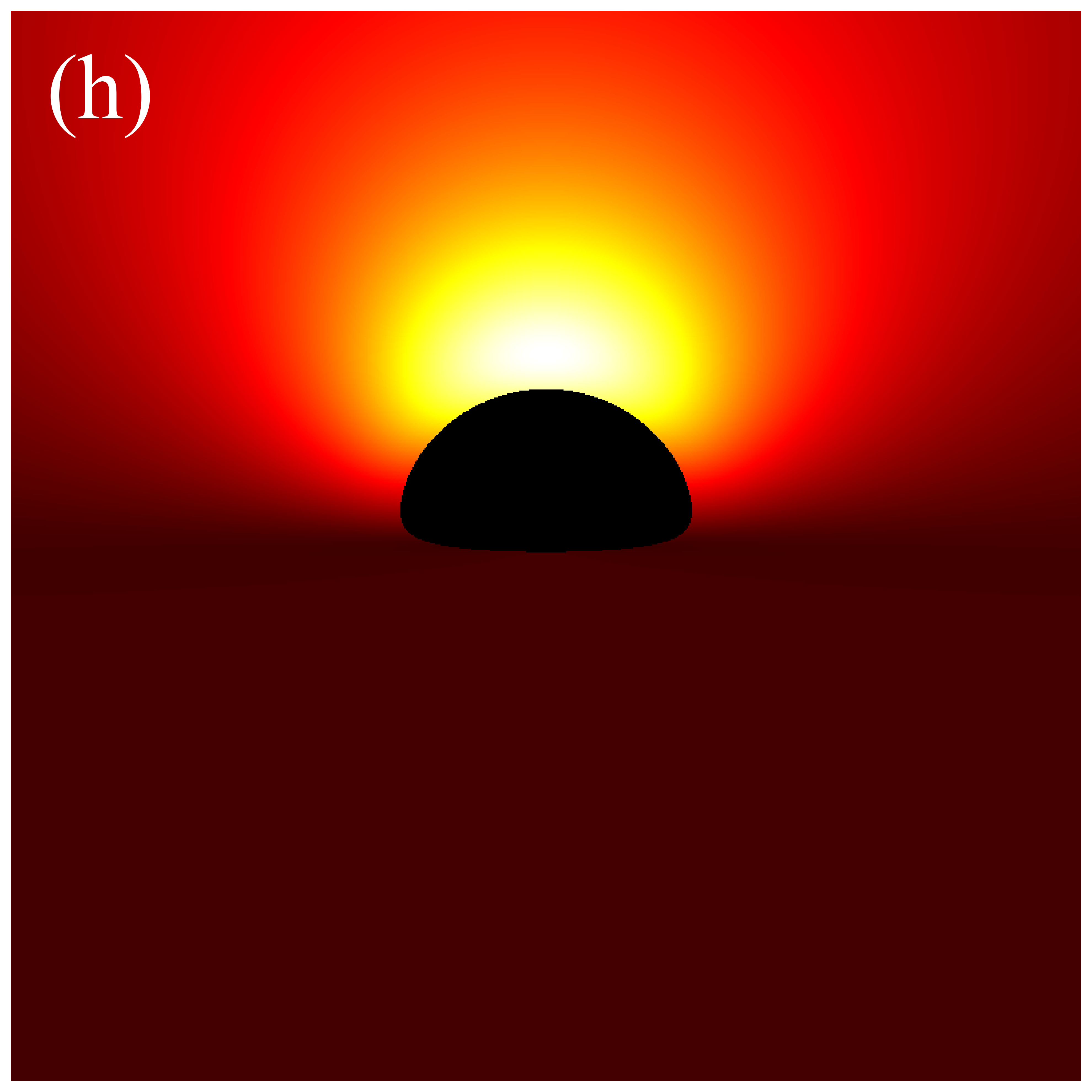}
\includegraphics[width=2.9cm]{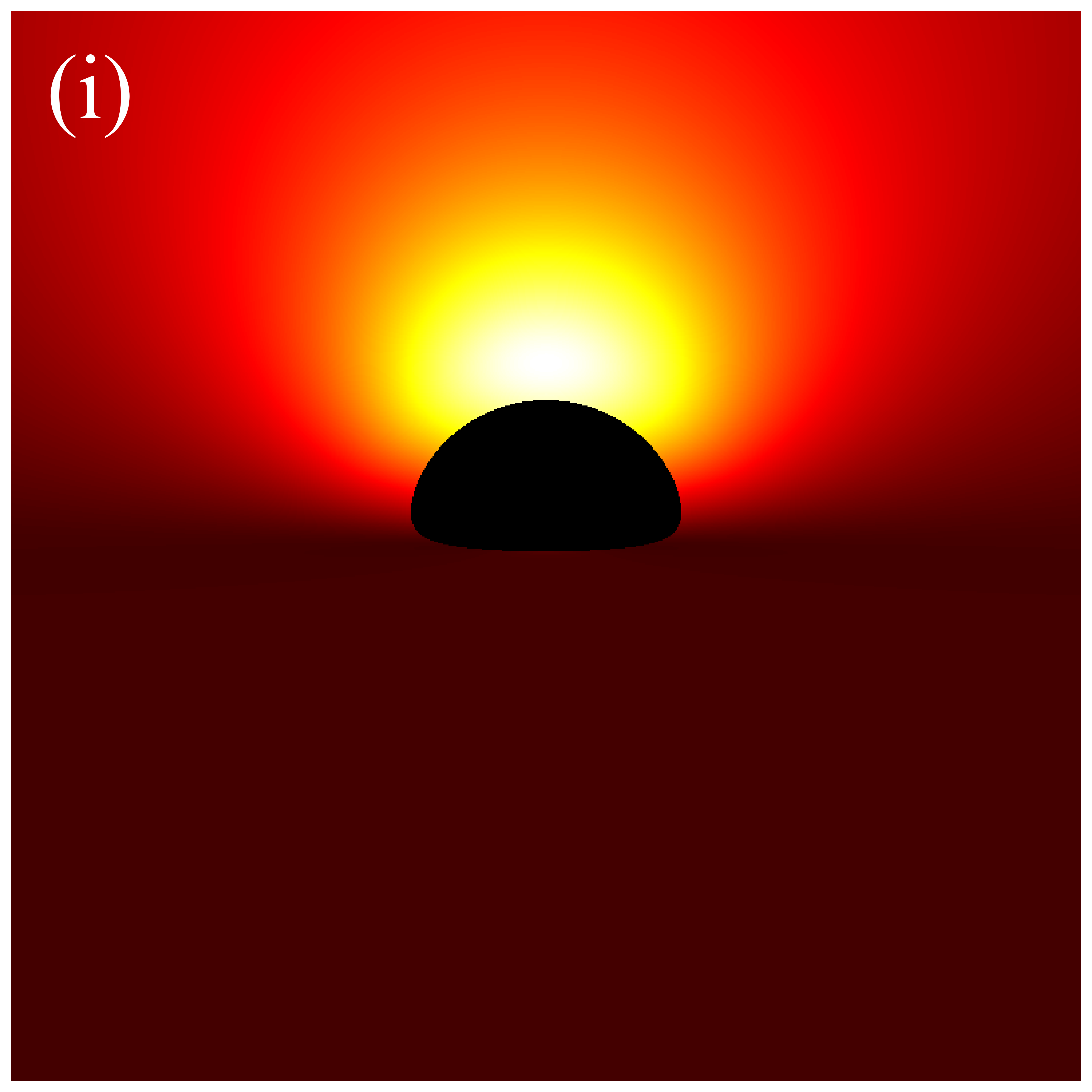}
\includegraphics[width=2.9cm]{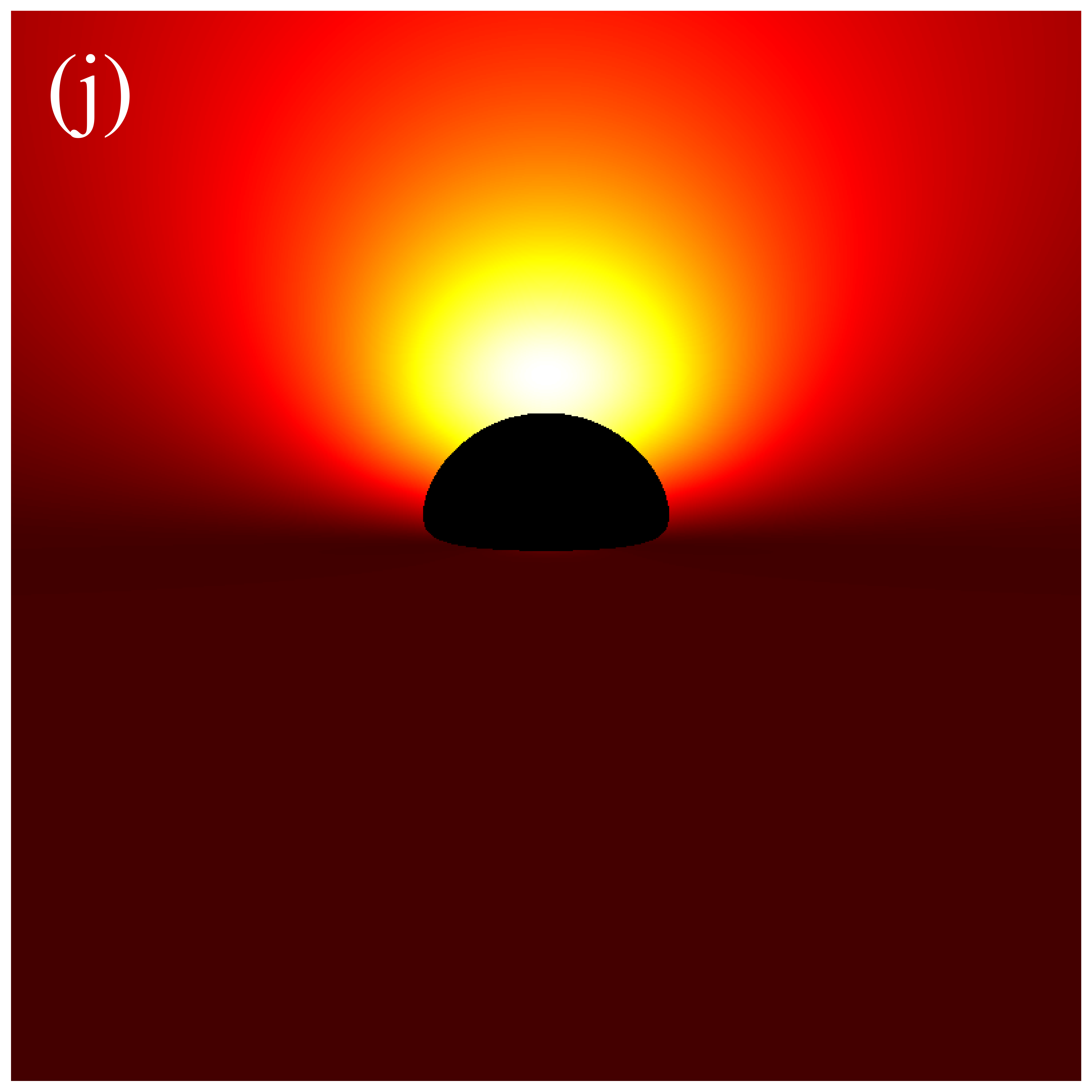}
\caption{Distribution of $\cos\delta$ values in the direct images of deformed Schwarzschild black holes across different parameter spaces. From left to right, the deformation parameters are $-7$, $-3$, $0$, $3$, and $7$. The first row corresponds to an observation inclination of $17^{\circ}$, while the second row corresponds to $85^{\circ}$. The values of $\cos\delta$ are represented using a continuous, linear color map transitioning from dark red to white, where dark red and white correspond to $\cos\delta=0$ and $\cos\delta=1$, respectively. It is evident that, due to gravitational lensing effects, light rays with smaller $\delta$ values are concentrated towards the upper part of the image.}}\label{fig8}
\end{figure*}

Next, we turn our attention to the image characteristics of deformed Schwarzschild black holes surrounded by an anisotropic accretion disk, and elucidate the impact of the accretion disk's projection effect on black hole images. The projection effect primarily arises from the introduction of the angle $\delta$ between the light rays and the normal to the accretion disk, making it essential to examine the evolution of $\delta$ with respect to the observation coordinates $(x,y)$. Figure 8 illustrates the distribution of $\cos\delta$ in the direct images of deformed Schwarzschild black holes. Here, dark red represents lower $\cos\delta$, which is associated with larger $\delta$, while white corresponds to higher $\cos\delta$ and smaller $\delta$, indicating that light rays are less deviated from the normal to the disk plane. It is observed that $\delta$ correlates with the deformation parameter: a smaller deformation parameter corresponds to a larger $\delta$, suggesting that the convergence effect of the deformed Schwarzschild black hole on light rays is more pronounced. This finding further supports the negative correlation between the curvature of the deformed Schwarzschild spacetime and the deformation parameter. Moreover, it is apparent that light rays originating from the upper half of the observation plane typically exhibit smaller $\delta$. This is due to gravitational lensing effects, which cause these light rays to be bent by the black hole and hit the accretion disk behind the black hole at an angle nearly parallel to the disk's normal (e.g., the red ray in figure 2). Consequently, this portion of the image satisfies $\kappa\approx 1$, and is thus almost unaffected by the projection effect. However, light rays with smaller $\delta$ are concentrated within a limited region, while the remaining portions of the observation plane generally feature rays hitting the accretion disk at larger $\delta$. This phenomenon becomes more pronounced at higher observation inclinations (e.g., the second row of figure 8 or the blue ray in figure 2). These observations indicate that the projection effect of an anisotropic accretion disk indeed has the potential to alter the observed characteristics of black hole images, particularly those of direct images, highlighting the need for further investigation.
\begin{figure*}%[tbph]
\center{
\includegraphics[width=3.5cm]{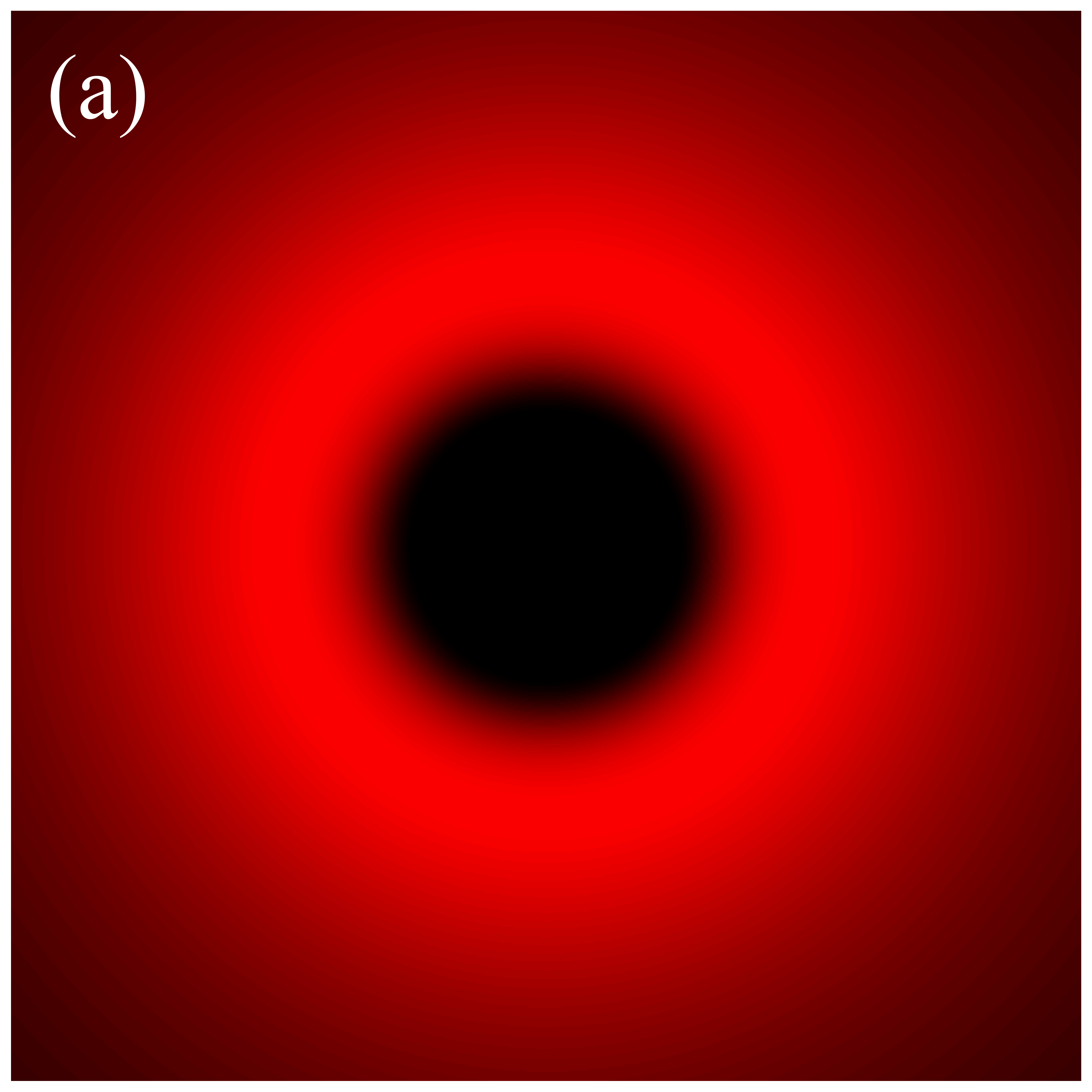}
\includegraphics[width=3.5cm]{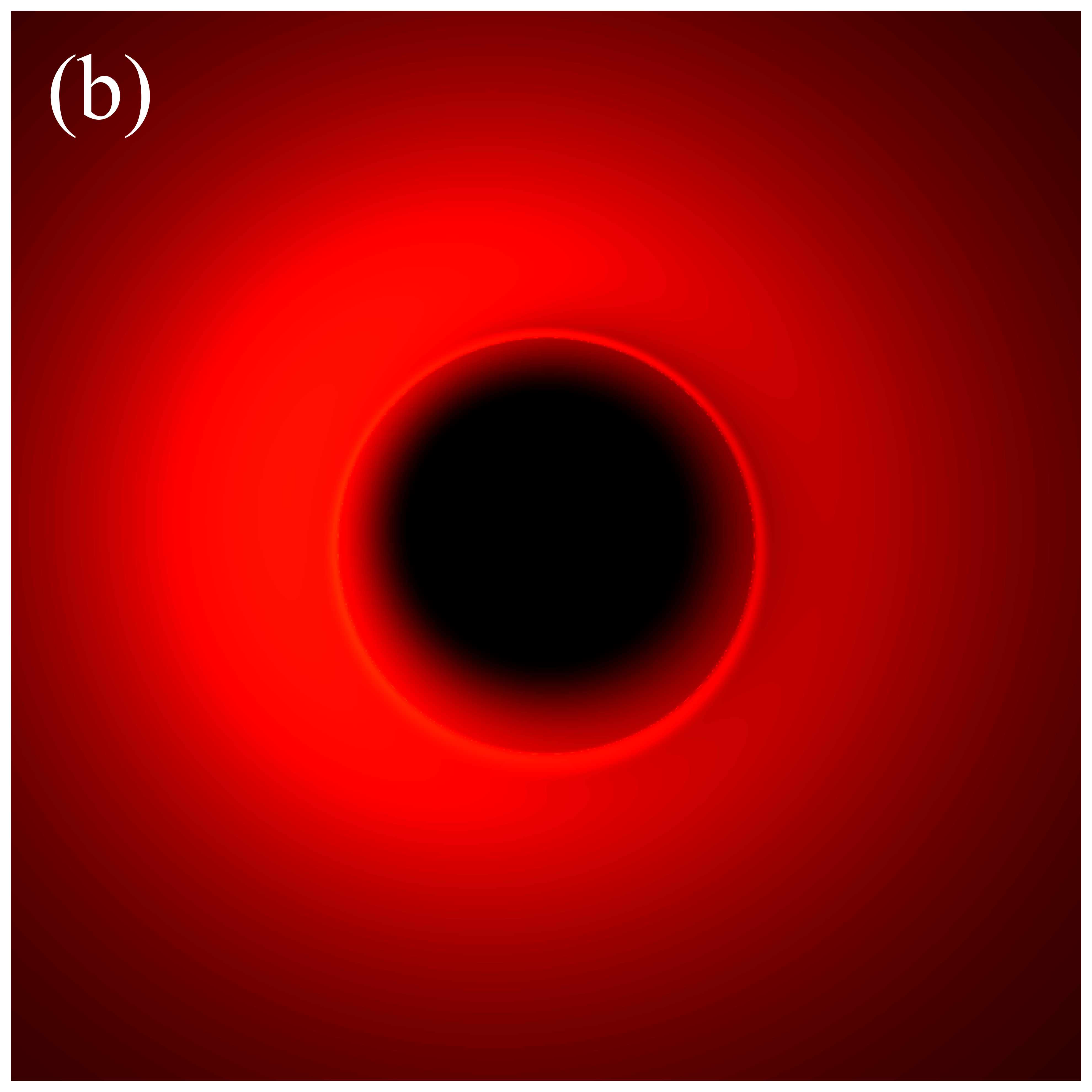}
\includegraphics[width=3.5cm]{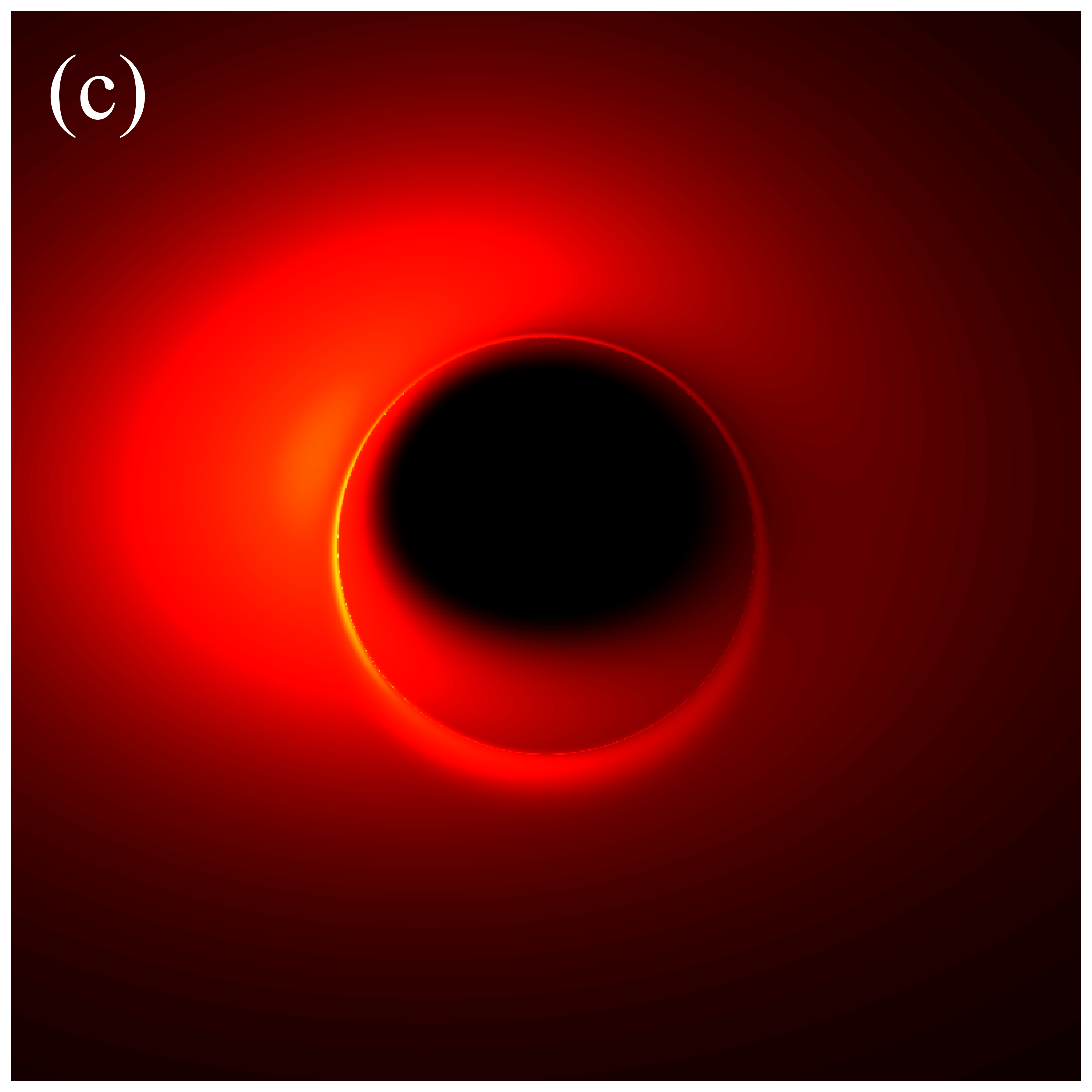}
\includegraphics[width=3.5cm]{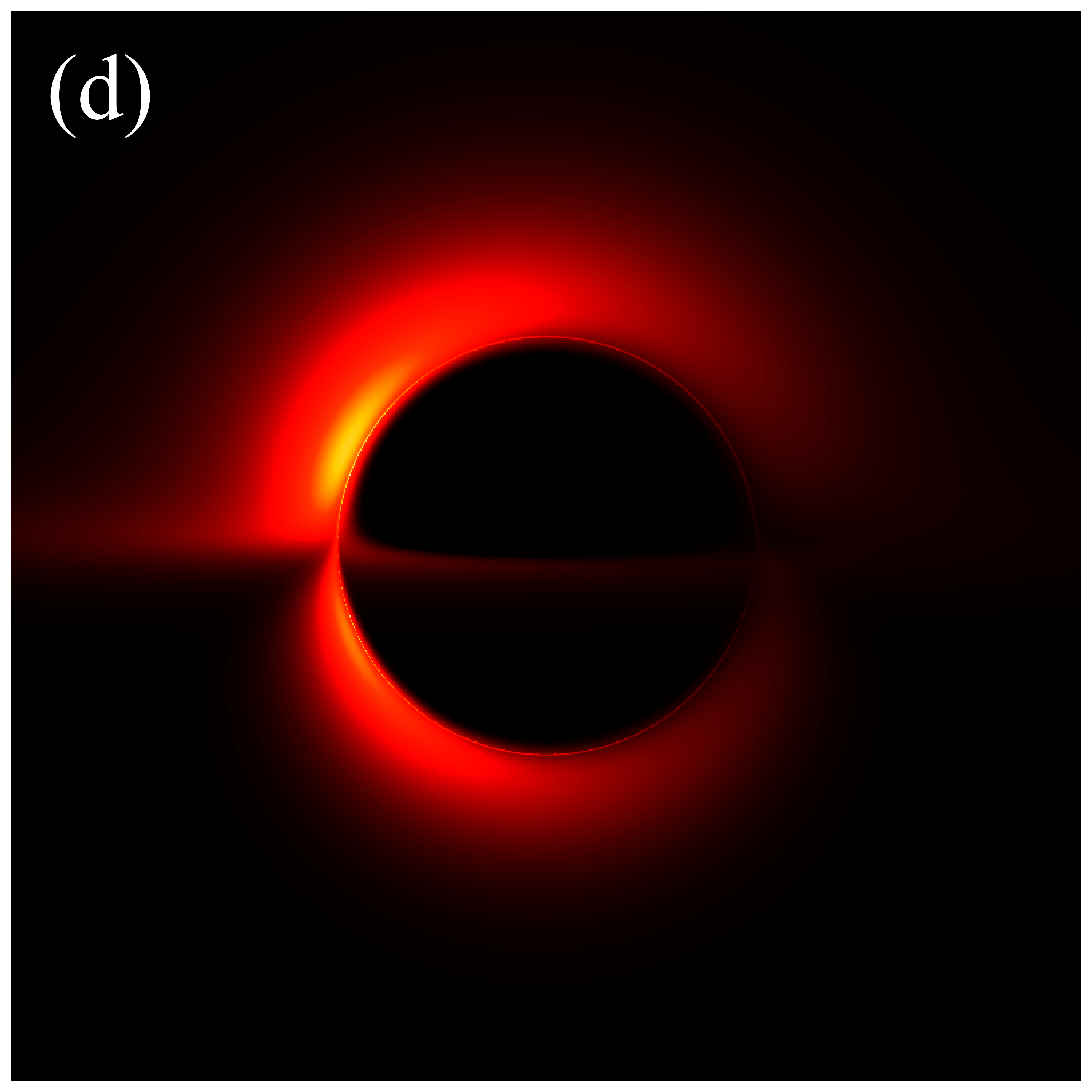}
\includegraphics[width=3.5cm]{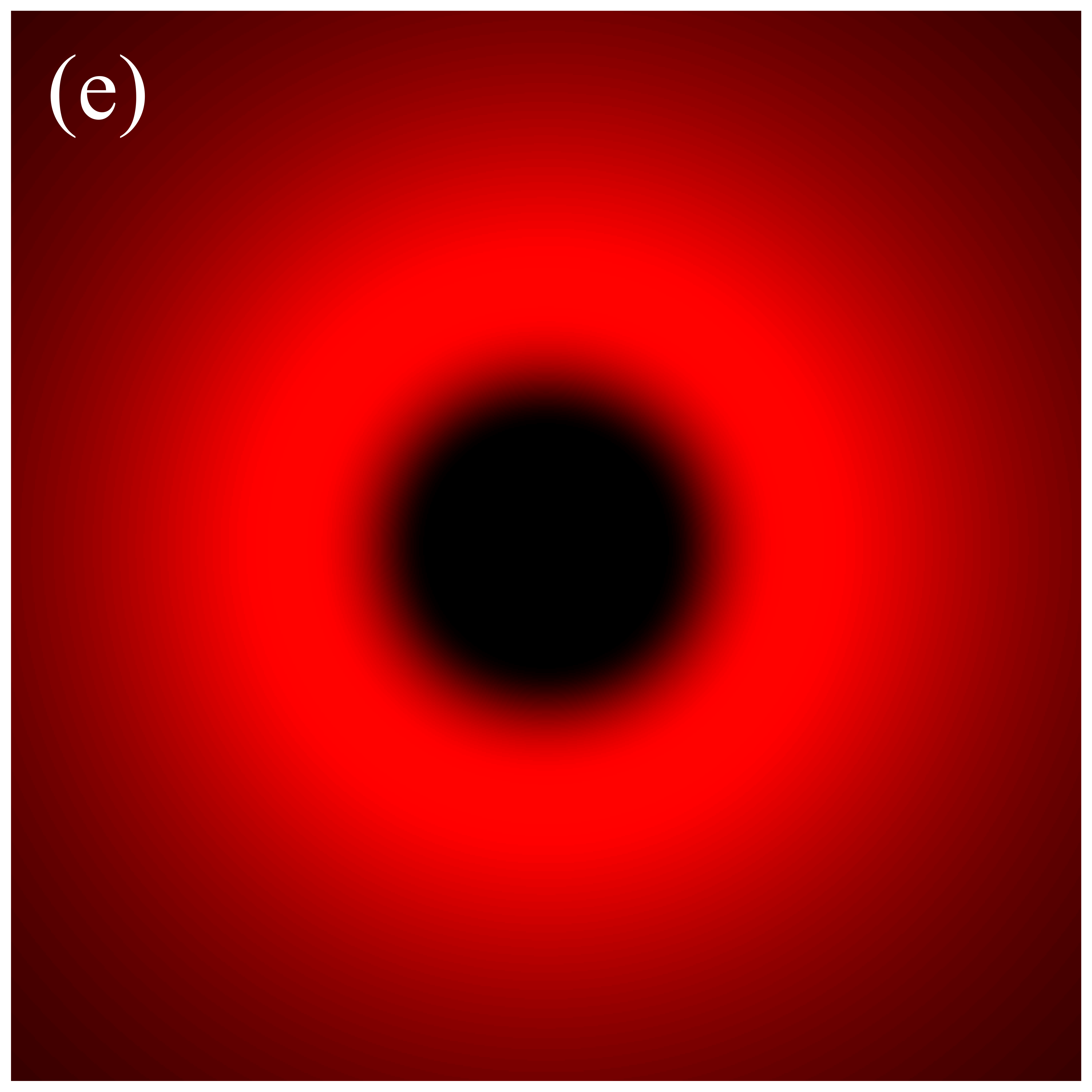}
\includegraphics[width=3.5cm]{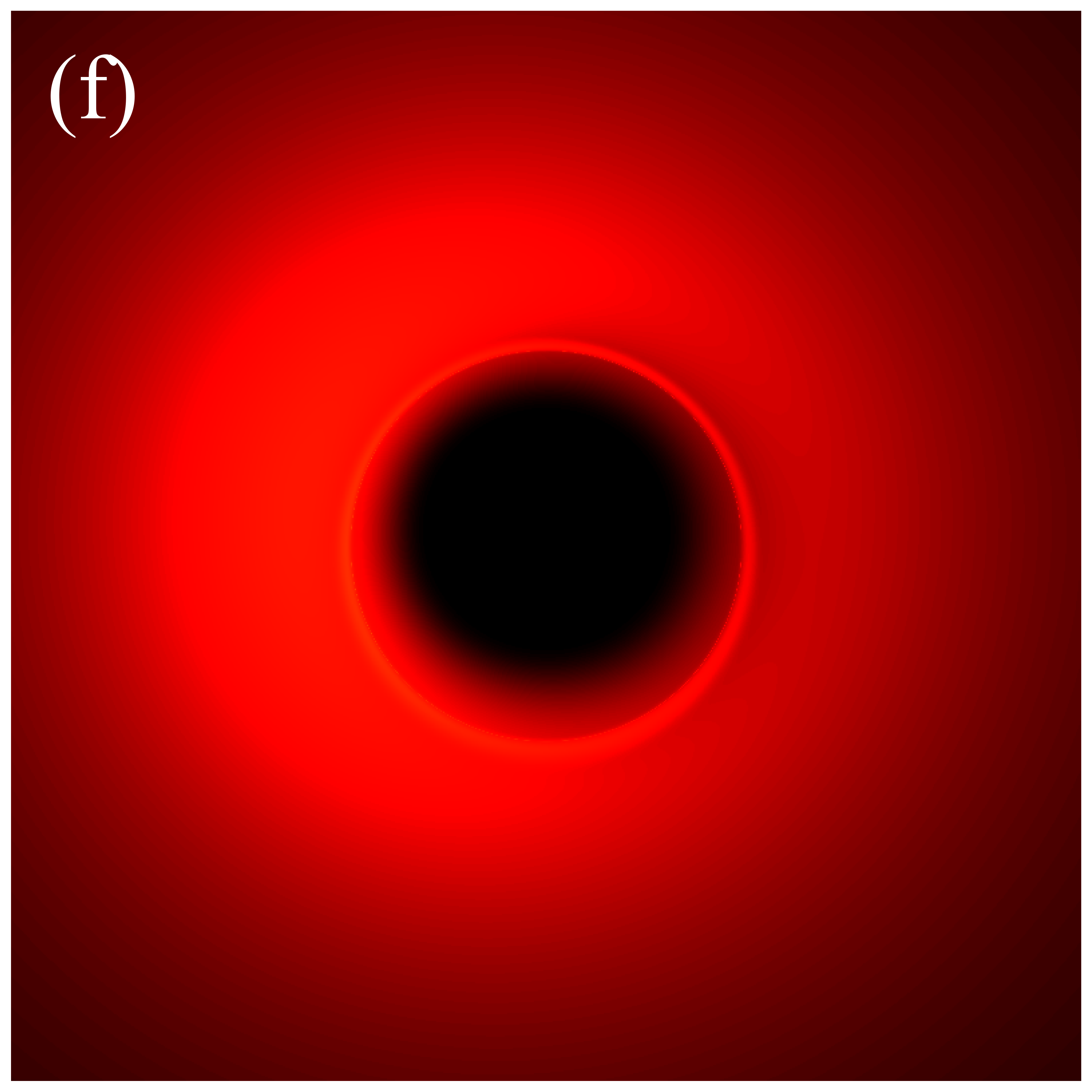}
\includegraphics[width=3.5cm]{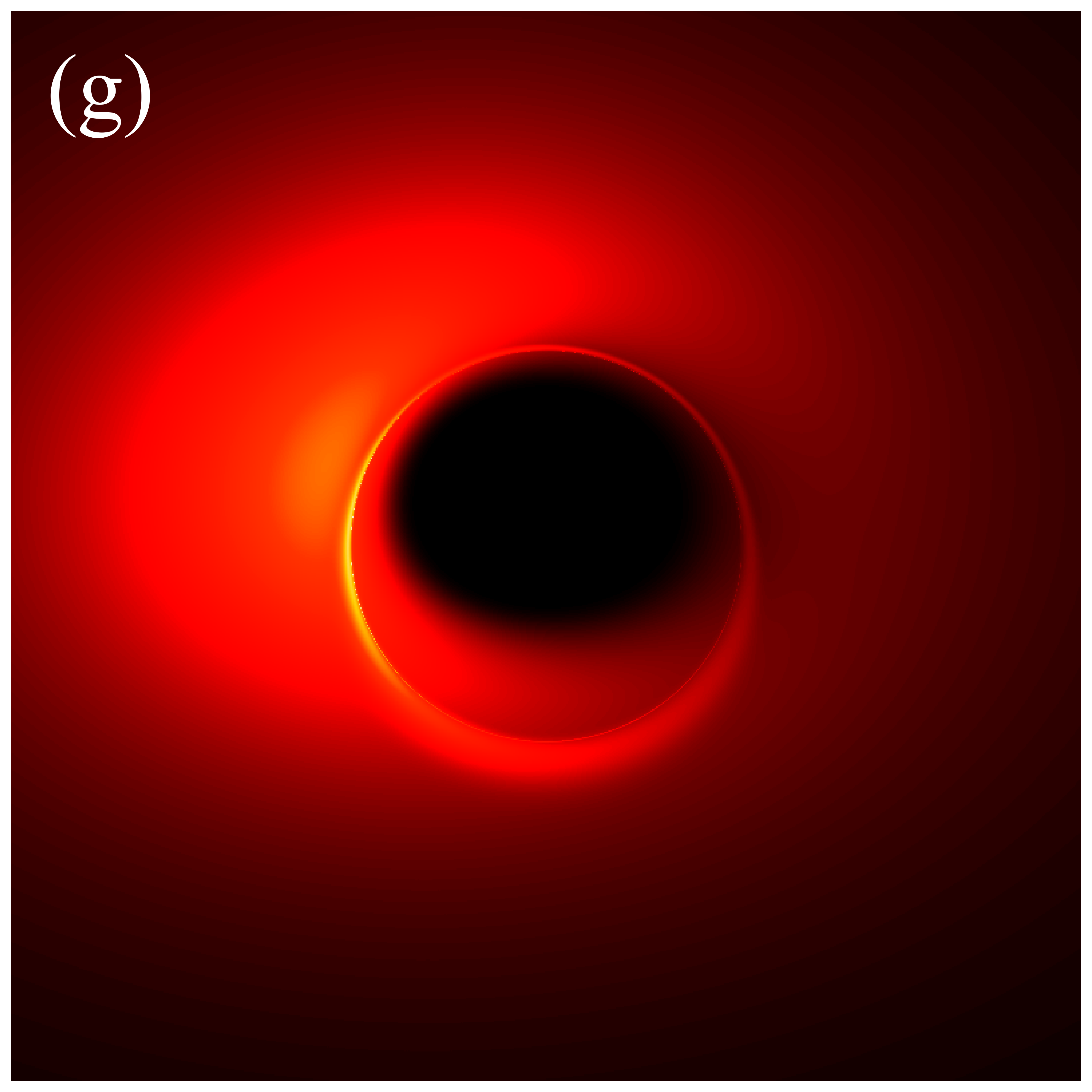}
\includegraphics[width=3.5cm]{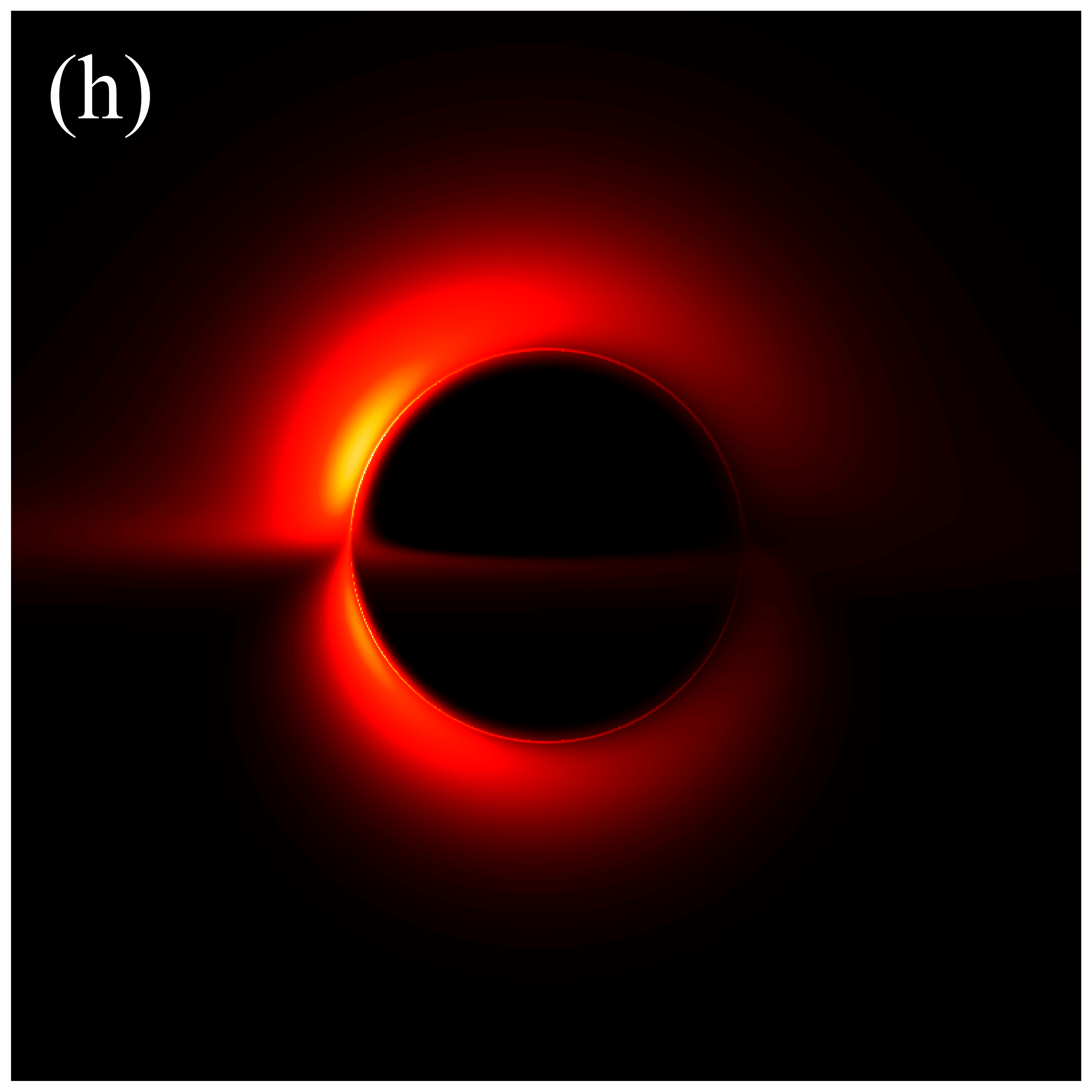}
\includegraphics[width=3.5cm]{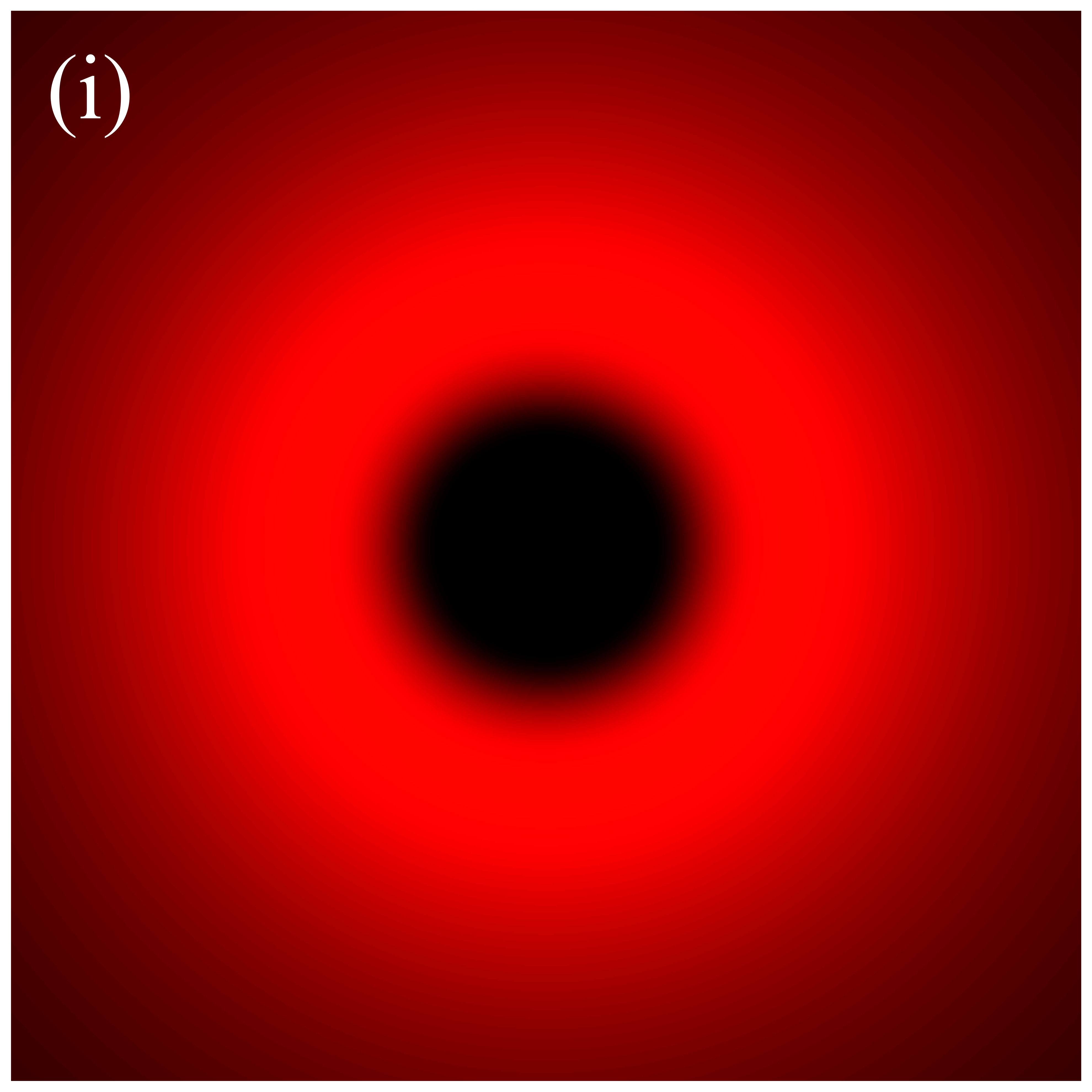}
\includegraphics[width=3.5cm]{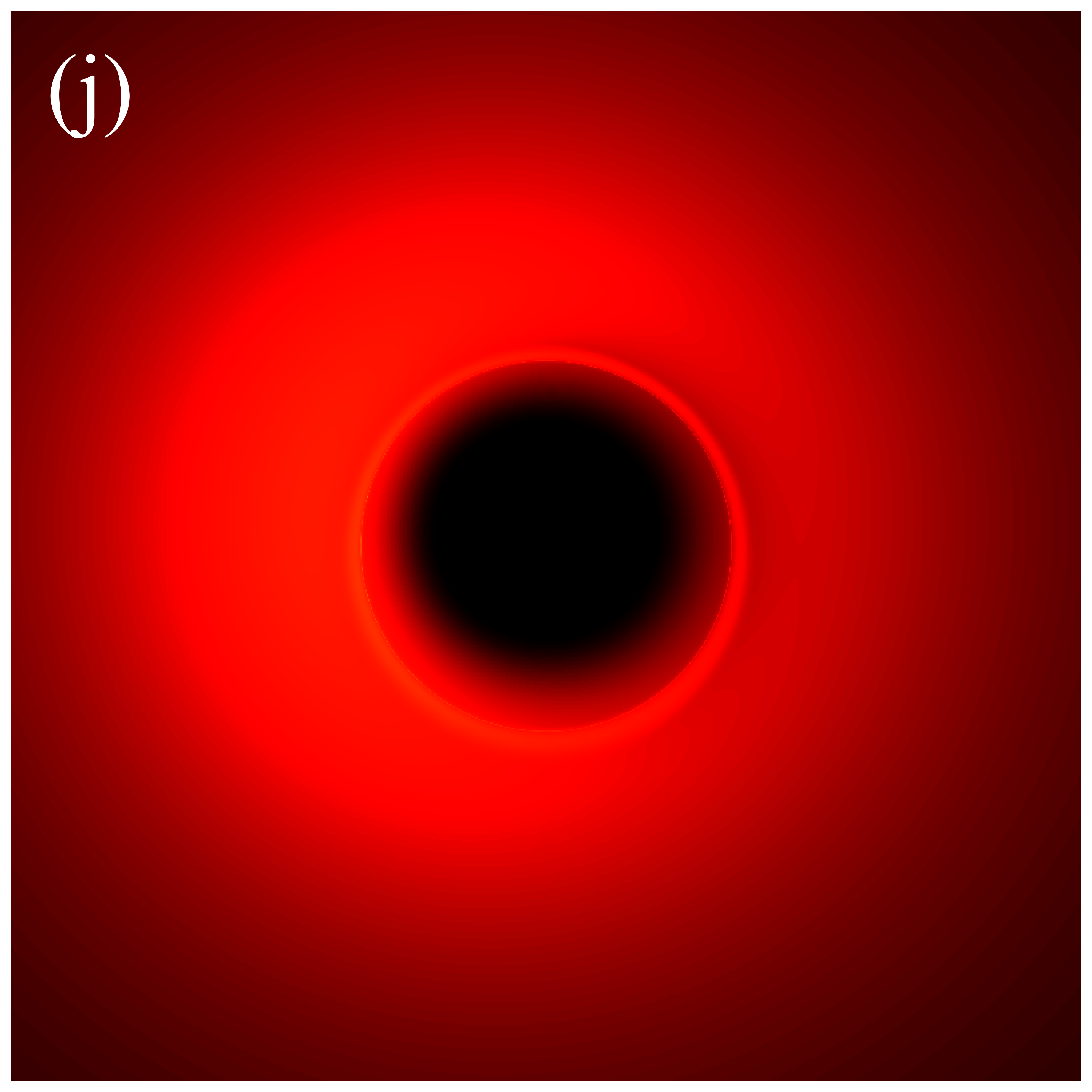}
\includegraphics[width=3.5cm]{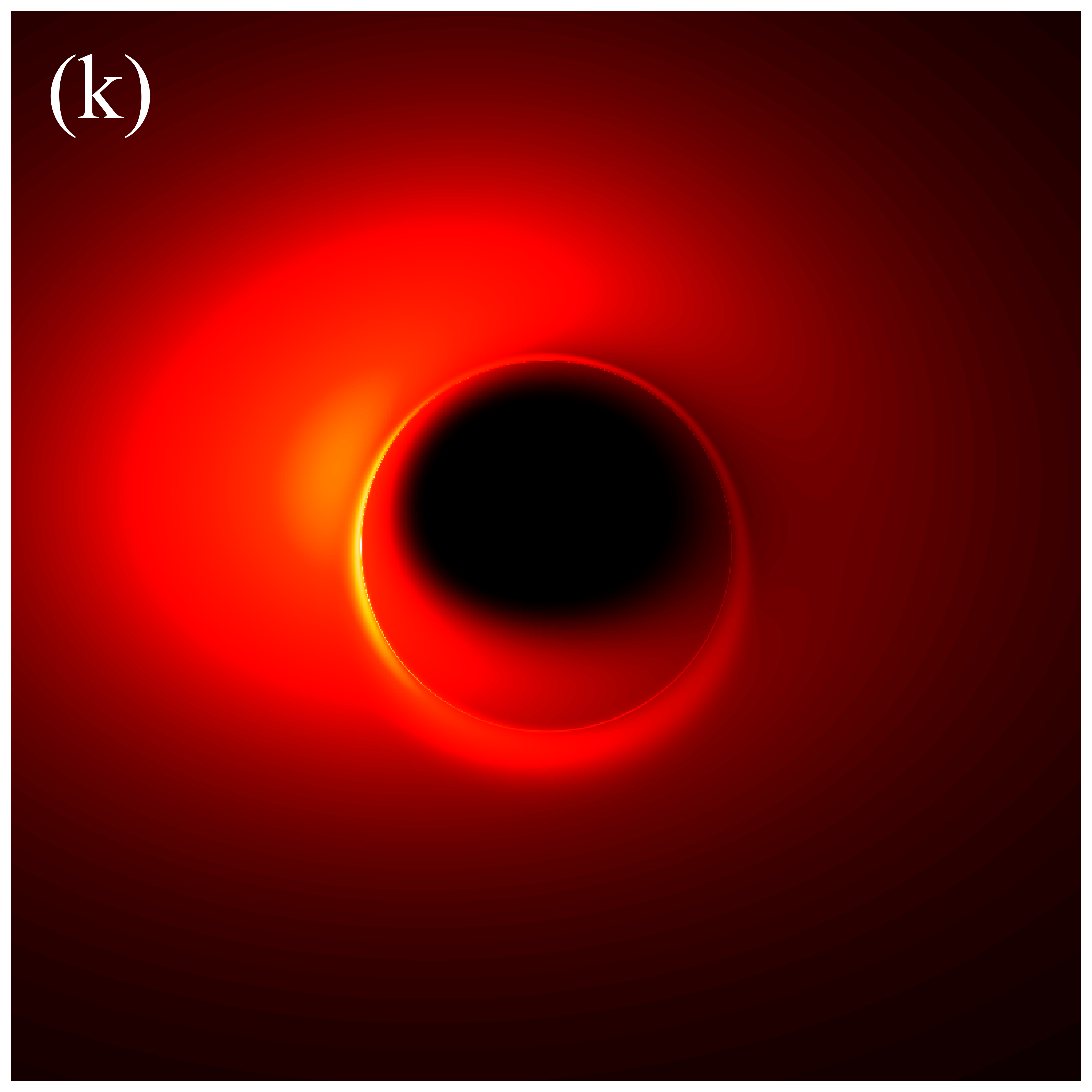}
\includegraphics[width=3.5cm]{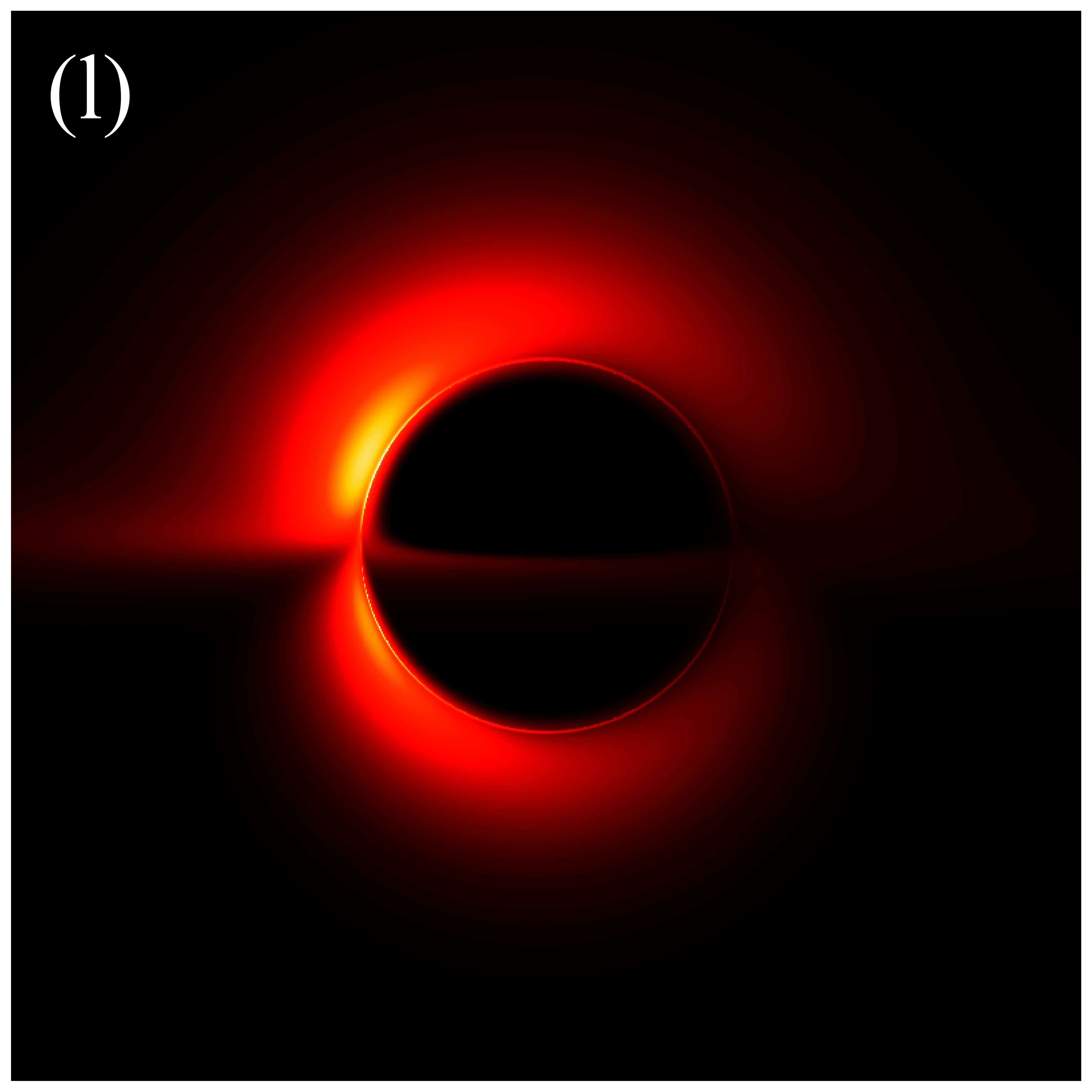}
\includegraphics[width=3.5cm]{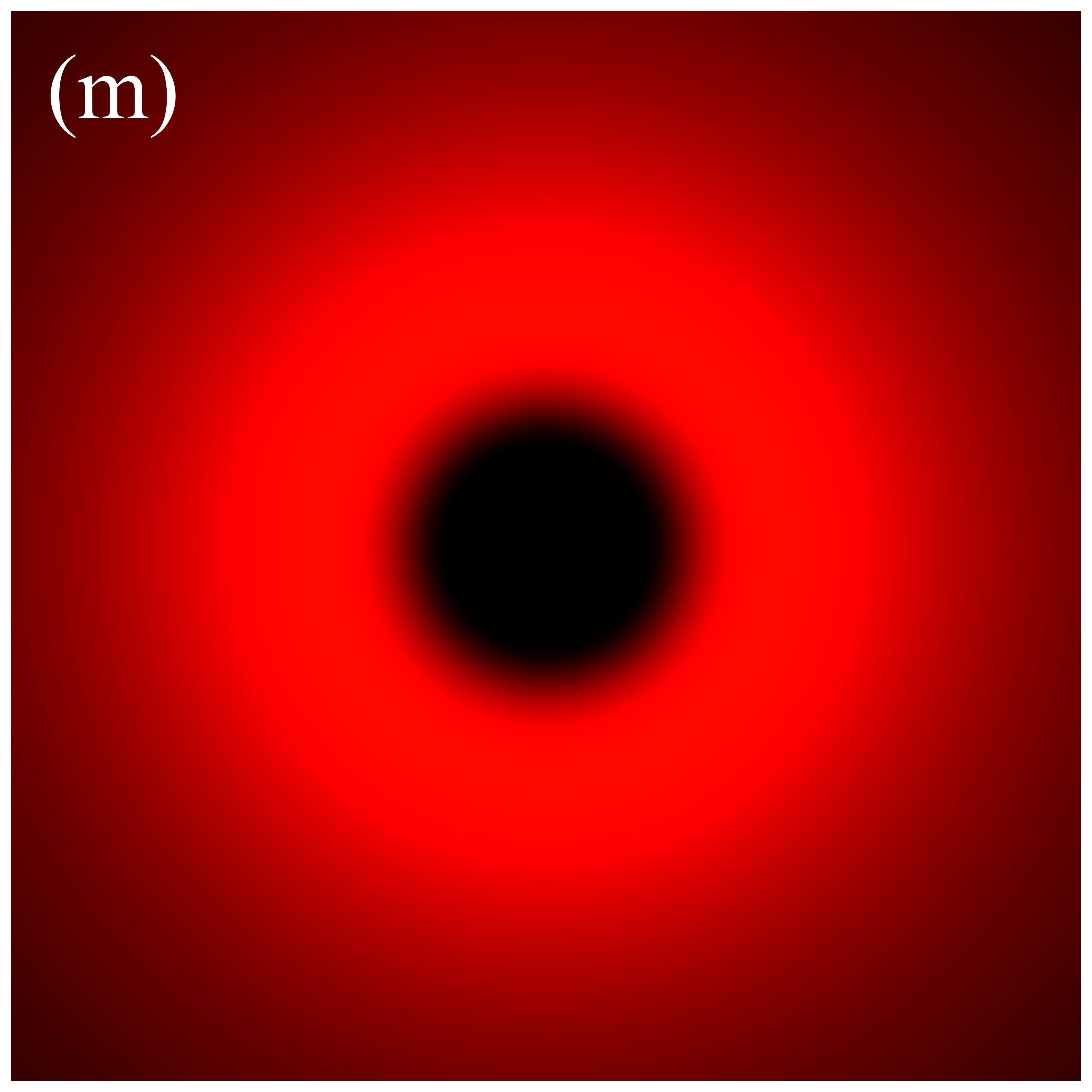}
\includegraphics[width=3.5cm]{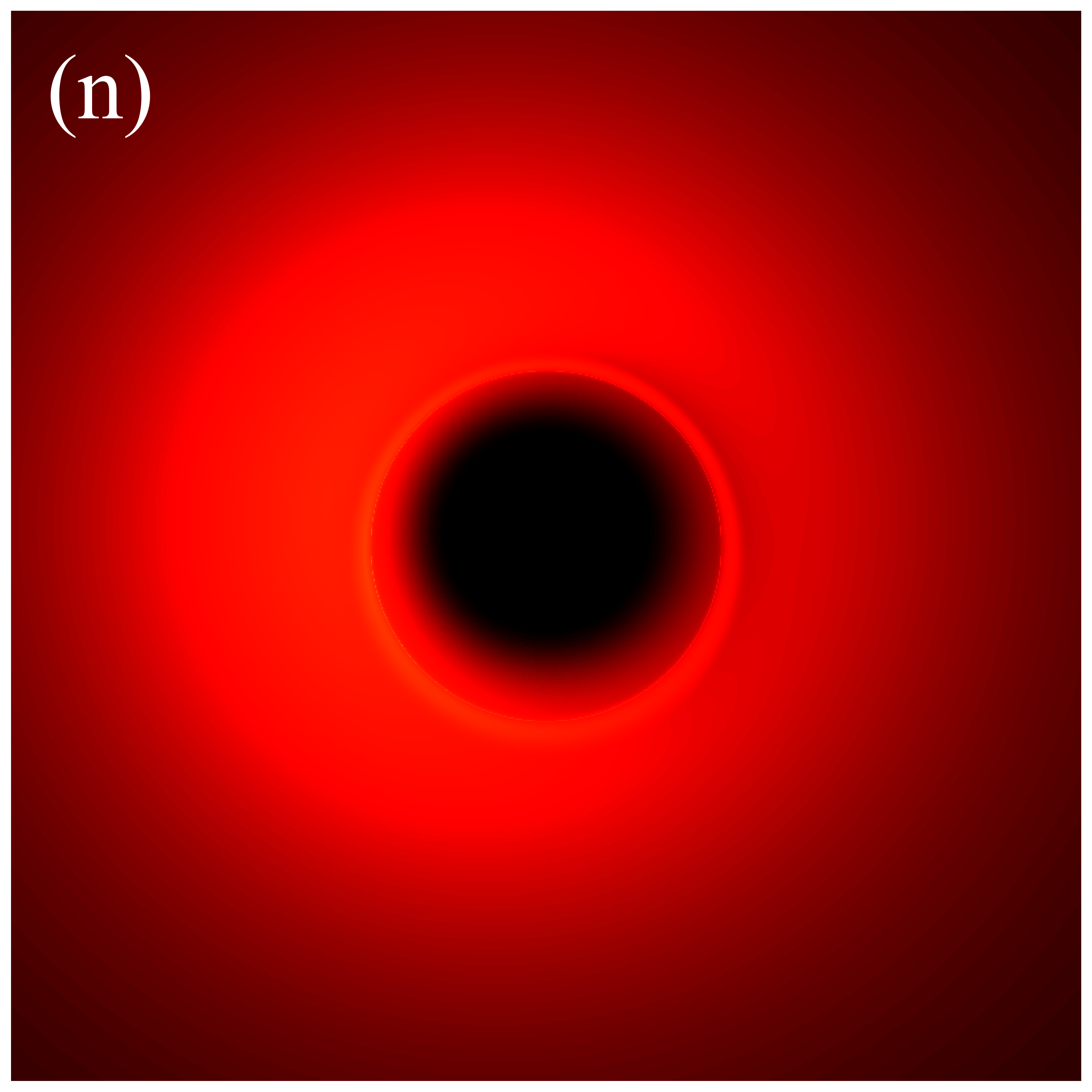}
\includegraphics[width=3.5cm]{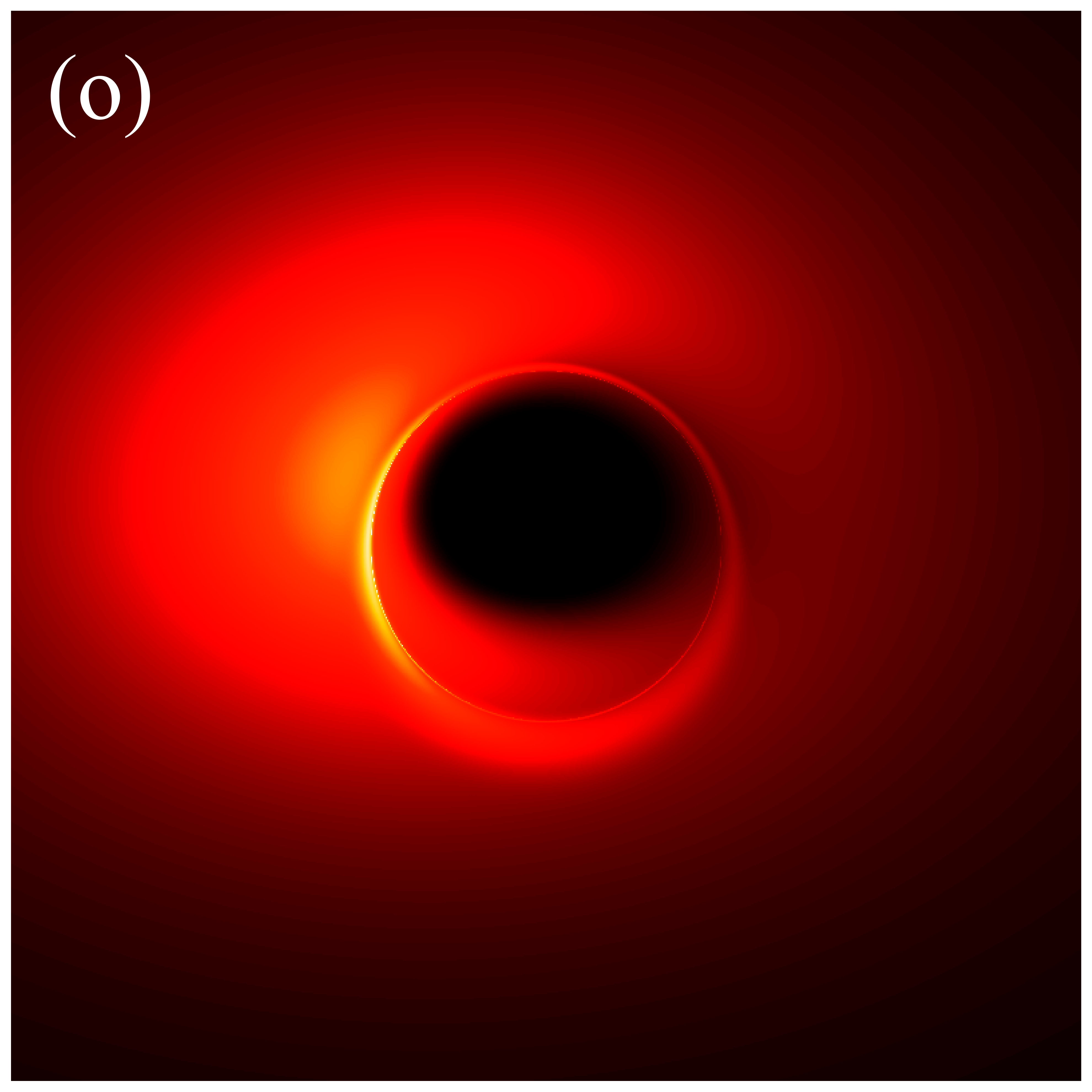}
\includegraphics[width=3.5cm]{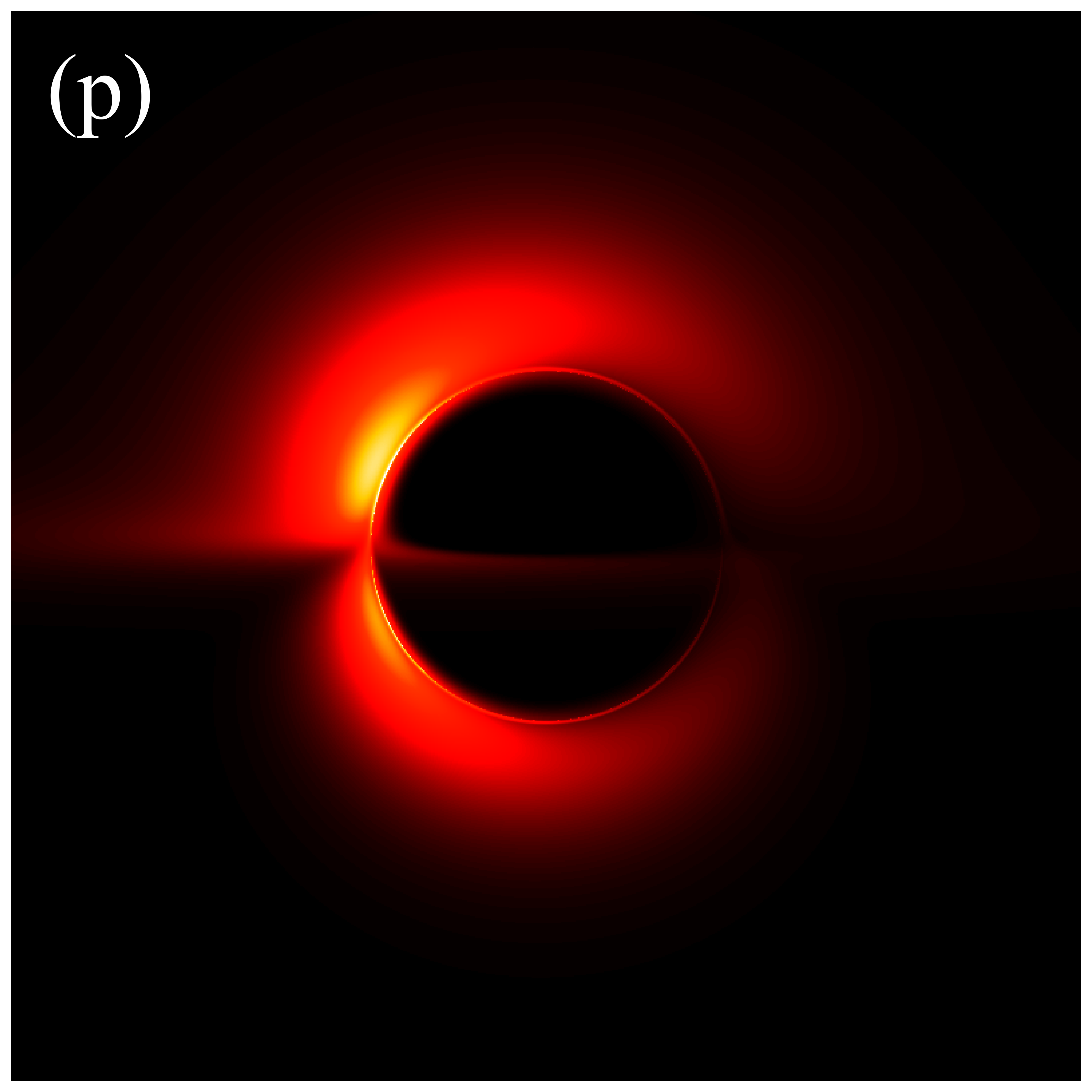}
\includegraphics[width=3.5cm]{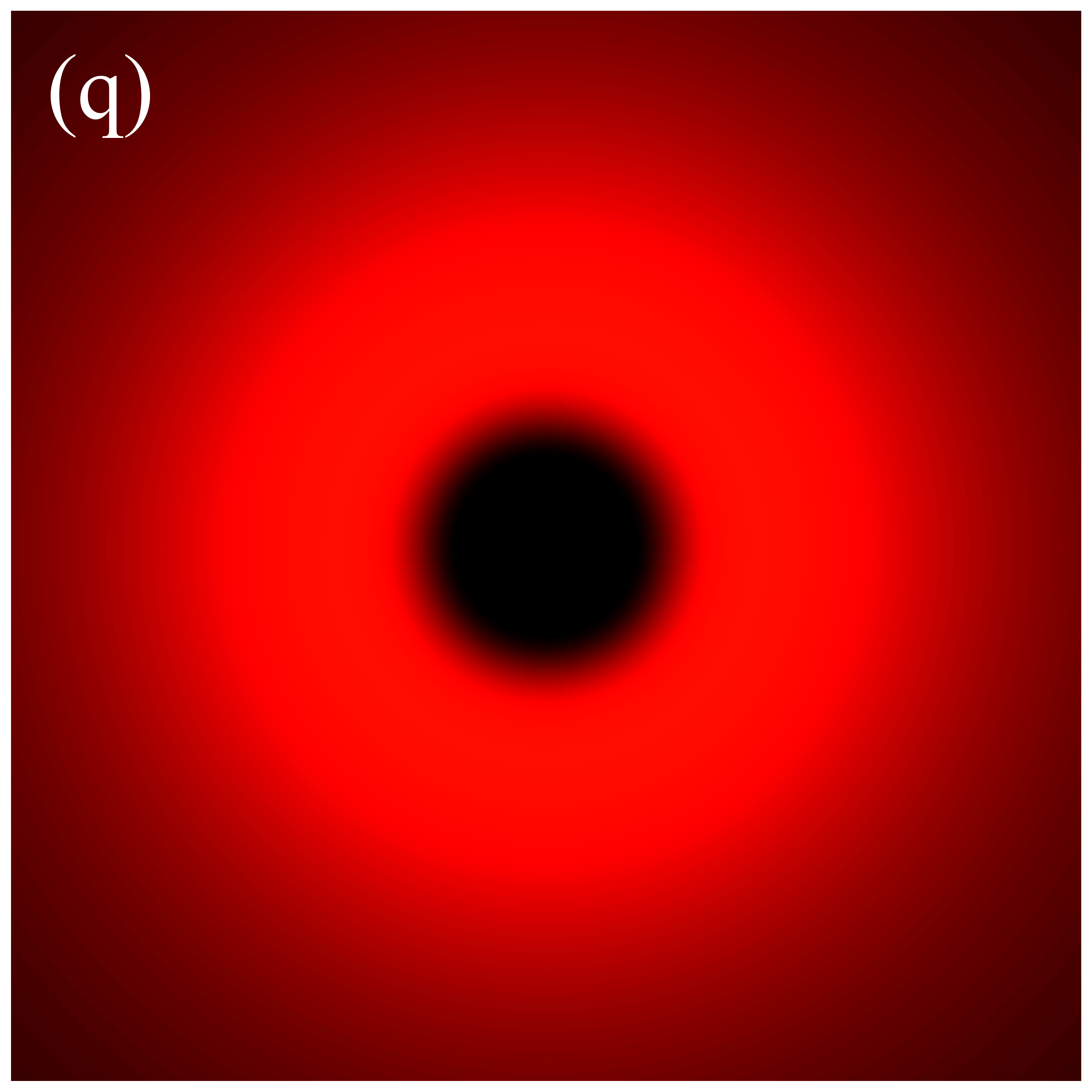}
\includegraphics[width=3.5cm]{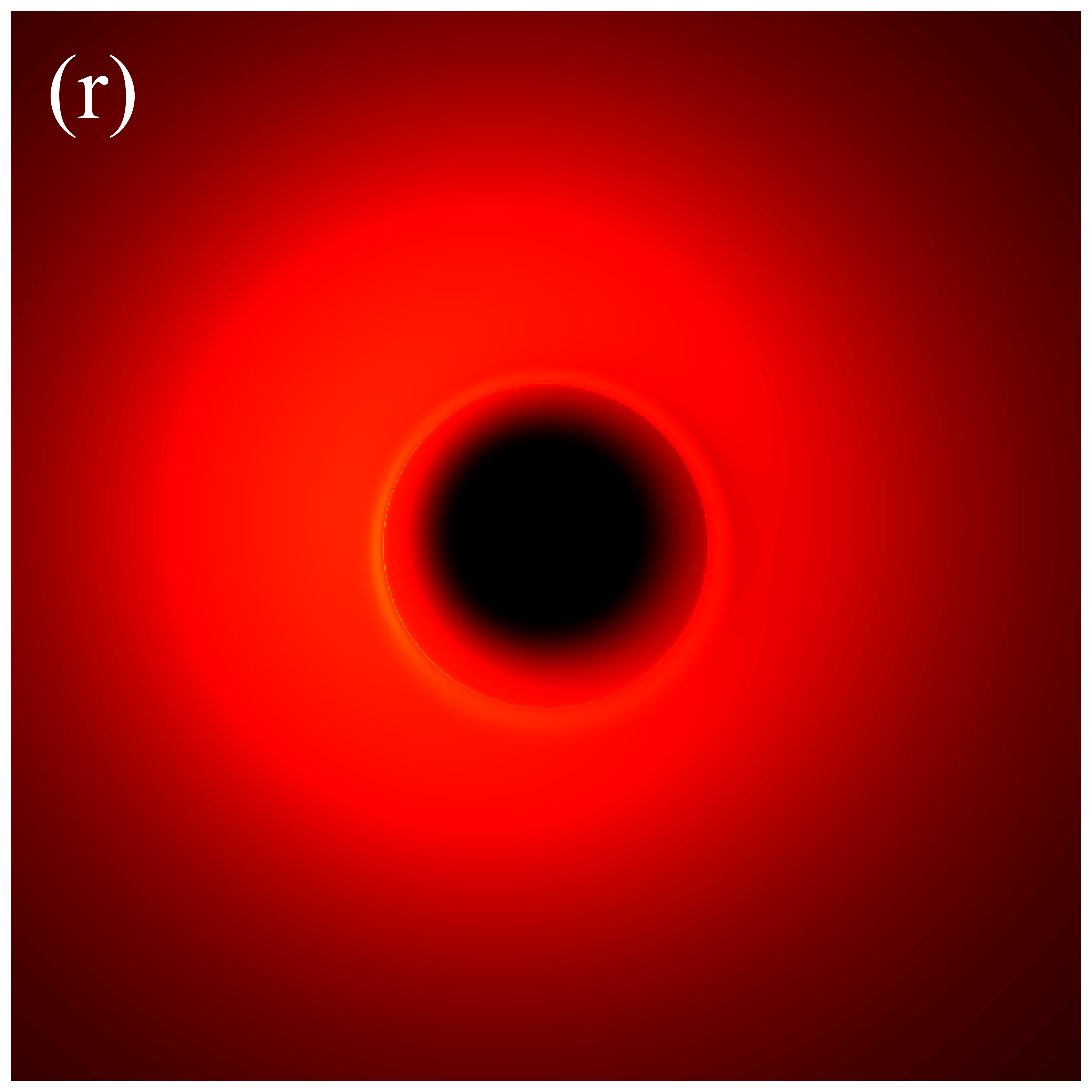}
\includegraphics[width=3.5cm]{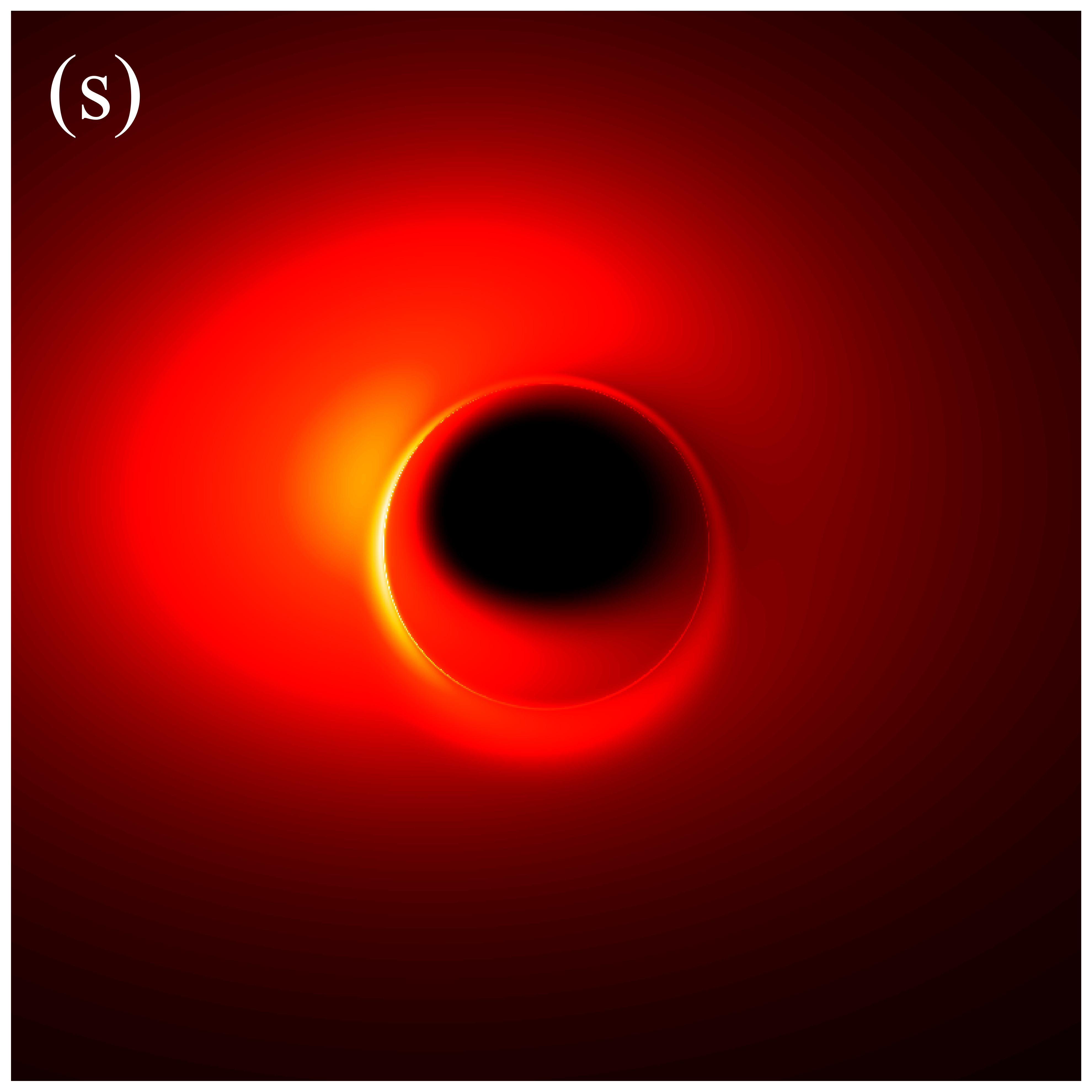}
\includegraphics[width=3.5cm]{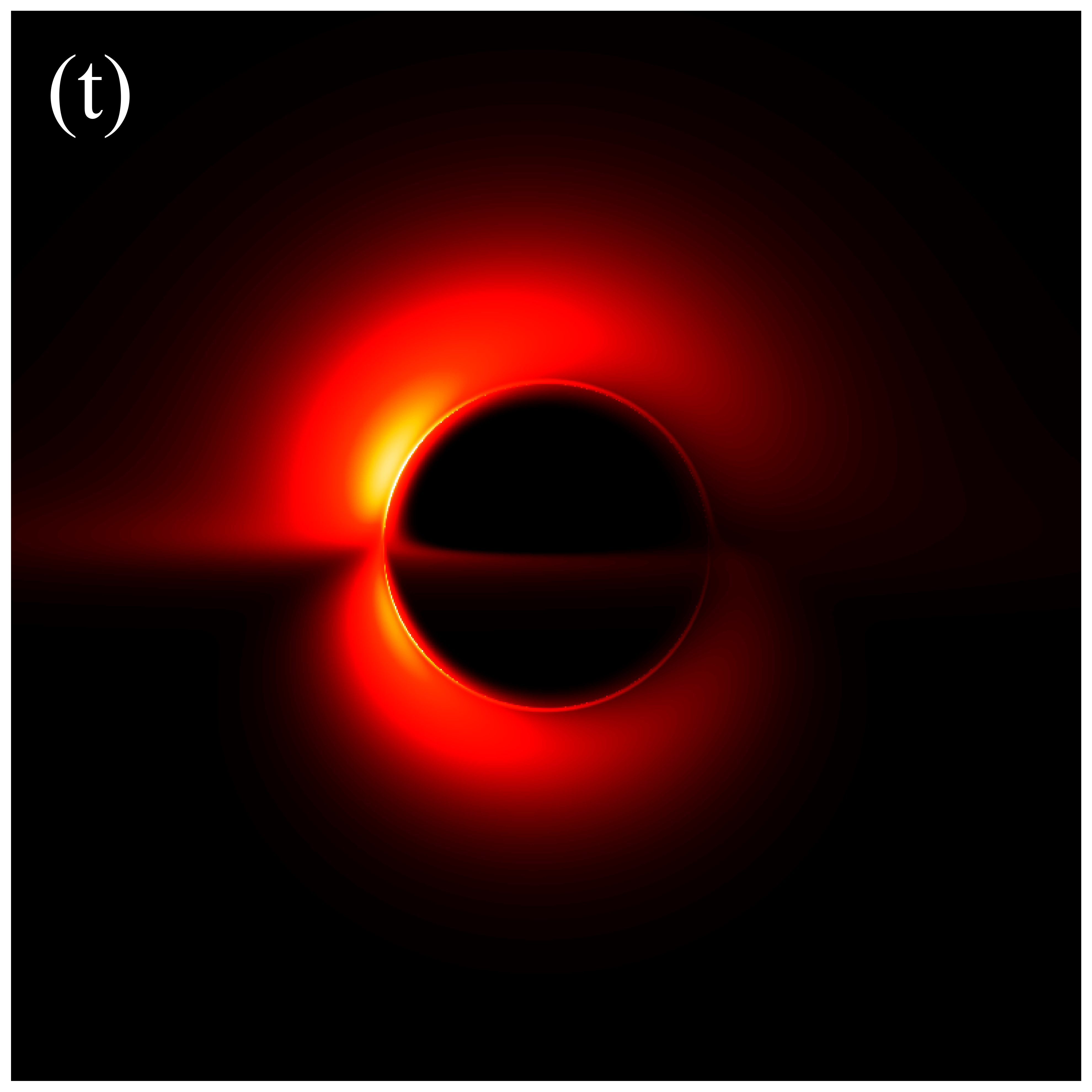}
\caption{Re-simulation of figure 6 incorporating the projection effect of an anisotropic accretion disk.}}\label{fig9}
\end{figure*}

Figures 9 and 10 present the results of re-simulations of figures 6 and 7, respectively, incorporating the projection effect from an anisotropic accretion disk. A comparison of figures 6 through 10 reveals that the projection effect significantly suppresses the image brightness, particularly affecting the intensity of bright spots and the eastern and western regions of the image. This suppression effect becomes more pronounced with increasing observation angle. This is anticipated because, as previously mentioned, light rays rarely cross the accretion disk perpendicularly during propagation, and with increasing observation inclination, the rays approach the disk nearly parallel to its plane, resulting in larger $\delta$ when hitting the disk. The projection effect diminishes the background brightness of image, leading to sharper and more discernible critical curves, as demonstrated in figures 6(s) and 9(s). This phenomenon arises because higher-order images are primarily contributed by light rays passing almost perpendicularly through the accretion disk, where the projection effect has minimal impact on their specific intensities. Conversely, light rays forming the direct image typically possess larger $\delta$, resulting in a significant reduction in their specific intensities. In summary, the projection effect exerts a more substantial influence on direct images compared to higher-order images, leading to reduced background brightness and more prominent higher-order subrings. Furthermore, it is important to note that in figures 9 and 10, the impacts of deformation parameter and observation inclination on image features such as the inner shadow, critical curve, and bright spots, as shown in figures 6 and 7, can still be accurately assessed.
\begin{figure*}%[tbph]
\center{
\includegraphics[width=3.5cm]{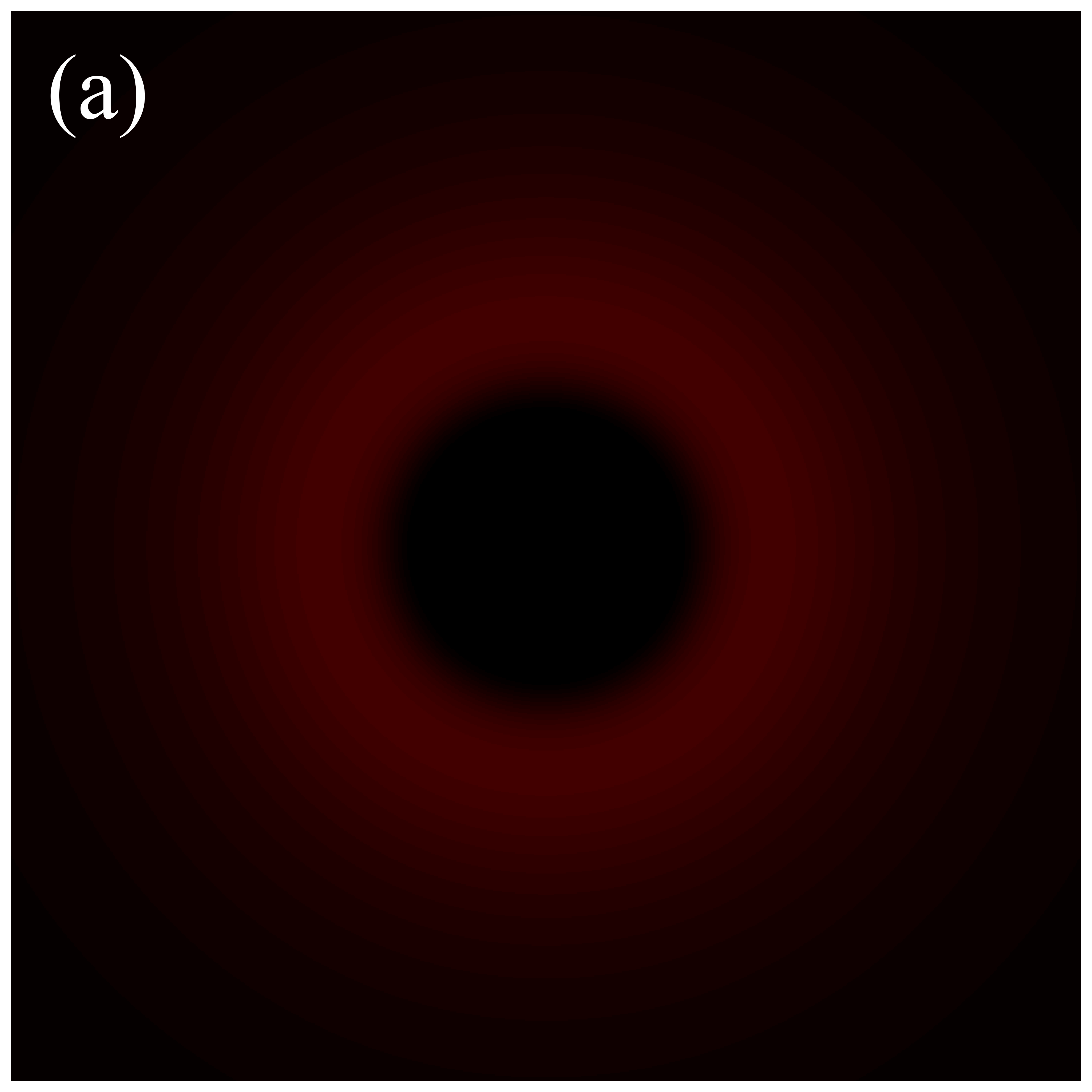}
\includegraphics[width=3.5cm]{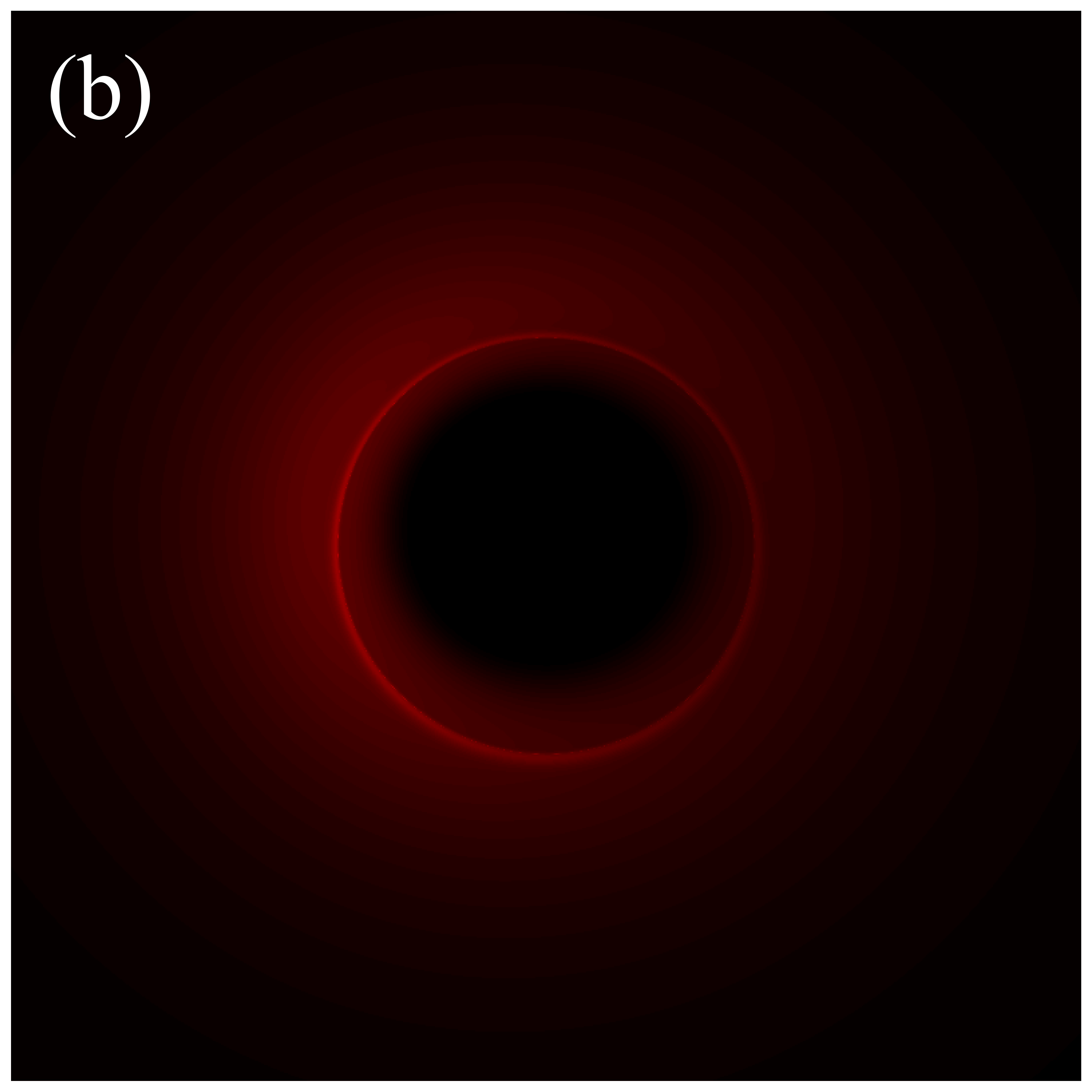}
\includegraphics[width=3.5cm]{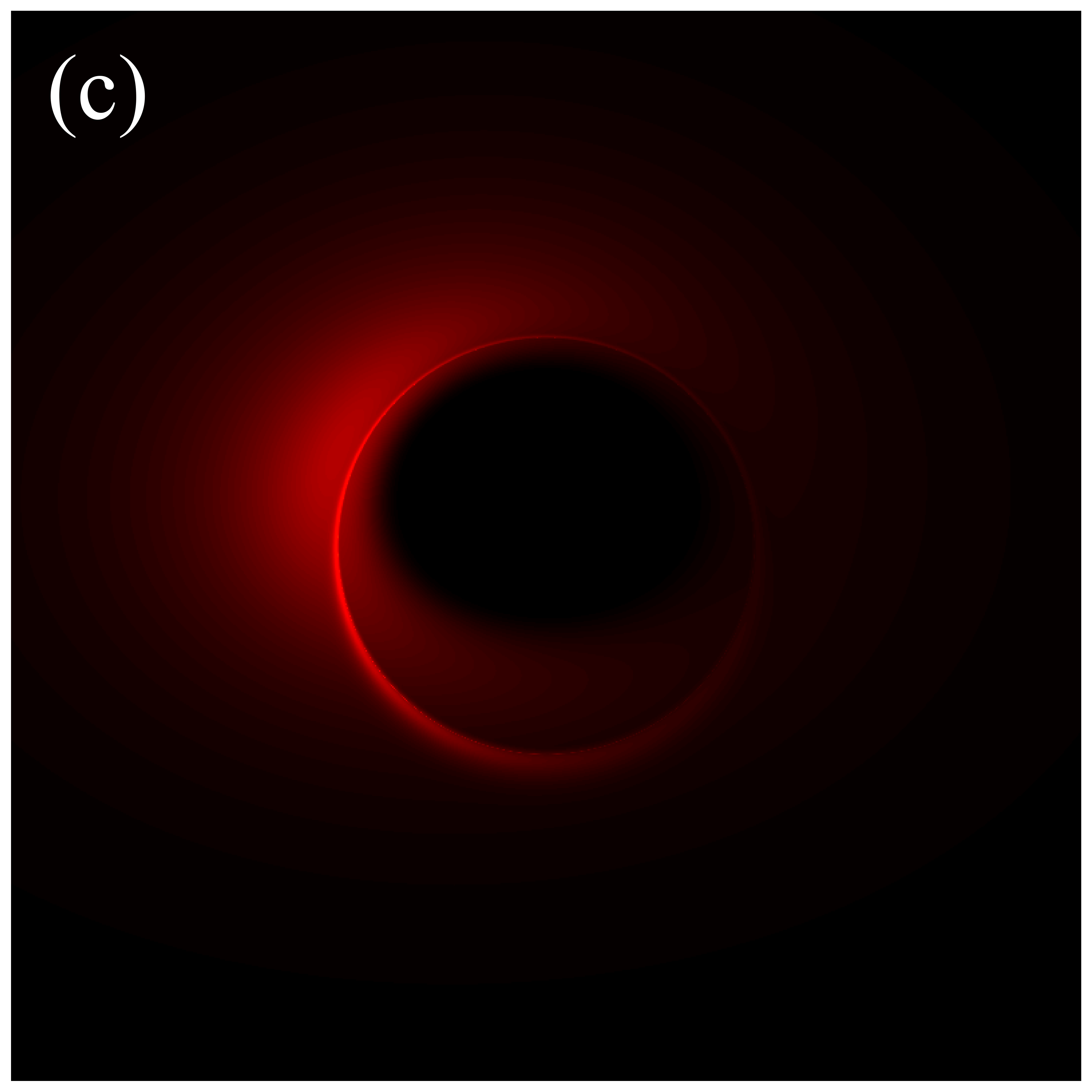}
\includegraphics[width=3.5cm]{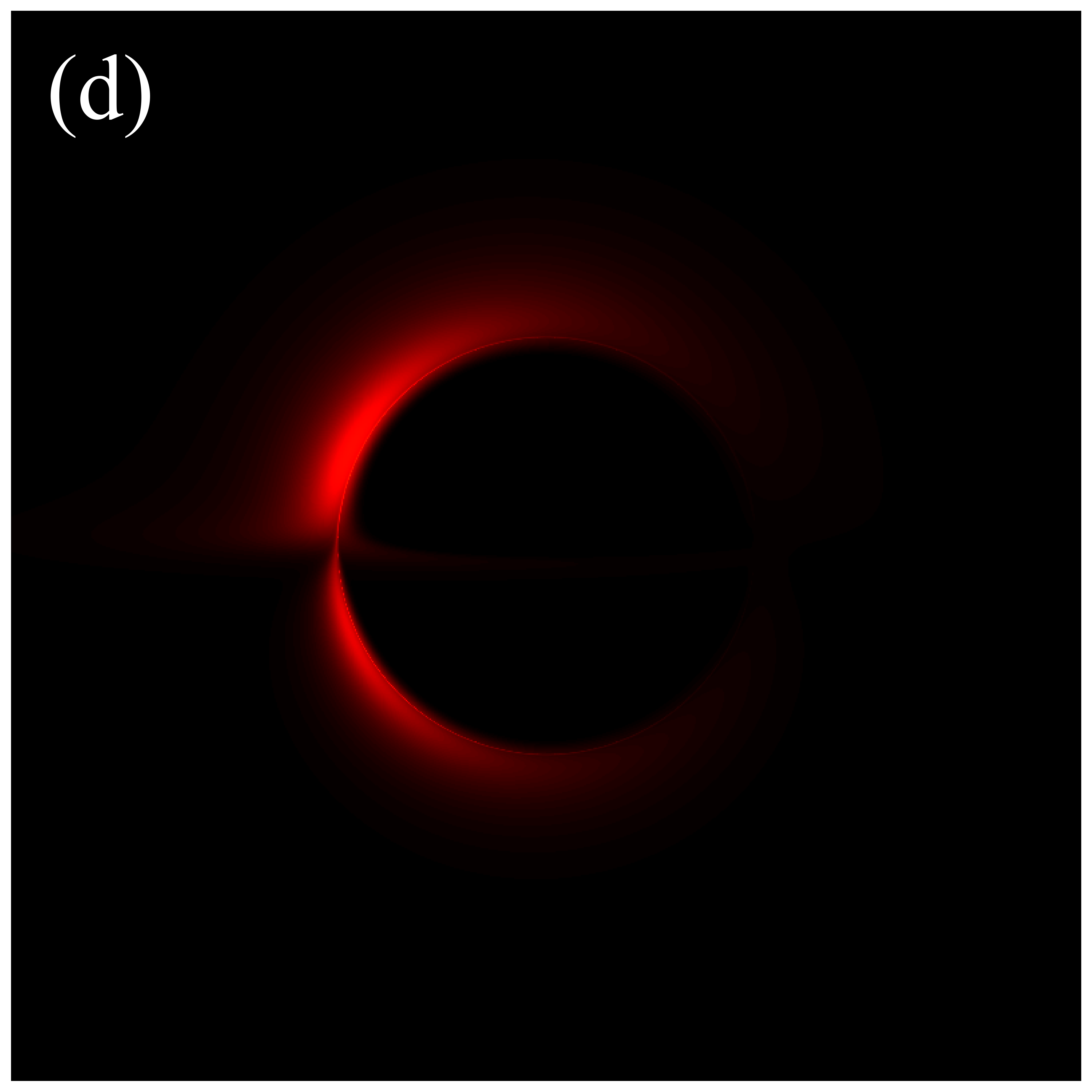}
\includegraphics[width=3.5cm]{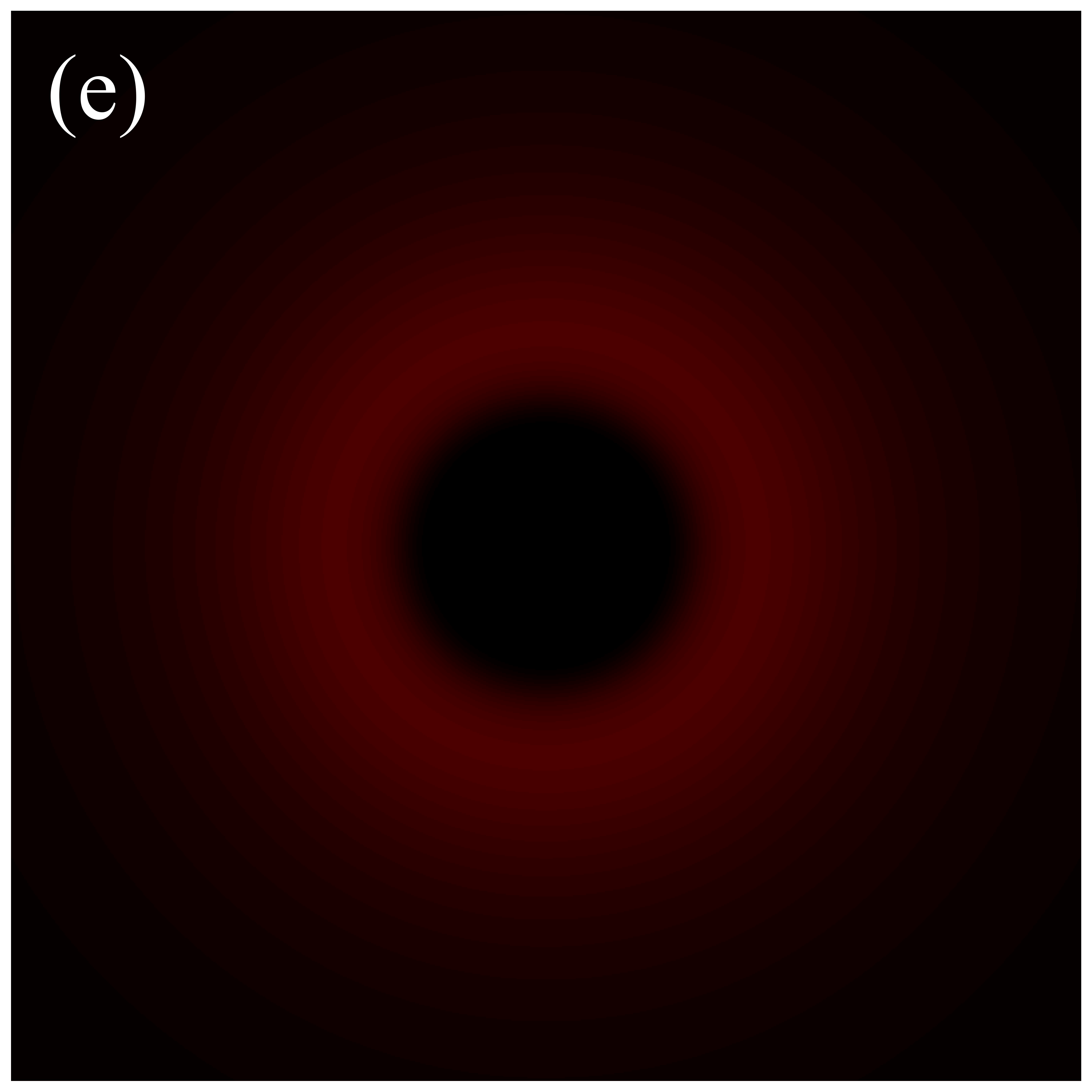}
\includegraphics[width=3.5cm]{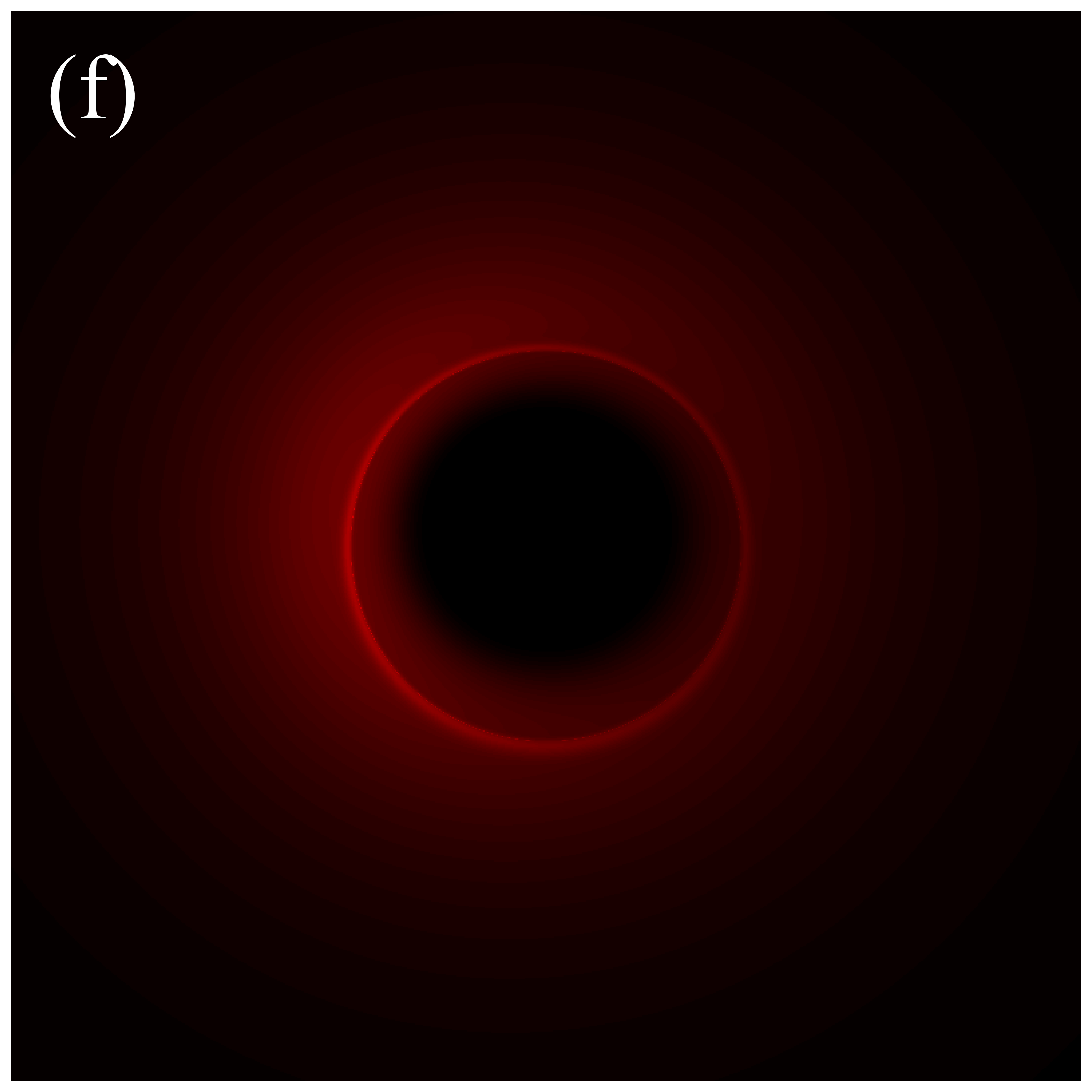}
\includegraphics[width=3.5cm]{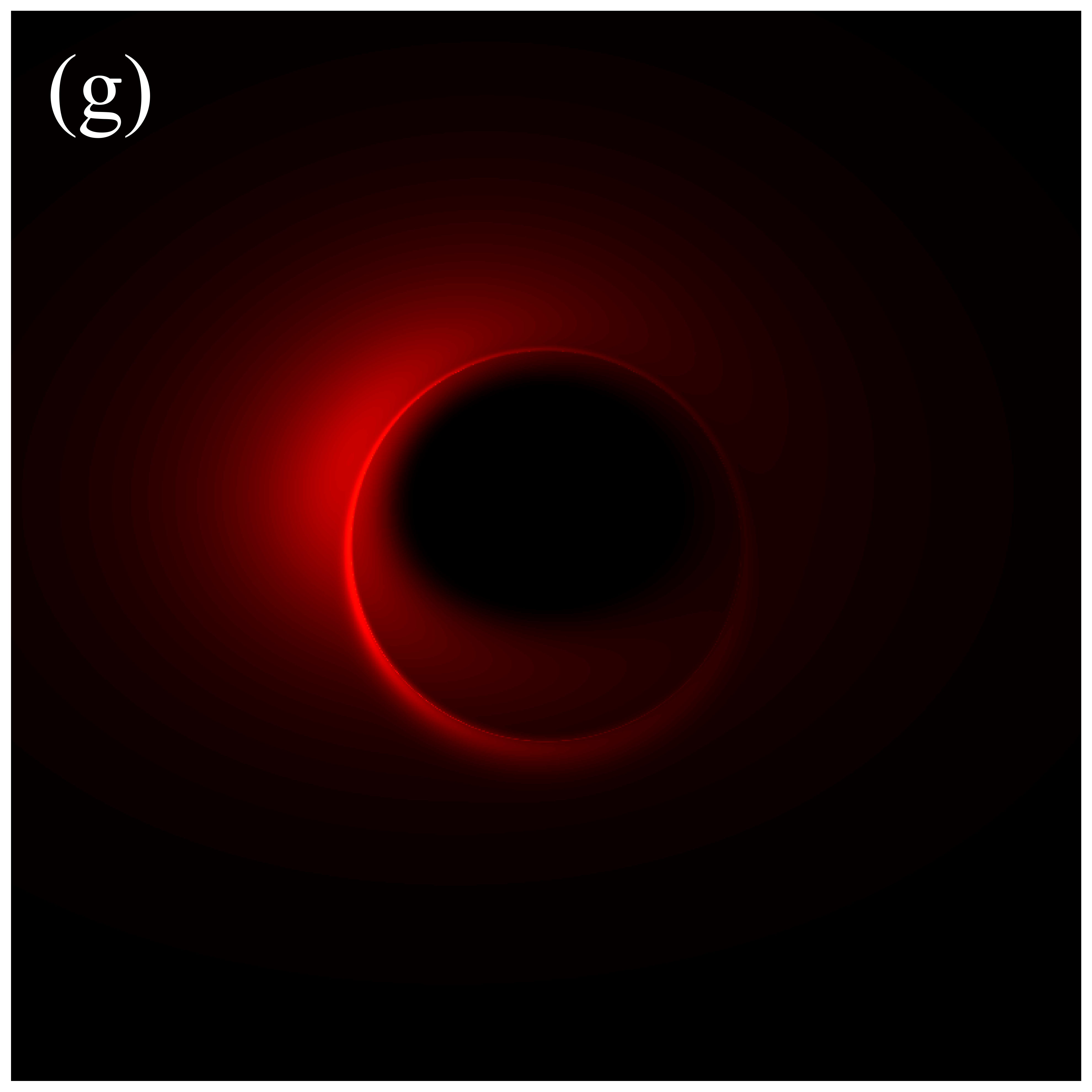}
\includegraphics[width=3.5cm]{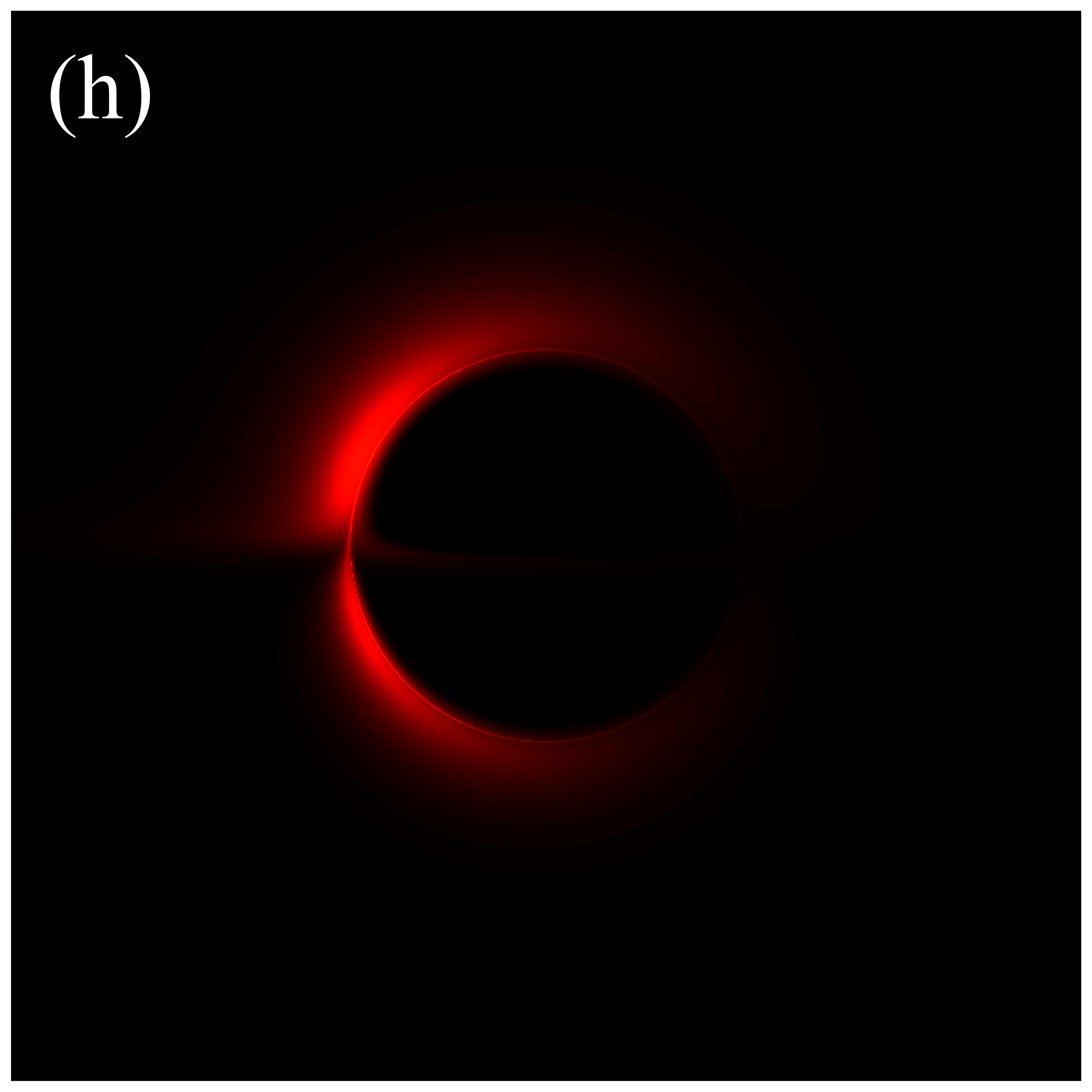}
\includegraphics[width=3.5cm]{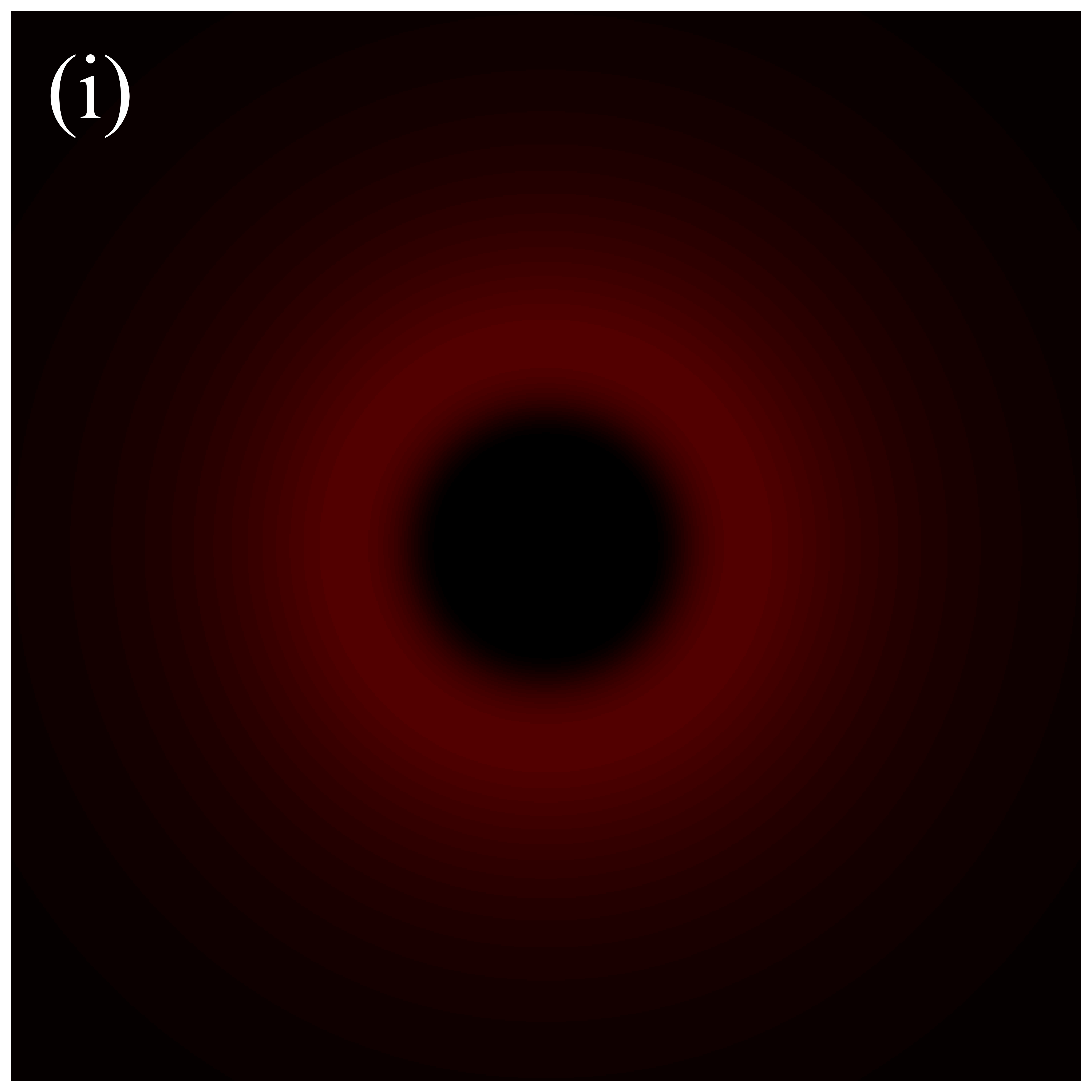}
\includegraphics[width=3.5cm]{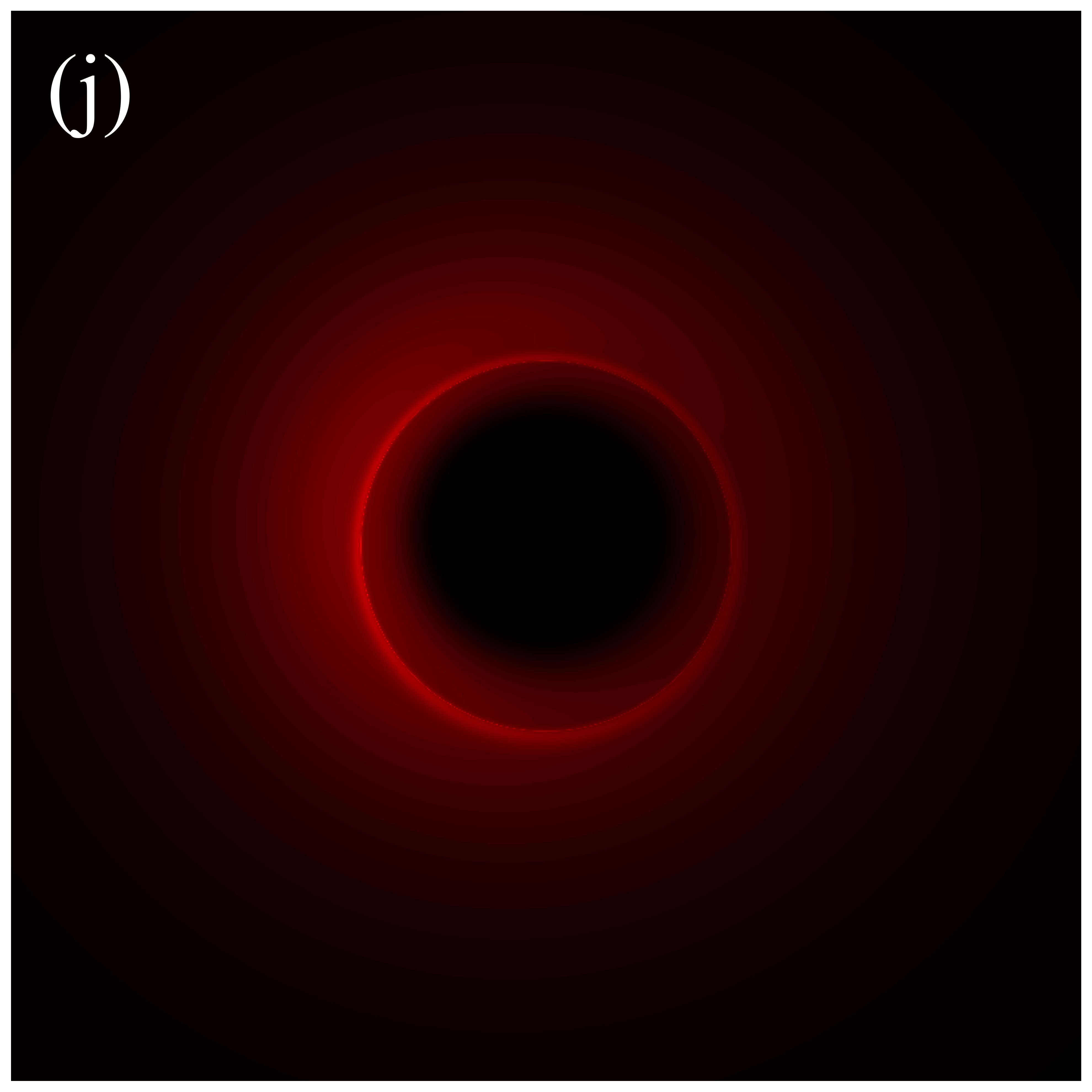}
\includegraphics[width=3.5cm]{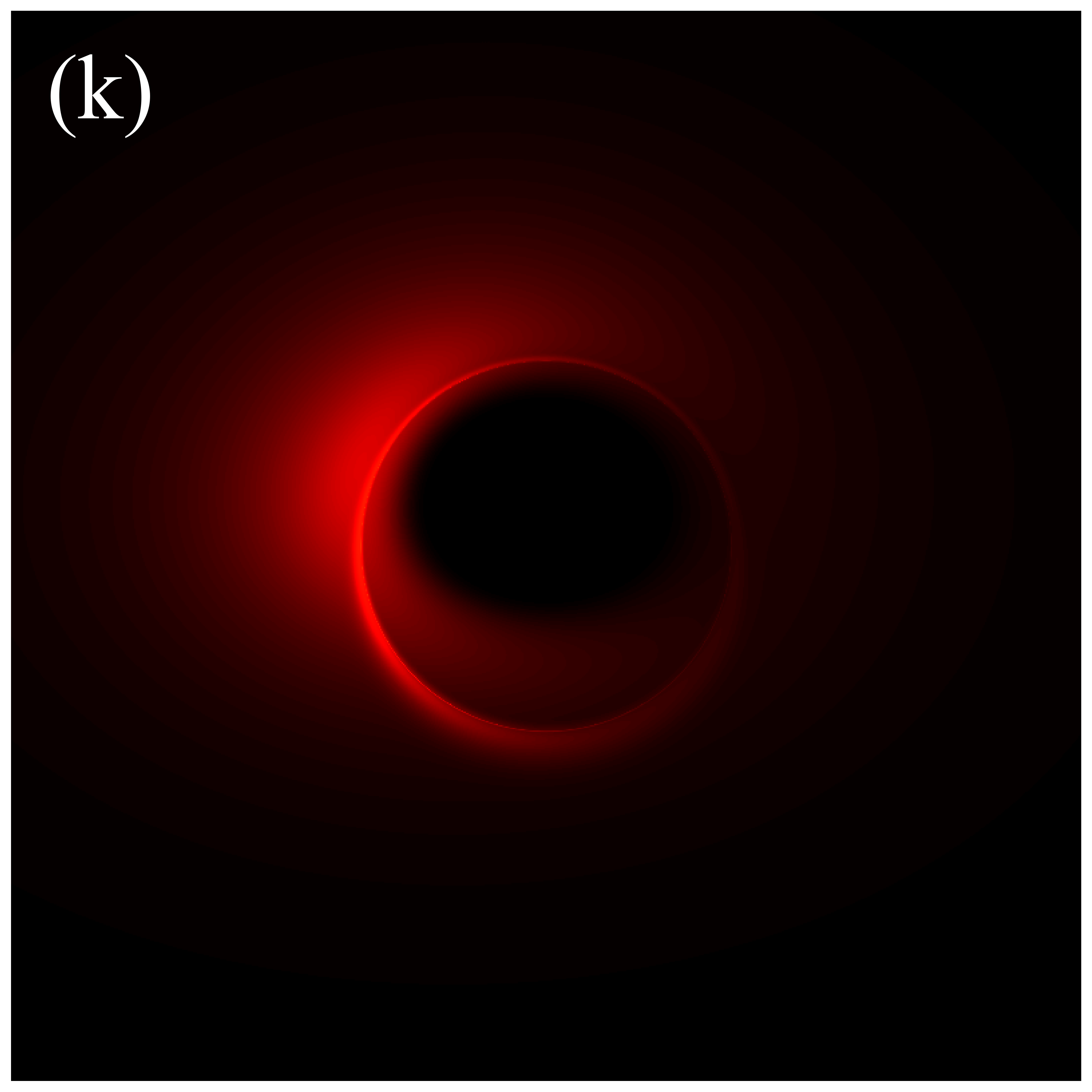}
\includegraphics[width=3.5cm]{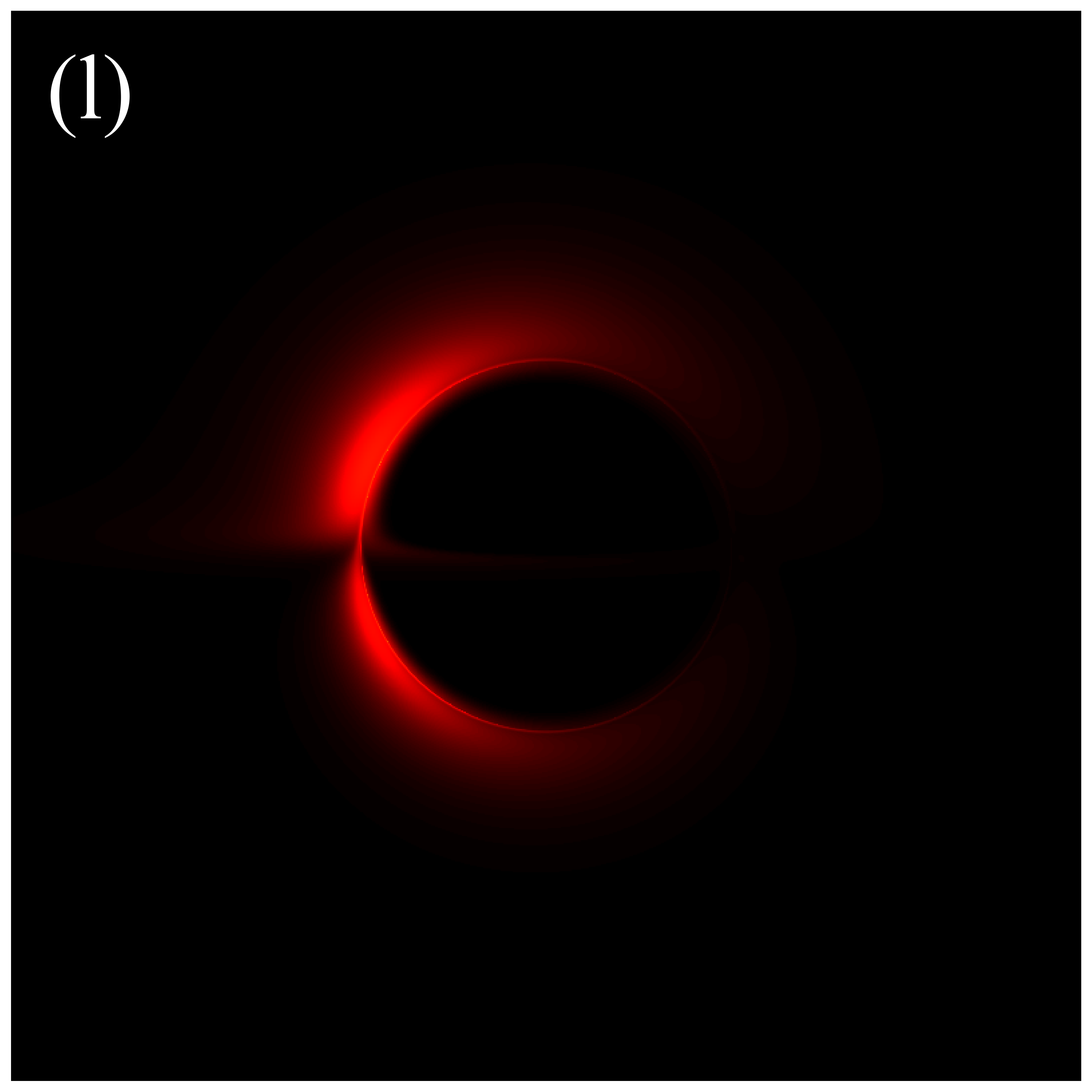}
\includegraphics[width=3.5cm]{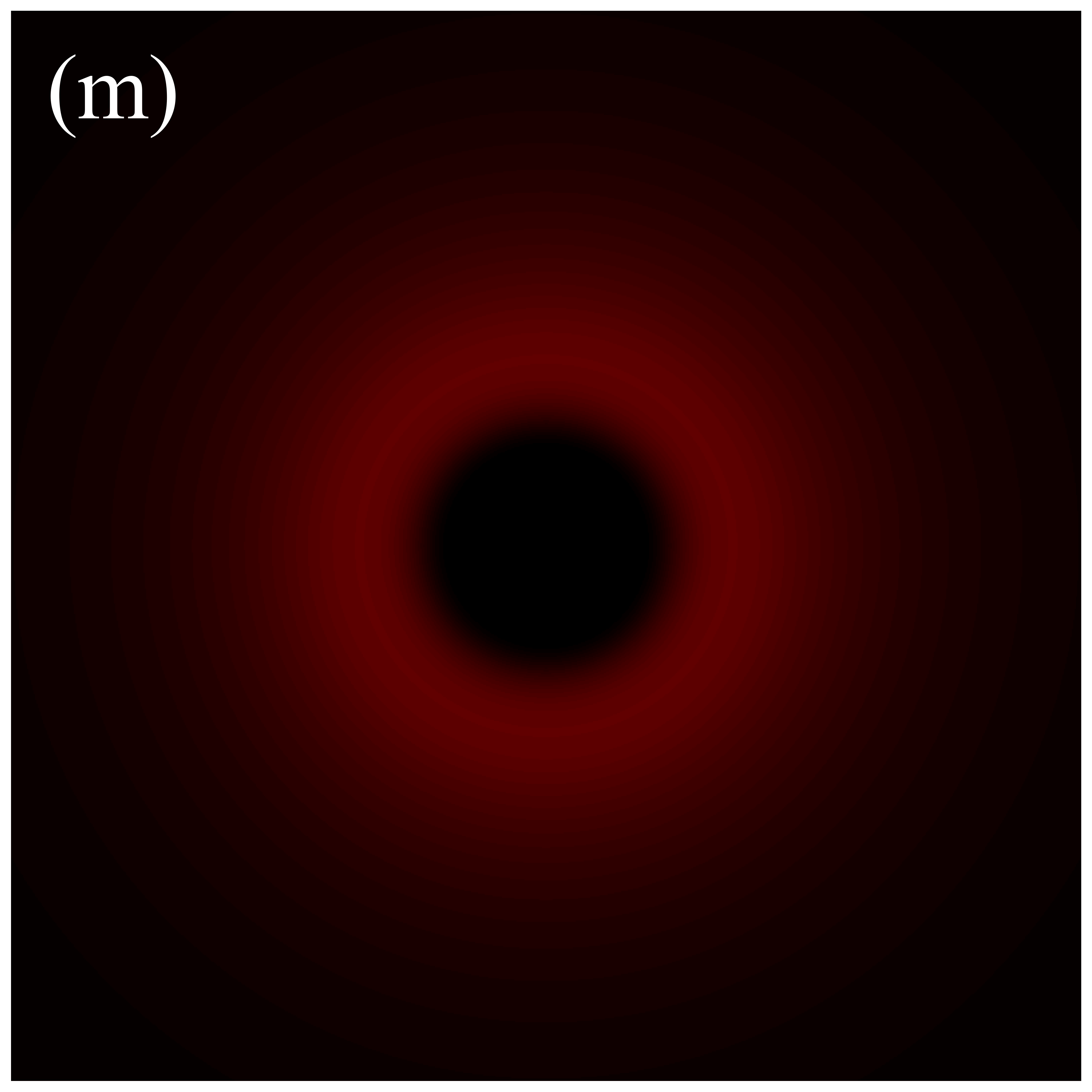}
\includegraphics[width=3.5cm]{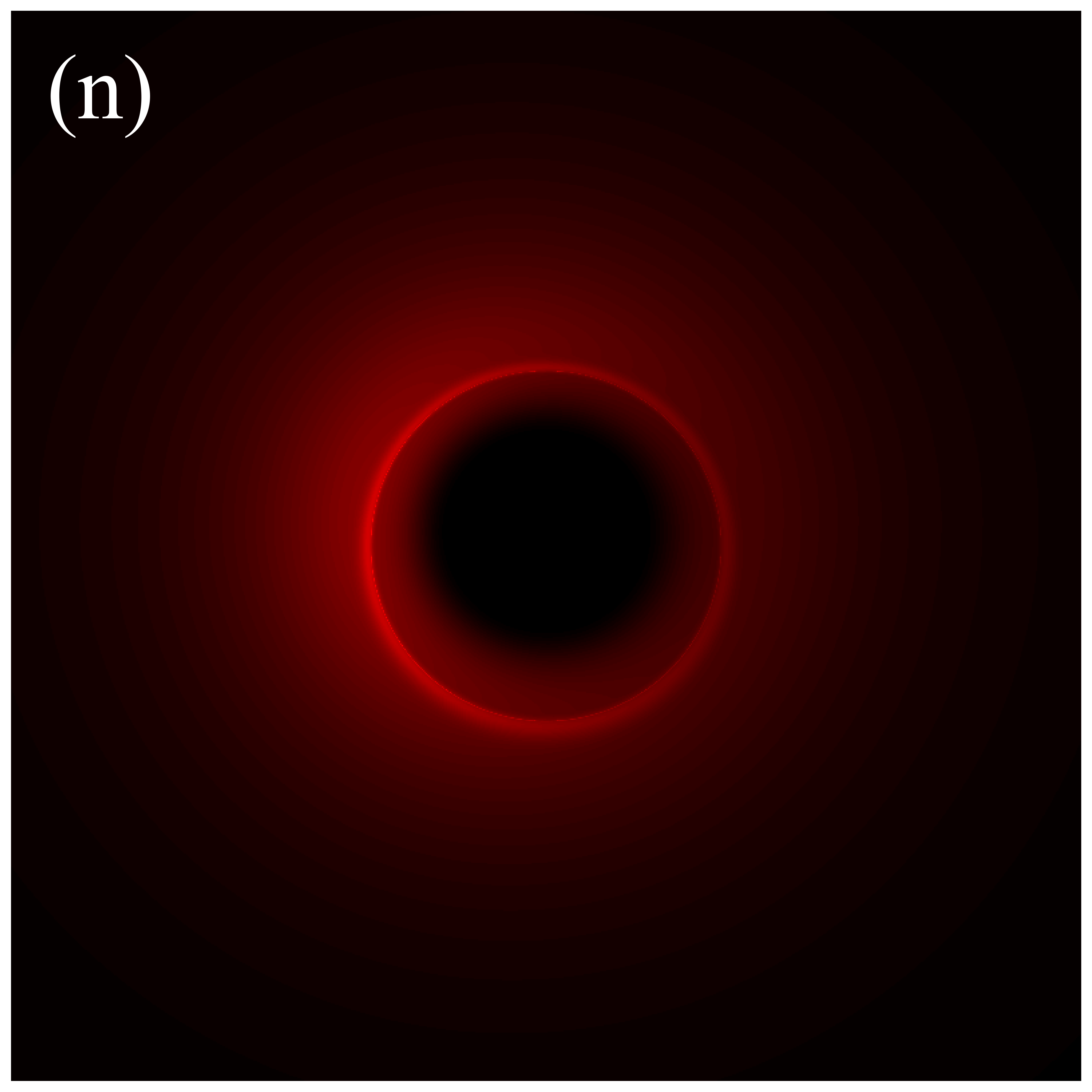}
\includegraphics[width=3.5cm]{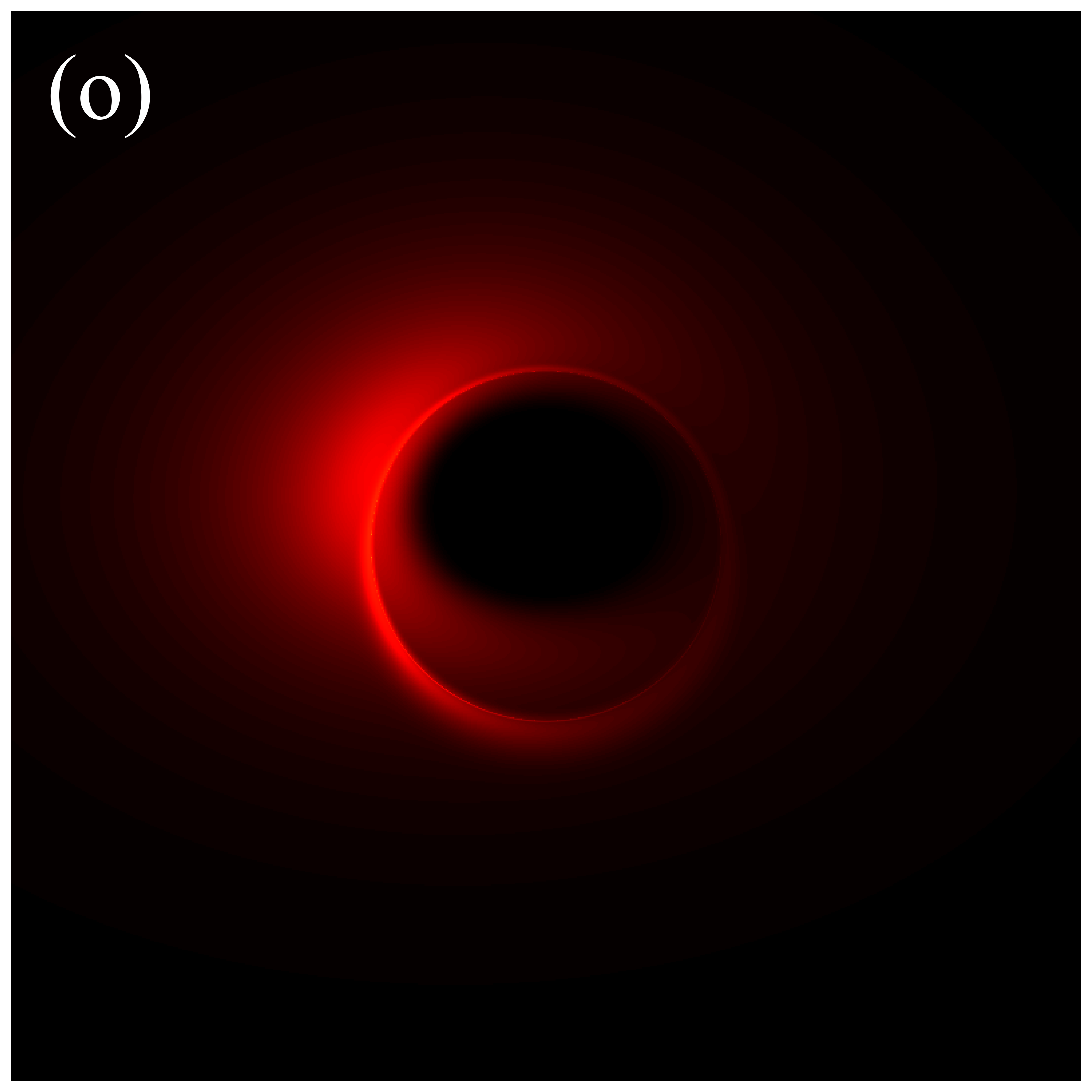}
\includegraphics[width=3.5cm]{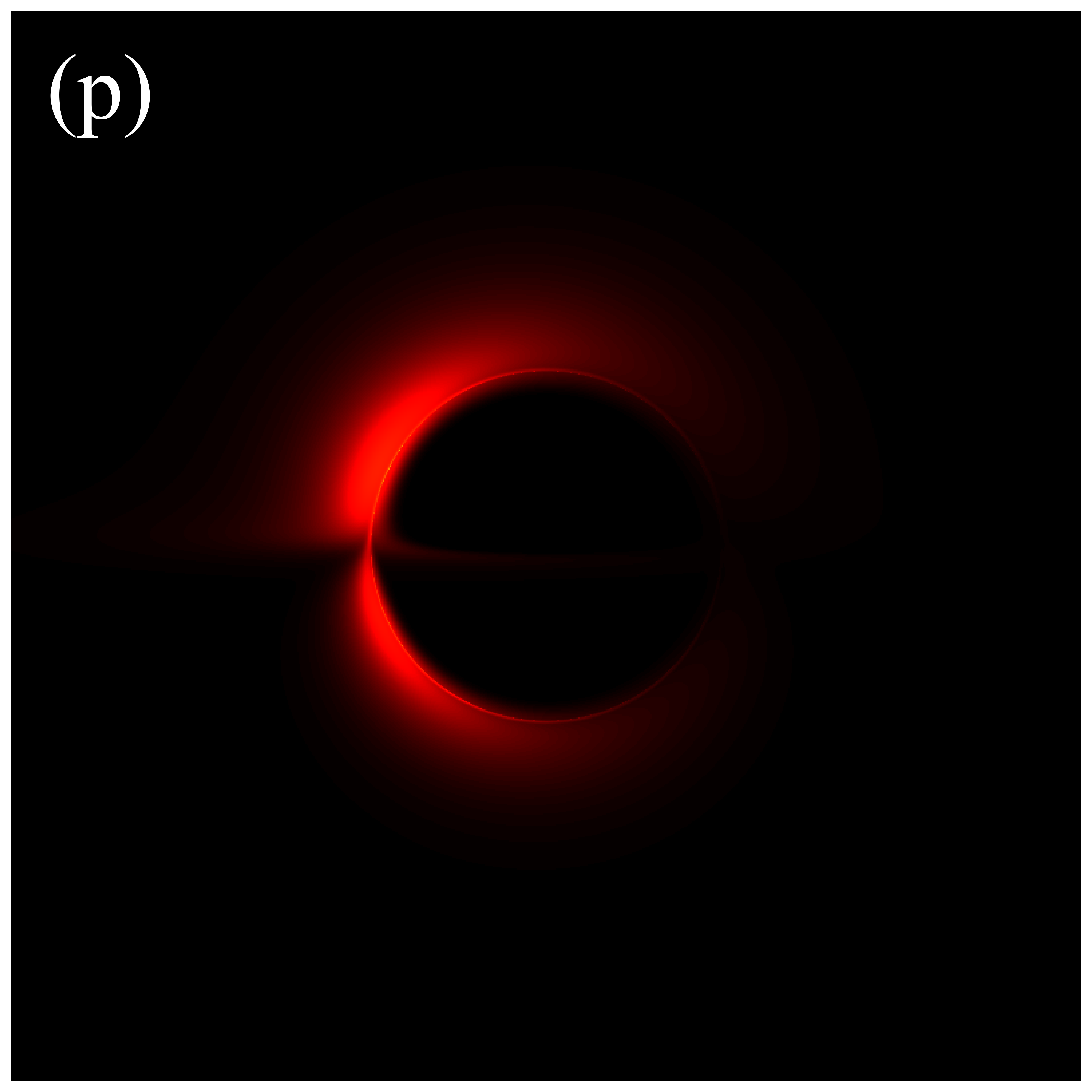}
\includegraphics[width=3.5cm]{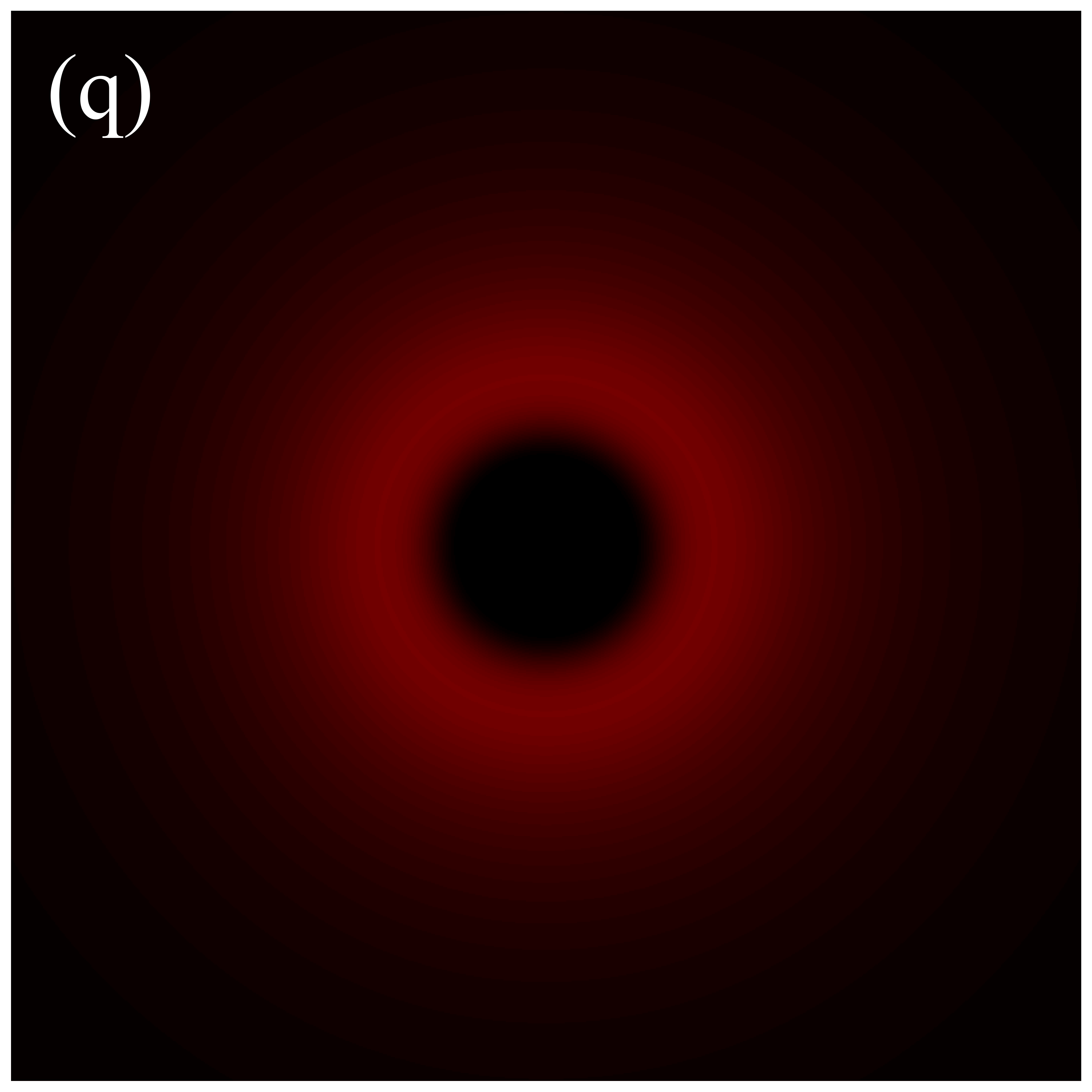}
\includegraphics[width=3.5cm]{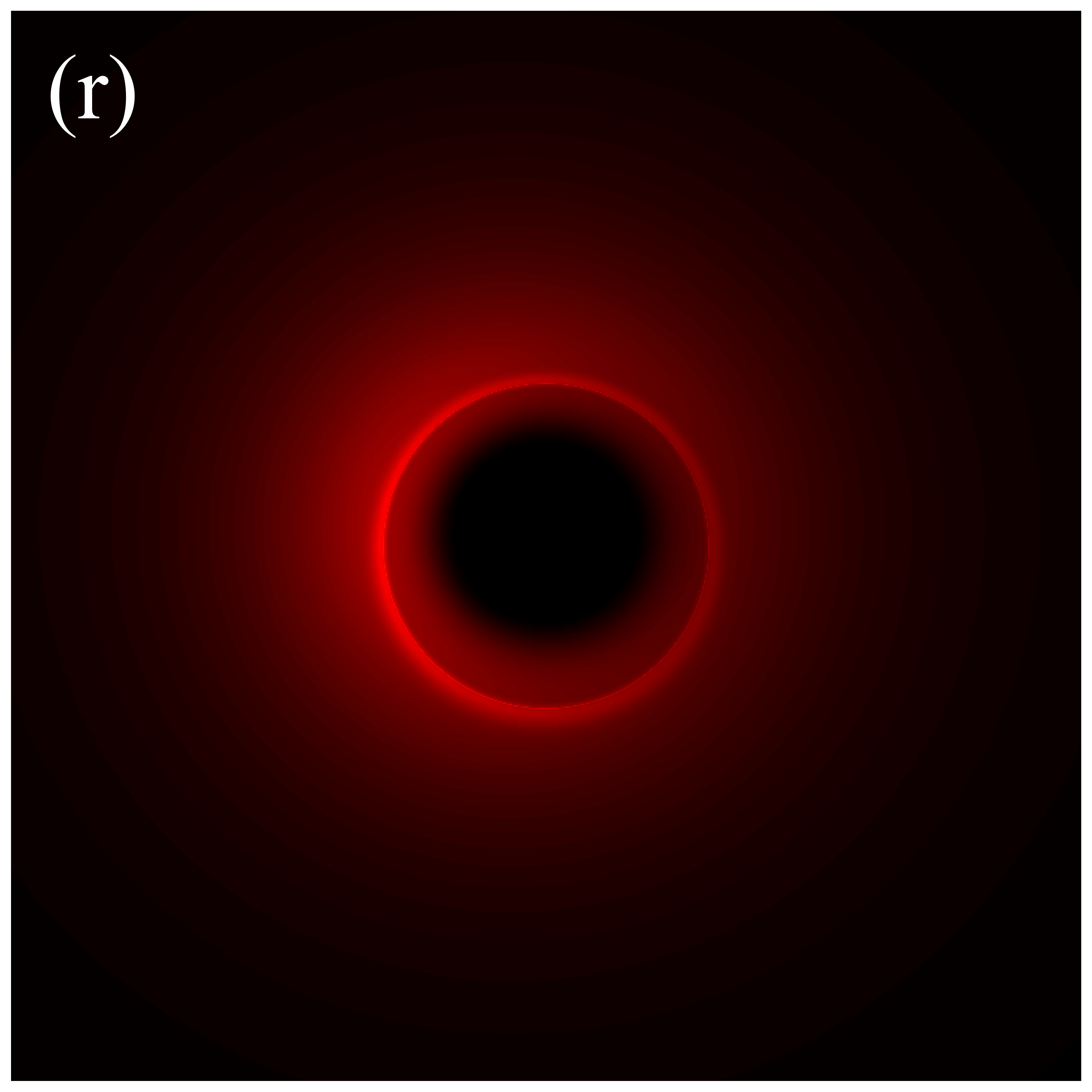}
\includegraphics[width=3.5cm]{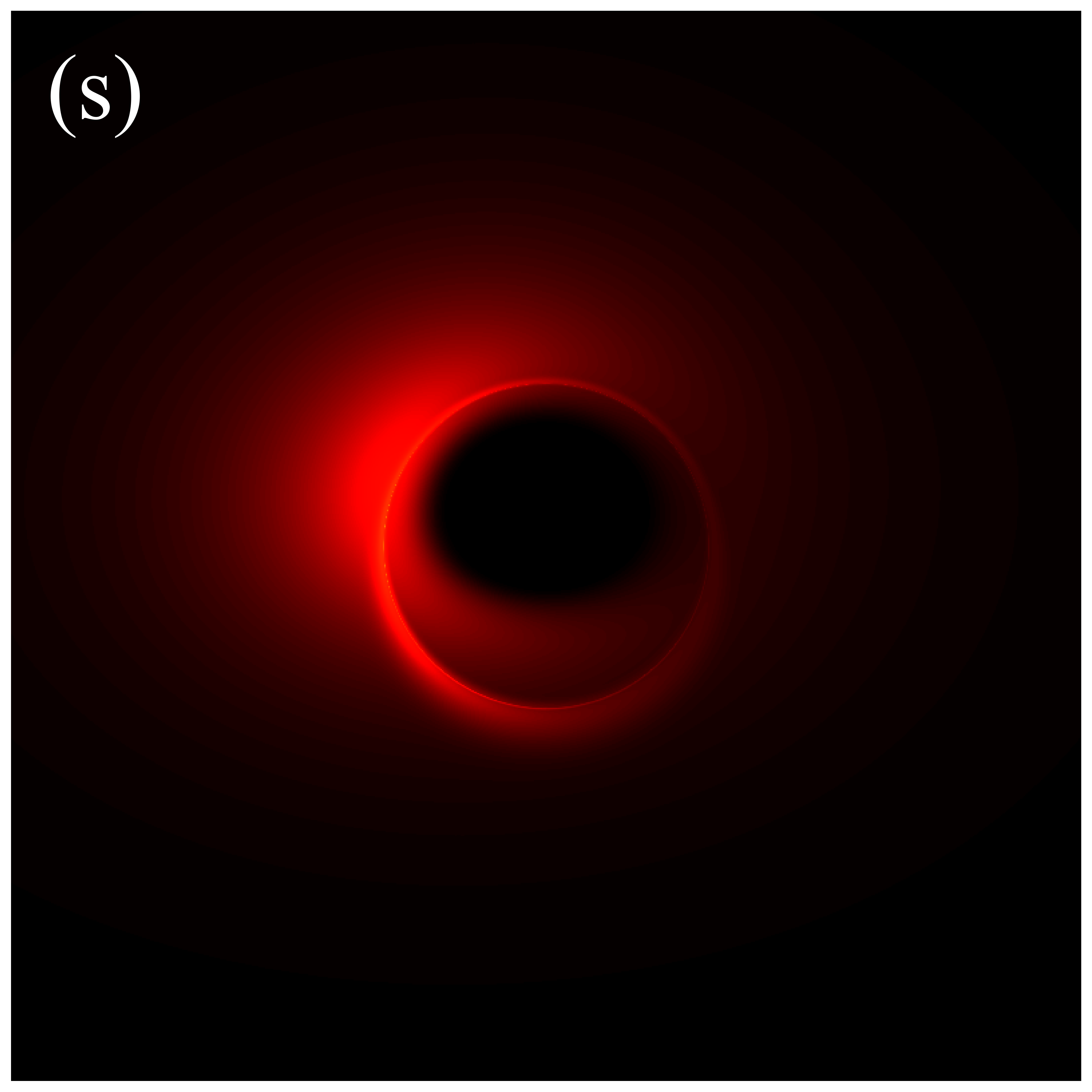}
\includegraphics[width=3.5cm]{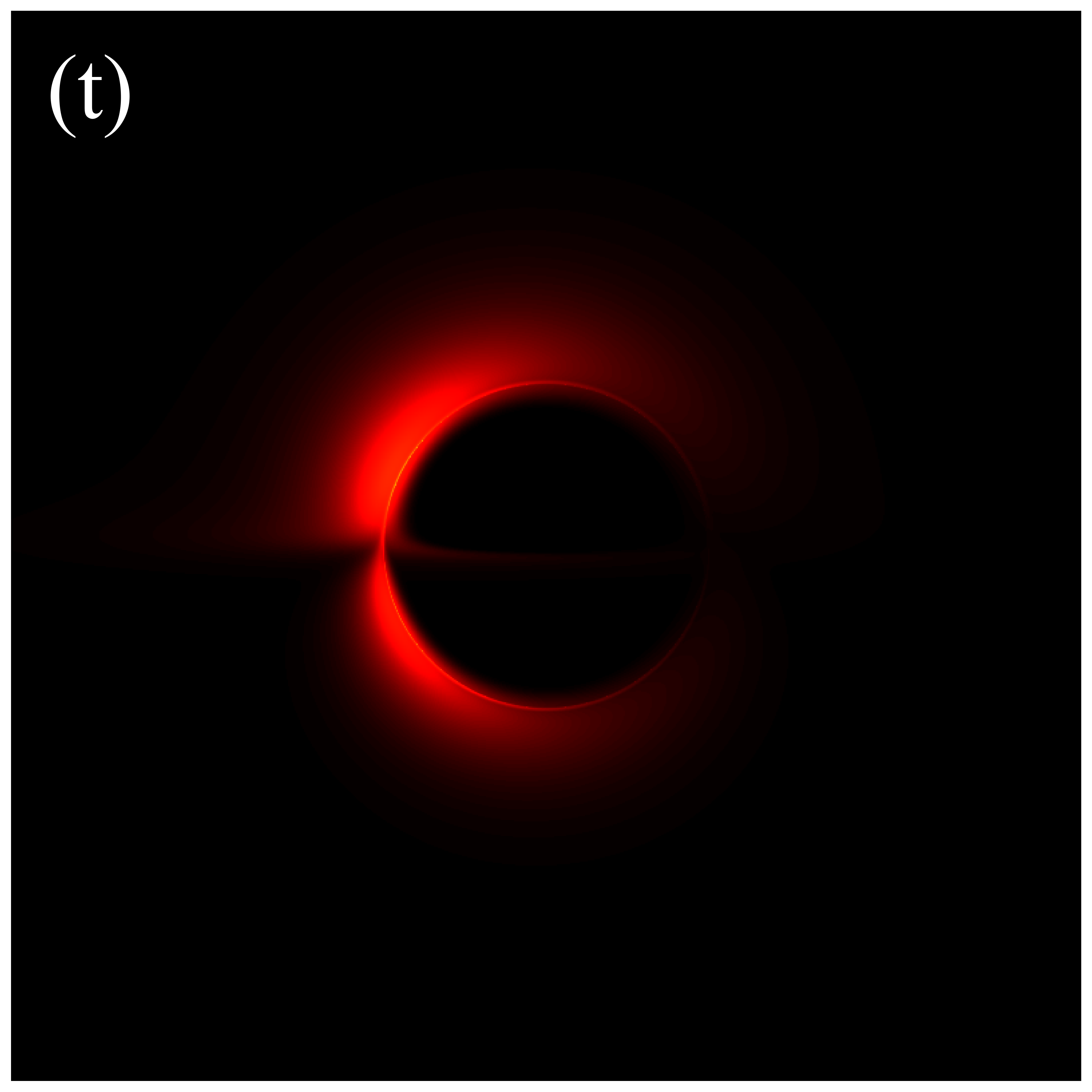}
\caption{Re-simulation of figure 7 incorporating the projection effect of an anisotropic accretion disk.}}\label{fig10}
\end{figure*}

To conduct a more detailed analysis of the projection effect on black hole images, we extracted the specific intensity distribution of light rays along the $x$-axis of the observation plane, as depicted in figure 11. The black solid line represents the scenario where $\kappa=1$, while the red dashed line corresponds to the case of $\kappa=\cos\delta$. When the observation inclination is $0^{\circ}$, the curve exhibits symmetry about $x=0$, with the two peaks representing the critical curve. Under these conditions, the reduction in specific intensity due to the projection effect is negligible. As the observation inclination increases, the peak positions remain constant, but the symmetry is disrupted, with the left and right peaks corresponding to Doppler blueshift and redshift, respectively. In this case, the projection effect increasingly attenuates the specific intensity, particularly in the slope and bulge near the Doppler blueshift peak. This attenuation enhances the prominence of the peaks, facilitating the precise extraction of geometric information from the critical curve. It is noteworthy that this suppression effect is more pronounced in the $86$ GHz simulations compared to the $230$ GHz images. Furthermore, it is observed that in all parameter spaces, the black solid line and the red dashed line overlap perfectly in the region where $I_{\textrm{obs}}=0$, indicating that the projection effect does not impact the size of the inner shadow.
\begin{figure*}%[tbph]
\center{
\includegraphics[width=4.7cm]{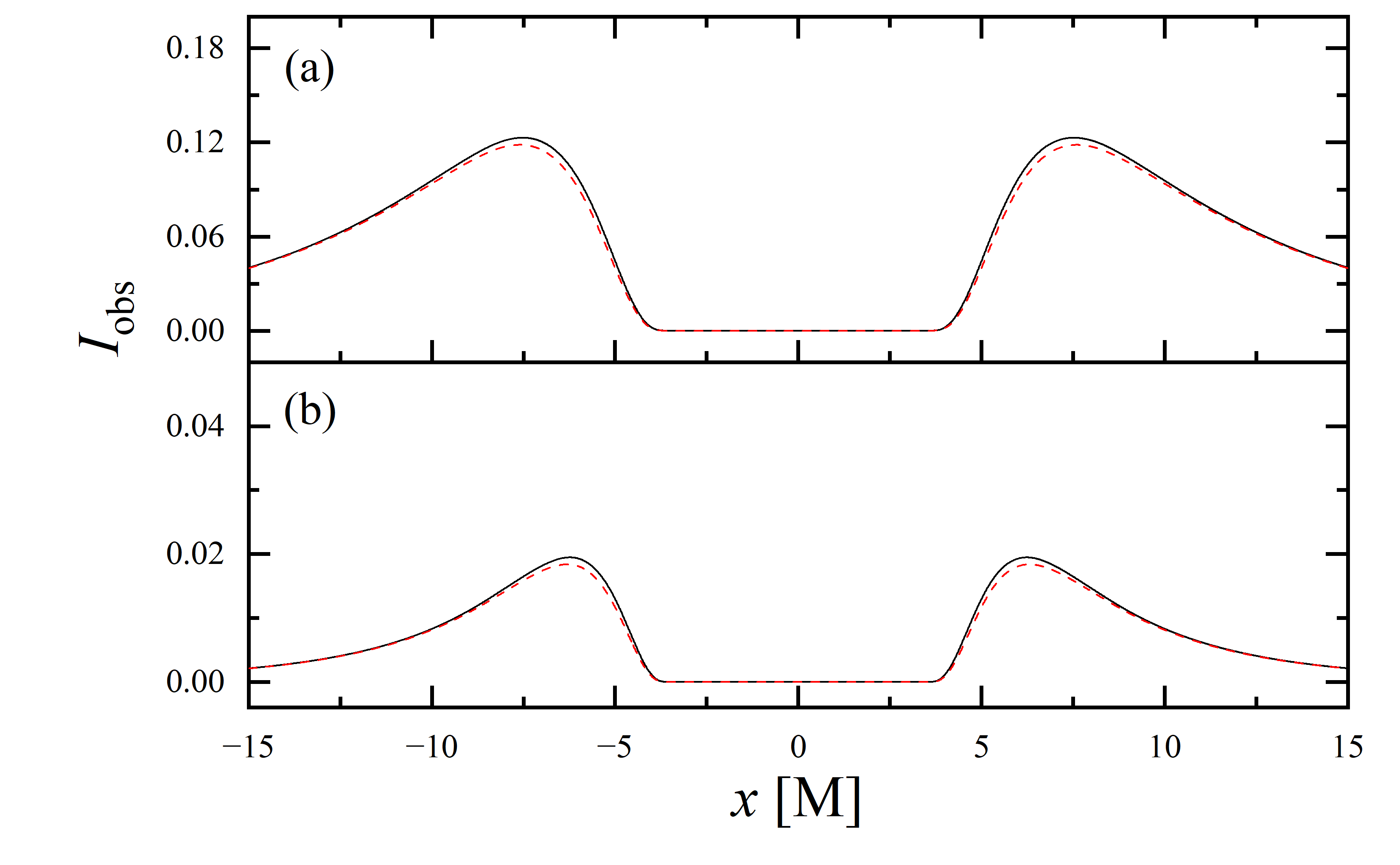}
\includegraphics[width=4.7cm]{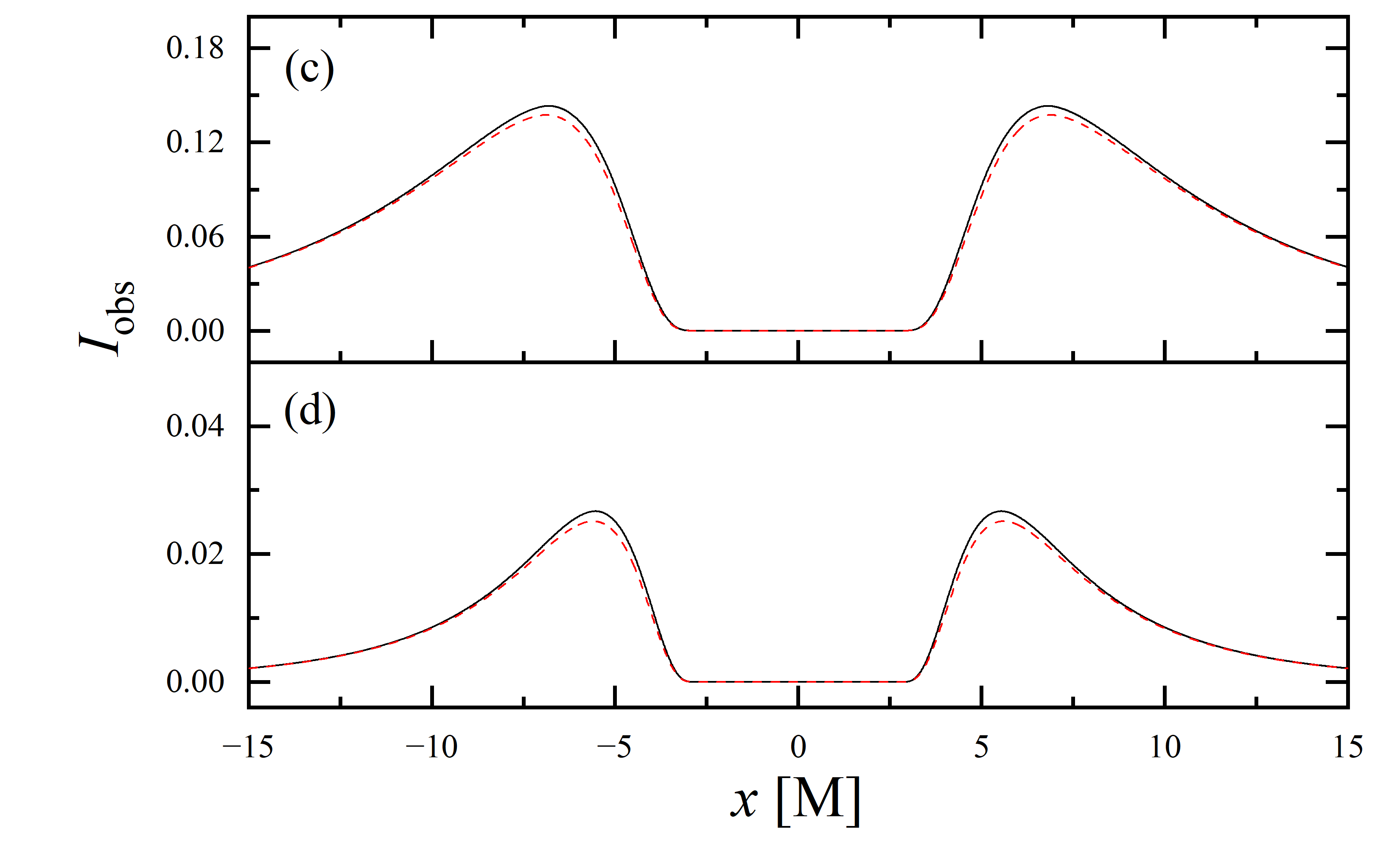}
\includegraphics[width=4.7cm]{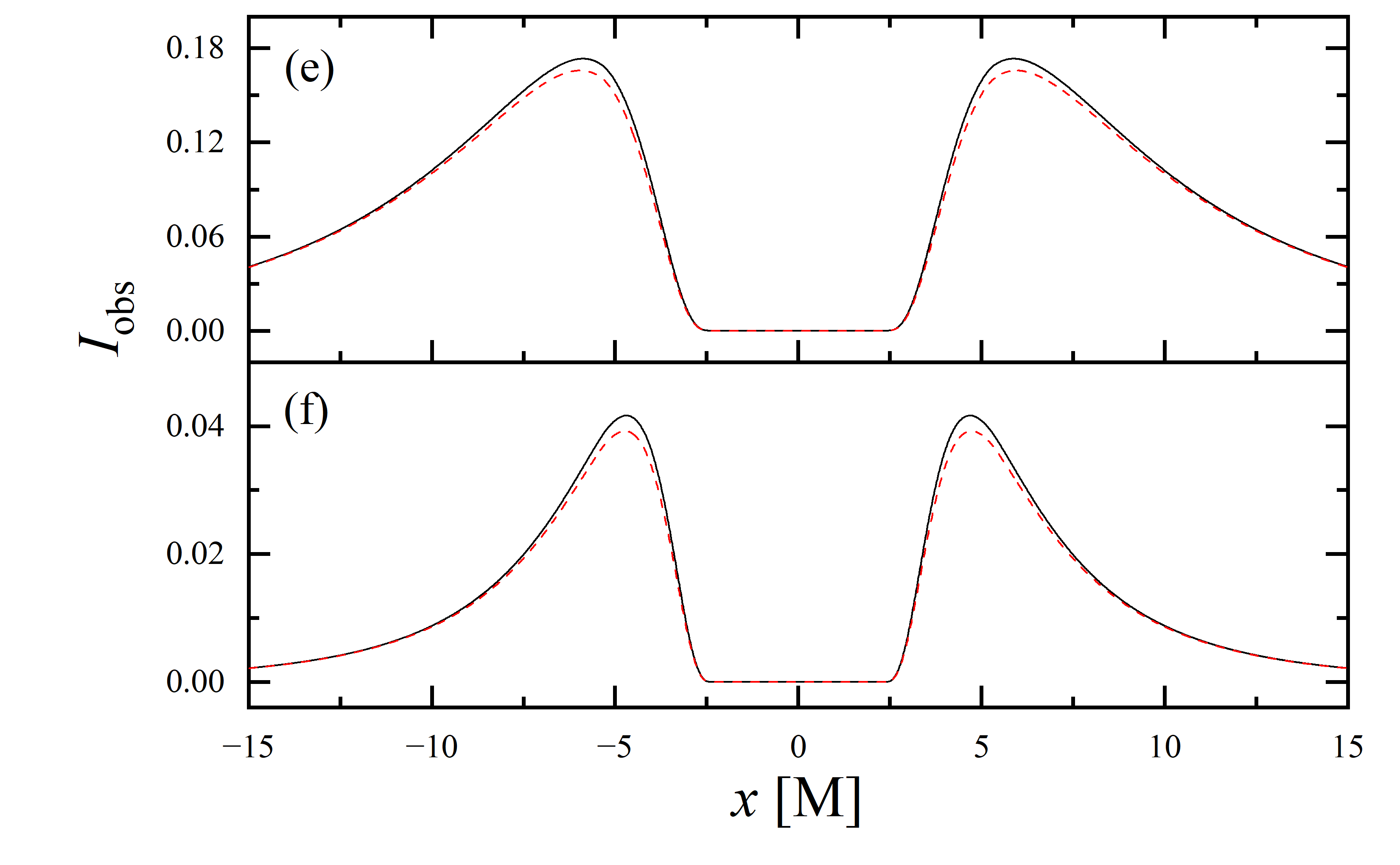}
\includegraphics[width=4.7cm]{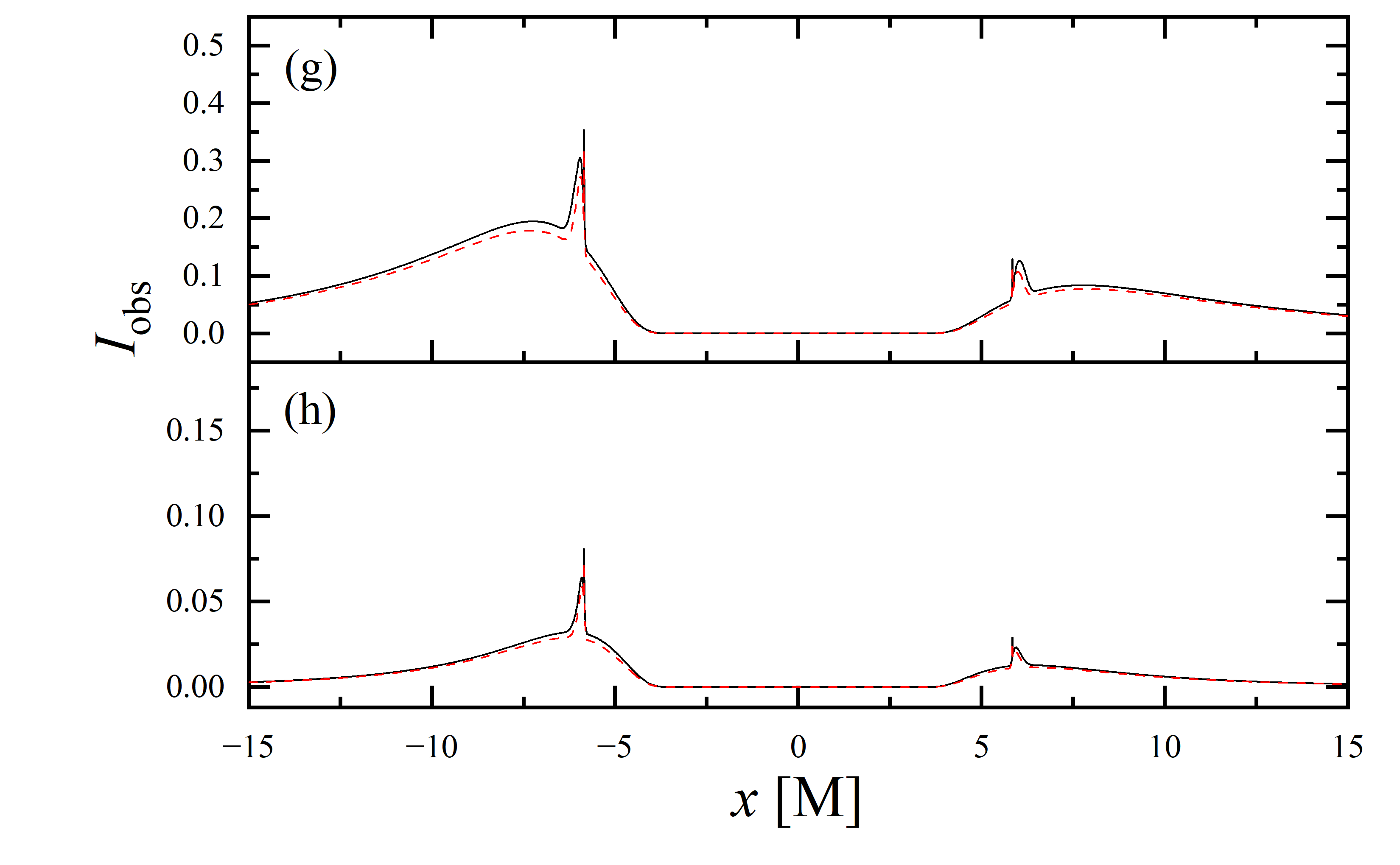}
\includegraphics[width=4.7cm]{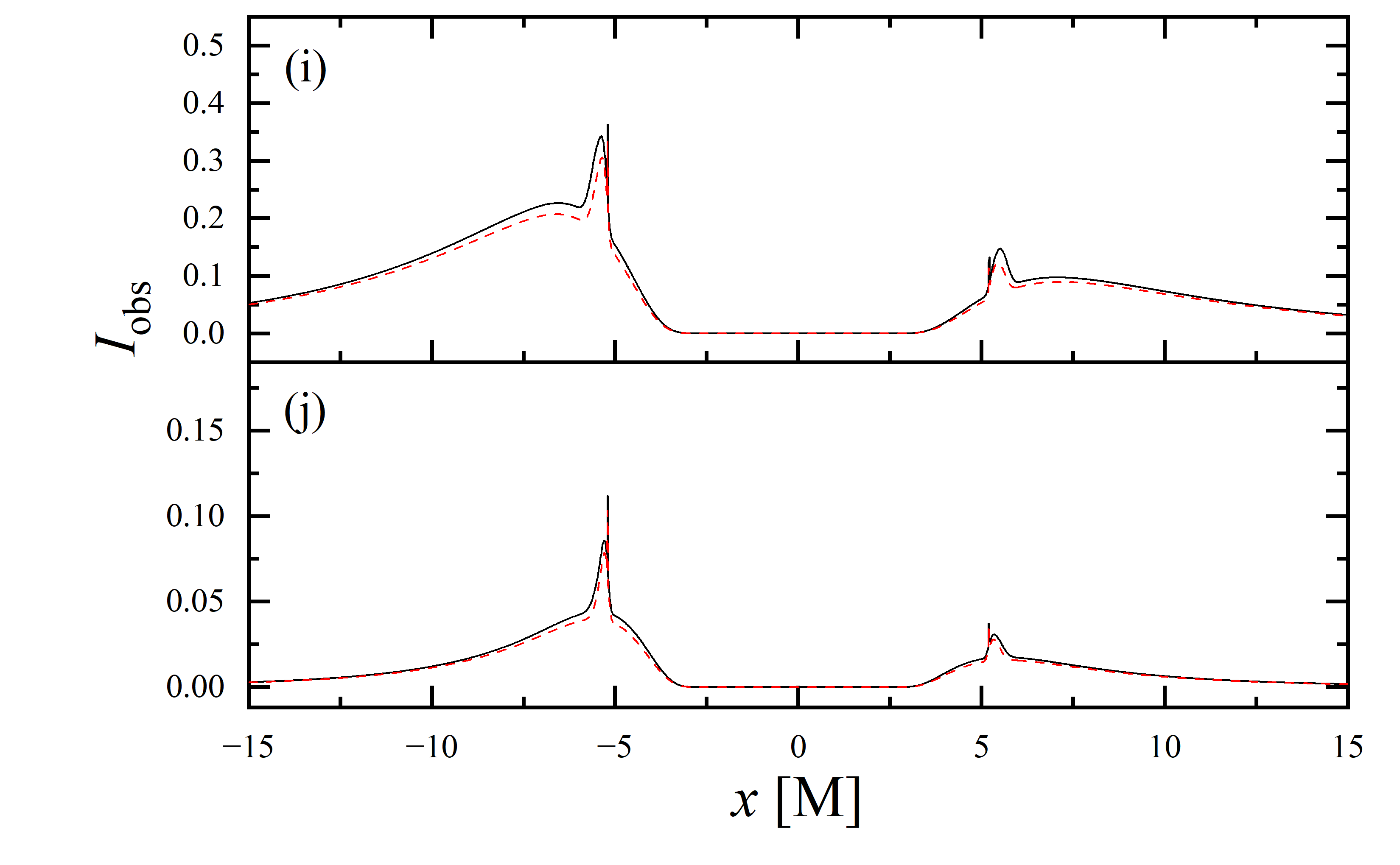}
\includegraphics[width=4.7cm]{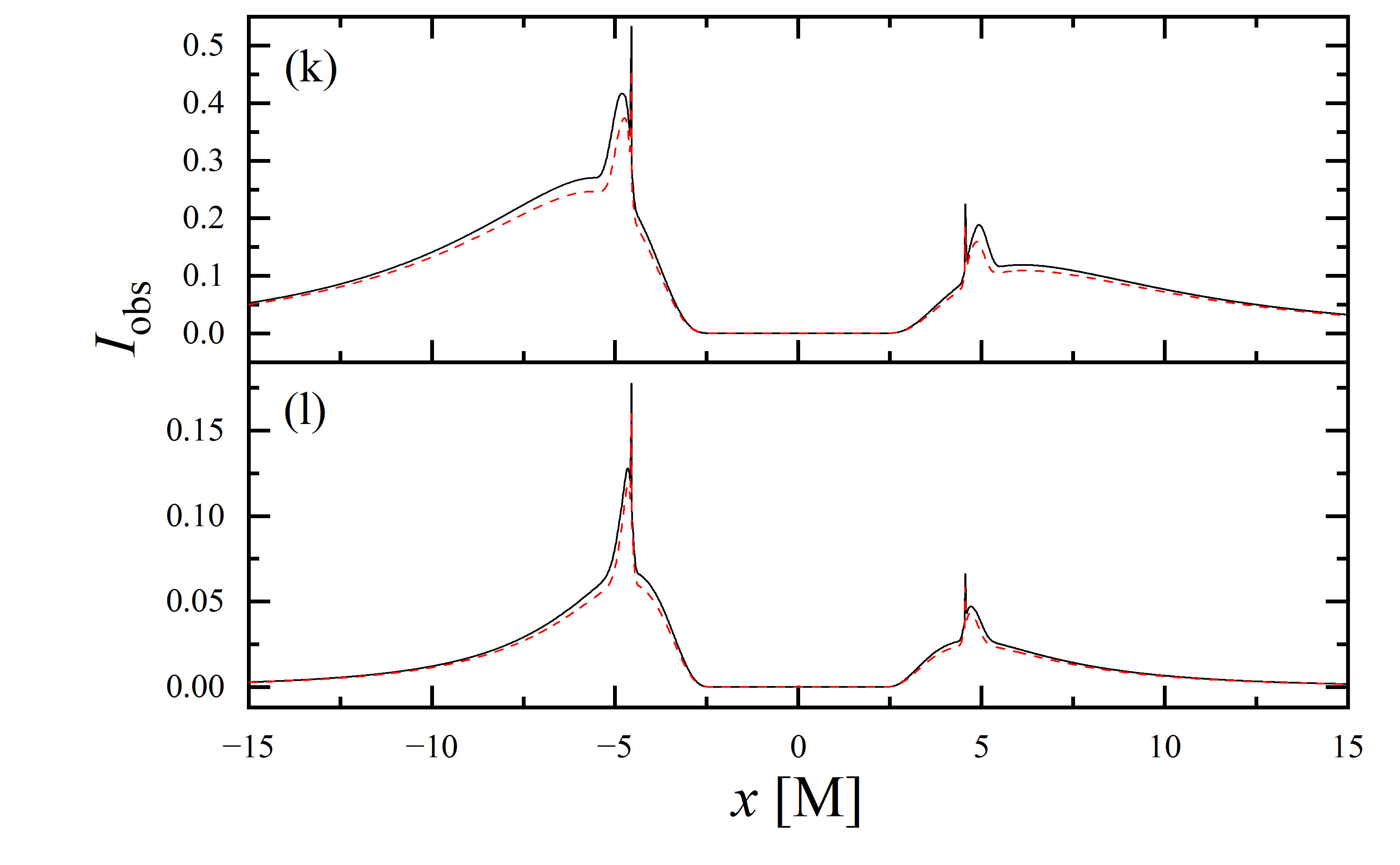}
\includegraphics[width=4.7cm]{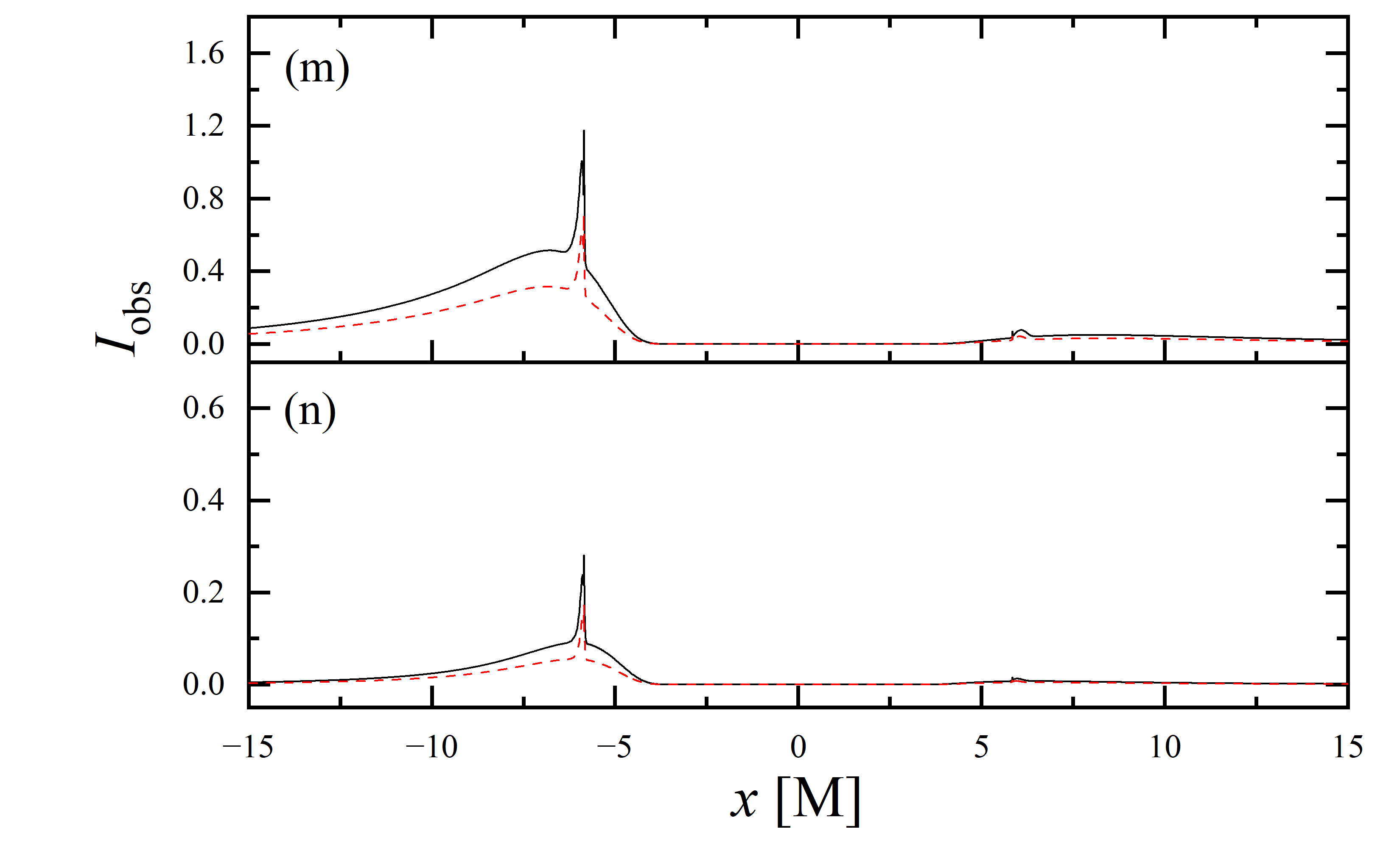}
\includegraphics[width=4.7cm]{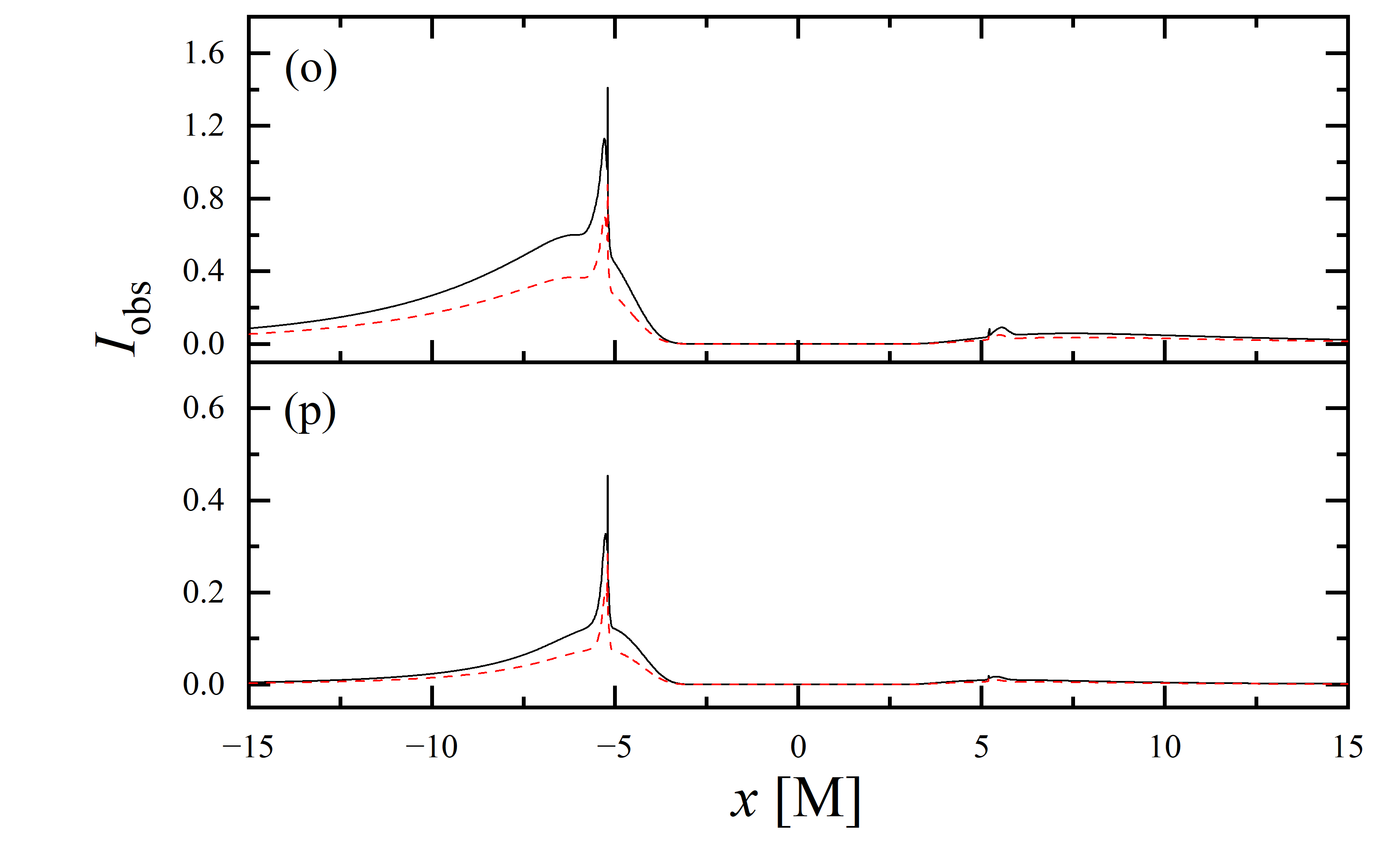}
\includegraphics[width=4.7cm]{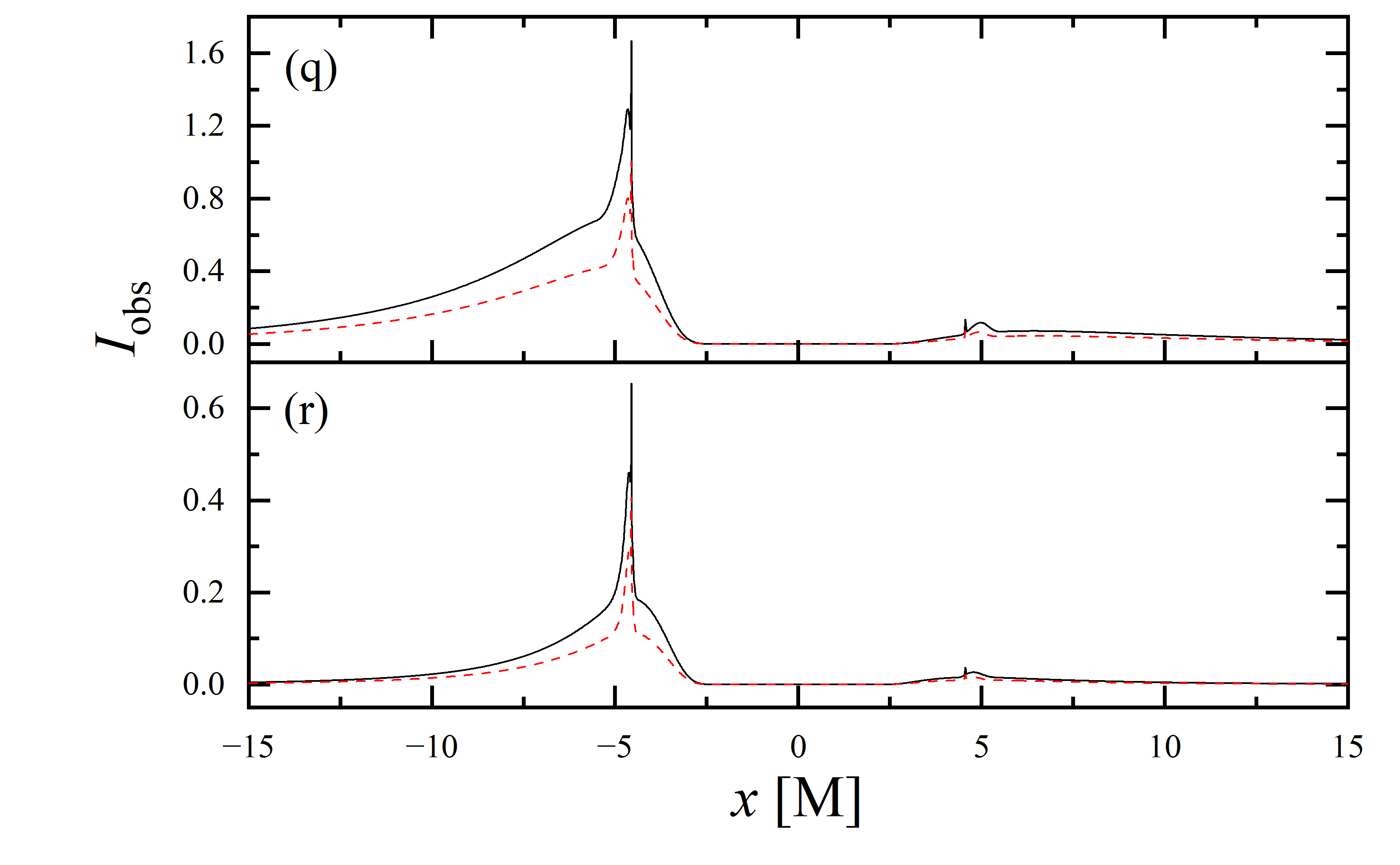}
\caption{Distribution of specific intensity along the $x$-axis in the observation plane across different parameter spaces. Each panel presents the results at $86$ GHz (top) and $230$ GHz (bottom), with the black solid line representing the isotropic accretion disk and the red dashed line representing the anisotropic accretion disk. From left to right, the values of deformation parameter are $-7$, $0$, and $7$. From top to bottom, the observation inclinations are $0^{\circ}$, $17^{\circ}$, and $50^{\circ}$. The projection effect is observed to suppress the specific intensity $I_{\textrm{obs}}$, with the suppression becoming more pronounced as the observation inclination increases.}}\label{fig11}
\end{figure*}

We also extracted the specific intensity distribution along the $y$-axis, as shown in figure 12. At an observation inclination of $0^{\circ}$, the curves are identical to those in figure 11, which is expected given the central symmetry of the images at this inclination. As the observation inclination increases, a pronounced suppression of the projection effect on the specific intensity of light rays becomes evident. Additionally, in the case of $\omega=50^{\circ}$, this suppression effect is more significant in the region where $y<0$ compared to where $y>0$. This difference is due to gravitational lensing, where light rays associated with the $y>0$ region have a smaller $\delta$ during their initial intersection with the accretion disk, while those corresponding to the $y<0$ region generally exhibit a larger $\delta$ during their first crossing of the disk. It is also observed that the projection effect does not change the size of the inner shadow along the $y$-axis.
\begin{figure*}%[tbph]
\center{
\includegraphics[width=4.7cm]{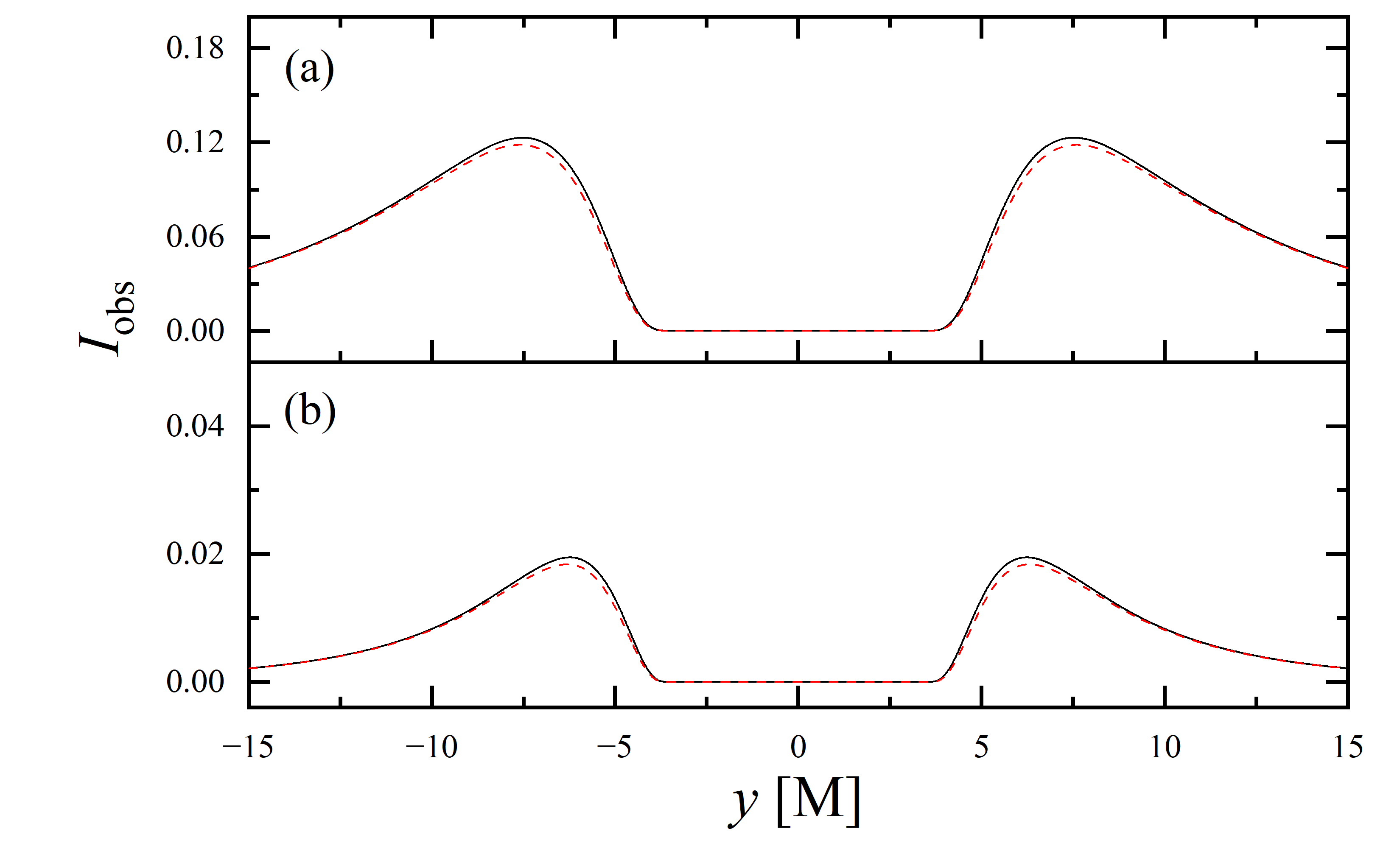}
\includegraphics[width=4.7cm]{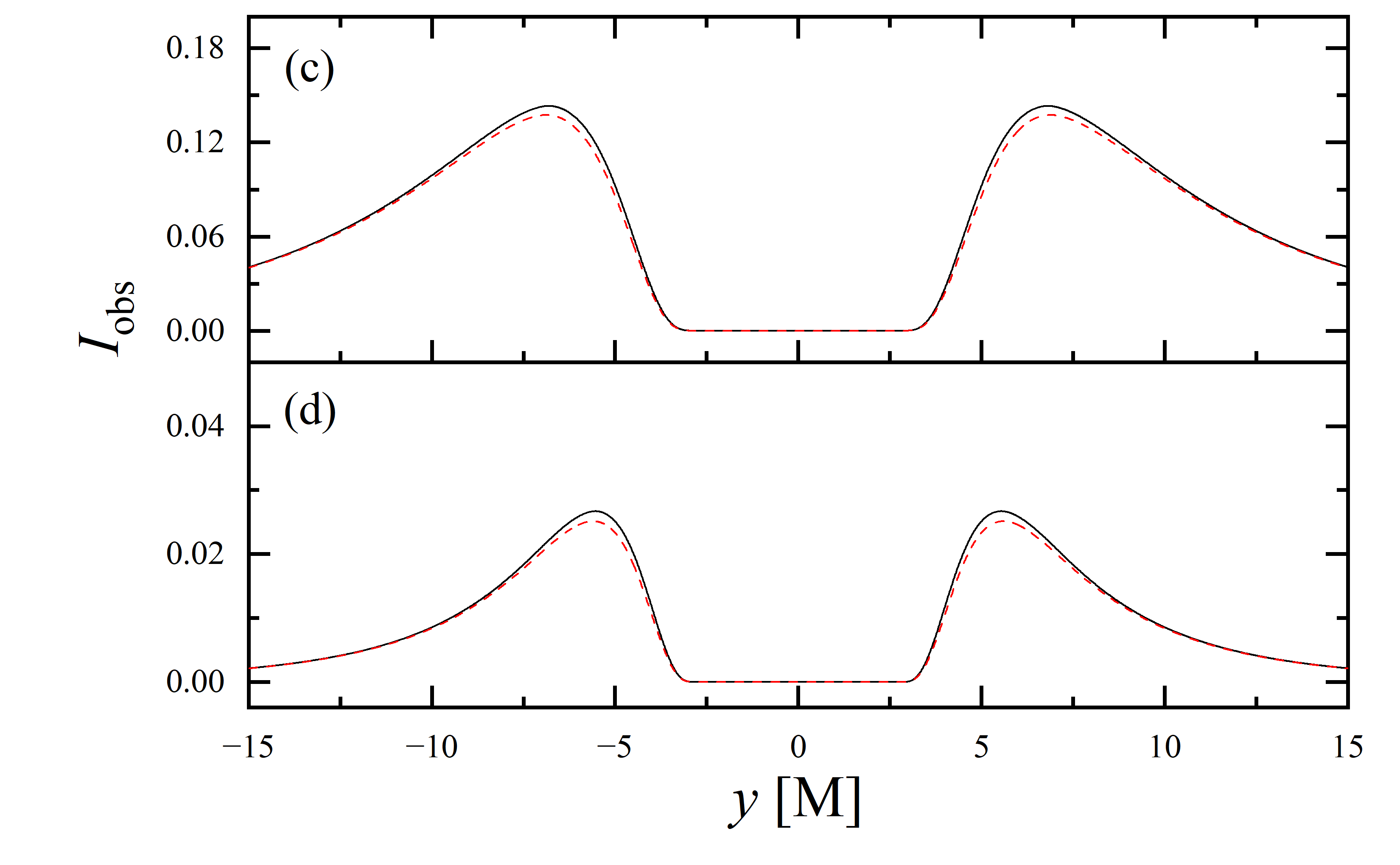}
\includegraphics[width=4.7cm]{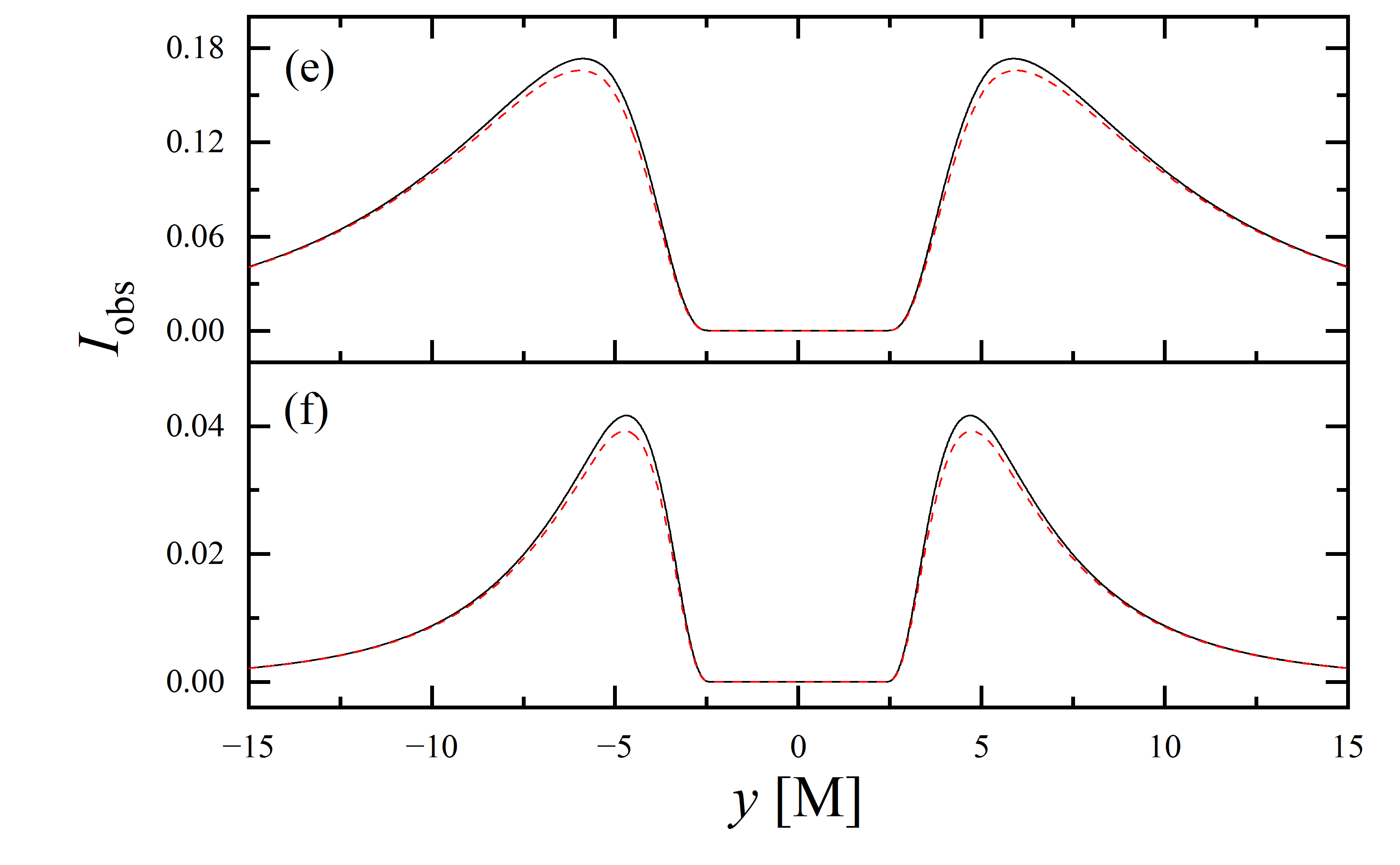}
\includegraphics[width=4.7cm]{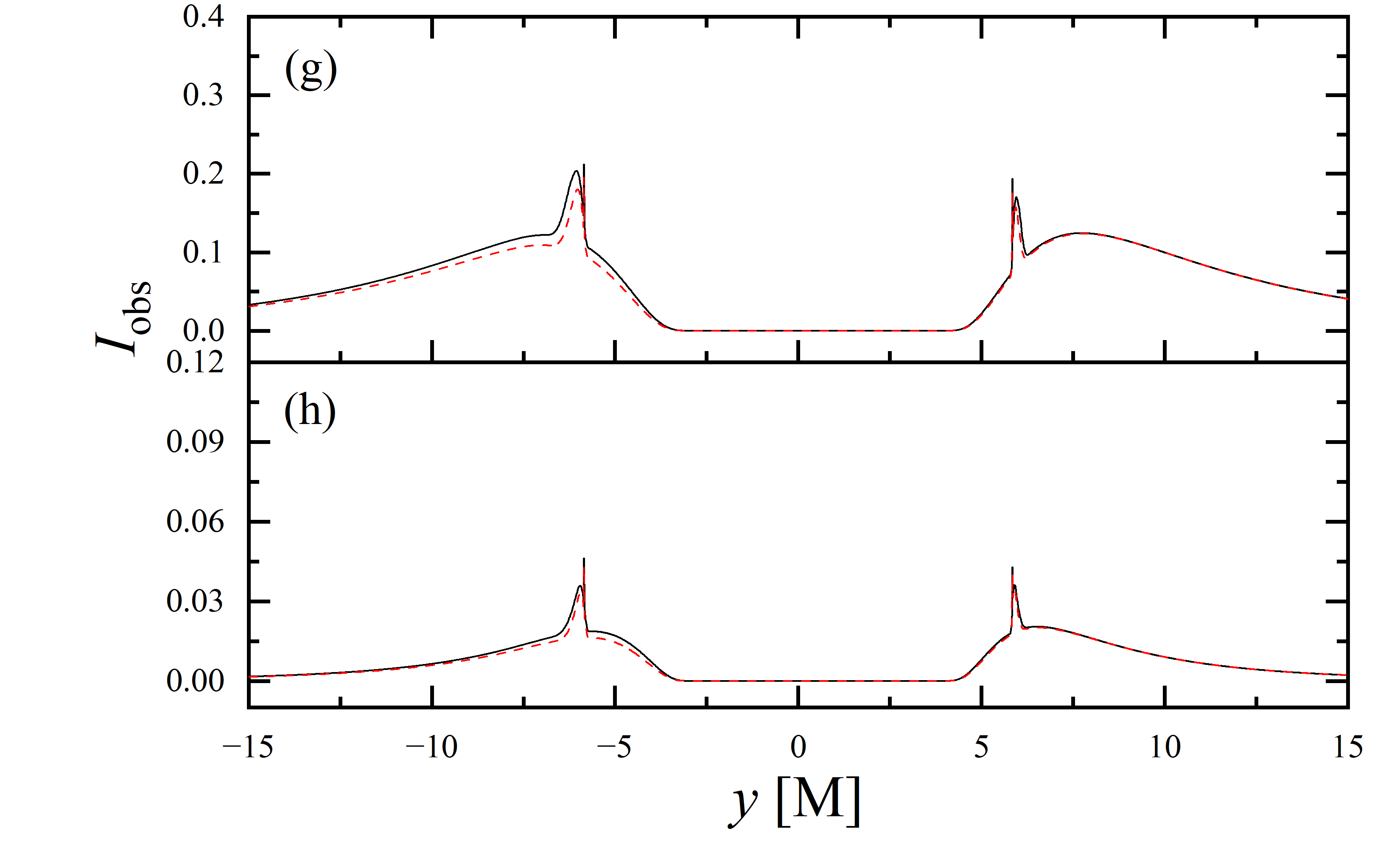}
\includegraphics[width=4.7cm]{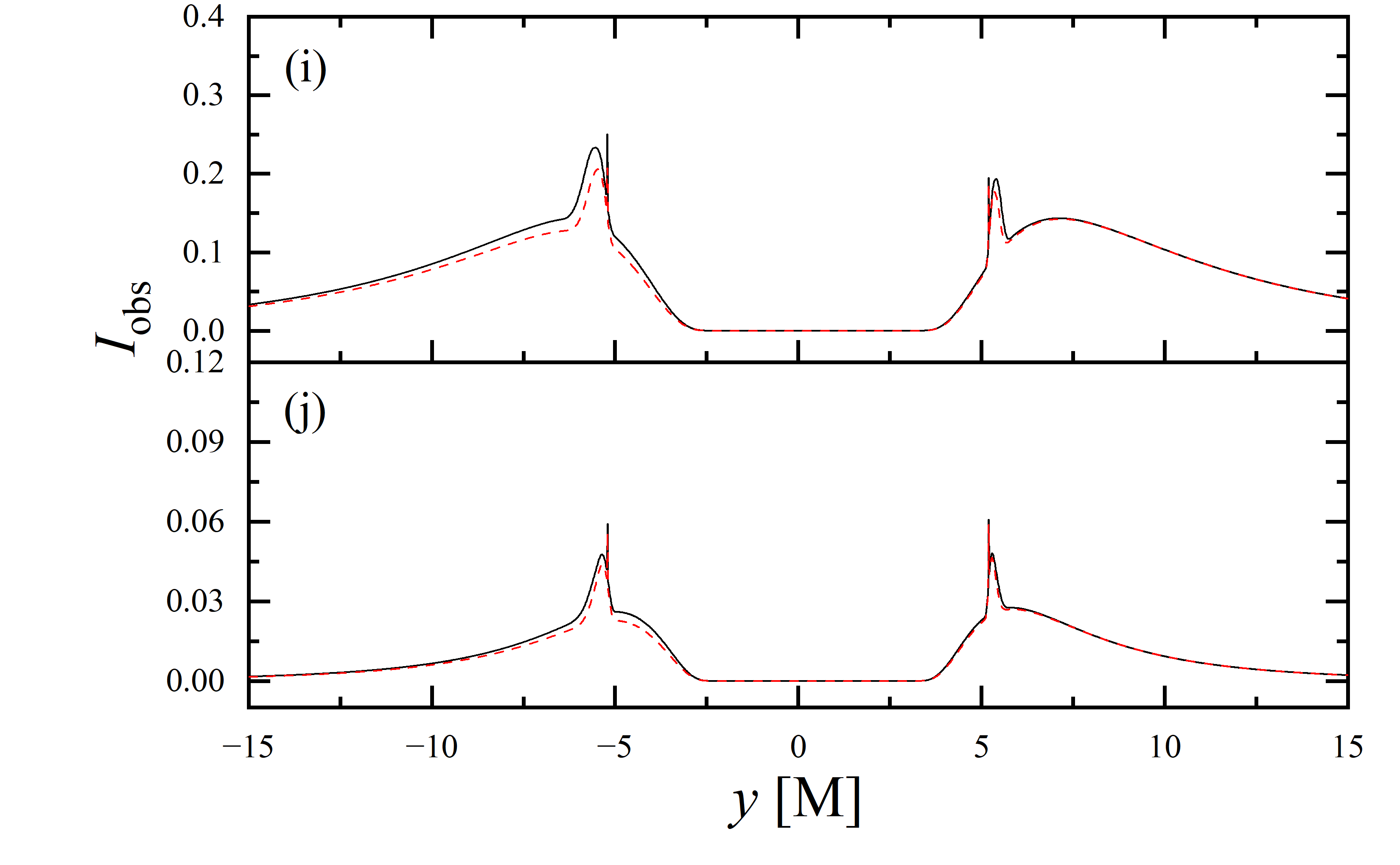}
\includegraphics[width=4.7cm]{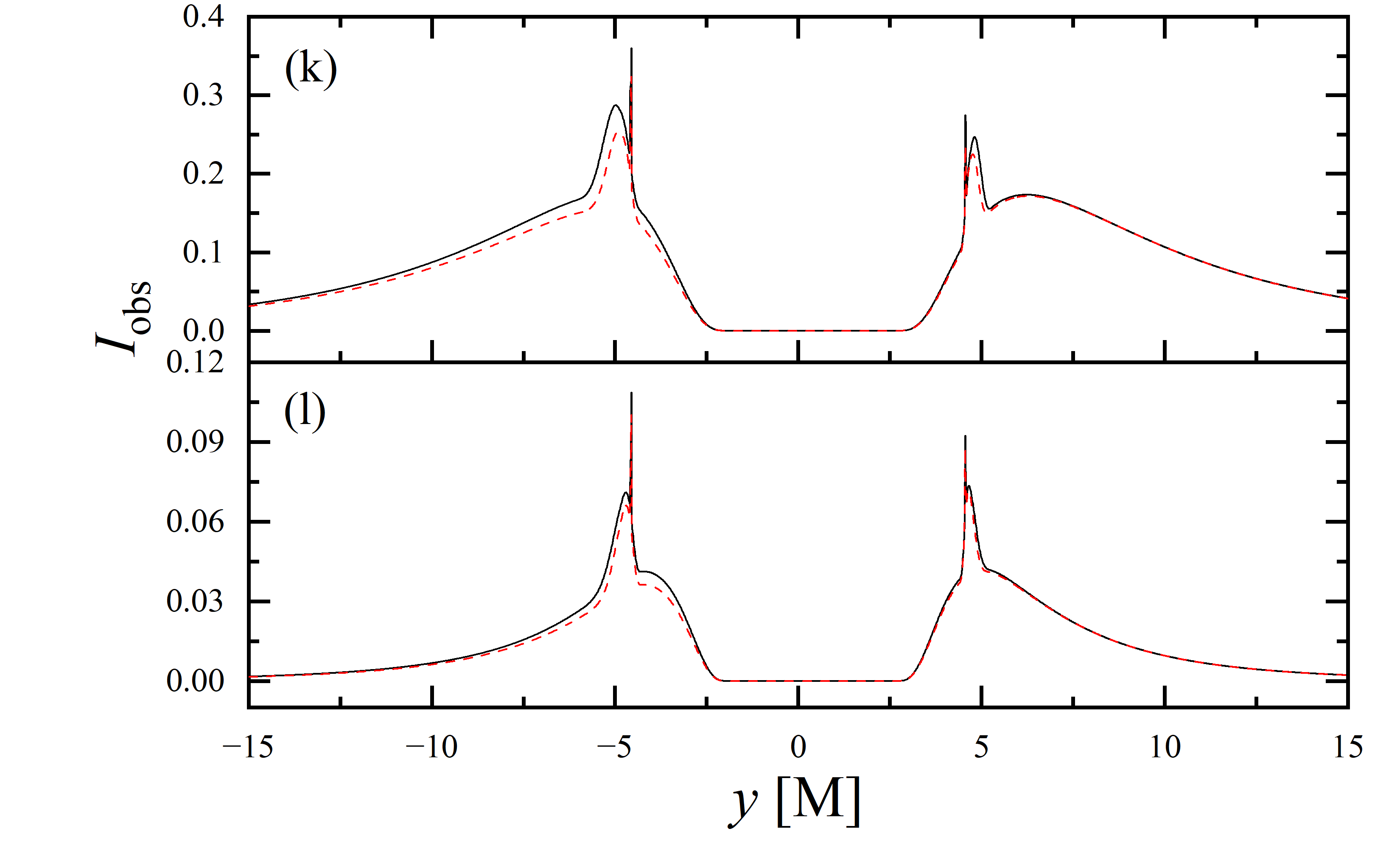}
\includegraphics[width=4.7cm]{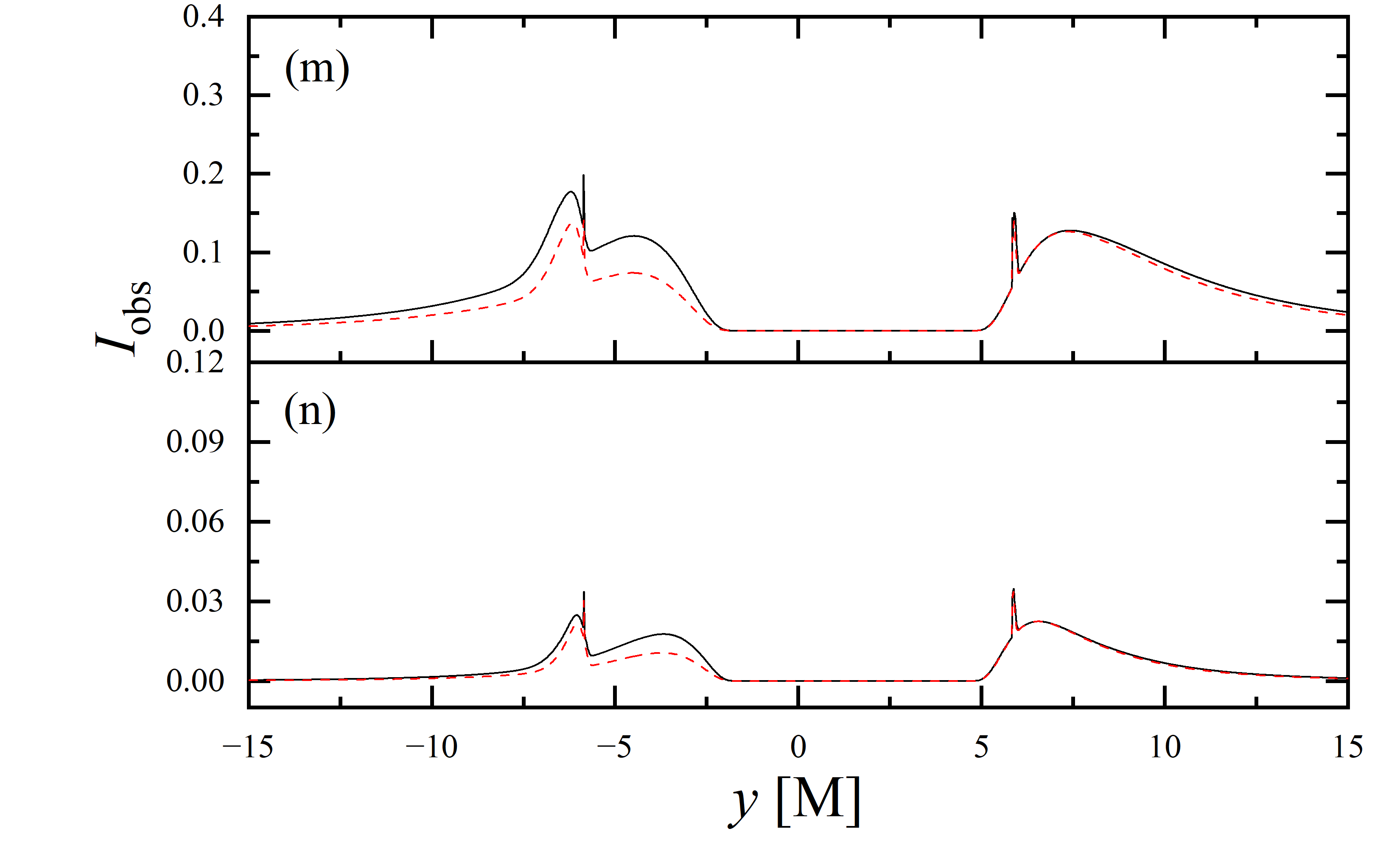}
\includegraphics[width=4.7cm]{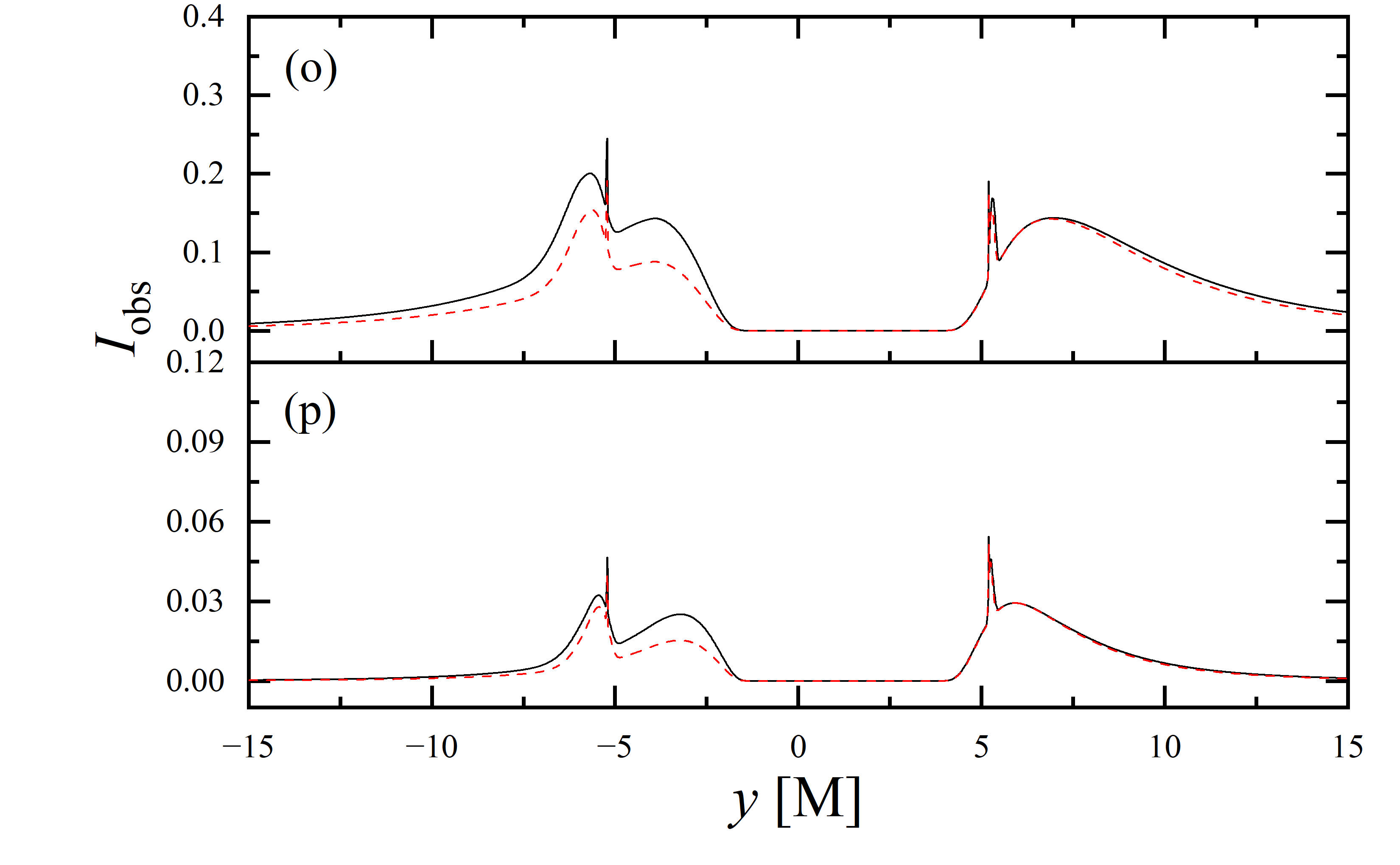}
\includegraphics[width=4.7cm]{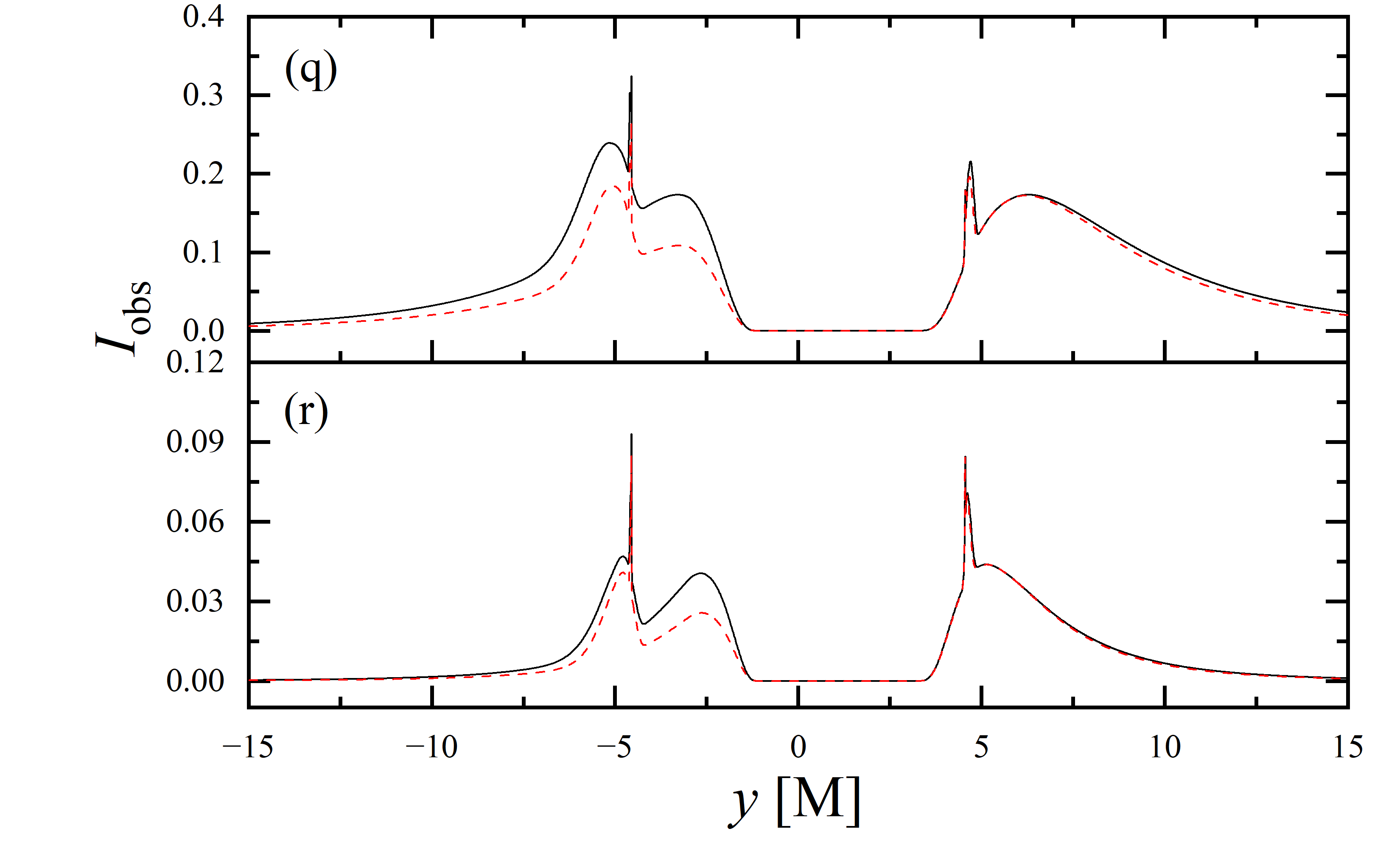}
\caption{Similar to figure 11, but for the distribution of specific intensity along the $y$-axis.}}\label{fig12}
\end{figure*}
\subsection{Parameter constraints by inner shadows}
The use of black hole images for constraining parameters has become a popular area of research, with many authors pioneering methods in this field \cite{Gralla et al. (2020),Hioki and Maeda (2009),Johannsen (2013b),Johannsen et al. (2016),Wei et al. (2019),Li et al. (2020),Farah et al. (2020),Afrin and Ghosh (2022),Cao et al. (2023)}. However, these constraining techniques rely on the precise extraction of the geometric information of the critical curve, a task that remains challenging given the current angular resolution of the EHT. Based on our previous findings, we can confirm that the features of the black hole's inner shadow, such as its contour and size, are unaffected by the projection effect and are instead determined solely by the intrinsic properties of the black hole and the observation angle. Consequently, the inner shadow naturally emerges as a viable alternative tool for parameter constraints.
\begin{figure*}%[tbph]
\center{
\includegraphics[width=7cm]{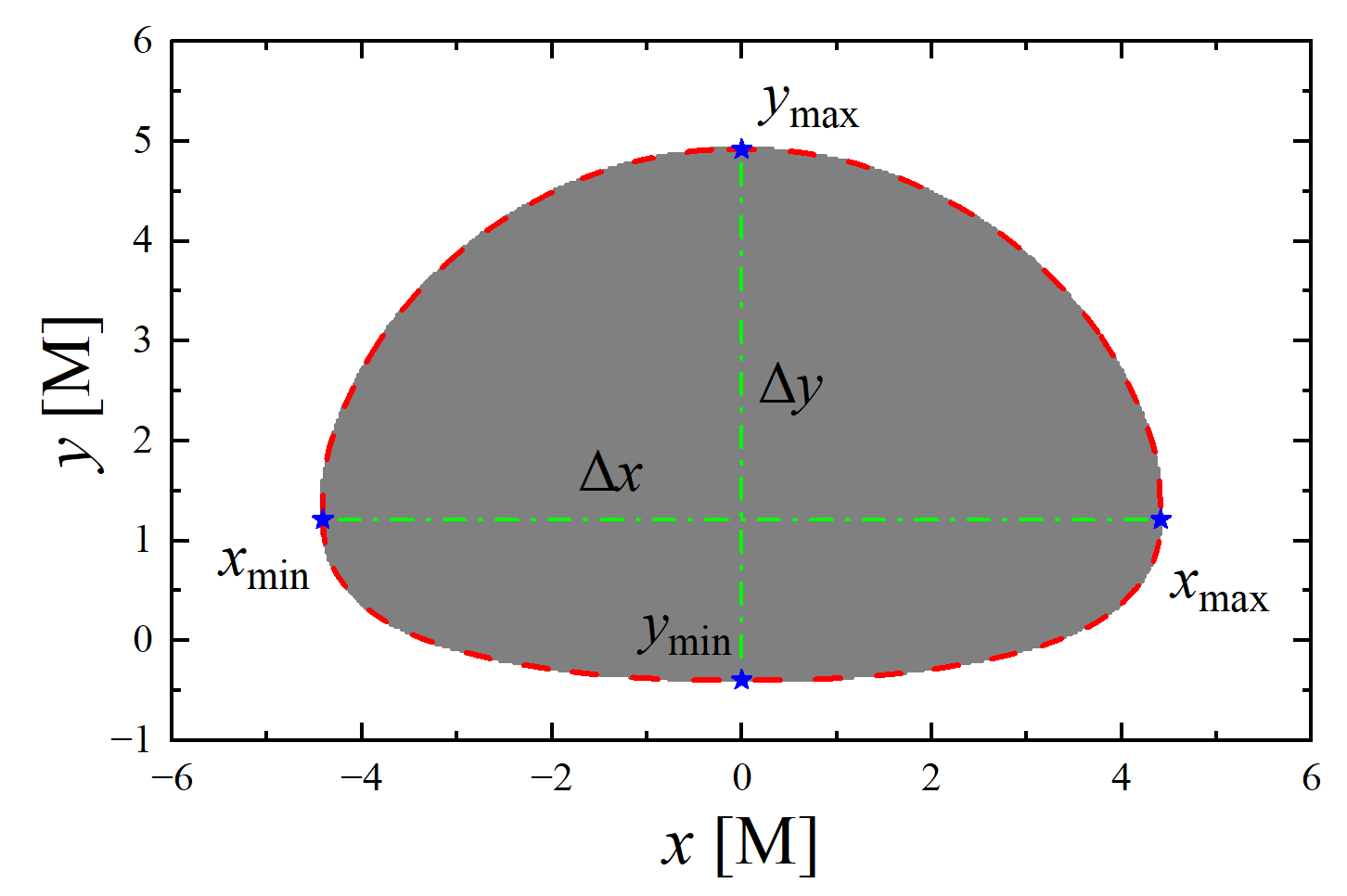}
\caption{Inner shadow of a deformed Schwarzschild black hole with an observation inclination of $80^{\circ}$ and a deformation parameter of $-5$. The image resolution is $500 \times 500$ pixels. The rightmost, leftmost, topmost, and bottommost points of the inner shadow profile are labeled as $x_{\textrm{max}}$, $x_{\textrm{min}}$, $y_{\textrm{max}}$, and $y_{\textrm{min}}$, respectively. We use $\Delta x = x_{\textrm{max}} - x_{\textrm{min}}$ and $\Delta y = y_{\textrm{max}} - y_{\textrm{min}}$ to constrain the black hole parameters and the observation angle.}}\label{fig13}
\end{figure*}

We propose that for spherically symmetric black holes, obtaining just two measurements of the black hole's inner shadow can uniquely determine the observation inclination and the black hole's parameter. Figure 13 illustrates the inner shadow of a deformed Schwarzschild black hole with an observation inclination of $\omega=80^{\circ}$ and a deformation parameter of $\varepsilon=-5$. Here, $x_{\textrm{max}}$, $x_{\textrm{min}}$, $y_{\textrm{max}}$, and $y_{\textrm{min}}$ correspond to the leftmost, rightmost, uppermost, and lowermost points of the inner shadow, respectively. The maximum horizontal distance of the inner shadow, $\Delta x$, defined mathematically as $x_{\textrm{max}}-x_{\textrm{min}}$, serves as the first observable used for parameter constraints. The second observable is the maximum vertical distance of the inner shadow, $\Delta y = y_{\textrm{max}}-y_{\textrm{min}}$. Clearly, both $\Delta x$ and $\Delta y$ are related to the deformation parameter, with $\Delta y$ being more sensitive to the observation angle than $\Delta x$. We simulated the inner shadow of deformed Schwarzschild black holes for different parameter spaces within an observation field of $6 \times 6$ M with a resolution of $1200 \times 1200$ pixels. The resulting contour maps of $\Delta x$ and $\Delta y$ with respect to observation angle and deformation parameter are shown in figure 14, where solid and dashed lines represent $\Delta x$ and $\Delta y$, respectively. It is evident that any given solid (dashed) line will intersect with any other given dashed (solid) line at most once. This implies that once $\Delta x$ and $\Delta y$ are measured, a unique pair of deformation parameter and observation inclination can be determined. It is also worth mentioning that we reasonably speculate that this method could be effective for any spherically symmetric black hole scenario characterized by one additional hair parameter beyond its mass.
\begin{figure*}%[tbph]
\center{
\includegraphics[width=6cm]{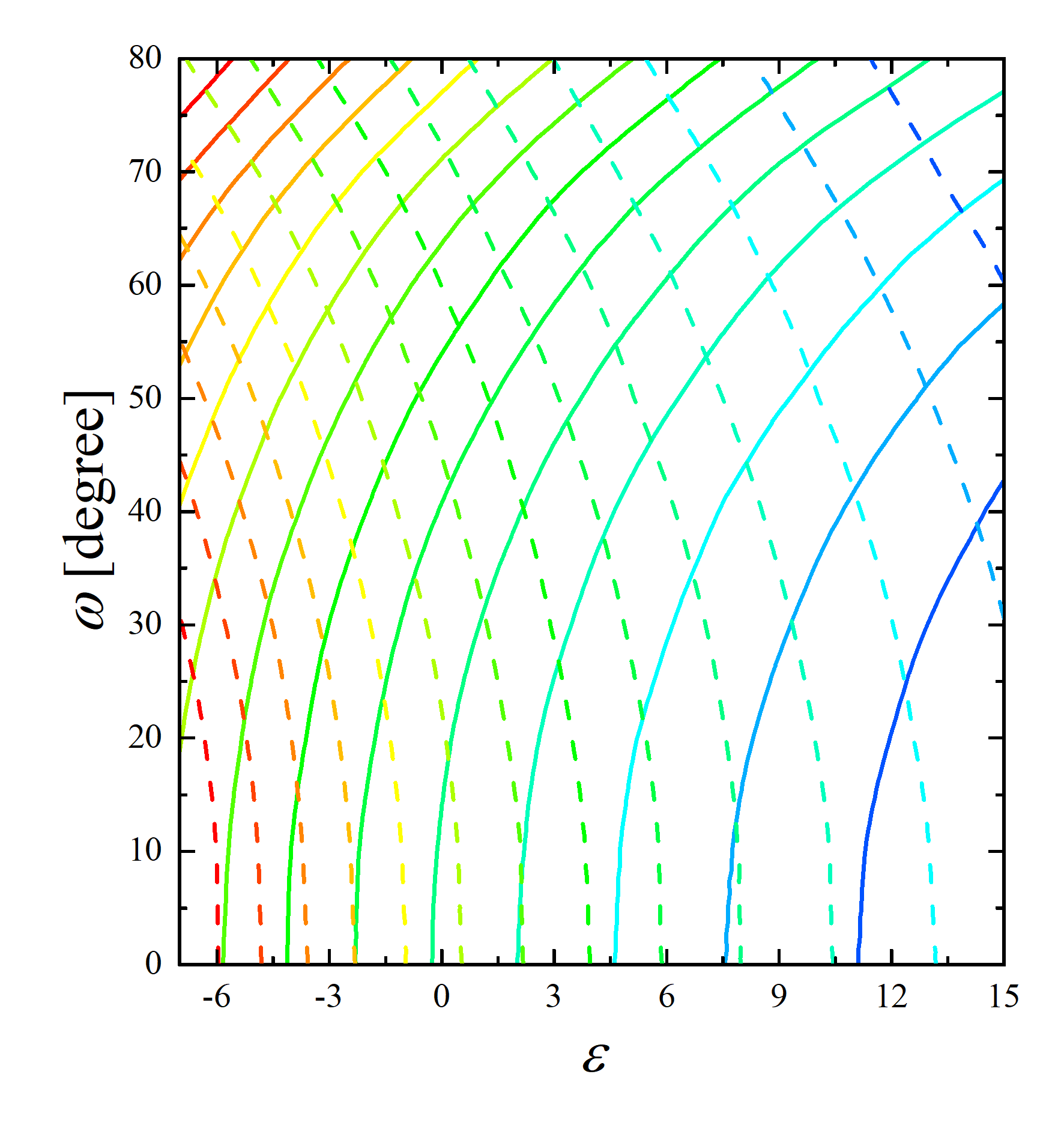}
\caption{Contour maps of $\Delta x$ (solid lines) and $\Delta y$ (dashed lines). Both $\Delta x$ and $\Delta y$ are represented by means of a linear, continuous color scale transitioning from blue to red. For $\Delta x$, blue and red correspond to values of $3.92$ and $9.36$, respectively, while for $\Delta y$, blue and red correspond to values of $3.35$ and $7.1$, respectively. Clearly, once $\Delta x$ and $\Delta y$ are obtained, the observation inclination and deformation parameter can be easily determined.}}\label{fig14}
\end{figure*}
\section{Conclusion}
The groundbreaking achievements of the EHT have opened new avenues for utilizing black hole images to study high-energy physics and gravitational theories, sparking widespread interest in the numerical simulations of black hole images within the scientific community. However, it is important to note that most current research overlooks the impact of the accretion disk projection effect on black hole images. In this paper, we employed a ray-tracing method to investigate the observational characteristics of deformed Schwarzschild black holes and revealed the influence of the anisotropic accretion disk projection effect on these images.

Our findings indicate that the size of the critical curve and inner shadow of deformed Schwarzschild black holes decreases as the deformation parameter increases, which is a consequence of the negative correlation between the deformation parameter and the black hole's gravitational field. Simultaneously, the brightness of the light spots and the critical curve in the image significantly diminishes with decreasing deformation parameter. These trends suggest that among oblate deformed Schwarzschild black holes, Schwarzschild black holes, and prolate deformed Schwarzschild black holes, the former exhibits the largest critical curve and inner shadow but the lowest brightness, whereas the latter displays the smallest critical curve and inner shadow but is the brightest. This relationship can be used to distinguish deformed Schwarzschild black holes from Schwarzschild black hole, providing a tool for testing the no-hair theorem and general relativity. Additionally, we proposed a novel method that simultaneously constrains both the deformation parameter and the observation angle by defining two observables: the maximum lengths of the inner shadow in the horizontal and vertical directions. This method has the potential to be extended to spherically symmetric black holes in other theories of gravity.

By comparing black hole images simulated under isotropic and anisotropic accretion disk conditions, we found that the projection effect does not alter the size of the black hole's inner shadow or critical curve. In other words, it does not impede the use of black hole images for studying gravitational theories. However, the projection effect does suppress the radiation intensity of anisotropic accretion disks, resulting in reduced image brightness. This suppression depends on both the observation inclination and frequency: increasing the observation angle or decreasing the frequency amplifies the suppression effect. It is also noteworthy that the projection effect tends to primarily suppress the specific intensity of direct emission, with a lesser impact on higher-order emissions. This suggests that the projection effect may enhance the visibility of higher-order bright rings in the image.

It is worth noting that the projection effect considered in this paper is modeled by $\kappa=\cos\delta$. In reality, the relevant expression is likely to be far more complex. Moreover, astrophysical black holes are generally described by the Kerr metric, which includes spin. Therefore, our next step will be to focus on how more realistic projection effect influence the observational signatures of rotating black holes.

\appendix
\section{Embedding diagrams of the deformed Schwarzschild spacetime}
Embedding diagrams provide a powerful tool for visualizing curved spacetime by mapping the black hole line element into Euclidean space \cite{Stuchlik and Hledik (1999),Hledik et al. (2006),Galison et al. (2024)}. This visualization technique allows us to intuitively grasp the geometry and deformation of spacetime around black holes, facilitating deeper insights into gravitational phenomena. Let's begin with a static spherically symmetric black hole described by eq. \eqref{1}. It is straightforward to obtain the spatial line element in the equatorial plane as
\begin{equation}\label{a1}
\textrm{d}l^{2} = g_{rr}\textrm{d}r^{2} + g_{\varphi \varphi}\textrm{d}\varphi^{2}.
\end{equation}
In Euclidean space, the line element of the cylindrical coordinates is written as
\begin{equation}\label{a2}
\textrm{d}l^{2} = \textrm{d}\rho^{2}+\textrm{d}z^{2}+\rho^{2}\textrm{d}\varphi^{2},
\end{equation}
further, we have
\begin{equation}\label{a3}
\textrm{d}l^{2} = \left[ 1+\left( \frac{\textrm{d}z}{\textrm{d}\rho} \right)^{2} \right]\textrm{d}\rho^{2}+\rho^{2}\textrm{d}\varphi^{2}.
\end{equation}
Embedding the black hole line element \eqref{a1} in Euclidean space gives $\rho^{2}=r^{2}$, $\textrm{d}\rho = \textrm{d}r$, and
\begin{equation}\label{a4}
g_{rr} = 1+\left(\frac{\textrm{d}z}{\textrm{d}\rho}\right)^{2}.
\end{equation}
Consequently, the relationship between $z$ and $\rho$ in the embedding diagram of the deformed Schwarzschild spacetime can be expressed as
\begin{equation}\label{a5}
\int \textrm{d}z = \int \sqrt{\frac{1+h(\rho)}{f(\rho)}-1}\textrm{d}\rho.
\end{equation}

\acknowledgments
This research has been supported by the National Natural Science Foundation of China [Grant Nos. 12403081 and 11973020].

\end{document}